\documentclass[preprint,11pt,colorlinks=true,allcolors=blue]{elsarticle}

\usepackage{orcidlink}
\usepackage[english]{babel}

\usepackage[a4paper,top=2cm,bottom=2cm,left=3cm,right=3cm,marginparwidth=1.75cm]{geometry}

\usepackage{amsmath, amssymb, amsthm}
\usepackage{graphicx}
\usepackage{subfig}
\usepackage{bm}
\usepackage{upgreek}
\usepackage{mathtools}

\theoremstyle{remark}

\newcommand{\mathsfbi}[1]{\mathit{\bm{\mathsf{#1}}}}
\newcommand{\bcdot}{\boldsymbol{\cdot}}
\newcommand{\matcomp}[1]{\mathit{#1}}
\newcommand{\bk}[1]{\left(#1\right)}
\newcommand{\bs}[1]{\left[#1\right]}

\newcommand{\rbasis}[2]{$\widehat{B}_{#1}^{#2}(\tilde{r},\theta)$}

\journal{Computer Physics Communications}

\begin{document}

\begin{frontmatter}
\title{Smooth Polar B-Splines with High-Order Regularity at the Origin}

\author{Peiyou Jiang$^a$\orcidlink{0000-0002-1295-6064}}
\author{Roman Hatzky$^a$\orcidlink{0000-0003-4616-9568}}
\author{Zhixin Lu$^a$\orcidlink{0000-0002-0557-9694}}
\author{Eric Sonnendr\"ucker$^{a,b}$\orcidlink{0000-0002-8340-7230}}
\author{Matthias Borchardt$^c$\orcidlink{0000-0003-4514-7391}}
\author{Ralf Kleiber$^c$\orcidlink{0000-0002-2261-2855}}
\author{Martin {Campos Pinto}$^a$\orcidlink{0000-0002-8915-1627}}
\author{and Ronald Remmerswaal$^a$\orcidlink{0000-0003-4394-2615}}

\affiliation{organization={Max Planck Institute for Plasma Physics},
            addressline={Boltzmannstr.~2}, 
            city={ Garching},
            postcode={85748}, 
            country={Germany}}

\affiliation{organization={Technical University of Munich, Department of Mathematics},
            addressline={Boltzmannstr.~3}, 
            city={ Garching},
            postcode={85748}, 
            country={Germany}}

\affiliation{organization={Max Planck Institute for Plasma Physics},
            addressline={Wendelsteinstr.~1}, 
            city={Greifswald},
            postcode={17491}, 
            country={Germany}}


\begin{abstract}
We introduce a smooth B-spline discretization in polar coordinates on the unit disc that corrects the loss of regularity present at the origin caused by the coordinate singularity in standard tensor-product B-spline formulations.
The method constructs ``smooth polar splines'' via a Galerkin projection of harmonic polar functions~$S_l^{-m}(r,\theta) \coloneqq r^l \sin(m\theta)$
and $S_l^{m}(r,\theta) \coloneqq r^l \cos(m\theta)$, derived from the polar representation of Cartesian monomials, onto the central tensor-product B-spline basis in the innermost radial region. 
The radial component reproduces $r^l$ exactly for $0 \le l \leq p$, where $p$ is the B-spline degree, satisfying the near-origin regularity condition. However, exact compatibility with $C^\infty$-regularity at the origin is recovered
only in the limit $\Delta\theta \to 0$, when the angular component resolves all
angular harmonics accurately.
The smooth polar splines are linear combinations of standard tensor-product B-splines and lie in the same function space, enabling mapping between the $C^\infty$-regular subspace and the original discretization space via an exact prolongation operator and a corresponding restriction operator acting on the discrete variables. 
They match standard tensor-product B-splines away from the origin, preserve orthogonality among the newly constructed origin‑centered basis functions, and maintain local support and sparse matrices. 
This smoothness and locality improve the conditioning of mass and stiffness matrices, conserve charge, and reduce statistical errors in particle-in-cell simulations near the origin, while eliminating spurious eigenvalues in eigenvalue problems. 
The approach provides a robust, high-order, and efficient adaptation of tensor-product B-splines for polar coordinates in physics simulations.
\end{abstract}

\begin{keyword}
B-splines \sep polar coordinates \sep  Galerkin finite-element method \sep  $C^\infty$-continuity (high-order continuity) \sep regularity at the origin \sep  particle-in-cell (PIC)
\end{keyword}
\end{frontmatter}

\section{Introduction}
\label{Sec_intro}

In simulations of toroidal fusion plasmas, such as those in tokamaks and stellarators, the poloidal cross-section is naturally represented in logical polar coordinates, which have a singular point at the magnetic axis. In this work, the magnetic axis coincides with the polar origin of the logical coordinate system, so that the coordinate singularity at $r=0$ corresponds exactly to the magnetic axis of the equilibrium. When the simulation domain includes the axis, the numerical discretization should respect regularity conditions; otherwise, nonphysical discontinuities in the solution or its derivatives can occur there.

In polar coordinates $(r,\theta)$, a poloidal Fourier decomposition of a regular function
\begin{equation}
\Phi(r,\theta) = \sum_{m\in\mathbb{Z}} \Phi_m(r) \, \exp({\mathrm{i} m \theta)}
\end{equation}
implies the regularity condition~\citep{Lewis1990constraints}
\begin{equation} \label{eq.crude_reg_cond}
\Phi_m(r) = \mathcal{O}(r^{|m|}) \quad \text{as} \quad r\to 0,
\end{equation}
where the exponent $|m|$ represents the minimal radial power needed to ensure regularity at $r=0$.

Early gyrokinetic particle simulations were developed to study microturbulence phenomena in magnetized plasmas~\citep{lee1983gyrokinetic}, focusing on the dynamics of small-scale fluctuations and their impact on transport.  
For strictly linear simulations of ion-temperature-gradient-driven modes (ITGs), the simulation domain typically excludes the region around the magnetic axis, since these modes are peaked near the strongest temperature gradients at mid-radius or in the outer plasma~\citep{Fivaz1998gygles,wan2012global}.  
In such cases, the regularity of the gradient of the electromagnetic potentials near the magnetic axis is irrelevant, as the instabilities have no significant presence in this region.  
However, the situation changes for nonlinear simulations of microturbulence, which include zonal flows and the nonlinear interactions associated with the radial spreading of turbulence~\citep{Lin98zonalflow,hatzky2002energy}.  
A major part of the zonal flow is the poloidally axisymmetric ($m{=}0$) Fourier component of the electromagnetic potentials, which remains finite at the axis and satisfies the regularity condition $\partial_r \Phi_0|_{r=0} = 0$.

Beyond microturbulence, significant advances in gyrokinetic models -- expanding from purely electrostatic to fully electromagnetic formulations~\citep{chen2007electromagnetic,mishchenko2014pullback,kraus2017gempic,hatzky2019reduction,lu2021development} -- have made it possible to simulate large-scale magnetohydrodynamic (MHD) modes~\citep{Nuehrenberg_JPP_2025}.  
Accurate treatment of axis regularity is essential for modes such as the global Alfv\'en eigenmode (GAE)~\citep{Mishchenko2008AlfvenicPIC} and the kink~\citep{cole2014fluid}, for which axis continuity and smoothness are crucial.
Consequently, a more rigorous treatment of the regularity conditions near the magnetic axis is needed.

In various gyrokinetic particle-in-cell (PIC) codes, polar tensor-product B-spline
discretizations implementing continuity at the axis for the electrostatic
potential~$\phi$ and parallel vector potential~$A_\parallel$ have been used.
This so-called ``unicity'' regularity condition -- meaning that the potentials are
continuous and single-valued at the magnetic axis, i.e.\ $C^0$-regularity -- has
been implemented in the \textsc{GYGLES}, \textsc{ORB5}, and \textsc{EUTERPE}
codes~\citep{Fivaz1998gygles,lanti2020orb5,kleiber2024euterpe}.
Recall that a function is of class $C^n$ if it is $n$~times differentiable and its
$n$-th derivative is continuous. Throughout this work, the term ``$C^n$-regularity''
will, unless explicitly stated otherwise, refer to regularity at the polar origin.

While $C^0$-level regularity aids in implementing the Laplacian operator (e.g.\ in the quasi-neutrality equation or parallel Amp\`ere's law), thereby preventing singular behaviour at the origin, it leaves the gradient of the potentials -- which appears in the equations of motion for the particles -- discontinuous there.  
Therefore, at least $C^1$-regularity, which ensures continuous first derivatives, is essential.

\citet{Toshniwal2017isogeometric} proposed a method for achieving 
$C^1$-regularity near the origin in polar coordinates within an isogeometric 
framework, using a polar B-spline base, where the geometry is exactly represented 
via the same spline mapping between a smooth logical domain and the physical domain. 
Several other numerical approaches have been developed to enforce $C^1$-regularity, which has been shown to be important for Vlasov--Poisson type simulations in \cite{zoni2019solving}. 
The structure-preserving hybrid code (\textsc{STRUPHY})~\citep{holderied2022magneto} builds on 
the ideas of \citet{Toshniwal2017isogeometric}, enabling implementation of the full de~Rham sequence. A related approach, employing projector-based techniques, is presented 
in~\citep{Guclu2025BrokenFEEC}. In our work, we do not adopt the isogeometric route 
because our physical domain is obtained from an MHD equilibrium solver, and modifying 
the existing geometry handling would require substantial changes to established code 
packages.

Here, we present a discretization of polar splines that achieves high-order
smoothness at the origin without modifying the existing computational geometry
framework.
The method enforces only $C^0$-regularity analytically at the origin, while
inducing an algebraic structure that encodes the modal constraints associated
with $C^\infty$-regularity.
For sufficiently high angular resolution, the numerical representation exhibits
smoothness consistent with $C^\infty$-regularity up to machine precision.
The method is formulated entirely within the tensor-product B-spline space,
thereby ensuring compatibility with existing codes and eliminating the loss of
differentiability that is characteristic of standard tensor-product splines at
the origin in polar coordinates.
By embedding the smooth basis into the tensor-product B-spline space, we enable exact prolongation and corresponding restriction operations on coefficient and load vectors between the smooth and standard formulations.
This allows us to retain charge- and current-assignment procedures of tensor-product B-splines, restricting them to the regularity subspace when required.  
Only the solver component must be modified to account for the regularity condition at the origin.  
The resulting coefficient vectors are then prolonged back to the tensor-product B-spline space, allowing field calculation routines to remain unchanged.

The rest of this article is organized as follows.
Section~\ref{sec.tensor-prod_B-spline} examines tensor-product B-splines in polar
coordinates.
Section~\ref{sec:general_info} provides general information about bases with
$C^n$-regularity at the origin.
Section~\ref{sec:c1scheme} derives two B-spline bases with
$C^\infty$-regularity at the origin, with the key idea presented in
Section~\ref{Sec.GalerkinCenter}.
Section~\ref{sec.theoryandanal} presents a numerical analysis of smooth polar
splines.
Section~\ref{sec:numerical_results_PIC} demonstrates the impact of the
high-order regularity method on PIC codes.
Finally, Section~\ref{sec:conclusion} concludes the paper, followed by the appendix.

\section{Tensor-product B-spline discretization}
\label{sec.tensor-prod_B-spline}

\subsection{Physical and logical domains with polar--Cartesian mapping}

We consider as physical domain the unit disc
\begin{equation} \label{eq.unit_disc}
  \Omega \coloneqq \{ (x,y) \in \mathbb{R}^2 \mid x^2 + y^2 \le 1 \}.
\end{equation}
For discretization, we employ polar coordinates $(r,\theta)$; thus the logical domain is
\begin{equation} 
\widehat{\Omega} \coloneqq [0,1] \times [0,2\uppi).
\end{equation}
The mapping $\boldsymbol{F}$ from polar to Cartesian coordinates is defined by
\begin{equation} \label{eq.polar_coord_trans}
\boldsymbol{F} : \widehat{\Omega} \longrightarrow \Omega \subset \mathbb{R}^2, 
\quad
\boldsymbol{F}(r,\theta) = (r\cos(\theta), r\sin(\theta)).
\end{equation}
For an arbitrary function $\tilde{f}(r,\theta)$, its Cartesian representation is defined for $(x,y) \neq (0,0)$ by
\begin{equation} \label{eq.polar_coord_trans_inverse}
f(x,y) = \tilde{f} \! \left( \boldsymbol{F}^{-1}(x,y) \right),
\end{equation}
where the inverse mapping $\boldsymbol{F}^{-1}$ from the physical to the logical domain is
\begin{equation}
  \boldsymbol{F}^{-1} : \Omega \longrightarrow \widehat{\Omega} \subset \mathbb{R}^2, 
\quad
\boldsymbol{F}^{-1}(x,y) = \left( \sqrt{x^2+y^2}, \mathrm{atan2}(y,x) \right),
\end{equation}
with $\mathrm{atan2}(y,x) \in [0,2\uppi)$. 
At the origin, $\theta$ is not uniquely defined, and $r=0$ constitutes a coordinate singularity.

\subsection{One-dimensional radial B-spline finite-element basis functions}
\label{Sec.1d_basis_radial}

We adopt an open, uniform knot vector over $r \in [0,1]$ with constant spacing
$
\Delta r \coloneqq 1 / n_{\mathrm{int}},
$
where $n_{\mathrm{int}}$ is the number of equal-length intervals. 
Interior knots have multiplicity one, while the knots at $r=0$ and $r=1$ have multiplicity $p{+}1$, where $p$ is the polynomial degree, 
which clamps the basis functions so that they interpolate the boundary control points and produce one-sided support at the boundaries.
This configuration facilitates regularity constraints at the origin (see Sec.~\ref{sec:c1scheme}) 
and boundary conditions at $r=1$ (see Sec.~\ref{Sec.galerkin_matrices}). 
Without special treatment, the repeated knots at $r=0$ reduce radial continuity at the boundary, 
and standard tensor-product bases (see Sec.~\ref{sec.tensor_product_basis}) generally fail to satisfy 
the polar regularity condition as $r \to 0$. 
Within each knot span, the spline basis functions are polynomial and therefore
$C^\infty$, while at a simple interior knot the continuity is $C^{p-1}$.

For $B_{i,p}(r)$, a B-spline of order $p{+}1$ and index $i$, 
the total number of basis functions is 
$
N_r \coloneqq n_{\mathrm{int}} + p,
$
where $n_{\mathrm{int}}$ denotes the number of intervals,
and the finite-dimensional subspace~$V_{r,h}$ is spanned by $N_r$ radial B-splines. 
Except for those adjacent to the boundary, the radial B-splines are identical in shape 
and can be obtained by translation across knot intervals.
A function $\phi_{r,h}(r) \in V_{r,h}$ can be expanded in this basis as
\begin{equation}
\phi_{r,h}(r) = \sum_{i=0}^{N_r-1} \phi_{r,i} B_{i,p}(r)
           = \boldsymbol{\phi}_r^{\mathrm T} \boldsymbol{B}_r(r),
\end{equation}
where $\boldsymbol{\phi}_r\in \mathbb{R}^{N_\mathrm{r}}$ and $\boldsymbol{B}_r$ are column vectors of length $N_r$ containing the coefficients $\phi_{r,i}$ and the linearly independent basis functions $B_{i,p}(r)$, respectively. 

An important property of radial B-splines is that they satisfy the partition of unity~\citep{deBoor2001splines}:
\begin{equation} \label{eq.unity_r}
\sum_{i=0}^{N_r-1} B_{i,p}(r) = 1, \qquad \forall r \in [0,1].
\end{equation}

\subsubsection{Approximation at the radial boundary}
\label{Sec.radial_boundary}

In the innermost interval at the left boundary of an open, uniform knot vector, 
exactly $p{+}1$ basis functions $B_{i,p}(r)$ are nonzero. 
For indices $i \le p$, these boundary B-splines are piecewise polynomials supported on 
$[0, \Delta r (i+1)]$. 
For $i < p$, the $B_{i,p}(r)$ have distinct shapes due to truncation by the boundary, 
whereas $B_{p,p}(r)$ has the same shape as the corresponding interior basis functions with $i > p$ when translated.  
It is convenient to introduce the normalized radial coordinate
$\tilde{r} \coloneqq r/\Delta r$,
which measures $r$ in units of the knot spacing. 
Explicit polynomial forms follow from the Cox--de~Boor recursion (App.~\ref{App.Bspline_rec}).

For a given $p$ and $i \le p$, the lowest power of $\tilde r$ appearing in
$B_{i,p}(\tilde r)$ on the first knot span $\tilde r \in [0,1]$ is $i$.
This implies that for radial B-splines we have
\begin{equation} \label{eq.boundary_poly_order}
 B_{i,p}(\tilde{r}) = \mathcal{O}(\tilde{r}^i) \quad \text{as} \quad \tilde{r} \to 0,
\end{equation}
which can be verified for example for cubic B-splines in the innermost interval $\tilde{r} \in [0, 1]$:
\begin{subequations}
\label{eq.Express_cubic}
\begin{align}
    B_{0,3}(\tilde{r}) &= 1 - 3\tilde{r} + 3\tilde{r}^2 - \tilde{r}^3 , \label{eq.Taylor0} \\
    B_{1,3}(\tilde{r}) &= 3\tilde{r} - \frac{9}{2} \tilde{r}^2 + \frac{7}{4} \tilde{r}^3 , \label{eq.Taylor1} \\
    B_{2,3}(\tilde{r}) &= \frac{3}{2}\tilde{r}^2 - \frac{11}{12} \tilde{r}^3 , \label{eq.Taylor2} \\
    B_{3,3}(\tilde{r}) &= \frac{1}{6}\tilde{r}^3 . \label{eq.Taylor3}
\end{align}
\end{subequations}

\begin{figure}
    \subfloat[Linear]{\includegraphics[width=0.3\textwidth]{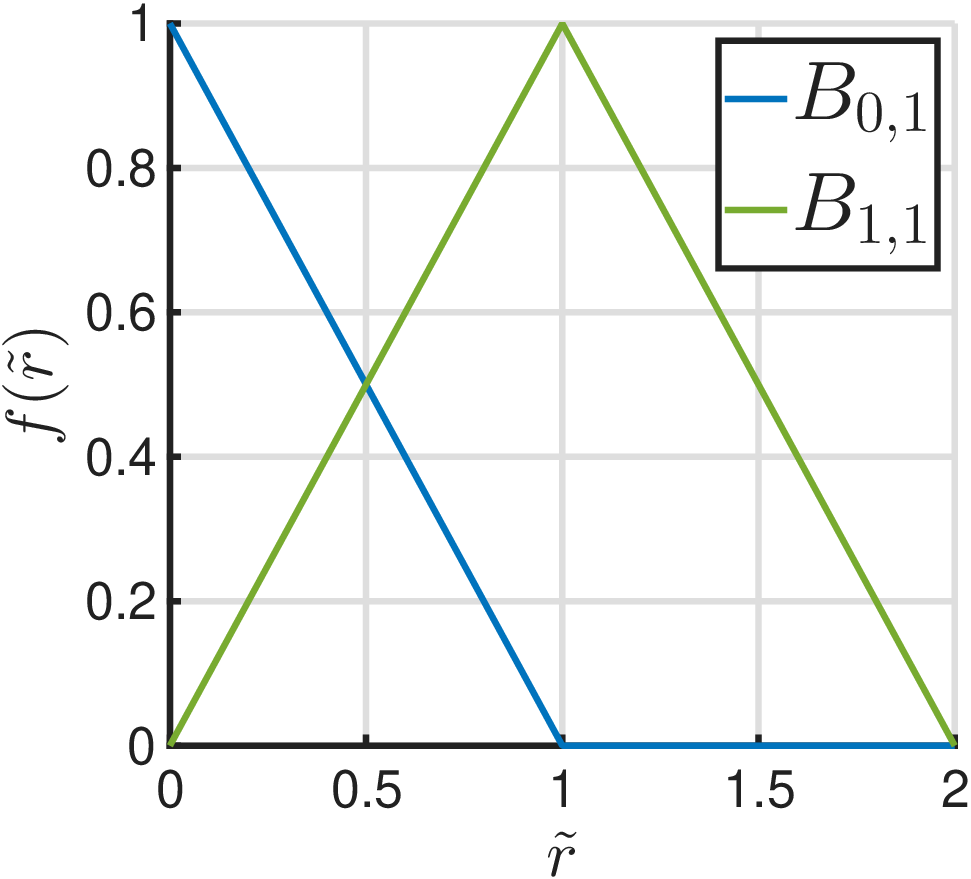}}
 \hfill 	
    \subfloat[Quadratic]{\includegraphics[width=0.3\textwidth]{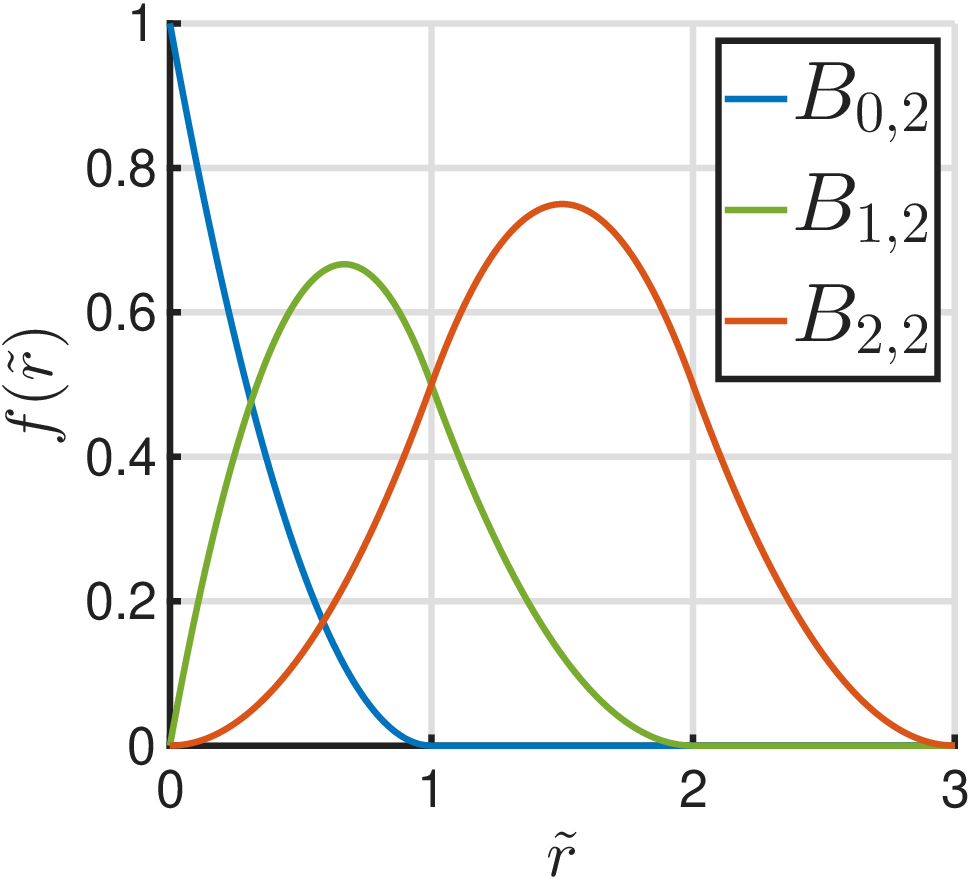}}
 \hfill	
    \subfloat[Cubic]{\includegraphics[width=0.3\textwidth]{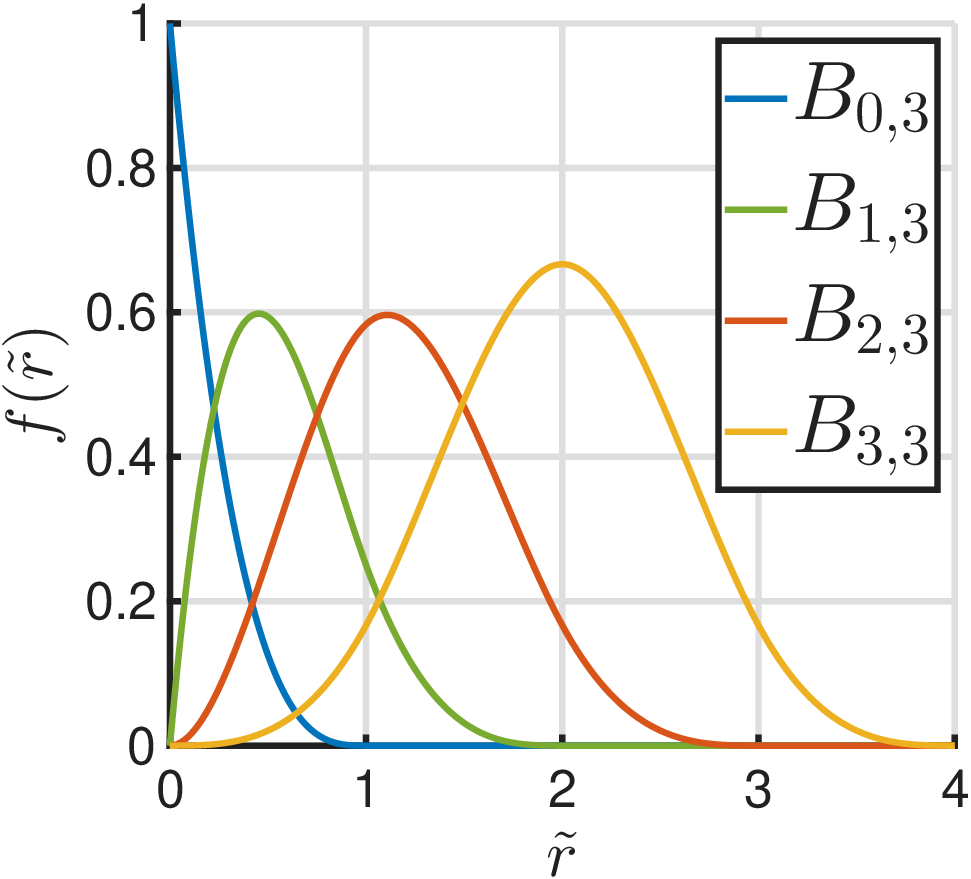}\label{fig.basis_cubic}} 
 \caption{Linear, quadratic, and cubic B-spline basis functions at the left boundary.}
 \label{fig.boundary_basis}
\end{figure}

For $0 \le n \leq p$ it follows that $\mathrm{d}^n/ \mathrm{d}r^n B_{i,p}(0) = 0$ for $n < i$. In particular,
\begin{equation}\label{eq.B0_r0_one}
    B_{0,p}(0) = 1 \quad \text{and} \quad B_{i,p}(0) = 0 \quad \text{for }  1 \le i \le p.
\end{equation}
Figure~\ref{fig.boundary_basis} illustrates this for the linear, quadratic, and cubic B-spline basis functions at the left boundary.

Hereafter we write $B_i(r)$ without the subscript $p$; when required, the polynomial degree~$p$ of the B-splines is stated in words.

\subsection{One-dimensional angular B-spline finite-element basis functions}
\label{Sec.1d_basis_angular}

\begin{figure}
    \subfloat[Quadratic ($p{=}2$)]{\includegraphics[width=0.45\textwidth]{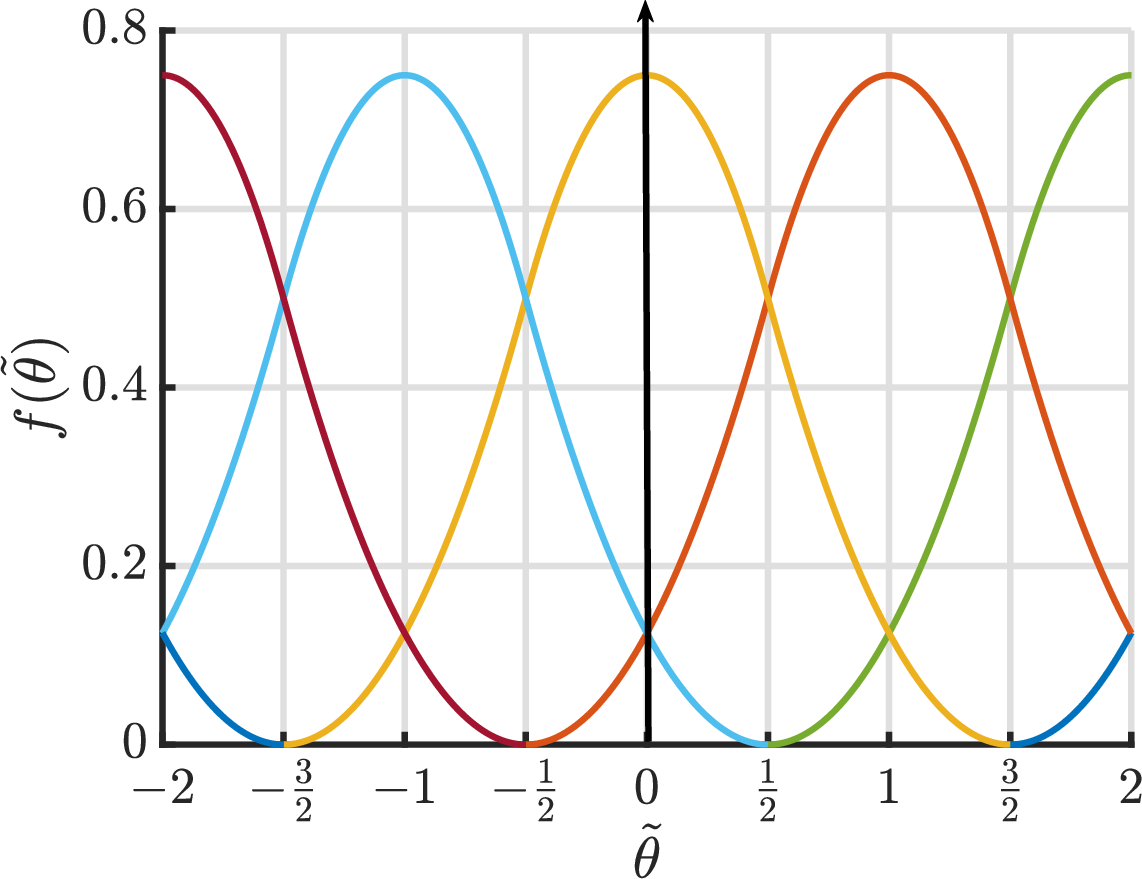}}
 \hfill	
    \subfloat[Cubic ($p{=}3$)]{\includegraphics[width=0.45\textwidth]{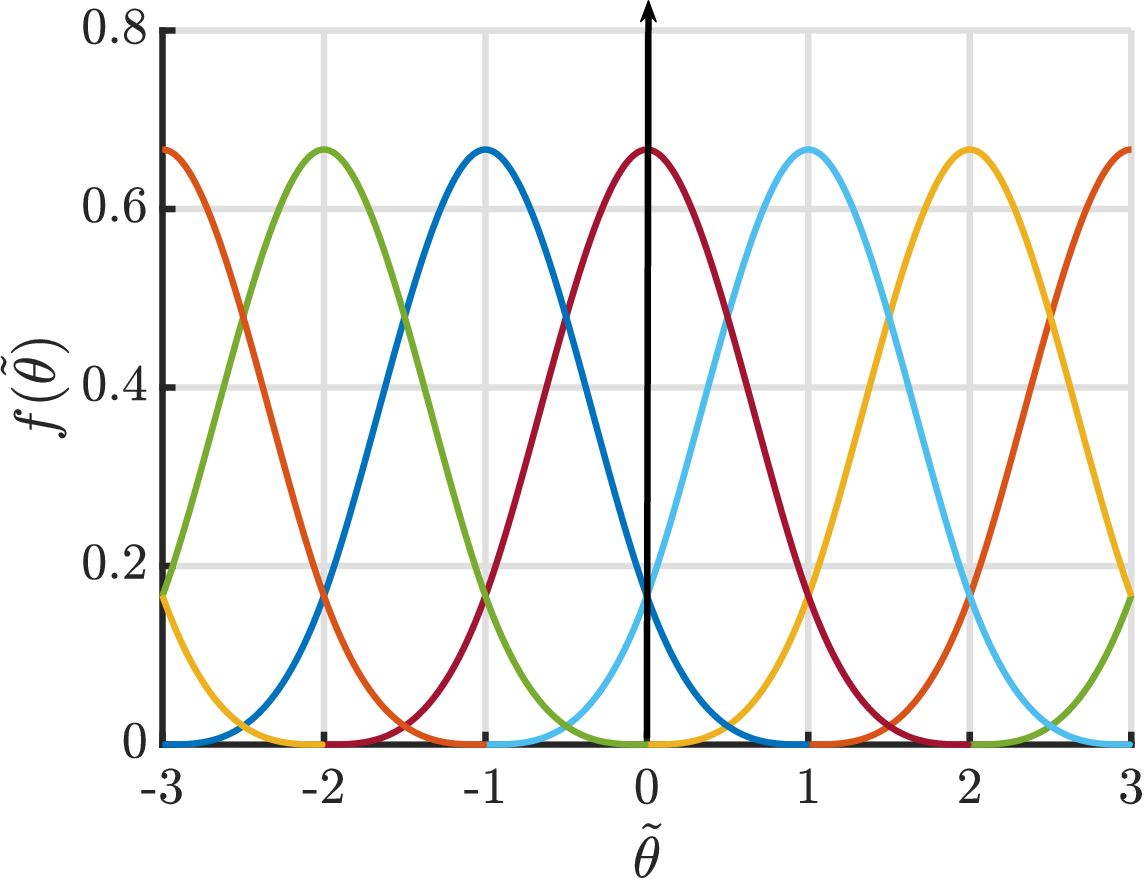}} 
 \caption{Quadratic and cubic periodic B-spline basis functions near $\theta=0$.}
 \label{fig.periodic_basis}
\end{figure}

We define the periodic B-spline basis on $\theta \in [0, 2\uppi)$ for fixed degree~$p$, with uniform spacing 
$\Delta \theta \coloneqq 2\uppi/N_\theta$, where $N_\theta$ denotes the number of angular basis functions. Thus, the finite-dimensional subspace~$V_{\theta,h}$ is spanned by
$N_\theta$ angular B-splines. The uniform periodic knot vector
$\theta_j \coloneqq j\Delta \theta$, where $j=0,1,\ldots,N_\theta-1$, is wrapped so that $\theta_{j+N_\theta} \equiv \theta_j \ (\mathrm{mod} \ 2\uppi)$. The first B-spline $B_0(\theta)$ is chosen such that its maximum occurs at $\theta=0$ (see Fig.~\ref{fig.periodic_basis}, where $\tilde{\theta} \coloneqq \theta / \Delta \theta$ is the normalized angular coordinate), 
and it is constructed to be even, $B_0(\theta) = B_0(-\theta)$. 
As part of the periodic spline basis, $B_0$ also satisfies $B_0(\theta + 2\uppi) = B_0(\theta)$.
The remaining basis functions are generated by periodic translation:
\begin{equation} \label{eq.Btheta_translate}
    B_j(\theta) \coloneqq B_0 \left(\theta - j \Delta \theta \right), 
    \qquad (1\leq j \leq N_\theta-1).
\end{equation}
As $B_0$ is chosen to be even, it follows that
\begin{equation}
    B_j(\theta) = B_{N_\theta-j}(2\uppi-\theta), 
    \qquad (1\leq j \leq N_\theta-1).
\end{equation}

A function $\phi_{\theta,h}(\theta) \in V_{\theta,h}$ can be expanded in this basis as
\begin{equation}
\phi_{\theta,h}(\theta) = \sum_{j=0}^{N_\theta-1} \phi_{\theta,j} B_j(\theta)
           = \boldsymbol{\phi}_\theta^{\mathrm T} \boldsymbol{B}_\theta(\theta),
\end{equation}
where $\boldsymbol{\phi}_\theta\in \mathbb{R}^{N_\theta}$ and $\boldsymbol{B}_\theta$ are column vectors of length $N_\theta$ containing the coefficients $\phi_{\theta,j}$ and the basis functions $B_j(\theta)$, respectively. 

By construction from uniform periodic translations of $B_0(\theta)$, the angular B-splines satisfy the partition of unity~\citep{deBoor2001splines}:
\begin{equation} \label{eq.unity_theta}
    \sum_{j=0}^{N_\theta-1} B_j(\theta) = 1, 
    \qquad \forall \theta \in [0,2\uppi).
\end{equation}

\subsection{Two-dimensional B-spline finite-element basis functions in polar coordinates}
\label{sec.tensor_product_basis}

For a two-dimensional problem in polar coordinates $(r,\theta)$ we define the B-spline basis functions
\begin{equation} \label{eq.tensor_Bspline}
    B_{i,j}(r,\theta) \coloneqq B_{r,i}(r) B_{\theta,j}(\theta) ,
\end{equation}
where $B_{r,i}(r)$ is a non-periodic radial basis function described in 
Sec.~\ref{Sec.1d_basis_radial} and $B_{\theta,j}(\theta)$ is a periodic angular basis 
function with periodicity $2\uppi$ described in Sec.~\ref{Sec.1d_basis_angular}.
The linearly independent two-dimensional B-splines $B_{i,j}$, defined by the product 
ansatz~\eqref{eq.tensor_Bspline}, are also called tensor-product B-splines. 
Thus, the finite-dimensional space~$V_h$ defined by the tensor product
\begin{equation}
 V_h \coloneqq \operatorname{span}\{ B_{r,i}(r) B_{\theta,j}(\theta) \mid
0 \le i \le N_r-1,\; 0 \le j \le N_\theta-1 \}
          = V_{r,h} \otimes V_{\theta,h},
\end{equation}
is spanned by
\begin{equation}\label{eq.DoF_original}
    N \coloneqq \dim V_h = N_r \, N_\theta
\end{equation}
tensor-product B-splines, where $N_r$ and $N_\theta$ denote the number of one-dimensional B-splines spanning $V_{r,h}$ and $V_{\theta,h}$, respectively.

Note that we restrict our discussion to the two-dimensional domain of a unit disc. By multiplying another periodic angular basis function~$B_{\varphi,k}(\varphi)$, we could easily cover the three-dimensional case of a torus.

When needed, we write the two-dimensional B-spline basis functions as a column vector
\begin{equation}\label{eq.Kronecker_product}
    \boldsymbol{B} \coloneqq \boldsymbol{B}_r \otimes \boldsymbol{B}_\theta ,
\end{equation}
where $\boldsymbol{B}_r$ and $\boldsymbol{B}_\theta$ are the column vectors for the radial and angular B-spline basis, respectively, and the operator $\otimes$ is the Kronecker product. It is defined as
\begin{subequations}
\begin{align}
       \boldsymbol{B}_r \otimes \boldsymbol{B}_\theta 
       &\coloneqq [B_{r,0} \boldsymbol{B}_\theta^\mathrm{T},\; B_{r,1} \boldsymbol{B}_\theta^\mathrm{T}, \ldots, B_{r,N_r-1} \boldsymbol{B}_\theta^\mathrm{T}]^\mathrm{T}, \\
       \label{kprime}
       B_{k'}(r,\theta) 
       &\coloneqq B_{r,i}(r) B_{\theta,j}(\theta), \quad
        (i = 0, \ldots, N_r-1,\; j = 0, \ldots, N_\theta-1,\; k' \coloneqq i N_\theta + j) ,
\end{align} 
\end{subequations}
where $\boldsymbol{B}_r \in V_{r,h}^{N_r}$, $\boldsymbol{B}_\theta \in V_{\theta,h}^{N_\theta}$, and $\boldsymbol{B} \in V_h^{N}$.

A function $\phi_h \in V_h$ can be expressed in the basis $\boldsymbol{B}$ by
\begin{equation} \label{eq.2d_bspline_phi_org}
     \phi_h(r,\theta) = \sum_{i=0}^{N_r-1} \sum_{j=0}^{N_\theta-1} \phi_{i,j} B_{i,j}(r,\theta),
\end{equation}
Let us denote the coefficients by $\boldsymbol{\phi}\in \mathbb{R}^{N}$, ordered such that the angular index $j$ varies fastest. We can rewrite Eq.~\eqref{eq.2d_bspline_phi_org} in accordance with Eq.~\eqref{kprime} with a single-index~$k'$ when the coefficient vector~$\boldsymbol{\phi}$ and the basis vector~$\boldsymbol{B}$ are adjusted correspondingly
\begin{equation} \label{eq.2d_bspline_phi_org_single}
     \phi_h(r,\theta) = \sum_{k'=0}^{N-1} \phi_{k'} B_{k'}(r,\theta) = \boldsymbol{\phi}^\mathrm{T} \boldsymbol{B}(r,\theta).
\end{equation}

For a better approximation, either the number of intervals in the radial and angular directions can be increased, or the polynomial degree of the B-splines can be raised. Note that no measures have been taken yet to enforce a specific $C^n$-regularity of the B-spline representation at the origin of polar coordinates used on the unit disc. We address this issue in Sec.~\ref{sec:regularity}.

For the tensor-product basis, the partition of unity follows from the partition of unity of the radial (see Eq.~\eqref{eq.unity_r}) and angular (see Eq.~\eqref{eq.unity_theta}) B-spline basis functions:
\begin{equation} \label{eq.unity_tensor}
    \sum_{k'=0}^{N-1} B_{k'}(r,\theta)
    = \sum_{i=0}^{N_r-1} \sum_{j=0}^{N_\theta-1} B_{r,i}(r) B_{\theta,j}(\theta)
    = \left( \sum_{i=0}^{N_r-1} B_{r,i}(r) \right) \left( \sum_{j=0}^{N_\theta-1} B_{\theta,j}(\theta) \right) = 1, \quad \forall(r,\theta)\in \Omega.
\end{equation}

\subsubsection{Galerkin finite-element formulation in polar coordinates}
\label{Sec.galerkin_matrices}

Having established the radial, angular, and tensor-product B-spline basis functions, we now turn to the question of how to use them to approximate a given function $f(r,\theta)$. A two-dimensional spline function~$\phi_h(r,\theta)$, defined by Eq.~\eqref{eq.2d_bspline_phi_org_single}, can be used for such approximations via various numerical schemes. Among possible approaches are the sampling method, the collocation method, and the finite-element method based on Galerkin orthogonality (see App.~\ref{App.trigono_approx}).
While the sampling and collocation schemes are of interest for their simplicity and ease of implementation, in the present work all computations are carried out exclusively using the finite-element method. This choice is motivated by its superior convergence properties and robustness, especially for higher-order B-splines. The remainder of this section outlines the formulation of the Galerkin finite-element method in polar coordinates, which forms the basis of our numerical implementation.

The $L^2$-Galerkin projection $\Pi : L^2(\Omega) \to V_h$, where $V_h \subset L^2(\Omega)$
is a finite-dimensional subspace, maps $f \in L^2(\Omega)$ to $\phi_h = \Pi[f] \in V_h$
uniquely characterized by the orthogonality condition
\begin{equation}\label{eq.Galerkin_ortho}
\int_\Omega \big(f(\boldsymbol{r}) - \Pi[f](\boldsymbol{r})\big)
B_k(\boldsymbol{r}) \, \mathrm{d}\boldsymbol{r} = 0,
\qquad \forall \, B_k \in V_h.
\end{equation}
That is, the discretization error (residual) $e_h \coloneqq f - \Pi[f]$ is orthogonal to
$V_h$ with respect to the $L^2(\Omega)$ inner product, a property commonly referred to
as Galerkin orthogonality. Here, $\boldsymbol{r}$ denotes the physical coordinate vector in the domain~$\Omega$
(independent of the choice of coordinate system). When expressed in polar coordinates, the integration measure becomes $\mathrm{d}\boldsymbol{r} = r\,\mathrm{d}r\,\mathrm{d}\theta$.

Moreover, $\Pi$ is understood as an operator acting on functions; its matrix
representation with respect to a chosen basis of $V_h$ yields the corresponding
coefficient vector of $\phi_h$.
Inserting \eqref{eq.2d_bspline_phi_org_single} in Eq.~\eqref{eq.Galerkin_ortho} gives
\begin{equation}
    \sum_{k'=0}^{N-1} \phi_{k'} \int_\Omega B_{k'}(\boldsymbol{r}) B_k(\boldsymbol{r}) \, \mathrm{d} \boldsymbol{r} = \int_\Omega f(\boldsymbol{r}) B_k(\boldsymbol{r}) \, \mathrm{d}\boldsymbol{r} ,
    \qquad (k = 0, \ldots, N-1)
\end{equation}
and leads to the linear system of equations
\begin{equation} \label{eq.mass_org}
    \mathsfbi{M} \boldsymbol{\phi} = \boldsymbol{f} ,
\end{equation}
where
\begin{align}
\label{eq.mass_matrix_org}
    \matcomp{m}_{k,k'} & \coloneqq \int_\Omega B_k(\boldsymbol{r}) B_{k'}(\boldsymbol{r}) \, \mathrm{d} \boldsymbol{r} , \\
\label{eq.RHS_f}
    f_k & \coloneqq \int_\Omega f(\boldsymbol{r}) B_k(\boldsymbol{r}) \, \mathrm{d} \boldsymbol{r}, \qquad (k,k' = 0, \ldots, N-1).
\end{align}
The single-indices are defined as $k\coloneqq i N_\theta + j$, $k'\coloneqq i' N_\theta + j'$, where the dual-indices are $i,i'= 0, \ldots, N_r-1$ and $j,j'= 0, \ldots, N_\theta-1$. Here, the $i$ (and $i'$) index refers to the radial B-spline basis functions $B_{r,i}(r)$, while the $j$ (and $j'$) index refers to the angular B-spline basis functions $B_{\theta,j}(\theta)$.
In finite-element literature, the matrix~$\mathsfbi{M}$ is known as the mass matrix, and the vector $\boldsymbol{f}$ as the load vector. The mass matrix $\mathsfbi{M}$ is symmetric and positive definite on~$V_h$ because the B-spline basis functions $B_k$ are linearly independent with nontrivial compact support. In linear algebra terms, $\mathsfbi{M}$ is the Gram matrix of the basis functions with respect to the $L^2$-inner product. 

The B-spline basis can also be employed to discretize partial differential equations within the finite-element method framework. 
We consider the linear boundary-value problem
\begin{subequations}
\label{eq.setofequations}
\begin{alignat}{3}
\label{eq.equation}
    \mathcal{L} \phi(\boldsymbol{r}) &= f(\boldsymbol{r})
    &&\quad \text{in}&& \quad \Omega , \\
\label{eq.boundary_condition}
    \phi &= 0 
    &&\quad \text{on}&& \quad \partial\Omega ,
\end{alignat}
\end{subequations}
where $\Omega$ denotes the region of the unit disc and $\partial \Omega$ represents its boundary.

The linear elliptic differential operator $\mathcal{L}$ is taken in divergence form as
\begin{subequations}
\label{eq.selfadj_oper}
\begin{equation}\label{eq.diff_oper}
    \mathcal{L} \phi(\boldsymbol{r}) \coloneqq 
    - \nabla \bcdot \left[ a(\boldsymbol{r}) \nabla \phi(\boldsymbol{r}) \right] 
    + c(\boldsymbol{r}) \phi(\boldsymbol{r}) ,
\end{equation}
where the real-valued coefficient functions $a(\boldsymbol{r})$ and $c(\boldsymbol{r})$ are prescribed and satisfy
\begin{equation} \label{eq.qp-funcs}
    a(\boldsymbol{r}) > 0, \qquad c(\boldsymbol{r}) \ge 0 .
\end{equation}
\end{subequations}

In this work, we restrict our attention to differential operators of the form~\eqref{eq.selfadj_oper} together with homogeneous Dirichlet boundary conditions~\eqref{eq.boundary_condition}. With these boundary conditions, the operator $\mathcal{L}$ is self-adjoint in the weak sense on $L^2(\Omega)$. Nevertheless, the same finite-element B-spline framework can be applied to more general coefficient functions, boundary conditions, and differential operators.

In the finite-element method, the weak form of Eq.~\eqref{eq.equation} is obtained by 
representing $\phi_h$ as in Eq.~\eqref{eq.2d_bspline_phi_org_single}, multiplying 
by the test basis function $B_k(\boldsymbol{r})$, and integrating over $\Omega$. 
Applying integration by parts and incorporating the boundary conditions yields:
\begin{multline} \label{eq.diffopt_splines}
    \sum_{k'=0}^{N-1} \phi_{k'} \left[ -\oint_{\partial \Omega} a(\boldsymbol{r}) B_k(\boldsymbol{r}) \nabla B_{k'}(\boldsymbol{r}) \bcdot \widehat{\boldsymbol{n}} \, \mathrm{d}l
          + \int_\Omega \left( a(\boldsymbol{r}) \nabla B_k(\boldsymbol{r}) \bcdot \nabla B_{k'}(\boldsymbol{r}) + c(\boldsymbol{r}) B_k(\boldsymbol{r}) B_{k'}(\boldsymbol{r}) \right) \mathrm{d} \boldsymbol{r} \right] \\
          = \int_\Omega f(\boldsymbol{r}) B_k(\boldsymbol{r}) \, \mathrm{d}\boldsymbol{r} ,
    \qquad (k = 0, \ldots, N-1),
\end{multline}
where $\widehat{\boldsymbol{n}}$ is the outward-pointing unit normal vector to $\partial\Omega$. The homogeneous Dirichlet boundary condition, Eq.~\eqref{eq.boundary_condition}, 
is enforced by omitting the boundary term in Eq.~\eqref{eq.diffopt_splines}. 
At the polar outer boundary $r = 1$, only the clamped radial B-spline 
$B_{r,N_r-1}(r)$ is nonzero. For homogeneous Dirichlet boundary conditions, 
the coefficients of these trial basis functions are set to zero. 
Consequently, all test functions $B_k(\boldsymbol{r})$ used in the weak form vanish 
on $\partial\Omega$, and the boundary integral is therefore identically zero.

Finally, we arrive at the following linear system of equations
\begin{equation} \label{eq.stiff_org}
    \mathsfbi{S} \boldsymbol{\phi} = \boldsymbol{f},
\end{equation}
where the symmetric stiffness matrix is defined as
\begin{equation} \label{eq.S_mat}
\begin{aligned}
       \matcomp{s}_{k,k'} \coloneqq {} & \int_{0}^{1} \int_{0}^{2\uppi} a(r,\theta)
          \frac{\mathrm{d}B_{r,i}(r)}{\mathrm{d}r} \frac{\mathrm{d}B_{r,i'}(r)}{\mathrm{d}r} 
          B_{\theta,j}(\theta) B_{\theta,j'}(\theta) \, r \, \mathrm{d}\theta \, \mathrm{d}r \\
       & + \int_{0}^{1} \int_{0}^{2\uppi} a(r,\theta) B_{r,i}(r) B_{r,i'}(r)
          \frac{1}{r} \frac{\mathrm{d}B_{\theta,j}(\theta)}{\mathrm{d}\theta} 
          \frac{\mathrm{d}B_{\theta,j'}(\theta)}{\mathrm{d}\theta} \, \mathrm{d}\theta \,\mathrm{d}r \\
       & + \int_{0}^{1} \int_{0}^{2\uppi} c(r,\theta) B_{r,i}(r) B_{r,i'}(r) B_{\theta,j}(\theta) B_{\theta,j'}(\theta) \,
          r \, \mathrm{d}\theta \, \mathrm{d}r.
\end{aligned}
\end{equation}
Here, the radial indices $i,i'$ run from $0$ to $N_r-2$ because the 
outer-boundary function $B_{r,N_r-1}(r)$, which is nonzero at $r=1$, is excluded 
from the trial and test spaces by the homogeneous Dirichlet condition. 
The angular indices $j,j'$ run over their full range $0,\ldots,N_\theta-1$.

The elements $\matcomp{s}_{k,k'}$ with dual radial indices $i {=} i' {=} 0$ cannot cancel the apparent $1/r$ singularity in the integrand in the second line of Eq.~\eqref{eq.S_mat}, since $B_{r,0}(0) B_{r,0}(0) = 1$ (see Eq.~\eqref{eq.B0_r0_one}) and we assume $a(r,\theta) \neq 0$. 
In contrast, elements for which $i = 1,\ldots,p$ or $i' = 1,\ldots,p$ satisfy $B_{r,i}(0) B_{r,i'}(0) = 0$, which removes the singularity. 
Numerical evaluation of Eq.~\eqref{eq.S_mat} using Gauss--Legendre quadrature yields finite values due to quadrature regularization, but the formulation remains inconsistent in the continuous weak form.
However, a mathematically consistent way to eliminate the pole is to impose $C^0$-regularity on the B-spline basis functions at the origin. 
This modification renders new basis functions~$\widehat{B}_0^0$ independent of~$\theta$, so that $\mathrm{d}\widehat{B}_0^0 / \mathrm{d}\theta = 0$ (see App.~\ref{App.C0-regularity}), and consequently removes the singular term exactly.

\section{General information about bases with $C^n$-regularity at the origin}
\label{sec:general_info}

\subsection{Regularity conditions for two-dimensional functions at the polar origin}
\label{sec:regularity}

In polar coordinates (see Eq.~\eqref{eq.polar_coord_trans}),
the coordinate singularity at the origin $r=0$ can cause a nonphysical
loss of regularity when a function or its derivatives are expressed
in these coordinates.
It has been shown in~\citep{eisen1991Spectra} that a function $f(x,y)$
is of class $C^n$ at $(0,0)$ if and only if its local expansion near the
origin agrees, up to order $n$, with that of a bivariate polynomial of
total degree $n$ in $(x,y)$. Consequently, if $f(x,y)$ is itself a
polynomial, then it satisfies this condition for all $n$ and is therefore
$C^\infty$ at $(0,0)$. Therefore, to enforce $C^\infty$ regularity at the
origin for a spline approximation in polar coordinates
(see Eq.~\eqref{eq.2d_bspline_phi_org_single}), it suffices to require that
$\phi_h\!\left( \boldsymbol{F}^{-1}(x,y) \right)$ coincide with a bivariate
polynomial in $(x,y)$ within an inner region of the grid, specified by
$r \leq \Delta r$, where $\Delta r$ denotes the radial grid spacing.

\subsection{Bivariate monomial basis}
\label{Sec.Cn_polynomial_space}

A bivariate polynomial in $(x,y)$ of total degree at most $n$, defined over 
the domain $\Omega$ (see Eq.~\eqref{eq.unit_disc}), can be written as
\begin{subequations}
\begin{equation}\label{eq.Cn_poly_xy}
    \mathcal{T}_n(x,y) 
    = \sum_{l=0}^{n} \sum_{p=0}^{l} t_l^p \,T_l^p(x,y),
\end{equation}
where the Cartesian monomial basis functions of total degree $l$ are defined by
\begin{equation}
    T_l^p(x,y) \coloneqq x^p y^{l-p}.
\end{equation}
\end{subequations}
Thus, the bivariate monomials~$T_l^p(x,y)$ of total degree $\leq n$ span the subspace of bivariate polynomials~$\mathbb{P}_n$, which has the dimension
\begin{equation} \label{eq.deg_freedom}
   N_n \coloneqq \dim \mathbb{P}_n = \frac{1}{2}(n+1)(n+2).
\end{equation}

Transforming Eq.~\eqref{eq.Cn_poly_xy} to polar coordinates via 
$(x,y) = (r\cos(\theta), r\sin(\theta))$ (see Eqs.~\eqref{eq.polar_coord_trans}) yields
\begin{subequations}
\begin{equation}\label{eq.Cn_poly_rtheta}
    \widetilde{\mathcal{T}}_n(r,\theta) 
    = \sum_{l=0}^{n} \sum_{p=0}^{l}
      t_l^p \, \widetilde{T}_l^p(r,\theta),
\end{equation}
where the polar-coordinate representation of the Cartesian monomials is
\begin{equation}
    \widetilde{T}_l^p(r,\theta) 
    \coloneqq r^l \cos^p(\theta) \sin^{l-p}(\theta).
\end{equation}
\end{subequations}
Since $\mathcal{T}_n(x,y)$ is a polynomial, it belongs to $C^\infty(\Omega)$ in Cartesian coordinates. Its polar representation $\widetilde{\mathcal{T}}_n(r,\theta)$ defines the same $C^\infty$ function on $\widehat{\Omega}$. At $r=0$, all terms with $l>0$ vanish, leaving $\widetilde{\mathcal{T}}_n(0,\theta) = t_0^0$ for all $\theta$, consistent with the continuity (and smoothness) of polynomials.

By construction, the Cartesian and polar representations agree everywhere:
\begin{equation}
\mathcal{T}_n(x,y) =
\begin{cases}
\widetilde{\mathcal{T}}_n\big( \boldsymbol{F}^{-1}(x,y) \big),
    & (x,y) \neq (0,0), \\[2mm]
t_0^0, 
    & (x,y) = (0,0).
\end{cases}
\end{equation}
Note that $\boldsymbol{F}^{-1}$ is not uniquely defined at $(0,0)$,
because $\theta$ is undefined there; however,
$\widetilde{\mathcal{T}}_n(r,\theta)$ is independent of $\theta$ for $r=0$,
so the constant value $t_0^0$ agrees with the Cartesian representation.

\subsection{Harmonic representation of the bivariate monomial basis}
\label{Sec.harmbasisfuncs}

Although the polar bivariate monomial basis functions~$\widetilde{T}_l^p$ are a straightforward choice with a simple radial part, they have the drawback that they are not orthogonal (see Sec.~\ref{Sec.orthonormal}) in either the radial or angular part. Therefore, we propose a more concise basis, called the harmonic polar basis, which spans the same polynomial space but is pairwise orthogonal in the angular part with respect to the $L^2$-inner product over $[0,2\uppi]$. We keep the radial part and rewrite the powers of $\sin(\theta)$ and $\cos(\theta)$ in~$\widetilde{T}_l^p$ as sums of the harmonic angular modes
\begin{equation}
    \label{eq.h_basis}
    h_{-m}(\theta) \coloneqq \sin(m\theta), \qquad
    h_0(\theta) \coloneqq 1, \qquad
    h_m(\theta) \coloneqq \cos(m\theta), \qquad m \in \mathbb{N}_0.
\end{equation}
Here, the harmonic index is signed: negative indices $-m$ correspond to sine
modes and non-negative indices $m$ to cosine modes.

We distinguish odd and even harmonic polar functions according to their angular parity. The odd harmonic polar functions are defined as
\begin{subequations}
\label{eq.smooth_basis}
\begin{equation}
    S_l^{-m}(r,\theta) \coloneqq r^l \sin(m\theta),
\end{equation}
and the even ones as
\begin{equation}
    S_l^{m}(r,\theta) \coloneqq r^l \cos(m\theta),
\end{equation}
\end{subequations}
where $l,m \in \mathbb{N}_0$ satisfy $m \le l$ and $m \equiv l \ (\mathrm{mod}\ 2)$.
Here, $m$ denotes the harmonic order, while the sign of the superscript
$\pm m$ distinguishes sine ($-$) and cosine ($+$) modes.
For every admissible pair $(l,m)$ with $m>0$, the harmonic polar basis contains precisely
two functions $S_l^{-m}$ and $S_l^{m}$.
For $m=0$, the sine term vanishes identically ($S_l^{-0}=0$) and only the cosine
term $S_l^{0}$ is present.

The harmonic polar basis functions~$S_l^{-m}$ and $S_l^m$ can be expressed as a linear combination of the basis functions~$\widetilde{T}_l^p$ and have the same cardinality as the polar monomial basis. We can express the polynomial $\widetilde{\mathcal{T}}_n$ by using the harmonic polar basis functions:
\begin{equation}\label{eq.bivar_monomials}
    \widetilde{\mathcal{T}}_n(r,\theta) =
    \sum_{l=0}^{n} 
    \sum_{\substack{m = -l \\ |m| \equiv l \ (\mathrm{mod}\ 2)}}^{l}
    s_l^m \, S_l^m(r,\theta).
\end{equation}

The parity condition $|m| \equiv l \ (\mathrm{mod}\ 2)$ means that the harmonic
order $|m|$ and the total degree $l$ are either both even or both odd.
This arises naturally when a homogeneous Cartesian monomial
$x^p y^{l-p}$ of total degree~$l$ is converted to polar coordinates and the
angular part $\cos^p(\theta)\sin^{\,l-p}(\theta)$ is expanded into harmonics.
The resulting terms $\cos(m\theta)$ and $\sin(m\theta)$ have harmonic orders
$|m|$ with the same parity as~$l$.
%
%
The “monomial parity” property \cite[Chapter~18.4]{Boyd2000} ensures that for each
nonzero harmonic order $|m|>0$, the harmonic polar basis contains both the sine term
$S_l^{-m}$ and the cosine term $S_l^{m}$.
The shorthand notation $S_l^{\pm m}$ denotes the pair of harmonic polar functions 
$\{S_l^{-m}, S_l^{m}\}$ associated with angular mode~$|m|$.
For $|m|>0$, this pair spans a two‑dimensional real subspace, whereas for $m=0$
only the cosine term $S_l^{0}$ is present.
The harmonic polar functions~$S_l^{\pm m}(r,\theta)$ form a basis of the polynomial
space $\mathbb{P}_n$.
Table~\ref{tab:tableau_Slm} displays an example of $S_l^{\pm m}$ for $n=3$.
\begin{table}[ht]
    \centering
    \begin{tabular}{|cccccccc|}
         \hline
          \phantom{$l=0$} \vline & $m=-3$ & $m=-2$ & $m=-1$ & $m=0$ & $m=\phantom{-}1$ & $m=\phantom{-}2$ & $m=\phantom{-}3$ \\
         \hline
          $l=0$ \vline &&&& $S_0^0$ &&& \\
          $l=1$ \vline &&& $S_1^{-1}$ && $S_1^1$ && \\
          $l=2$ \vline && $S_2^{-2}$ && $S_2^0$ && $S_2^2$ & \\
          $l=3$ \vline &$S_3^{-3}$ && $S_3^{-1}$ && $S_3^1$ && $S_3^3$  \\
         \hline
    \end{tabular}
    \caption{Tableau of the harmonic polar functions ~$S_l^{\pm m}$ of degree at most
    $n=3$. Negative values of $m$ correspond to sine modes and positive values to cosine modes. The index~$l$ labels rows and the index~$m$ labels columns. Admissible index pairs satisfy $0 \le l \le n$ and $|m| \le l$, with the parity condition $|m| \equiv l \pmod{2}$.} 
    \label{tab:tableau_Slm}
\end{table}

Finally, we rewrite the harmonic polar functions $S_l^m$ in a product form
separating radial and angular parts (see Eq.~\eqref{eq.smooth_basis}):
\begin{equation} \label{eq.gen_harmon_ansatz_l}
    S_l^m(r,\theta) = R_l(r) h_m(\theta),
\end{equation}
where the radial part is a monomial of degree~$l$
\begin{equation} \label{eq.R_l_coeff}
    R_l(r) \coloneqq r^l.
\end{equation}
An important observation is that, although the harmonic polar functions $S_l^{m}(r,\theta)$
are well defined for arbitrary index pairs $(l,m)$, 
only the admissible ones can be expressed as Cartesian polynomials, in the sense that:
\begin{equation} \label{eq.Slm_smoothness}
S_l^{m}(\boldsymbol{F}^{-1}(x,y)) \in C^\infty(\Omega) 
\quad \iff \quad |m| \le l \ \text{and}\ |m| \equiv l \ (\mathrm{mod}\,2),
\ \text{with } l \in \mathbb{N}_0,\ m \in \mathbb{Z}.
\end{equation}
This characterization of $C^\infty$-polar modes, which was also derived in \cite{Lewis1990constraints}, will be central in our construction of smooth polar splines (see Sec.~\ref{sec:c1scheme}).

\subsection{Mutual $L^2$-orthogonality of the basis functions~$S_l^{\pm m}$ over $\widehat{\Omega}$}
\label{Sec.orthonormal}

In numerical methods such as Galerkin discretization, using monomials $r^l$ as basis functions often leads to ill-conditioned systems. As $l$ increases, normalized radial monomials become increasingly parallel in the $L^2$-inner product: the inner product of two normalized monomials of degrees $l-1$ and $l$ tends toward 1. This amplifies round-off errors and degrades the convergence of iterative solvers.

Orthogonalized bases, such as Zernike polynomials~\citep{Zernike1934,Niu2022}, avoid these problems: they are mutually orthogonal, reduce cross‑coupling in the mass matrix, and often yield matrices that are well‑conditioned. Orthogonal functions separate scales cleanly, so higher‑order modes do not contaminate lower‑order behavior, and coefficients can be computed stably. As a result, orthogonal bases improve numerical stability, accuracy, and efficiency.

On the unit disc, the continuous $L^2$-inner product in polar coordinates is defined by
\begin{equation} \label{eq.L2_scalar_prod}
   \left\langle \phi_h , \psi_h \right\rangle \coloneqq \int_0^1 \int_{0}^{2\uppi} \phi_h(r,\theta) \, \psi_h(r,\theta) \, r \, \mathrm{d} \theta \, \mathrm{d} r,
\end{equation}
where $\phi_h$ and $\psi_h$ are assumed to be real-valued. Substituting 
$\phi_h(r,\theta) = \sum_{k=0}^{N-1} \phi_k B_k(r,\theta)$ and
$\psi_h(r,\theta) = \sum_{k'=0}^{N-1} \psi_{k'} B_{k'}(r,\theta)$  
into Eq.~\eqref{eq.L2_scalar_prod} yields
\begin{equation}\label{eq.Mscalar_prod}
\langle \phi_h, \psi_h \rangle
= \sum_{k=0}^{N-1} \sum_{k'=0}^{N-1} \phi_k \psi_{k'}
\int_0^1 \!\!\int_0^{2\uppi} B_k(r,\theta) B_{k'}(r,\theta) \, r \, \mathrm{d}\theta \,\mathrm{d}r
= \boldsymbol{\phi}^\mathrm{T} \mathsfbi{M} \boldsymbol{\psi},
\end{equation}
where $\mathsfbi{M}$ is the mass matrix for the basis $\boldsymbol{B}$, as defined in Eq.~\eqref{eq.mass_matrix_org}. 
Hence, the continuous $L^2$-inner product corresponds to the $M$-inner product in coefficient space.

The $L^2$-norm induced by Eq.~\eqref{eq.L2_scalar_prod} is
\begin{equation} \label{eq.L2_norm}
   \lVert \phi_h\rVert_{L^2} = \sqrt{\left\langle \phi_h , \phi_h \right\rangle} 
   = \sqrt{ \boldsymbol{\phi}^\mathrm{T} \mathsfbi{M} \boldsymbol{\phi} },
\end{equation}
and will be used in the normalization step after orthogonalization.

The harmonics~$h_m$ (see Eq.~\eqref{eq.h_basis}) are mutually orthogonal with respect to the $L^2$-inner product,
\begin{equation} \label{eq.h_orthogonality}
    \left\langle h_m, h_{m'} \right\rangle_\theta \coloneqq \int_0^{2\uppi} h_m(\theta) h_{m'}(\theta) \, \mathrm{d}\theta \propto \delta_{m,m'}.
\end{equation}
Thus, no orthogonalization is required between basis functions with different $m$ values. 
For basis functions sharing the same $m$ index, however, orthogonality must be enforced in the radial direction: the monomial radial parts $R_l$ are not orthogonal with respect to the radial $L^2$-inner product,
\begin{equation} \label{eq.innerprod_r}
\langle R_l, R_{l'} \rangle_r \coloneqq \int_0^1 R_l(r) R_{l'}(r) \, r \, \mathrm{d}r = \frac{1}{l+l'+2},
\end{equation}
and therefore require orthogonalization.

We orthogonalize the radial parts $R_l$ within each $m$-column of tableau~\ref{tab:tableau_Slm} (i.e.\ for fixed harmonic index $m$). 
In this case, the radial parts $R_l$ take a generalized form $\overline{R}_l^m$ that depends on both indices $l$ and $m$ (see App.~\ref{App.smooth_constraint}). This mutual orthogonalization can be stated as
\begin{equation}
    \left\langle \overline{R}_l^{m}, \overline{R}_{l'}^{m} \right\rangle_r \propto \delta_{l,l'}.
\end{equation}
By performing Gram--Schmidt orthogonalization of the radial part along each $m$ column of tableau~\ref{tab:tableau_Slm}, the orthogonalized functions $\overline{S}_l^{m}$ are the Zernike polynomials~$Z_l^m$. They form a complete orthogonal basis with respect to the full $L^2$-inner product~\eqref{eq.L2_scalar_prod} on the unit disc.

Finally, after radial orthogonalization, each basis function is normalized to unit $L^2$-norm using Eq.~\eqref{eq.L2_norm}:
\begin{equation}
    \widehat{S}_l^{\pm m}(r,\theta) = \frac{\overline{S}_l^{\pm m}}{\lVert \overline{S}_l^{\pm m} \rVert_{L^2}},
\end{equation}
yielding the orthonormal set $\widehat{\boldsymbol{S}}_l^{\pm m}(r,\theta)$.

The orthogonalization procedure for smooth center-splines is discussed in 
Sec.~\ref{Sec.orthonormal_spline}.

\section{B-spline bases with smoothness at the origin}
\label{sec:c1scheme}

\subsection*{Overview of the smooth polar spline construction}

Our goal is to design a subspace of the tensor-product B-splines
that preserves the algebraic $C^\infty$-regularity structure of harmonic 
polar functions, encoded by the admissible index pairs $(l,m)$ (see Sec.~\ref{sec:general_info}),
as well as the high-order approximation properties of B-splines.
This will enable efficient approximation of functions with steep local 
gradients, while ensuring smoothness at the polar origin.

The basic idea is to find a suitable approximation of the harmonic polar 
functions~$S^m_l$ by tensor-product B-splines in the innermost radial region,
and to use admissible index pairs $(l,m)$ which correspond to $C^\infty$-regularity (see Eq.~\eqref{eq.Slm_smoothness}) and are bounded by the degree~$p$ of the B-splines. 

By retaining fewer independent angular degrees of freedom close to the origin, 
this approach ensures exact $C^\infty$-regularity as the angular 
discretization is refined ($\Delta\theta \rightarrow 0$), and, for finite angular grids, it preserves the smoothness structure of harmonic polar functions.
In particular, our approximation is exact for radial monomials~$r^l$ for $l \le p$
on the innermost radial interval $[0,\Delta r]$, and it preserves the orthogonality
property of the harmonic angular modes $h_m(\theta)$ on $[0,2\pi)$.

For clarity, we briefly summarize the construction of the smooth polar-spline
basis:
\begin{enumerate}
  \item We start from the family of separable harmonic polar functions
  $S_l^m(r,\theta)$, obtained from the harmonic representation of the bivariate
  polynomial space $\mathbb P_p$ in polar coordinates as characterized in
  Sec.~\ref{Sec.harmbasisfuncs}. These functions are projected by a Galerkin
  $L^2$-projection onto the tensor-product B-spline space on the center
  subdomain, yielding the corresponding center-spline functions~$\widetilde B_l^m$. The projection is performed for all index pairs $(l,m)$ in a
  tensor-product index set $\mathcal I_{\mathrm c}
  = \mathcal I_{\mathrm c}(p,N_\theta)$ of cardinality
  $|\mathcal I_{\mathrm c}| = (p+1)N_\theta$, corresponding to a full
  tensor-product tableau using all $N_\theta$ angular degrees of freedom (see Tab.~\ref{tab:tableau_Blm}).

  \item To preserve the smoothness structure at the origin, the resulting center-spline functions~$\widetilde B_l^m$ are restricted to an admissible index set
  $\widetilde{\mathcal I}_{\mathrm c} \subseteq \mathcal I_{\mathrm c}$ of
  cardinality $|\widetilde{\mathcal I}_{\mathrm c}| = (p+1)(p+2)/2$. This reduced
  index set corresponds to the first $p+1$ rows of the tableau
  associated with $C^\infty$-regularity at the origin  (see Tab.~\ref{tab:tableau_Slm}).
\end{enumerate}

Owing to the separable structure of the harmonic polar functions and the tensor-product
B-spline basis, the Galerkin projection admits a factorization into independent
radial and angular components. As a consequence, the modal index pairs $(l,m)$
are preserved under the Galerkin $L^2$-projection, and distinct modes do
not interact algebraically, even at finite angular resolution. 
Exact compatibility with $C^\infty$-regularity at the polar origin is recovered in
the limit $\Delta\theta \to 0$, as the discrete angular space converges to the
continuous harmonic representation.

\subsection{Galerkin $L^2$-projection onto the center subspace~$V_h(\widehat{\Omega}_\mathrm{c})$}
\label{Sec.GalerkinCenter}

We define the center subdomain of the logical polar-coordinate domain
\begin{equation}
    \widehat{\Omega}_\mathrm{c} \coloneqq [0,\Delta r] \times [0,2\uppi),
    \qquad \widehat{\Omega}_\mathrm{c} \subset \widehat{\Omega}.
\end{equation}
Moreover, we define the center subspace $V_h(\widehat{\Omega}_\mathrm{c})$ as the tensor‑product
space spanned by the B-splines whose support intersects the subdomain $\widehat{\Omega}_\mathrm{c}$:
\begin{equation}
  V_h(\widehat{\Omega}_\mathrm{c}) \coloneqq V_{r,h}([0, \Delta r]) \otimes V_{\theta,h} = \operatorname{span}\!\left\{
        B_{i,j} = B_{r,i} \otimes B_{\theta,j} \;\middle|\;
        0 \le i \le p,\; 0 \le j \le N_\theta - 1
    \right\},
\end{equation}
where $V_{r,h}([0,\Delta r])$ denotes the restriction of the radial subspace $V_{r,h}$
to the innermost radial interval $[0,\Delta r]$, spanned by the $p{+}1$ radial 
B-splines $B_{r,i}(r)$ with support intersecting $[0,\Delta r]$ (see Sec.~\ref{Sec.radial_boundary}), and $V_{\theta,h}$
is the angular finite‑element subspace spanned by the $N_\theta$ periodic angular 
B-splines $B_{\theta,j}(\theta)$ (see Sec.~\ref{Sec.1d_basis_angular}). 
The subspace~$V_h(\widehat{\Omega}_\mathrm{c})$ has dimension 
\begin{equation} \label{eq.Vh_Omega_dim}
  N_\mathrm{c} \coloneqq \dim V_h(\widehat{\Omega}_\mathrm{c}) = (p+1)\, N_\theta
\end{equation}
and is spanned by the center tensor-product B-spline basis functions
$\boldsymbol{B}_{\mathrm{c}}$ whose support intersects
$\widehat{\Omega}_{\mathrm{c}}$.

Here, the functions $S_l^m = r^l h_m(\theta)$ are considered simply as
separable functions indexed by $(l,m)$, without enforcing the smoothness-related
index restrictions introduced in Sec.~\ref{Sec.harmbasisfuncs}. The index set is
instead determined solely by the polynomial degree~$p$ of the B-splines and the
angular resolution~$N_\theta$ (see below).

We define the Galerkin $L^2$-projection~$\Pi_\mathrm{c}$ by
\begin{equation} \label{eq.Galerkin_op_full}
    \Pi_\mathrm{c} : V(\widehat{\Omega}_\mathrm{c}) \to V_h(\widehat{\Omega}_\mathrm{c}),
    \qquad
    V(\widehat{\Omega}_{\mathrm c}) \coloneqq 
L^2\!\left(\widehat{\Omega},
1_{\widehat{\Omega}_{\mathrm c}}\, r\,\mathrm{d} r\,\mathrm{d}\theta\right).
\end{equation}
With this choice of measure, the space $V(\widehat{\Omega}_{\mathrm c})$ is
naturally identified with the weighted space
$L^2(\widehat{\Omega}_{\mathrm c},\, r\,\mathrm{d}r\,\mathrm{d}\theta)$.
It maps a function $f \in V(\widehat{\Omega}_\mathrm{c})$ to its discrete representative
$\Pi_\mathrm{c}[f] \in V_h(\widehat{\Omega}_\mathrm{c})$,
characterized uniquely by the orthogonality condition
\begin{equation} \label{eq.Galerkin_c}
    \int_{\widehat{\Omega}_\mathrm{c}} \left( f(r,\theta) - \Pi_\mathrm{c}[f](r,\theta) \right)
    B_k(r,\theta) \, r \,\mathrm{d}r \, \mathrm{d}\theta = 0,
    \quad \forall \, B_k \in V_h(\widehat{\Omega}_\mathrm{c}).
\end{equation}
In particular, for the harmonic polar functions~$S_l^m(r,\theta)$, we denote their Galerkin projections by
$\widetilde{B}_l^m(r,\theta) \coloneqq \Pi_\mathrm{c}[S_l^m] \in V_h(\widehat{\Omega}_\mathrm{c})$.
The center-spline functions $\widetilde{B}_l^m(r,\theta)$ are expressed as linear combinations of the 
$B_{k'} \in V_h(\widehat{\Omega}_\mathrm{c})$:
\begin{equation} \label{eq.2d_B_lk_lincomb}
     \widetilde{B}_l^m(r,\theta) = \sum_{k'=0}^{N_\mathrm{c}-1} (\boldsymbol{P}_l^m)_{k'}  B_{k'}(r,\theta),
\end{equation}
where $\boldsymbol{P}_l^m\in \mathbb{R}^{N_\mathrm{c}}$ is the coefficient vector
associated with the index pair $(l,m)$.

We define the maximal angular mode number as
\begin{equation} \label{eq.m_max}
    m_\mathrm{max} \coloneqq \lfloor N_\theta/2 \rfloor .
\end{equation}
Accordingly, all harmonic functions~$h_m$ with $|m| \le m_\mathrm{max}$ can be represented in the angular finite‑element space in the Galerkin
$L^2$ sense (see Sec.~\ref{Sec.angular_basis}).
For even $N_\theta$, the mode $m = m_\mathrm{max}$ corresponds to the Nyquist frequency and is representable only in its cosine form
$h_{m_\mathrm{max}} = \cos(m_\mathrm{max}\theta)$; the corresponding sine mode
$h_{-m_\mathrm{max}}$ is absent.
Moreover, we require that the harmonics~$h_m(\theta)$ with $|m| \le p$ be resolved, which is ensured by choosing at least
\begin{equation} \label{eq.theta_condition_p}
    N_\theta \ge N_{\theta,\mathrm{min}} = 2p + 1
\end{equation}
angular B-splines (see Eq.~\eqref{eq.theta_condition}). 
Thus, we consider the truncated index set
\begin{equation}
    \mathcal{I}_{\mathrm{c}} \coloneqq \{(l,m)\mid 0 \le l \le p,\ |m| \le m_{\max} \},
\end{equation}
where, for even $N_\theta$, the sine mode at $|m|=m_{\max}$ is excluded.
In particular, for all $N_\theta$ we have $|\mathcal{I}_{\mathrm c}| = (p+1)N_\theta = N_{\mathrm c}$.

Each function $S_l^m$ factors into a radial and an angular part 
(see Eq.~\eqref{eq.gen_harmon_ansatz_l}), and the tensor-product B-spline basis 
$\boldsymbol{B}_{\mathrm{c}}$ shares the same separable structure 
(see Eq.~\eqref{eq.tensor_Bspline}). Since the weight $r$ in the polar $L^2$-inner
product depends only on the radial coordinate (see Eq.~\eqref{eq.Galerkin_c}), the associated mass matrix preserves this tensor-product structure. Thus, the Galerkin $L^2$-projection
problem (see Eq.~\eqref{eq.Galerkin_op_full}) factorizes exactly, in the
tensor‑product sense, into independent one‑dimensional radial and angular
$L^2$-projection problems:
\begin{equation} \label{eq.galerkin_tensorproduct}
    \Pi_\mathrm{c} = \Pi_{r,\mathrm{c}} \otimes \Pi_{\theta,\mathrm{c}},
\end{equation}
where~$\Pi_{r,\mathrm{c}}$ and $\Pi_{\theta,\mathrm{c}}$ are defined by
Eqs.~\eqref{eq.Galerkin_op_r} and \eqref{eq.Galerkin_op_theta}.
The final coefficient vectors~$\boldsymbol{P}_l^m$ are then obtained as the
Kronecker products of the corresponding radial and angular coefficient vectors, 
as detailed in Sec.~\ref{sec.polsplinebasis}.

\subsection{Radial part of the center-splines} \label{Sec.radial_basis}

The radial Galerkin $L^2$-projection
\begin{equation} \label{eq.Galerkin_op_r}
    \Pi_{r,\mathrm{c}} : V_r([0,\Delta r]) \to V_{r,h}([0,\Delta r]),
        \qquad V_r([0,\Delta r]) \coloneqq L^2\bigl([0,\Delta r],\, r\,\mathrm{d} r\bigr),
\end{equation}
maps a function $f \in V_r([0,\Delta r])$ to its discrete representative
$\Pi_{r,\mathrm{c}}[f] \in V_{r,h}([0,\Delta r])$,
characterized uniquely by the orthogonality condition
\begin{equation} \label{eq.FEM_r}
\int_0^{\Delta r} \big( f(r) - \Pi_{r,\mathrm{c}} [f](r) \big)\,
B_{r,i}(r) \, r \, \mathrm{d}r = 0,
\quad \forall \, B_{r,i} \in V_{r,h}([0,\Delta r]).
\end{equation}

In particular, for the radial monomials $R_l(r)=r^l$ with $0 \le l \le p$,
we denote their Galerkin projections by
$
\widetilde{R}_l(r) \coloneqq \Pi_{r,\mathrm{c}}[R_l] \in V_{r,h}([0,\Delta r]).
$
The discrete functions $\widetilde{R}_l(r)$ are expressed in the radial
B-spline basis of degree~$p$ as
\begin{equation} \label{eq.tilde_B_r_l}
    \widetilde{R}_l(r) =
    \left( \boldsymbol{c}_{r,l} \right)^\mathrm{T}
    \boldsymbol{B}_{r,\mathrm{c}}(r),  \qquad 0 \le l \le p, 
\end{equation}
where $\boldsymbol{c}_{r,l}\in \mathbb{R}^{p+1}$ is the coefficient vector
associated with the index $l$ and $\boldsymbol{B}_{r,\mathrm{c}}$ denotes the $(p{+}1)$-component column vector of center radial B-spline basis functions, evaluated on their full support (see Sec.~\ref{Sec.radial_boundary}).

Since $r^l$ for $0 \le l \le p$ admits an exact representation in
$V_{r,h}([0,\Delta r])$ on $[0,\Delta r]$ (see App.~\ref{App.exact_poly_r}),
we have $\widetilde{R}_l(r) = r^l$ on $[0,\Delta r]$.
Thus, the center radial B-splines span the polynomial space
$\mathbb{P}_{r,p}([0,\Delta r])$, and any monomial $r^l$ with
$0 \le l \le p$ lies in $V_{r,h}([0,\Delta r])$:
\begin{equation}  \label{eq.Gal_r_invari}
\Pi_{r,\mathrm{c}}[r^l] = r^l, \qquad 0 \le l \le p .
\end{equation}
Linear independence of the radial monomials, together with their inclusion in
$V_{r,h}([0,\Delta r])$, implies exact preservation of the monomials so that no mixing between different polynomial degrees occurs in the Galerkin $L^2$-projection.

\begin{figure}[htbp]
\centering
\includegraphics[width=0.55\linewidth]{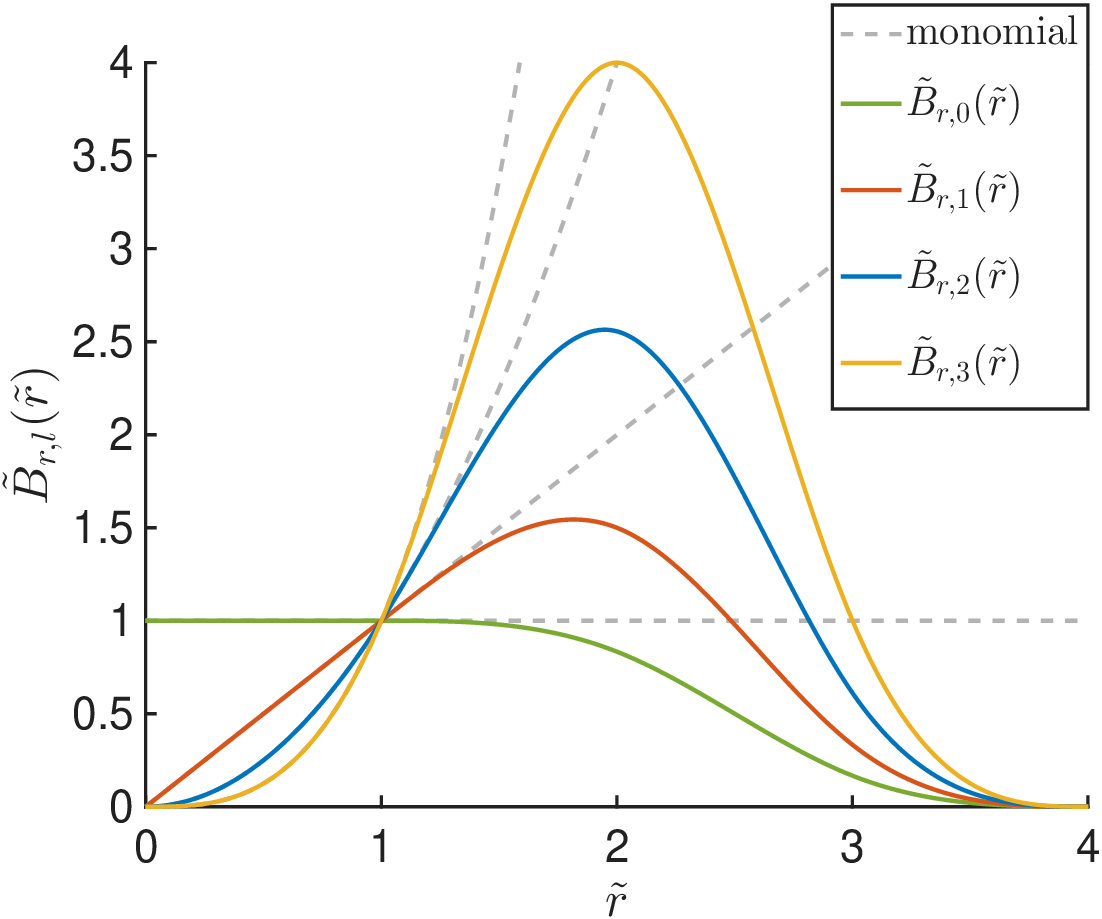}
\caption{Radial parts $\widetilde{R}_l$ of the center-spline basis functions (colored lines) expressed 
as linear combinations of cubic B-spline basis functions, reproducing the monomials in the innermost radial interval $\tilde{r} \in [0,1]$. 
Dashed lines indicate the monomials extended to $\tilde{r} \in [1,4]$.}
\label{fig.monomial_bspline}
\end{figure}

Evaluating the resulting B-spline combinations~\eqref{eq.tilde_B_r_l} on the full support of the $p{+}1$ innermost radial basis functions (e.g.\ $\tilde r \in [0,4]$ for cubic B-splines, where $\tilde{r} \coloneqq r / \Delta r$ is the normalized coordinate) defines the radial part of the center-splines (see Eq.~\eqref{m_eq}):
\begin{equation}\label{eq.radial_monomial_spline}
     \widetilde{\boldsymbol{B}}_{r,\mathrm{c}}(r) \coloneqq [\boldsymbol{c}_{r,0},\ldots,\boldsymbol{c}_{r,p}]^{\mathrm T} \boldsymbol{B}_{r,\mathrm{c}}(r).
\end{equation}
Outside the interval $\tilde{r} \in [0,1]$, the B-spline is generally not equal 
to the original monomial~$\tilde{r}^l$, and it retains the spline continuity $C^{p-1}$ 
at the knot (see Fig.~\ref{fig.monomial_bspline}).

\subsection{Angular part of the center-splines} \label{Sec.angular_basis}

The angular Galerkin $L^2$-projection
\begin{equation} \label{eq.Galerkin_op_theta}
    \Pi_{\theta,\mathrm{c}} : V_\theta \to V_{\theta,h}
        \qquad V_\theta \coloneqq L^2([0,2\uppi)),
\end{equation}
maps a function $f \in V_\theta$ to its discrete representative
$\Pi_{\theta,\mathrm{c}}[f] \in V_{\theta,h}$,
characterized uniquely by the orthogonality condition
\begin{equation} \label{eq.FEM_theta}
\langle  \left( f(\theta) - \Pi_{\theta,\mathrm{c}}[f](\theta) \right),
v_h \rangle_\theta = 0,
\quad \forall \, v_h \in V_{\theta,h}.
\end{equation}

In particular, for the harmonics~$h_m(\theta)$, we denote their
Galerkin projections by
$
\widetilde{h}_m(\theta) \coloneqq \Pi_{\theta,\mathrm{c}}[h_m]
\in V_{\theta,h}.
$
In practice, the Galerkin condition~\eqref{eq.FEM_theta} is enforced by choosing
$v_h = B_{\theta,j}(\theta)$ for $0\le j < N_\theta$ .
The discrete functions~$\widetilde{h}_m(\theta)$ are expressed (see App.~\ref{App.trigono_approx}) in the angular B-spline basis as
\begin{equation}\label{eq.tilde_B_theta_m}
    \widetilde{h}_m(\theta) = \left( \boldsymbol{c}_{\theta,m} \right)^\mathrm{T} \boldsymbol{B}_\theta(\theta), \qquad m \in \mathcal{I}_{\theta,\mathrm{c}},
\end{equation}
where $\boldsymbol{c}_{\theta,m} \in \mathbb{R}^{N_\theta}$
is the angular coefficient vector of the $m$-th harmonic, which can be
computed analytically (see App.~\ref{App.angularmodecoeff}), and
$\boldsymbol{B}_\theta(\theta)$ denotes the periodic angular B-spline basis
of degree~$p$, defined over $N_\theta$ uniformly spaced angular intervals
covering $[0,2\uppi)$ (see Section~\ref{Sec.1d_basis_angular}).
Here, we consider the truncated index set
\begin{equation}
     \mathcal{I}_{\theta,\mathrm{c}} \coloneqq \{ m \mid  |m| \le m_{\max} \},
\end{equation}
where $m_{\max}$ is defined in Eq.~\eqref{eq.m_max}, and again with the understanding that, for even $N_\theta$, the Nyquist sine mode is omitted.
In particular, $|\mathcal{I}_{\theta,\mathrm{c}}| = N_\theta$, ensuring that each
continuous harmonic mode~$h_m$ with $m \in \mathcal{I}_{\theta,\mathrm{c}}$ is
represented without aliasing in coefficient space, i.e.\ without mixing of
distinct harmonic functions~$h_m$ in the discrete periodic B-spline representation
(see App.~\ref{App.trigono_approx}).

Since $\widetilde{h}_m, \widetilde{h}_{m'} \in V_{\theta,h}$, their continuous
$L^2$-inner product is exactly represented by the $M_\theta$-inner product in
coefficient space (see Eq.~\eqref{eq.Mscalar_prod}). Owing to the circulant structure of $\mathsfbi{M}_\theta$ given by Eq.~\eqref{eq.M_theta}, assuming exact quadrature in its assembly, and the orthogonality relation
\eqref{eq:Mtheta_ortho_general} for the complex coefficient vectors
$\overline{\boldsymbol{c}}_{\theta,m}$ for Fourier modes, and upon restricting to the truncated
index set $\mathcal I_{\mathrm{c},\theta}$, we obtain
\begin{equation} \label{eq.h_tilde_ortho}
\langle \widetilde{h}_m, \widetilde{h}_{m'} \rangle_\theta
\propto \delta_{m,m'}, \qquad
m,m' \in \mathcal{I}_{\theta,\mathrm{c}}.
\end{equation}
By choosing the test function $v_h = \widetilde{h}_{m'} \in V_{\theta,h}$ in the Galerkin
orthogonality condition~\eqref{eq.FEM_theta}, we obtain the identity
\begin{equation} \label{eq.m_mixing}
\langle h_m, \widetilde{h}_{m'} \rangle_\theta
=
\langle \widetilde{h}_m, \widetilde{h}_{m'} \rangle_\theta,
\qquad m,m' \in \mathcal{I}_{\theta,\mathrm{c}}.
\end{equation}
The right-hand side can be interpreted as the matrix representation of the
Galerkin $L^2$-projection~$\Pi_{\theta,\mathrm{c}}$ in the discrete harmonic
basis; this matrix is diagonal on the truncated index set
$\mathcal I_{\mathrm c,\theta}$ (see Eq.~\eqref{eq.h_tilde_ortho}). As a result, we obtain the bi‑orthogonality between the continuous harmonics $h_m$ and their Galerkin projections $\widetilde{h}_{m}$,
\begin{equation} \label{eq.no_m_mixing}
\langle h_m, \widetilde{h}_{m'} \rangle_\theta
\propto \delta_{m,m'},
\qquad m,m' \in \mathcal{I}_{\theta,\mathrm{c}}.
\end{equation}
Thus, the action of the Galerkin $L^2$-projection~$\Pi_{\theta,\mathrm{c}}$ is
mode‑wise decoupled on $\mathcal I_{\mathrm c,\theta}$, and the harmonic index
$m$ is preserved under the angular Galerkin mapping from the continuous
harmonics $h_m$ to their discrete representatives $\widetilde{h}_m$.

On the truncated index set $\mathcal{I}_{\theta,\mathrm{c}}$, the continuous
harmonics $h_m$  (see Eq.~\eqref{eq.h_orthogonality}) and their Galerkin projections $\widetilde{h}_m$ (see Eq.~\eqref{eq.h_tilde_ortho}) form orthogonal
families in $L^2$, with Galerkin orthogonality inducing a bi‑orthogonality relation
between the two (see Eq.~\eqref{eq.no_m_mixing}). Moreover, Galerkin orthogonality implies a bi‑orthogonality relation (see Eq.~\eqref{eq.no_m_mixing}) between these two families.
As a result, harmonics~$h_m$ corresponding to distinct indices~$m$
cannot couple algebraically under the angular Galerkin $L^2$-projection into the discrete approximation, even at finite angular resolution.
Although a finite angular resolution limits the pointwise accuracy of the angular
approximation (leading to an $L^2$-approximation error of order $\mathcal{O}((\Delta \theta)^{p+1})$ for each fixed harmonic $h_m$), the mode‑wise decoupling of the Galerkin projection ensures that
the discrete solution remains compatible with the regularity requirements at the
polar origin.

When we restrict the Galerkin projection~\eqref{eq.Galerkin_op_theta} to
\begin{equation}
    \Pi_{\mathcal{I}_{\theta,\mathrm{c}}}: V_{\mathcal{I}_{\theta,\mathrm c}} \to
    V_{\theta,h},
\end{equation}
where
$
V_{\mathcal{I}_{\theta,\mathrm{c}}}
\coloneqq \operatorname{span}\{ h_m \mid m\in \mathcal{I}_{\theta,\mathrm{c}} \},
$
the projector maps each one‑dimensional modal subspace
$V_\theta^m=\operatorname{span}\{h_m\}$
into the corresponding discrete subspace
$V_{\theta,h}^m=\operatorname{span}\{\widetilde h_m\}$.
It therefore admits the algebraic direct‑sum representation
\begin{equation} \label{eq.no_Galerkin_coupling}
  \Pi_{\mathcal{I}_{\theta,\mathrm{c}}}
  =
  \bigoplus_{m \in \mathcal{I}_{\theta,\mathrm{c}}} \Pi_{\theta,\mathrm{c}}^m,
  \qquad
  \Pi_{\theta,\mathrm{c}}^m : V_\theta^m \to V_{\theta,h}^m.
\end{equation}
The direct sum is taken over mutually
$L^2$-orthogonal modal subspaces $V_\theta^m$,
yielding an orthogonal decomposition of the Galerkin projector
$\Pi_{\mathcal{I}_{\theta,\mathrm{c}}}$ with respect to the modal decomposition of
the space $V_{\mathcal{I}_{\theta,\mathrm{c}}}$.

An important special case is given by the constant harmonic $h_0 = 1$, which
belongs exactly to the angular B‑spline space due to the
partition‑of‑unity property (see Eq.~\eqref{eq.vecofones}). Consequently,
\begin{equation} \label{eq.h_theta0_invari}
    \Pi_{\theta,\mathrm{c}}(h_0) = h_0.
\end{equation}

\subsection{Smooth polar-spline basis via tensor-product B-splines}
\label{sec.polsplinebasis}

Using Eqs.~\eqref{eq.tilde_B_r_l} and~\eqref{eq.tilde_B_theta_m}, we construct each center-spline basis function as the tensor product of its radial and angular components (see Eq.~\eqref{eq.2d_B_lk_lincomb}):
\begin{equation} \label{eq.B_lk_spline}
     \widetilde{B}_l^m(r,\theta) = \widetilde{R}_l(r) \, \widetilde{h}_m(\theta) 
     = \big(\boldsymbol{P}_l^m\big)^\mathrm{T} \boldsymbol{B}_\mathrm{c}(r,\theta),
     \quad (l,m) \in \mathcal{I}_{\mathrm{c}}.
\end{equation}
In the final smooth center‑spline basis defined in Eq.~\eqref{eq.smooth_center_basis}, the index pairs $(l,m)$ will be further restricted due to regularity constraints at the origin. The coefficient vector
\begin{equation}
    \boldsymbol{P}_l^m \coloneqq \boldsymbol{c}_{r,l} \otimes \boldsymbol{c}_{\theta,m}
\end{equation}
is given by the Kronecker product of the radial and angular coefficient vectors.  
This representation expresses the basis function $\widetilde{B}_l^m$ in terms of the center tensor-product B-spline basis
\begin{equation}
    \boldsymbol{B}_\mathrm{c}(r,\theta) \coloneqq \boldsymbol{B}_{r,\mathrm{c}}(r) \otimes \boldsymbol{B}_\theta(\theta).
\end{equation}

An important special case is given by the set of axisymmetric basis functions
$S_l^0(r)$ for $0 \le l \le p$, which, due to the exact preservation of radial monomials~$r^l$ (see Eq.~\eqref{eq.Gal_r_invari}) and of the constant harmonic~$h_0$ (see Eq.~\eqref{eq.h_theta0_invari}), are exactly reproduced by the center-splines
$\widetilde{B}_l^0(r)$ on $[0,\Delta r]$:
\begin{equation}
    \Pi_\mathrm{c}(S_l^0) = S_l^0, \qquad 0 \le l \le p .
\end{equation}
The invariance of the lowest‑order axisymmetric mode $S_0^0$ under the projection $\Pi_\mathrm{c}$ implies exact
$C^0$‑regularity at the origin, with a single $\theta$‑independent degree of
freedom retained there (see Sec.~\ref{Sec.Cn_regularity}).

As a consequence of the exact preservation of radial monomials under the
radial Galerkin projection (see Eq.~\eqref{eq.Gal_r_invari}) and the mode-wise orthogonal decomposition of the
angular Galerkin projector (see Eq.~\eqref{eq.no_Galerkin_coupling}), together with
the tensor-product structure of both the separable modes
$
S_l^m(r,\theta) = r^l h_m(\theta)
$
and the center B-spline basis, and the exact factorization
$
\Pi_{\mathrm c} = \Pi_{r,\mathrm c} \otimes \Pi_{\theta,\mathrm c}
$ (see Eq.~\eqref{eq.galerkin_tensorproduct}),
the Galerkin projection preserves the ordered modal indices $(l,m)$. In
particular, for each $(l,m)$ the one-dimensional space
$\operatorname{span}\{ S_l^m \}$ is mapped onto the corresponding discrete
subspace $\operatorname{span}\{ \widetilde B_l^m \}$, and no algebraic coupling occurs between modes with distinct radial or angular indices.

The algebraic modal decoupling (not necessarily orthogonal) permits a corresponding algebraic decomposition
\begin{equation}
V_{h}(\widehat{\Omega}_\mathrm{c}) = \widetilde{V}_h(\widehat{\Omega}_\mathrm{c})
\oplus W_h(\widehat{\Omega}_\mathrm{c}),
\end{equation}
where $\widetilde{V}_h(\widehat{\Omega}_\mathrm{c})
\coloneqq
\operatorname{span}\{ \widetilde B_l^m \mid (l,m)\in \widetilde{\mathcal I}_{\mathrm c}\}$ denotes the subspace compatible
with $C^\infty$-regularity at the origin in the continuous angular limit, and
$W_h(\widehat{\Omega}_\mathrm{c})$ contains the remaining irregular modes.
The subspace $\widetilde{V}_h(\widehat{\Omega}_\mathrm{c})$ is selected a posteriori by the admissible index pairs $(l,m) \in \widetilde{\mathcal{I}}_{\mathrm{c}}$ (see Eq.~\eqref{eq.Slm_smoothness}), where
\begin{equation}
    \widetilde{\mathcal{I}}_{\mathrm{c}}
    \coloneqq
    \{(l,m)\mid 0 \le l \le p,\ |m| \le l,\ |m| \equiv l \ (\mathrm{mod}\ 2)\},
    \qquad \widetilde{\mathcal{I}}_\mathrm{c} \subseteq \mathcal{I}_\mathrm{c}.
\end{equation}
 
We assemble the smooth center-spline basis functions $\widetilde{\boldsymbol{B}}_\mathrm{c}$ spanning the subspace~$\widetilde{V}_h(\widehat{\Omega}_\mathrm{c})$
(see Eq.~\eqref{eq.B_lk_spline}) by collecting all $\widetilde{B}_l^m$ with $(l,m) \in \widetilde{\mathcal{I}}_{\mathrm{c}}$ into a single column vector :
\begin{equation} \label{eq.smooth_center_basis}
    \widetilde{\boldsymbol{B}}_\mathrm{c}(r,\theta) = \mathsfbi{P}_\mathrm{c}^\mathrm{T} \boldsymbol{B}_\mathrm{c}(r,\theta),
\end{equation}
with
\begin{align}
    \widetilde{\boldsymbol{B}}_\mathrm{c}(r,\theta) & \coloneqq [\widetilde{B}_l^m(r,\theta)], \\
    \label{eq.P_lm_matrix}
    \mathsfbi{P}_\mathrm{c} & \coloneqq \left[\boldsymbol{P}_l^m \right],
     \qquad \quad (l,m) \in \widetilde{\mathcal{I}}_{\mathrm{c}}.
\end{align}
Here, $[\widetilde{B}_l^m(r,\theta)]$ denotes the column vector whose entries are the basis functions $\widetilde{B}_l^m$ ordered according to the admissible pairs $(l,m) \in \widetilde{\mathcal{I}}_{\mathrm{c}}$, and $[\boldsymbol{P}_l^m]$ denotes the matrix whose columns are the coefficient vectors $\boldsymbol{P}_l^m$.

An example of this indexing and arrangement is shown in Table~\ref{tab:tableau_Blm}, which lists all smooth center-spline basis functions $\widetilde{B}_l^m$ (shown in black) for the cubic B-spline case $p{=}3$.

\begin{table}[ht]
    \centering
    \begin{tabular}{|cccccccccc|}
         \hline
          \phantom{$l=0$} \vline & $m=-4$ & $m=-3$ & $m=-2$ & $m=-1$ & $m=0$ & $m=\phantom{-}1$ & $m=\phantom{-}2$ & $m=\phantom{-}3$ & $m=\phantom{-}4$ \\
         \hline
         \rule{0pt}{16pt}%
         $l=0$ \vline & \color{black!30}{$\widetilde{B}_0^{-4}$} & \color{black!30}{$\widetilde{B}_0^{-3}$} & \color{black!30}{$\widetilde{B}_0^{-2}$} & \color{black!30}{$\widetilde{B}_0^{-1}$} & $\widetilde{B}_0^0$ &\color{black!30}{$\widetilde{B}_0^1$} &\color{black!30}{$\widetilde{B}_0^2$} & \color{black!30}{$\widetilde{B}_0^3$} & \color{black!30}{$\widetilde{B}_0^4$} \\[6pt]
          $l=1$ \vline & \color{black!30}{$\widetilde{B}_1^{-4}$} & \color{black!30}{$\widetilde{B}_1^{-3}$} & \color{black!30}{$\widetilde{B}_1^{-2}$} & $\widetilde{B}_1^{-1}$ & \color{black!30}{$\widetilde{B}_1^0$}& $\widetilde{B}_1^1$ &\color{black!30}{$\widetilde{B}_1^2$} & \color{black!30}{$\widetilde{B}_1^3$} & \color{black!30}{$\widetilde{B}_1^4$} \\[6pt]
          $l=2$ \vline & \color{black!30}{$\widetilde{B}_2^{-4}$} & \color{black!30}{$\widetilde{B}_2^{-3}$} & $\widetilde{B}_2^{-2}$ & \color{black!30}{$\widetilde{B}_2^{-1}$} & $\widetilde{B}_2^0$ & \color{black!30}{$\widetilde{B}_2^1$} & $\widetilde{B}_2^2$ & \color{black!30}{$\widetilde{B}_2^3$} & \color{black!30}{$\widetilde{B}_2^4$} \\[6pt]
          $l=3$ \vline & \color{black!30}{$\widetilde{B}_3^{-4}$} & $\widetilde{B}_3^{-3}$ & \color{black!30}{$\widetilde{B}_3^{-2}$} & $\widetilde{B}_3^{-1}$ & \color{black!30}{$\widetilde{B}_3^0$} & $\widetilde{B}_3^1$ &\color{black!30}{$\widetilde{B}_3^2$} & $\widetilde{B}_3^3$ & \color{black!30}{$\widetilde{B}_3^4$} \\[6pt]
         \hline
    \end{tabular}
\caption{Tableau of the $N_{\mathrm{c}} = 36$ center‑spline basis functions
$\widetilde{B}_l^m$, with index pairs $(l,m) \in \mathcal{I}_\mathrm{c}$ (shown in gray and black), for the cubic B‑spline case $p=3$ using $N_\theta = 9$ angular B‑splines.
Among these, exactly $\widetilde{N}_{\mathrm{c}} = 10$ basis functions satisfy the smoothness constraints. The radial polynomial-degree index satisfies $0 \le l \le p$ (rows), and the harmonic index satisfies $|m| \le m_{\mathrm{max}}$ (columns).
The admissible index pairs $(l,m) \in \widetilde{\mathcal{I}}_\mathrm{c}$ (shown in black) yield smooth center‑spline basis functions that are compatible with $C^\infty$-regularity at the origin in the continuous angular limit.}
\label{tab:tableau_Blm}
\end{table}

For each admissible index pair $(l,m)\in \widetilde{\mathcal{I}}_{\mathrm{c}}$, the
$L^2$-projection of $S_l^m$ onto $V_h(\widehat{\Omega}_{\mathrm c})$ does not vanish.
Assuming that the associated coefficient matrix $\mathsfbi{P}_\mathrm{c}$ (see
Eq.~\eqref{eq.P_lm_matrix}) has full column rank, the set of basis functions
$\{\widetilde{B}_l^m\}$ for $(l,m) \in \widetilde{\mathcal{I}}_{\mathrm{c}}$ is linearly independent. Consequently, the Galerkin $L^2$‑projection~$\Pi_{\widetilde{\mathcal{I}}_{\mathrm c}}$, restricted to the
bivariate polynomial space~$\mathbb P_p$ expressed in the admissible harmonic polar basis
(see Eq.~\eqref{eq.bivar_monomials}), is injective and maps $V_{\widetilde{\mathcal{I}}_{\mathrm c}}(\widehat{\Omega}_\mathrm{c})$ bijectively onto
$\widetilde V_h(\widehat\Omega_c)$. The subspace $\widetilde{V}_h(\widehat{\Omega}_\mathrm{c})$ spanned by the basis functions $\widetilde{\boldsymbol{B}}_\mathrm{c}$ is therefore isomorphic to the bivariate polynomial space~$\mathbb P_p$
(see Eq.~\eqref{eq.deg_freedom}), i.e.\
\begin{equation} \label{eq.deg_freedom_Blm}
    \widetilde{N}_\mathrm{c} \coloneqq \dim \widetilde{V}_h(\widehat{\Omega}_\mathrm{c}) = \frac{(p+1)(p+2)}{2}.
\end{equation}

Finally, the smooth polar-spline basis is defined globally by
\begin{equation}\label{eq.Bspline_regularity}
    \widetilde{\boldsymbol{B}}(r,\theta) \ \coloneqq\  \mathsfbi{P}^{\mathrm T} \boldsymbol{B}(r,\theta),
\end{equation}
where $\mathsfbi{P}$ has the block‑diagonal form
\begin{equation}\label{eq.restriction_op}
    \mathsfbi{P} \coloneqq \left[
    \begin{array}{c|c}
       \mathsfbi{P}_\mathrm{c} & \boldsymbol{0} \\
        \hline
        \boldsymbol{0} & \mathsfbi{I}
    \end{array}
    \right],
\end{equation}
and $\mathsfbi{I}$ is the identity matrix. The matrix $\mathsfbi{P}^\mathrm{T}$ has full row rank and is therefore a surjective mapping from the coefficient space spanned by the basis functions $\boldsymbol{B}$ to the subspace spanned by $\widetilde{\boldsymbol{B}}$. Moreover, $\mathsfbi{P}$ has full column rank, making it an injective mapping from the coefficient space of $\widetilde{\boldsymbol{B}}$ to $\boldsymbol{B}$.

\subsection{Imposing $C^\infty$-regularity at the polar origin in the finite-element method} 
\label{Sec.projection}

\subsubsection{Subspace~$\widetilde{V}_h$ spanned by the smooth polar-spline basis}

Let $V_h$ be the finite-element space spanned by the tensor-product B-spline basis functions 
$\boldsymbol{B}(r,\theta)$ and let $\widetilde{V}_h$ be the subspace spanned by the smooth polar-spline basis functions $\widetilde{\boldsymbol{B}}(r,\theta)$, which satisfy the regularity condition at $r = 0$ ensuring smoothness with respect to Cartesian coordinates. The dimension of $\widetilde{V}_h$ is
\begin{equation}\label{eq.DoF_Zernike}
   \widetilde{N} \coloneqq \dim \widetilde{V}_h
   = N - N_\mathrm{c} + \widetilde{N}_\mathrm{c}
   = \left[ N_r - (p+1) \right] N_\theta
     + \frac{(p+1)(p+2)}{2},
\end{equation}
where $N$, $N_\mathrm{c}$, and $\widetilde{N}_\mathrm{c}$ are defined in Eqs.~\eqref{eq.DoF_original}, \eqref{eq.Vh_Omega_dim}, and \eqref{eq.deg_freedom_Blm}.

The construction of the smooth center-spline basis assumes that the harmonics~$h_m(\theta)$ be minimally resolved, which is ensured by choosing at least $N_{\theta,\mathrm{min}} = 2p + 1$ angular B-splines (see Eq.~\eqref{eq.theta_condition_p}). Under this assumption, the number of degrees of freedom associated with the innermost radial interval is reduced from $(p + 1) N_\theta$ to $(p + 1)(p + 2)/2$, so that one obtains $\widetilde{N} \le N$. In the purely radial case $m=0$ only, corresponding to $N_{\theta,\mathrm{min}} = 1$, no reduction occurs and $\widetilde{N} = N$. In all cases,
\begin{equation}
    \widetilde{V}_h \subseteq V_h ,
\end{equation}
because each smooth polar-spline basis function can be expressed exactly as a linear combination of tensor-product B-splines (see Sec.~\ref{sec.polsplinebasis}).

Furthermore, using Eq.~\eqref{eq.Bspline_regularity}, any function 
$\widetilde{\phi}_h \in \widetilde{V}_h$ can be expressed as
\begin{equation} \label{eq.2d_bspline_phi_tilde}
\widetilde{\phi}_h(r,\theta) = \widetilde{\boldsymbol{\phi}}^{\mathrm T} 
\widetilde{\boldsymbol{B}}(r,\theta) 
= \widetilde{\boldsymbol{\phi}}^{\mathrm T} \mathsfbi{P}^{\mathrm T} \boldsymbol{B}(r,\theta) 
= (\mathsfbi{P} \widetilde{\boldsymbol{\phi}})^{\mathrm T} \boldsymbol{B}(r,\theta) ,
\end{equation}
which shows that the coefficients of $\widetilde{\boldsymbol{\phi}}$ in the 
$\boldsymbol{B}(r,\theta)$ basis are naturally defined by
\begin{equation} \label{eq.tildephi_phi_trans}
    \boldsymbol{\phi}_{\mathrm{reg}} \coloneqq \mathsfbi{P} \widetilde{\boldsymbol{\phi}}.
\end{equation}

The mapping $\mathsfbi{P}:\mathbb{R}^{\widetilde{N}} \to \mathbb{R}^N$ defined by 
$\boldsymbol{\phi}_{\mathrm{reg}} = \mathsfbi{P} \widetilde{\boldsymbol{\phi}}$ is injective, and 
$\boldsymbol{\phi}_{\mathrm{reg}} \in \mathcal{R}(\mathsfbi{P}) \subset \mathbb{R}^N$, where 
$\mathcal{R}(\mathsfbi{P}) \coloneqq \{ \mathsfbi{P}\widetilde{\boldsymbol{\phi}} \;|\; 
\widetilde{\boldsymbol{\phi}} \in \mathbb{R}^{\widetilde{N}} \}$ denotes the range of $\mathsfbi{P}$. 
The coefficient vectors satisfy $\boldsymbol{\phi}, \boldsymbol{\phi}_{\mathrm{reg}} \in \mathbb{R}^N$ 
and $\widetilde{\boldsymbol{\phi}} \in \mathbb{R}^{\widetilde{N}}$. 

Finally, we write Eq.~\eqref{eq.2d_bspline_phi_tilde} more explicitly as
\begin{equation}\label{eq.phiapprox_Cp}
    \widetilde{\phi}_h(r,\theta) =
    \sum_{l=0}^{p} 
    \sum_{\substack{m = -l \\ |m| \equiv l \ (\mathrm{mod}\ 2)}}^{l}
    \widetilde{b}_l^m \widetilde{B}_{l}^m(r,\theta)
    + \sum_{i=p+1}^{N_r-1} \sum_{j=0}^{N_\theta-1}
    \widetilde{\phi}_{i,j} \widetilde{B}_{i,j}(r,\theta) ,
\end{equation}
where the coefficient vector is
$
\widetilde{\boldsymbol{\phi}} =
\big[ \widetilde{b}_0^0, \widetilde{b}_1^{-1},\widetilde{b}_1^{1},\dots, \widetilde{b}_p^p, \dots, \widetilde{\phi}_{i,j}, \dots \big]^{\mathrm T},
$
and $p$ denotes the polynomial degree of the B-splines.

\subsubsection{Transfer of coefficient and load vectors}

When projecting a function $\phi_h(\boldsymbol{r}) \in V_h$ onto the subspace $\widetilde{V}_h$, 
we apply the Galerkin orthogonality condition (see Eq.~\eqref{eq.Galerkin_ortho})
\begin{equation} \label{eq.Galerkin_ortho_tilde}
    \int_\Omega \big( \phi_h(\boldsymbol{r}) - \widetilde{\phi}_h(\boldsymbol{r}) \big) 
    \widetilde{B}_k(\boldsymbol{r}) \, \mathrm{d} \boldsymbol{r} = 0,
    \qquad \forall \, \widetilde{B}_k \in \widetilde{V}_h .
\end{equation}
This yields the relation between the coefficient vectors
\begin{equation} \label{eq.phi_tildephi_trans}
   \widetilde{\mathsfbi{M}} \widetilde{\boldsymbol{\phi}} 
   = \mathsfbi{P}^{\mathrm T} \mathsfbi{M} \boldsymbol{\phi}
\quad \implies \quad
\widetilde{\boldsymbol{\phi}} = \widetilde{\mathsfbi{M}}^{-1} \mathsfbi{P}^{\mathrm T} \mathsfbi{M} \boldsymbol{\phi} ,
\end{equation}
where we have used Eqs.~\eqref{eq.mass_org} and~\eqref{eq.Bspline_regularity}. 
The associated mass matrix $\widetilde{\mathsfbi{M}}$ in $\widetilde{V}_h$ is defined by
\begin{equation} \label{eq.mass_matrix_org_reg}
    \widetilde{\matcomp{m}}_{k,k'} \coloneqq \int_\Omega \widetilde{B}_k(\boldsymbol{r}) 
    \widetilde{B}_{k'}(\boldsymbol{r}) \, \mathrm{d} \boldsymbol{r}, 
    \qquad (k,k' = 0, \ldots,\widetilde{N}-1) ,
\end{equation}
and is symmetric positive definite since the basis functions $\widetilde{B}_k$ are linearly independent. Inserting Eq.~\eqref{eq.Bspline_regularity} into Eq.~\eqref{eq.mass_matrix_org_reg} gives the relation between the mass matrix in the two spaces:
\begin{equation}\label{eq.mass_regularity_trans}
    \widetilde{\mathsfbi{M}} = \mathsfbi{P}^{\mathrm T} \mathsfbi{M}  \mathsfbi{P}.
\end{equation}

\begin{figure}[htbp]
\centering
\includegraphics[width=0.55\linewidth]{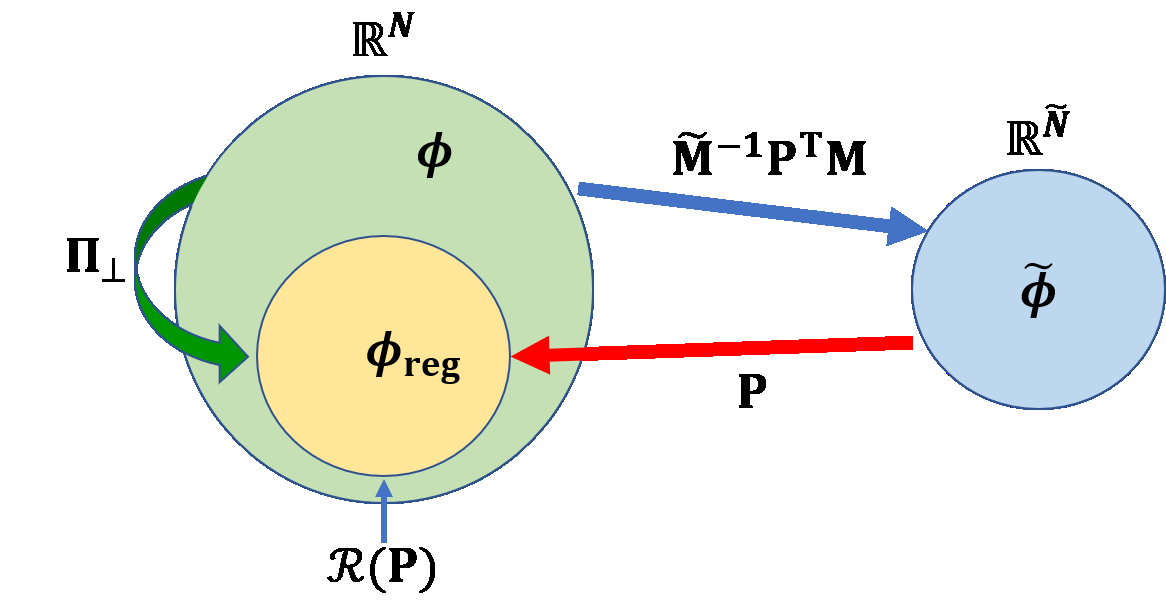}
\caption{
Coefficient spaces $\mathbb{R}^N$ of the tensor-product B-splines (green) and 
$\mathbb{R}^{\widetilde{N}}$ of the smooth polar splines (blue). 
The $C^\infty$-regular  subspace $\mathcal{R}(\mathsfbi{P}) \subset \mathbb{R}^N$ (yellow) 
contains coefficient vectors $\mathsfbi{\phi}_{\mathrm{reg}}$ that are 
$C^\infty$-regular at the polar origin.  
The $L^2$-orthogonal projection
$\widetilde{\mathsfbi{M}}^{-1} \mathsfbi{P}^{\mathrm T} \mathsfbi{M}$ (blue arrow) 
maps $\mathsfbi{\phi} \in \mathbb{R}^N$ onto 
$\widetilde{\mathsfbi{\phi}} \in \mathbb{R}^{\widetilde{N}}$ in the smooth polar-spline space.  
The prolongation $\mathsfbi{P}$ (red arrow) maps 
$\widetilde{\mathsfbi{\phi}}$ into 
$\mathsfbi{\phi}_{\mathrm{reg}} \in \mathcal{R}(\mathsfbi{P})$.  
The projector $\mathsfbi{\Pi}_\perp$ given by Eq.~\eqref{eq.Pi_perp_operator} (green arrow) maps any coefficient vector 
$\mathsfbi{\phi} \in \mathbb{R}^N$ to its $C^\infty$-regular (at the polar origin) part 
$\mathsfbi{\phi}_{\mathrm{reg}}$.
}
\label{fig.projection_space}
\end{figure}

Figure~\ref{fig.projection_space} illustrates the coefficient spaces of the tensor-product B-spline space $\mathbb{R}^N$ (green) and the smooth polar-spline space $\mathbb{R}^{\widetilde{N}}$ (blue). Coefficient vectors are transferred from $\mathbb{R}^N$ to $\mathbb{R}^{\widetilde{N}}$ (blue arrow) via the $L^2$-projection $\widetilde{\mathsfbi{M}}^{-1} \mathsfbi{P}^{\mathrm T} \mathsfbi{M}$ enforcing $C^\infty$-regularity at the origin (see Eq.~\eqref{eq.phi_tildephi_trans}), which projects out the modes that violate smoothness at the origin. In the opposite direction, vectors are transferred from $\mathbb{R}^{\widetilde{N}}$ to $\mathbb{R}^N$ (red arrow) by the prolongation (embedding) operator $\mathsfbi{P}$ (see Eq.~\eqref{eq.tildephi_phi_trans}), yielding $\mathcal{R}(\mathsfbi{P})$ (yellow), the subspace of all coefficient vectors that are $C^\infty$-regular at the polar origin.

When projecting a function $f(\boldsymbol{r}) \in L^2(\Omega)$
onto the subspace $\widetilde{V}_h$, we obtain, using the Galerkin
orthogonality condition, the linear system
\begin{equation} \label{eq.mass_tilde}
\widetilde{\mathsfbi{M}}\,\widetilde{\boldsymbol{\phi}}
= \widetilde{\boldsymbol{f}},
\end{equation}
where the mass matrix is given by Eq.~\eqref{eq.mass_matrix_org_reg} and
\begin{equation}
\label{eq.RHS_f_reg}
\tilde{f}_{k'}
\coloneqq \int_\Omega f(\boldsymbol{r}) \widetilde{B}_{k'}(\boldsymbol{r})
\, \mathrm{d}\boldsymbol{r},
\qquad k' = 0, \ldots, \widetilde{N}-1 .
\end{equation}
Inserting Eq.~\eqref{eq.Bspline_regularity} into Eq.~\eqref{eq.RHS_f_reg}, the corresponding transformation reads
\begin{equation} \label{eq.f_regularity_trans}
\widetilde{\boldsymbol{f}}
= \mathsfbi{P}^{\mathrm T} \boldsymbol{f},
\end{equation}
which shows that $\mathsfbi{P}^\mathrm{T}$ acts as the discrete restriction operator for the load vector~$\boldsymbol{f}$.

\subsubsection{Galerkin discretization of the differential operator in the smooth polar-spline space}

Instead of discretizing Eqs.~\eqref{eq.setofequations} in the tensor-product space, we apply the Galerkin discretization of the operator~$\mathcal{L}$ in the smooth polar-spline space, i.e.\ we use the basis functions~$\widetilde{\boldsymbol{B}}$. This results in 
\begin{equation} \label{eq.L_regularity}
    \widetilde{\mathsfbi{S}} \widetilde{\boldsymbol{\phi}} = \widetilde{\boldsymbol{f}},
\end{equation}
where
\begin{equation}
\label{eq.stiff_matrix_regularity}
    \widetilde{\matcomp{s}}_{k,k'} \coloneqq \int_\Omega \widetilde{B}_k(\boldsymbol{r}) \mathcal{L} \widetilde{B}_{k'}(\boldsymbol{r}) \, \mathrm{d} \boldsymbol{r}, \qquad (k,k' = 0, \ldots, \widetilde{N}-1).
\end{equation}
Inserting Eq.~\eqref{eq.Bspline_regularity} into Eq.~\eqref{eq.stiff_matrix_regularity} gives the relation between the stiffness matrix in the two spaces:
\begin{equation}\label{eq.stiff_regularity_trans}
    \widetilde{\mathsfbi{S}} = \mathsfbi{P}^{\mathrm T} \mathsfbi{S}  \mathsfbi{P}.
\end{equation}

\begin{figure}[htbp]
\centering
\includegraphics[width=0.35\linewidth]{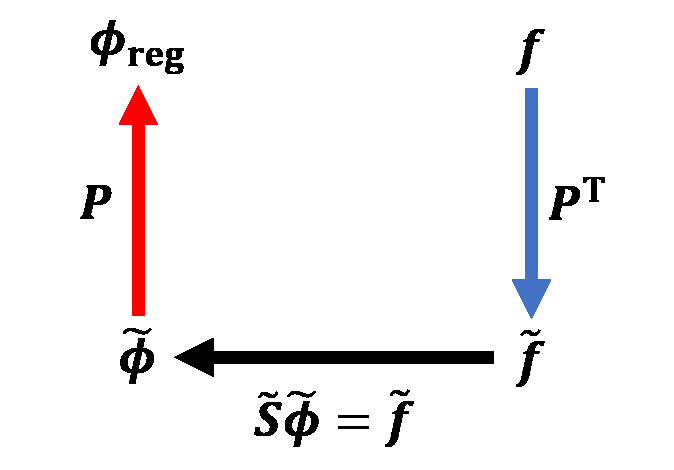}
\caption{The load vector $\boldsymbol{f}$ is calculated in the original tensor-product B-spline basis and then transformed to the smooth polar-spline basis, in which the matrix equation is solved. The resulting coefficient vector~$\widetilde{\boldsymbol{\phi}}$ is subsequently mapped back to the original basis.}
\label{fig.projection_equation}
\end{figure}

For practical reasons, working in the tensor-product space may be easier, since all basis functions share the same structure, which simplifies the implementation. Moreover, if one is working with legacy code, it can be advantageous to change only the linear system solver while keeping the rest of the code unchanged. Thus, we first calculate the stiffness matrix~$\mathsfbi{S}$ and the load vector~$\boldsymbol{f}$ in the original basis $\boldsymbol{B}$, transform them to the regularized basis~$\widetilde{\boldsymbol{B}}$ using Eqs.~\eqref{eq.stiff_regularity_trans} and \eqref{eq.f_regularity_trans}, and then solve for $\widetilde{\boldsymbol{\phi}}$ by Eq.~\eqref{eq.L_regularity}. Finally, the solution vector $\widetilde{\boldsymbol{\phi}}$ is mapped back to the original basis~$\boldsymbol{B}$ using the embedding given in Eq.~\eqref{eq.tildephi_phi_trans}. The scheme is visualized in Fig.~\ref{fig.projection_equation}.

\subsubsection{Construction of a $C^\infty$-regularity filter for the coefficient vector}
\label{sec:regularity_filter}

By concatenating the operators $\widetilde{\mathsfbi{M}}^{-1} \mathsfbi{P}^{\mathrm T} \mathsfbi{M}$ and $\mathsfbi{P}$ (see Eqs.~\eqref{eq.phi_tildephi_trans} and~\eqref{eq.tildephi_phi_trans}), we obtain
\begin{equation} \label{eq.projection}
    \boldsymbol{\phi}_\mathrm{reg} 
    = \mathsfbi{P} \widetilde{\boldsymbol{\phi}} 
    = \mathsfbi{P} \widetilde{\mathsfbi{M}}^{-1} \mathsfbi{P}^{\mathrm T} \mathsfbi{M} \boldsymbol{\phi} 
    = \mathsfbi{\Pi}_\perp \boldsymbol{\phi},
\end{equation}
where the projection operator is defined as
\begin{equation} \label{eq.Pi_perp_operator}
    \mathsfbi{\Pi}_\perp \ \coloneqq \ \mathsfbi{P} \widetilde{\mathsfbi{M}}^{-1} \mathsfbi{P}^{\mathrm T} \mathsfbi{M},
\end{equation}
which maps a coefficient vector $\boldsymbol{\phi}$ into the $C^\infty$-regular  subspace $\mathcal{R}(\mathsfbi{P})$ (see Fig.~\ref{fig.projection_space}).

To show that $\mathsfbi{\Pi}_\perp$ is an $M$-orthogonal projector, we first prove that it is idempotent, i.e.\ $\mathsfbi{\Pi}_\perp \mathsfbi{\Pi}_\perp = \mathsfbi{\Pi}_\perp$:
\begin{equation}
     \mathsfbi{\Pi}_\perp \mathsfbi{\Pi}_\perp 
     = \mathsfbi{P} \widetilde{\mathsfbi{M}}^{-1} \big(\mathsfbi{P}^{\mathrm T} \mathsfbi{M} \mathsfbi{P}\big) \widetilde{\mathsfbi{M}}^{-1} \mathsfbi{P}^{\mathrm T} \mathsfbi{M} 
     =  \mathsfbi{P} \widetilde{\mathsfbi{M}}^{-1} \widetilde{\mathsfbi{M}} \widetilde{\mathsfbi{M}}^{-1} \mathsfbi{P}^{\mathrm T} \mathsfbi{M} 
     =  \mathsfbi{\Pi}_\perp,
\end{equation}
where we have used Eq.~\eqref{eq.mass_regularity_trans}.

Next, we prove that it is $M$-symmetric, i.e.\ $\mathsfbi{\Pi}_\perp^\mathrm{T} \mathsfbi{M} = \mathsfbi{M} \mathsfbi{\Pi}_\perp$.  
Since both $\mathsfbi{M}$ and $\widetilde{\mathsfbi{M}}$ are symmetric, we have $(\widetilde{\mathsfbi{M}}^{-1})^\mathrm{T} = \widetilde{\mathsfbi{M}}^{-1}$ and $\mathsfbi{M}^\mathrm{T} = \mathsfbi{M}$:
\begin{equation}
\mathsfbi{\Pi}_\perp^\mathrm{T} 
= \big( \mathsfbi{P} \widetilde{\mathsfbi{M}}^{-1} \mathsfbi{P}^\mathrm{T} \mathsfbi{M} \big)^\mathrm{T}
= \mathsfbi{M} \mathsfbi{P} \widetilde{\mathsfbi{M}}^{-1} \mathsfbi{P}^\mathrm{T}.
\end{equation}
Multiplying by $\mathsfbi{M}$ on the right, we obtain:
\begin{equation}
\mathsfbi{\Pi}_\perp^\mathrm{T} \mathsfbi{M} 
= \mathsfbi{M} \mathsfbi{P} \widetilde{\mathsfbi{M}}^{-1} \mathsfbi{P}^\mathrm{T} \mathsfbi{M}
= \mathsfbi{M} \mathsfbi{\Pi}_\perp.
\end{equation}

The coefficient space $\mathbb{R}^N$ admits the $M$‑orthogonal decomposition
$\mathbb{R}^N = \mathcal{R}(\mathsfbi{P}) \oplus W_h$ with respect to the
$M$‑inner product (see Eq.~\eqref{eq.Mscalar_prod}),
where $\mathcal{R}(\mathsfbi{P})$ denotes the subspace of coefficient vectors
whose associated spline functions are $C^\infty$‑regular at the polar origin,
and $W_h$ is its $M$‑orthogonal complement. By definition,
$
W_h \ \coloneqq\ \{ \boldsymbol{w}_h \mid (\mathsfbi{M} \boldsymbol{\phi}_\mathrm{reg}) \bcdot \boldsymbol{w}_h = 0, \ 
\forall \, \boldsymbol{\phi}_\mathrm{reg} \in \mathcal{R}(\mathsfbi{P}) \},
$
i.e.\ every coefficient vector $\boldsymbol{w}_h \in W_h$ is $M$-orthogonal to every regular coefficient vector $\boldsymbol{\phi}_\mathrm{reg} \in \mathcal{R}(\mathsfbi{P})$.

The projector $\mathsfbi{\Pi}_\perp$ represents, in coefficient space, the
$L^2$-orthogonal projector from the finite-element space $V_h$ onto the
$C^\infty$-regular polar-spline subspace $\widetilde{V}_h$.
This subspace corresponds to $\mathcal{R}(\mathsfbi{P}) \subset \mathbb{R}^N$,
and orthogonality is understood with respect to the continuous $L^2$-inner
product on the unit disc, which is represented exactly in coefficient space
by the mass matrix $\mathsfbi{M}$.
Its kernel is the $L^2$‑orthogonal complement $W_h \subset V_h$.
In other words, $\mathsfbi{\Pi}_\perp$ maps any spline function to its
$C^\infty$‑regular component at the polar origin in the $L^2$ sense, removing
all modes that violate smoothness at the axis.
Consequently, as shown in Eq.~\eqref{eq.projection}, applying $\mathsfbi{\Pi}_\perp$
acts as a $C^\infty$‑regularity filter in both coefficient space and function space.

\subsubsection{Orthogonalization and normalization of the smooth center-splines}
\label{Sec.orthonormal_spline}

In Sec.~\ref{Sec.orthonormal}, we noted that an orthonormal basis has numerical advantages.
Taking orthonormalization into account (see App.~\ref{App.M_ortho_radial_coeff}), 
the radial parts of the orthonormal smooth center-splines are
\begin{equation}\label{eq.hat_tilde_B_r_l_m}
    \widehat{R}_{r,l}^m(r) = \left( \widehat{\boldsymbol{c}}_{r,l}^m \right)^\mathrm{T} \boldsymbol{B}_{r,\mathrm{c}}(r),
\end{equation}
where $\widehat{\boldsymbol{c}}_{r,l}^m$ denotes the $M_r$-orthonormalized coefficient vector of the radial part, i.e.\ orthonormal with respect to the
radial $L^2$-inner product defined in Eq.~\eqref{eq.innerprod_r}. Because the orthogonalization is performed separately for each fixed $m$-column, the orthonormalized coefficients $\widehat{\boldsymbol{c}}_{r,l}^m$ depend on both 
the row index $l$ and the column index $m$. The orthonormalization does not introduce coupling between different angular modes. In contrast, the original coefficients $\boldsymbol{c}_{r,l}$, defined in Eq.~\eqref{eq.tilde_B_r_l}, depend only on~$l$.

For cubic B-splines, Eq.~\eqref{eq.hat_tilde_B_r_l_m} becomes  
(for quadratic B-splines, see App.~\ref{App.quad_splines}):
\begin{equation}\label{eq.cubic_C3_radial_projection}
    \begin{bmatrix}
     \widehat{B}_{r,0}^0 \\[6pt] \widehat{B}_{r,2}^0 \\[6pt] \widehat{B}_{r,1}^1 \\[6pt] 
     \widehat{B}_{r,3}^1 \\[6pt] \widehat{B}_{r,2}^2 \\[6pt] \widehat{B}_{r,3}^3
    \end{bmatrix}
    =
   \mathrm{diag}\!\left(
   \begin{bmatrix}
   4\sqrt{\frac{21}{853}} \\[4pt] 
   4\sqrt{\frac{7}{8637878057}} \\[4pt]
   2\sqrt{\frac{70}{14431}} \\[4pt] 
   6\sqrt{\frac{210}{10052014067}} \\[4pt] 
   2\sqrt{\frac{42}{22277}} \\[4pt] 
   3\sqrt{\frac{35}{302}} 
    \end{bmatrix}
    \right)
    \begin{bmatrix}
       1  & 1 & 1 & 1 \\[6pt]
       -11029  &  -11029 & -7617 & 7737 \\[6pt]
       0  & 1 & 3 & 6 \\[6pt]
       0  & -17175 & -5725 & 2981 \\[6pt]
       0  & 0 & 2 & 11 \\[6pt]
       0  & 0 & 0 & 1
    \end{bmatrix}
    \begin{bmatrix}
     B_{r,0} \\[6pt] B_{r,1} \\[6pt] B_{r,2} \\[6pt] B_{r,3}
    \end{bmatrix}.
\end{equation}

Note that $\widehat{B}_{r,0}^0 = 4\sqrt{\frac{21}{853}} (B_{r,0} + B_{r,1} + B_{r,2} + B_{r,3})$ 
guarantees charge conservation (see Sec.~\ref{Sec.charge_cons_regularity_spline}).  
In addition, for the highest index~$p$ we have $\widehat{B}_{r,p}^p \propto B_{r,p}$.

\begin{figure}[htbp]
\centering
\includegraphics[width=0.55\linewidth]{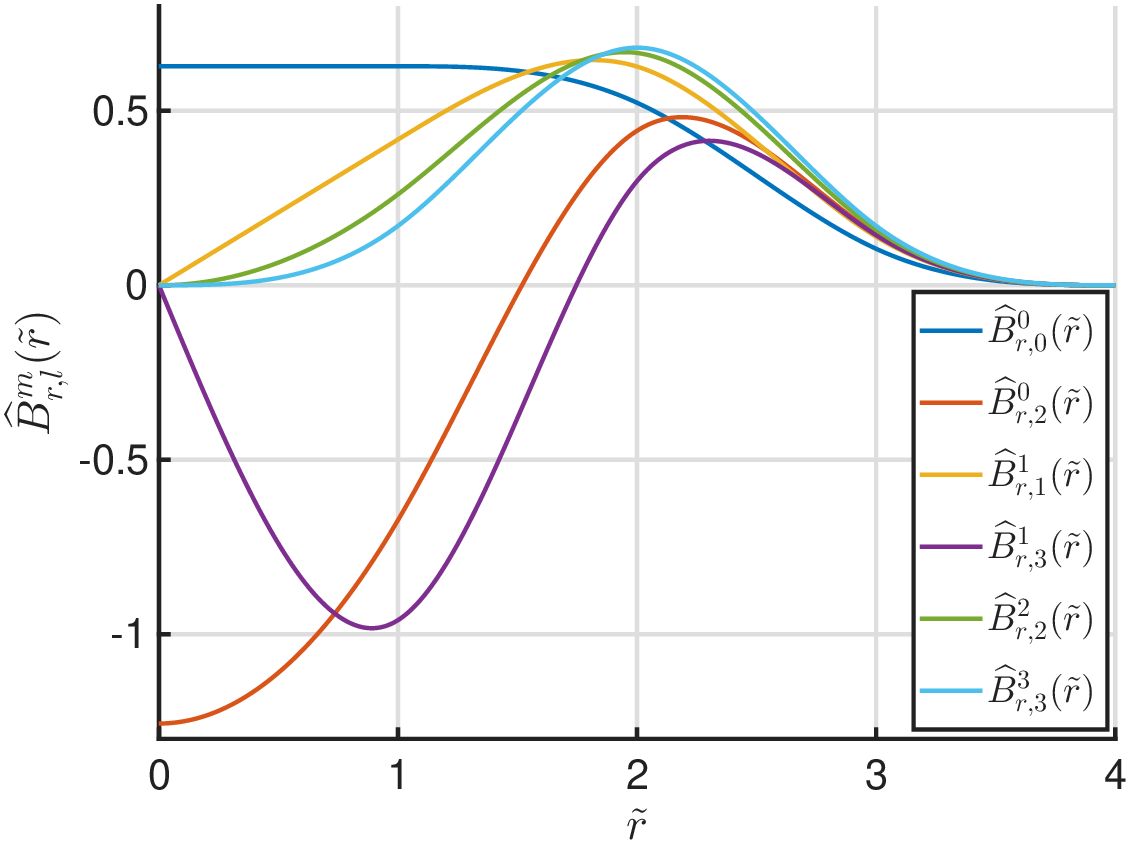}
\caption{Normalized radial basis functions~$\widehat{R}_{r,l}^m(r)$ for cubic B-splines. The functions $\widehat{B}_{r,0}^0$ and $\widehat{B}_{r,2}^0$, as well as
$\widehat{B}_{r,1}^1$ and $\widehat{B}_{r,3}^1$, are orthogonal with respect to the radial $L^2$-inner product (see Eq.~\eqref{eq.innerprod_r}).}
\label{fig.radial_orthogonal}
\end{figure}

The coefficient vectors~$\boldsymbol{c}_{\theta,m}$ of the angular part 
$\widetilde{h}_m$ are $M_\theta$-orthogonal to each other (see Eq.~\eqref{eq.h_tilde_ortho}).  
After normalization, the angular part becomes
\begin{equation}\label{eq.tilde_B_theta_m_norm}
    \widehat{h}_m(\theta) 
    = \left( \widehat{\boldsymbol{c}}_{\theta,m} \right)^\mathrm{T} \boldsymbol{B}_\theta(\theta),
\end{equation}
where the normalized coefficient vector is
\begin{equation}\label{eq.c_theta_m_norm}
    \widehat{\boldsymbol{c}}_{\theta,m} \coloneqq 
    \frac{\boldsymbol{c}_{\theta,m}}
    {\sqrt{(\boldsymbol{c}_{\theta,m})^\mathrm{T} \mathsfbi{M}_\theta \boldsymbol{c}_{\theta,m}}},
\end{equation}
with $\mathsfbi{M}_\theta$ given by Eq.~\eqref{eq.M_theta}, ensuring consistency 
with the $L^2$-metric in Sec.~\ref{Sec.orthonormal}.

The full $L^2$-orthonormal smooth center-spline basis functions~$\widehat{B}_l^m$ are the tensor product 
of the radial and angular parts (see Eqs.~\eqref{eq.hat_tilde_B_r_l_m} and~\eqref{eq.tilde_B_theta_m_norm}):
\begin{equation} \label{eq.B_lk_spline_norm}
     \widehat{B}_l^m(r,\theta) 
     = \widehat{R}_{r,l}^m(r) \, \widehat{h}_m(\theta) 
     = (\widehat{\boldsymbol{P}}_l^m)^\mathrm{T} \boldsymbol{B}_\mathrm{c}(r,\theta),
\end{equation}
where the normalized coefficient vector is given by
\begin{equation}
    \widehat{\boldsymbol{P}}_l^m \coloneqq \widehat{\boldsymbol{c}}_{r,l}^m \otimes \widehat{\boldsymbol{c}}_{\theta,m}.
\end{equation}

\begin{figure*}[ht]
\centering
    \subfloat[$\widehat{B}_0^0(\tilde{r},\theta)$]{\includegraphics[width = 0.4\linewidth]{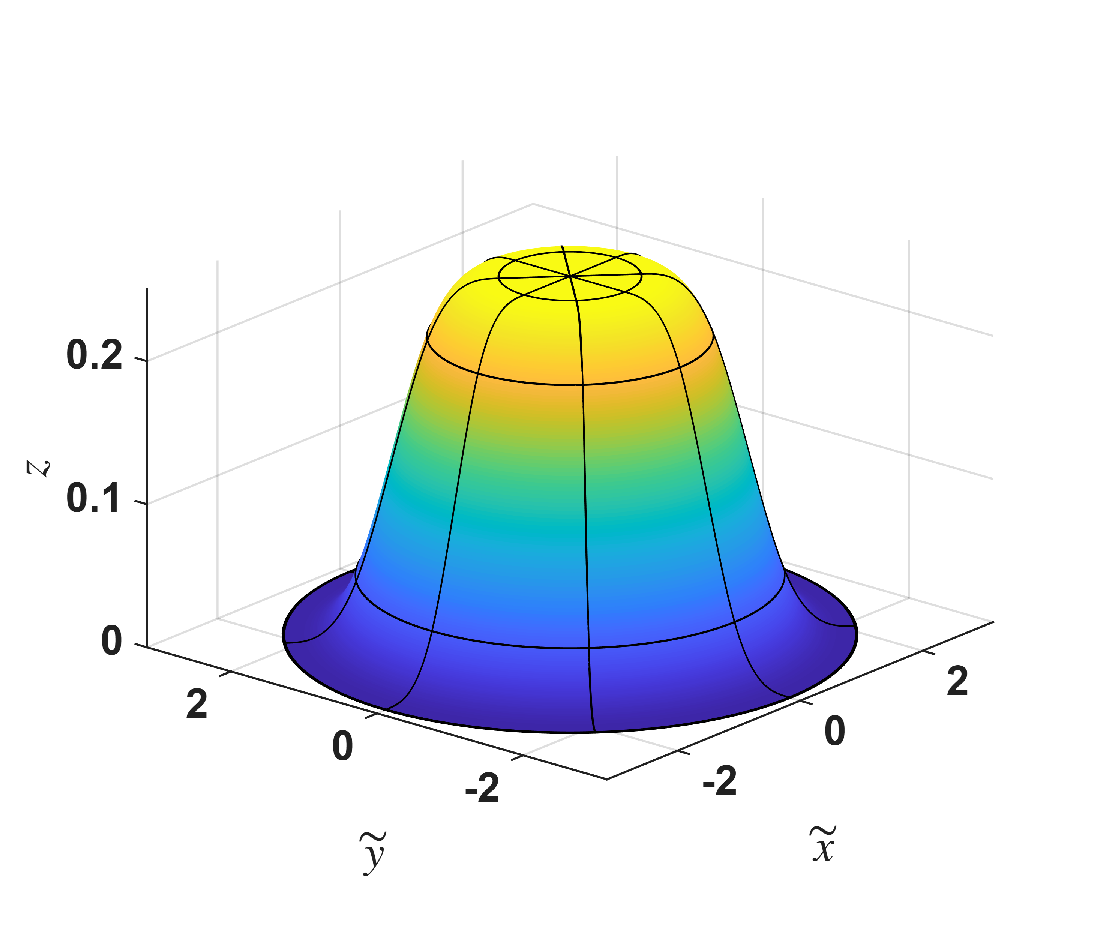}} \\
	\subfloat[$\widehat{B}_{1}^{-1}(\tilde{r},\theta)$]{\includegraphics[width = 0.4\linewidth]{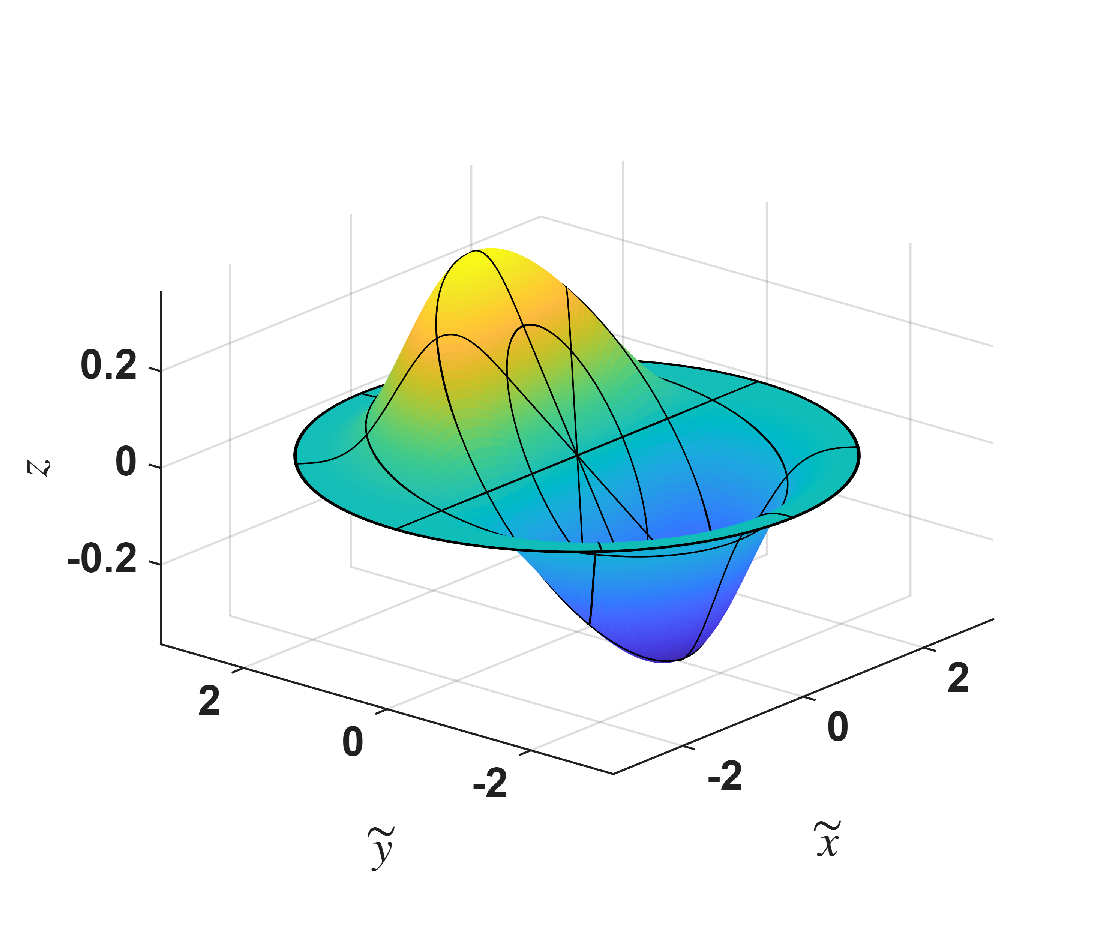}}
	\subfloat[$\widehat{B}_1^1(\tilde{r},\theta)$]{\includegraphics[width = 0.4\linewidth]{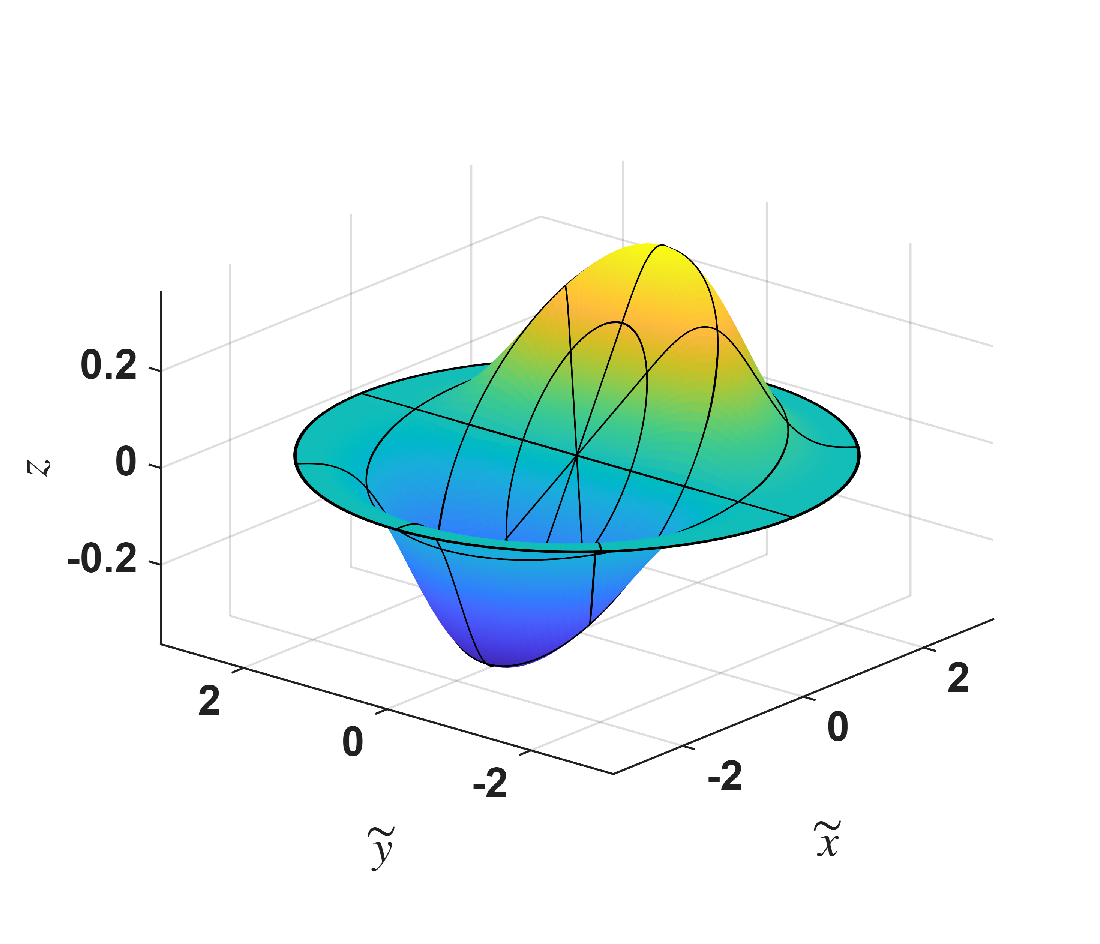}}
\caption{Examples of $\widehat{B}_0^0(\tilde{r},\theta)$, $\widehat{B}_1^{-1}(\tilde{r},\theta)$, 
and $\widehat{B}_1^1(\tilde{r},\theta)$, with the radial part given by cubic B-splines and the 
angular part by normalized harmonic functions~$\widehat{h}_m$. 
Closed curves at fixed $\tilde{r}$ indicate boundaries of the four radial intervals.}
\label{fig.cubic_basis_c3_1}
\end{figure*}

\begin{figure*}[ht]
\centering
	\subfloat[\rbasis{2}{-2}]{\includegraphics[width = 0.24\linewidth]{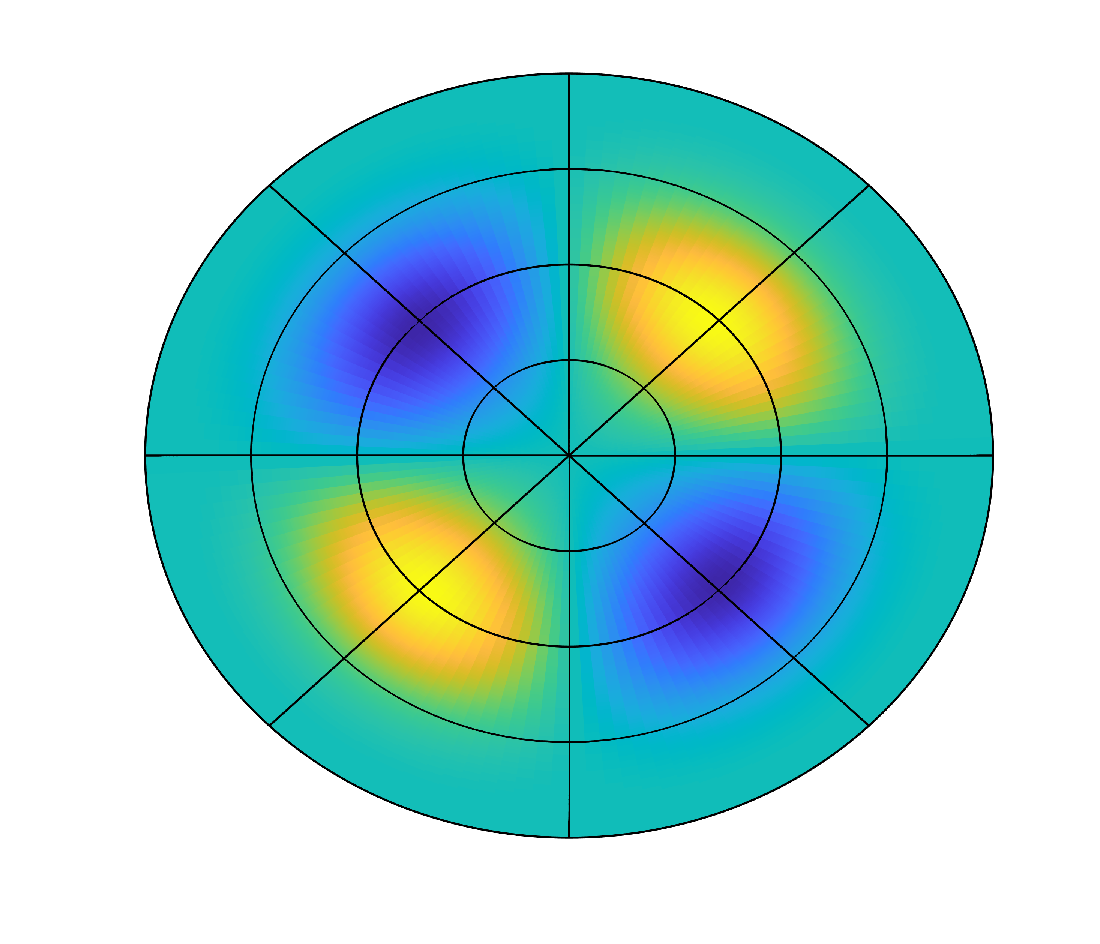}}
	\subfloat[\rbasis{2}{0}]{\includegraphics[width = 0.24\linewidth]{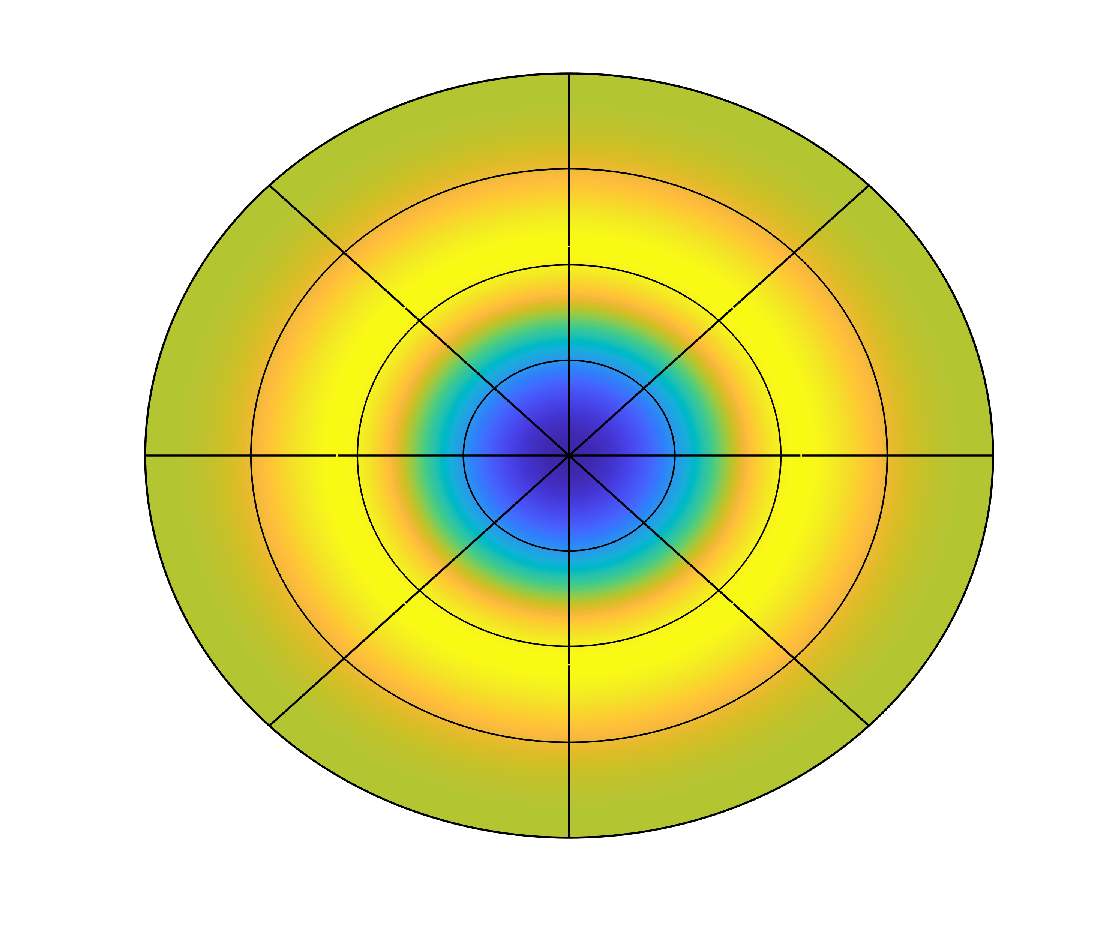}}
    \subfloat[\rbasis{2}{2}]{\includegraphics[width = 0.24\linewidth]{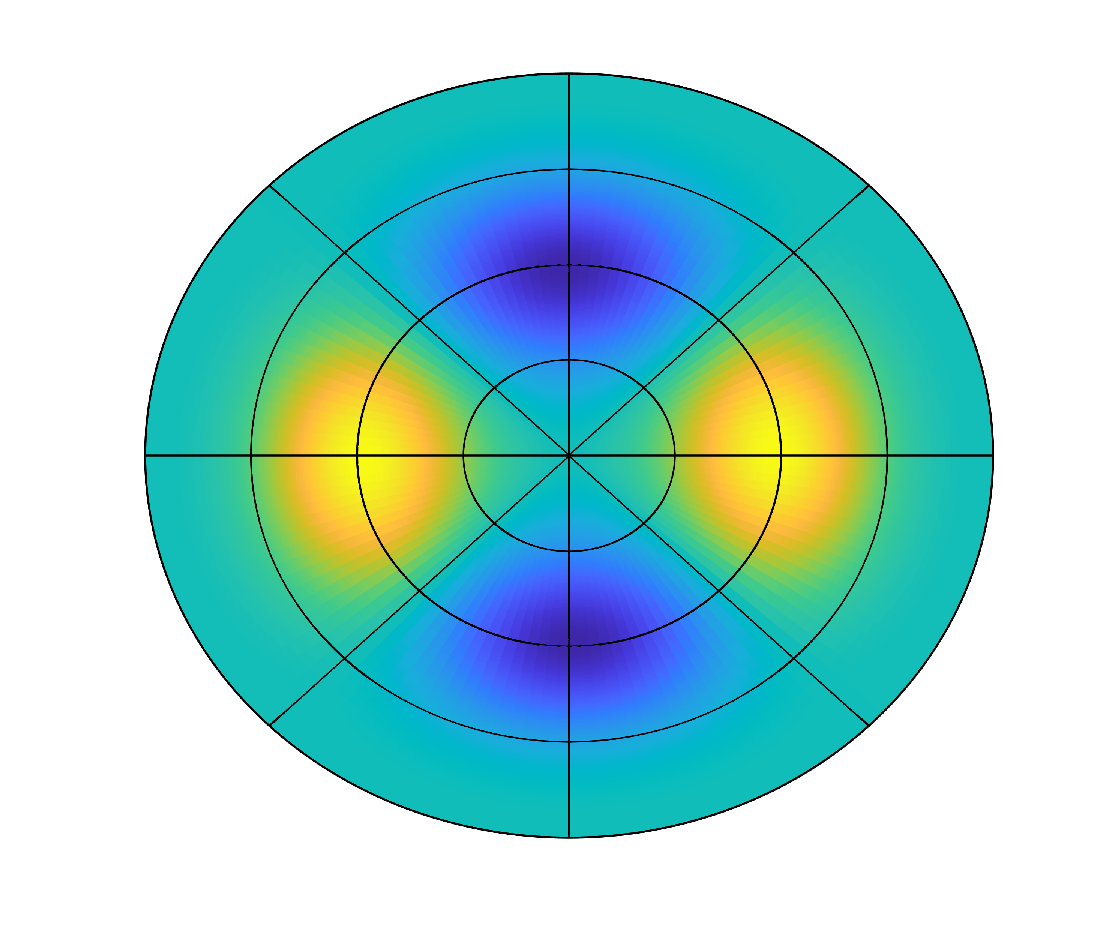}} \\
    \subfloat[\rbasis{3}{-3}]{\includegraphics[width = 0.24\linewidth]{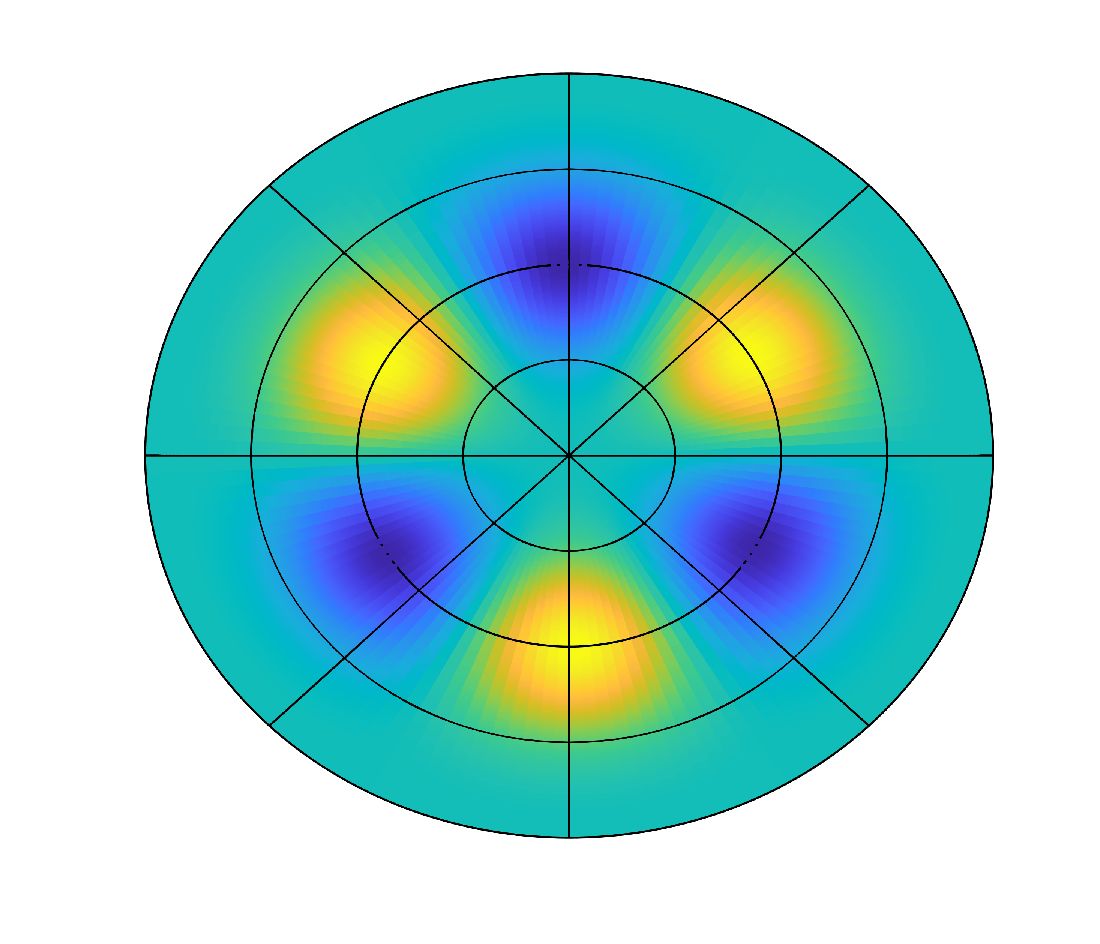}}
	\subfloat[\rbasis{3}{-1}]{\includegraphics[width = 0.24\linewidth]{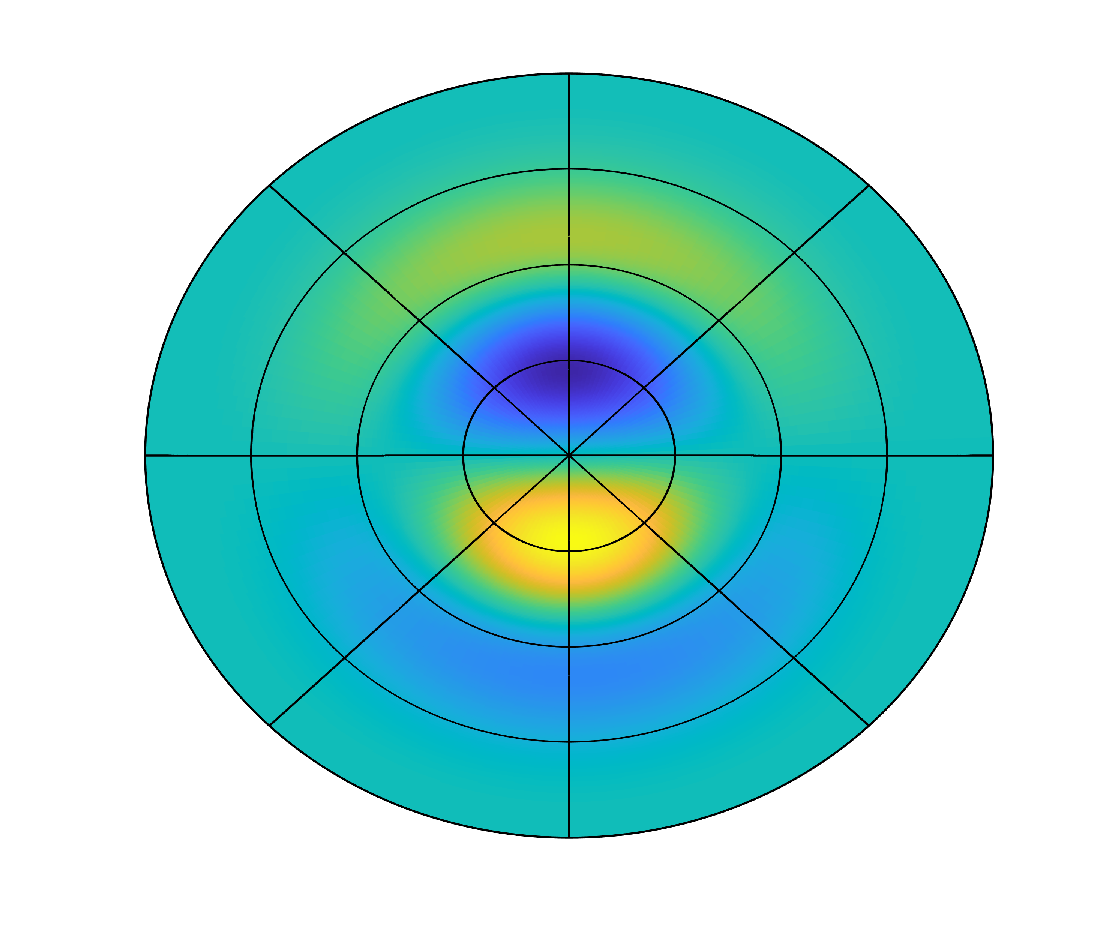}}
	\subfloat[\rbasis{3}{1}]{\includegraphics[width = 0.24\linewidth]{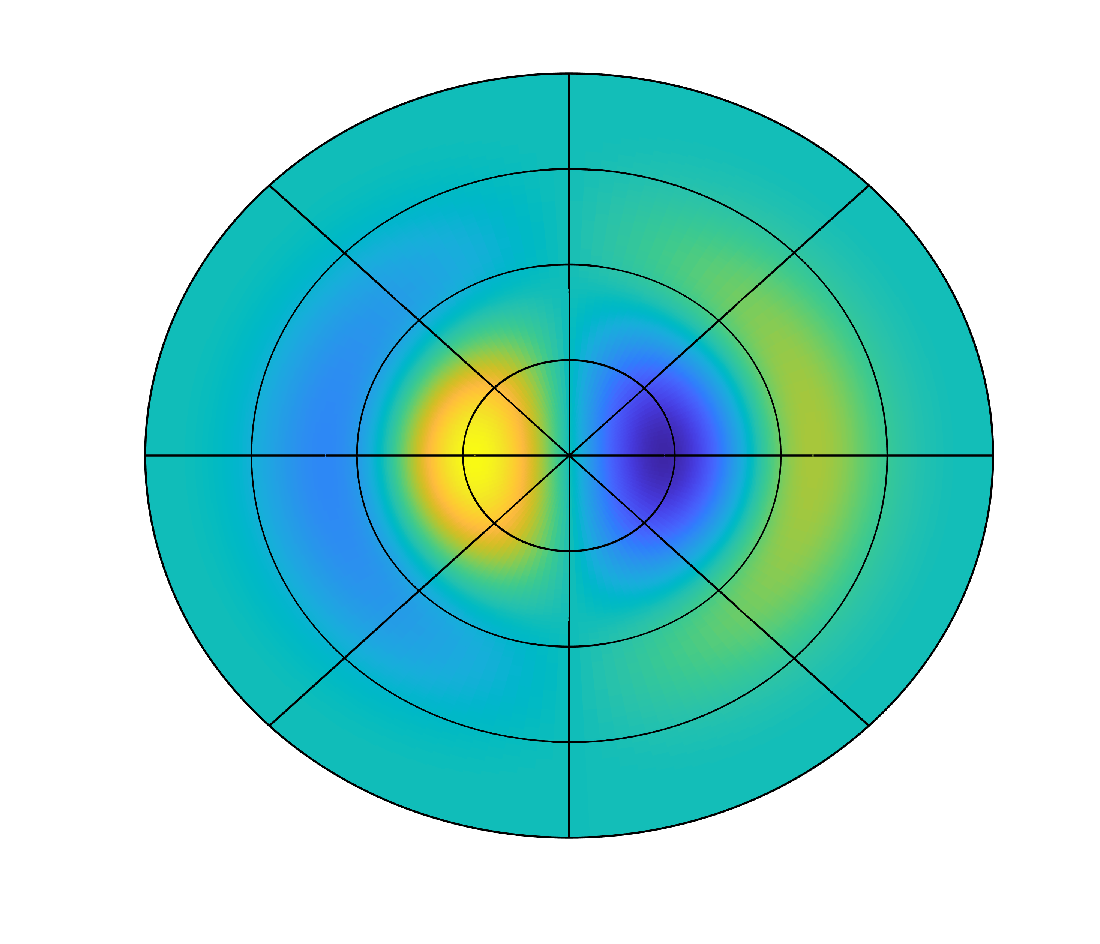}} 
    \subfloat[\rbasis{3}{3}]{\includegraphics[width = 0.24\linewidth]{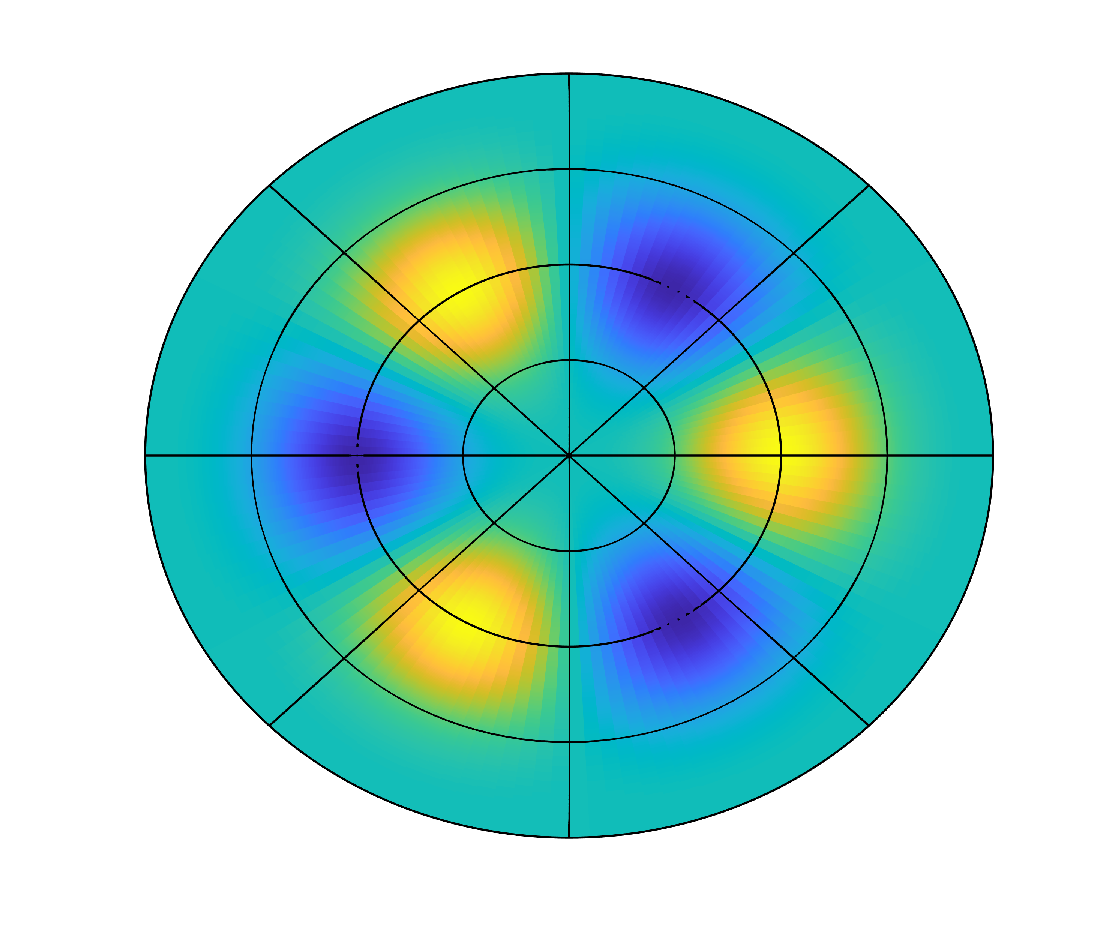}}
    \caption{Contour plots of $\widehat{B}_l^m(\tilde{r},\theta)$ for $0 \leq l \leq 3$, $-l \leq m \leq l$, 
    $|m| \equiv l \ (\mathrm{mod}\ 2)$. Radial part: cubic B-splines; angular part: normalized harmonic functions $\widehat{h}_m$. Circles at fixed $\tilde{r}$ mark boundaries of the four innermost radial intervals.}
\label{fig.cubic_basis_c3_2}
\end{figure*}

Figures~\ref{fig.cubic_basis_c3_1} and~\ref{fig.cubic_basis_c3_2} show the orthonormal smooth 
center-spline basis functions with $C^\infty$-regularity in the limit $N_\theta \to \infty$, using 
cubic B-splines in the radial part and normalized harmonic functions~$\widehat{h}_m(\theta)$ 
in the angular part. Quadratic B-spline representations for $C^2$-regularity are given in 
App.~\ref{App.quad_splines}.

\subsection{Imposing $C^n$-regularity ($n<p$) at the polar origin using smooth center-splines}
\label{Sec.Cn_regularity}

So far, we chose the maximum order of $C^\infty$-regularity. In this section, we consider a lower $C^n$-regularity $n<p$. To implement the $C^n$-regularity at the origin, only the radial parts of the smooth center-spline basis functions with leading monomial degree $l \le n$ are constrained. That is, in the construction step we retain only those radial monomial components $r^l$ with $l \le n$, and ignore terms of $\mathcal{O}(r^{n+1})$ for the purpose of enforcing the reduced regularity. Thus, for the underlying B-splines in the innermost interval, contributions of degree higher than $n$ can be omitted when enforcing the reduced regularity. These truncated Taylor expansions are used solely to identify which regularity constraints are enforced at the origin.

For example, if we want to achieve $C^2$-regularity for the cubic B-spline basis, we have for the innermost interval $\tilde{r} \in [0,1]$
\begin{subequations}
\label{eq.Taylor}
\begin{align}
    B_{r,0}(\tilde{r}) &= 1 - 3\tilde{r} + 3\tilde{r}^2 +  \mathcal{O}(\tilde{r}^3) \label{eq.cubic_C2_Taylor0} , \\
    B_{r,1}(\tilde{r}) &= 3\tilde{r} -\frac{9}{2} \tilde{r}^2+  \mathcal{O}(\tilde{r}^3) \label{eq.cubic_C2_Taylor1} , \\
    B_{r,2}(\tilde{r}) &= \frac{3}{2}\tilde{r}^2+  \mathcal{O}(\tilde{r}^3) \label{eq.cubic_C2_Taylor2}, \\
    B_{r,3}(\tilde{r}) &= \mathcal{O}(\tilde{r}^3) \label{eq.cubic_C2_Taylor3}.
\end{align}
\end{subequations}
Ignoring contributions of~$\mathcal{O}(r^3)$ and using matrix notation we achieve
\begin{equation} \label{eq.mat_C2}
    \left[
    \begin{array}{c}
       B_{r,0}  \\[6pt] B_{r,1} \\[6pt] B_{r,2} \\
    \end{array}
    \right]
    =
    \left[
    \begin{array}{c c c c}
       1  & -3 & 3  \\[6pt]
       0  &  3 & -\frac{9}{2} \\[6pt]
       0  & 0 & \frac{3}{2} \\
    \end{array}
    \right]
    \left[
    \begin{array}{c}
       1  \\[6pt] \tilde{r} \\[6pt] \tilde{r}^2 \\
    \end{array}
    \right] .
\end{equation}
The matrix is a $(n+1) \times (n+1)$ leading principal submatrix of the $(p{+}1) \times (p{+}1)$ matrix~$\mathsfbi{D}$ used in Eq.~\eqref{eq.cubic_C3_radial_matrix}, obtained by deleting the last $(p{-}n)$ columns and rows corresponding to the highest polynomial degrees. Inverting Eq.~\eqref{eq.mat_C2} results in
\begin{equation}
    \left[
    \begin{array}{c}
     1 \\[6pt] \tilde{r} \\[6pt] \tilde{r}^2 \\
    \end{array}
    \right]
    =
    \left[
    \begin{array}{c c c}
       1  & 1 & 1 \\[6pt]
       0  &  \frac{1}{3} & 1  \\[6pt]
       0  & 0 & \frac{2}{3}  \\
    \end{array}
    \right]
    \left[
    \begin{array}{c}
     B_{r,0} \\[6pt] B_{r,1} \\[6pt] B_{r,2}  \\
    \end{array}
    \right] .
\end{equation}
Note that the submatrix is identical with the matrix we obtain by deleting the last column and row of the $(p{+}1) \times (p{+}1)$ matrix $\mathsfbi{D}^{-1}$ in Eq.~\eqref{eq.cubic_C3_m_matrix}. This follows because $\mathsfbi{D}$ is upper triangular: for such matrices, the inverse of a leading principal block is itself the corresponding block of the inverse -- a property of block triangular inversion.

In general, we can obtain a $l \times l$ submatrix from $\mathsfbi{D}$ by deleting the last $(p{-}l{+}1)$ columns and rows (highest indices). Its inverse can be obtained by deleting similarly the last $(p{-}l{+}1)$ columns and rows of the $\mathsfbi{D}^{-1}$ matrix. As a result, we exclude with each deleted column and row of the $\mathsfbi{D}^{-1}$ matrix the contribution of another B-spline basis function to the monomials in the innermost interval. By doing so we reduce the imposed regularity order at the origin.

In the angular direction, we reduce the angular basis functions~$\widetilde{h}_m(\theta)$ to the ones which are still needed, i.e.\ $|m| \le n$ (see Sec.~\ref{Sec.angular_basis}). The angular basis functions themselves, which are B-spline representations of the harmonics, are not modified.

Following Eq.~\eqref{eq.B_lk_spline_norm} with the reduced $\widehat{R}_{r,l}^m(r)$ and full $\widehat{h}_m(\theta)$ we can construct the reduced smooth center-spline basis functions~$\widehat{\boldsymbol{B}}_\mathrm{c}$ with a degree equal or smaller than~$p$. Such a reduced basis implements the envisaged $C^n$-regularity at the origin. However, in practical discretizations, the $C^n$-regularity will be exact only in the limit $r \rightarrow 0$, because unconstrained higher-degree B-spline parts can still appear in the innermost radial region. In addition, B-splines~$B_{i,j}$ for $i>n$ and $i \le p$ are not fixed by the regularity condition when contributing to the innermost interval. To avoid this issue, it is generally safest to implement the maximal regularity of $C^\infty \equiv C^p$ determined by the degree $p$ of the B-splines.

Examples for $C^0$- and $C^1$-regularity can be found in App.~\ref{App.C0andC1examples}.

\subsection{Spectral method applied to the angular direction of the basis}
\label{Sec.spectral}
As we have discussed in Sec.~\ref{Sec.angular_basis}, the B-spline basis can only approximate the harmonics~$h_m$ that are part of the smooth center-spline basis functions~$\widehat{B}_l^m$. Hence, we consider a hybrid basis $\breve{\boldsymbol{B}}$ which consists of the radial part of the smooth center-splines~$\widetilde{B}_{r,\mathrm{c}}$ and the harmonics $\breve{\boldsymbol{B}}_{\theta,\mathrm{c}}$ in the angular direction, defined as
\begin{equation}\label{eq.angular_mon_spline_harm}
     \breve{\boldsymbol{B}}_{\theta,\mathrm{c}} \coloneqq [\widehat{h}_{-p},\ldots,\widehat{h}_0,\ldots,\widehat{h}_p],
\end{equation}
where $\breve{\boldsymbol{B}}_{\theta,\mathrm{c}}$ contains $2p{+}1$ orthonormal angular functions. The normalized harmonics are
\begin{equation}\label{eq.norm_h_basis}
    \widehat{h}_{-m} \coloneqq \frac{1}{\sqrt{\uppi}} \sin(m\theta), \qquad
    \widehat{h}_0 \coloneqq \frac{1}{\sqrt{2\uppi}} , \qquad
    \widehat{h}_m  \coloneqq \frac{1}{\sqrt{\uppi}} \cos(m\theta)
\end{equation}
for $0 \le m \le p$, and they are orthogonal to each other over $[0,2\uppi]$. A specific angular basis function
\begin{equation}\label{eq.breve_B_theta_m}
    \breve{h}_m(\theta) = \left(\breve{\boldsymbol{c}}_{\theta,m}\right)^\mathrm{T} \breve{\boldsymbol{B}}_{\theta,\mathrm{c}}(\theta)
\end{equation}
is selected by its canonical coefficient vector
\begin{equation}
    \left(\breve{\boldsymbol{c}}_{\theta,m}\right)^\mathrm{T} = (0,\ldots,1,\ldots,0),
\end{equation}
where the 1 is in the $(m{+}p{+}1)$-th position and all other entries are 0.
Note that the angular basis outside the inner region may contain a spectrum of harmonics with $|m| \le \breve{N}_\theta$:
\begin{equation}
     \breve{\boldsymbol{B}}_\theta \coloneqq [\widehat{h}_{-\breve{N}_\theta},\ldots,\widehat{h}_0,\ldots,\widehat{h}_{\breve{N}_\theta}], \qquad (\breve{N}_\theta \ge p).
\end{equation}

Using Eqs.~\eqref{eq.hat_tilde_B_r_l_m} and~\eqref{eq.breve_B_theta_m}, we construct a smooth center-spline basis function
\begin{equation} \label{eq._breve_B_lk_spline}
     \widetilde{\breve{B}}_l^m(r,\theta) = \widehat{R}_{r,l}^m(r) \, \breve{h}_m(\theta) =  \left( \breve{\boldsymbol{P}}_l^m \right)^\mathrm{T} \breve{\boldsymbol{B}}_\mathrm{c}(r,\theta),
\end{equation}
where $\widehat{R}_{r,l}^m$ denotes the radial part of the smooth center-spline basis functions.
The coefficient vector is
\begin{equation}
    \breve{\boldsymbol{P}}_l^m \coloneqq \widehat{\boldsymbol{c}}_{r,l}^m \otimes \breve{\boldsymbol{c}}_{\theta,m},
\end{equation}
and it describes $\widetilde{\breve{B}}_l^m$ using hybrid, smooth tensor-product B-spline basis functions
\begin{equation}
    \breve{\boldsymbol{B}}_\mathrm{c}(r,\theta) \coloneqq \boldsymbol{B}_{r,\mathrm{c}}(r) \otimes \breve{\boldsymbol{B}}_{\theta,\mathrm{c}}(\theta).
\end{equation}
Such a hybrid basis represents, in the innermost radial interval, the harmonic polar functions ~$S_l^m$ exact up to machine precision.

A function $\breve{\phi}_h \in \widetilde{\breve{V}}_h$ can be expressed in the basis $\widetilde{\breve{\boldsymbol{B}}}$ spanning $\widetilde{\breve{V}}_h$ (see Eq.~\eqref{eq.phiapprox_Cp}) by
\begin{equation}\label{eq.phiapprox_Cp_spec}
   \breve{\phi}_h(r,\theta) = \sum_{l=0}^{p} \sum_{\substack{m = -l \\ |m| \equiv l(\mathrm{mod}\ 2)}}^l \widetilde{\breve{b}}_l^m \widetilde{\breve{B}}_{l}^m(r,\theta) + \sum_{i=p+1}^{N_r-1} \sum_{j=-\breve{N}_\theta}^{\breve{N}_\theta}\breve{\phi}_{i,j}\breve{B}_{i,j}(r,\theta).
\end{equation}

Although the hybrid basis appears to be well suited for implementing 
$C^p$‑regularity on the unit disc, there are several caveats. 
First, for tokamaks and stellarators, the mass matrix in the $\theta$‑coupling 
is generally dense, because the Fourier‑type harmonics have global support 
and are coupled through the $\theta$‑dependent Jacobian.
Second, resolving steep local gradients may require a large number of basis 
functions to achieve reasonable convergence to the target function. 
Third, the numerical evaluation of harmonic functions~$h_m$ is typically more 
expensive than that of low‑order polynomials. 
For these reasons, we favor the smooth polar-spline basis over the hybrid basis.

The hybrid basis can equivalently be formulated by replacing the real-valued harmonic functions~$h_m$ with complex exponentials of the form $\exp(\pm \mathrm{i} m\theta)$.

\section{Numerical analysis of smooth polar splines}
\label{sec.theoryandanal}

\subsection{Basis function normalization and matrix condition numbers}

In this section, we examine the condition numbers of the mass and stiffness matrices, 
as these directly influence the convergence rate of iterative solvers. 
The larger the condition number, the more iterations an iterative solver typically requires to achieve a given accuracy. 

For a symmetric positive definite matrix~$\mathsfbi{A}$, which includes the mass and stiffness matrices in the discretizations considered here, 
the condition number~$\kappa(\mathsfbi{A})$ is defined as the ratio of its largest to smallest eigenvalue:
\begin{equation}
\kappa(\mathsfbi{A}) = \frac{\lambda_{\max}(\mathsfbi{A})}{\lambda_{\min}(\mathsfbi{A})}.
\end{equation}

The diagonal entries of the mass matrix in Eq.~\eqref{eq.mass_matrix_org} represent the squared $L^2$-norms (see Eq.~\eqref{eq.L2_norm}) of the basis functions (tensor‑product B-splines or smooth polar splines). When the basis functions are normalized, these diagonal entries become unity. Equal diagonal entries are advantageous for improving a matrix’s condition number.

A purely algebraic procedure can instead rescale the system matrix, achieving the same effect as explicit basis normalization while being more general. In this approach, the scaled system reads
\begin{subequations}
 \label{eq.Jacobi_precond}
\begin{equation} 
   \mathsfbi{D} \mathsfbi{A} \widetilde{\boldsymbol{\phi}} = 
   \mathsfbi{D} \widetilde{\boldsymbol{f}},
\end{equation}
where
\begin{equation}
   \mathsfbi{D} \coloneqq \mathrm{diag}^{-1}(\mathsfbi{A}),
\end{equation}   
\end{subequations}
and $\mathrm{diag}^{-1}(\mathsfbi{A})$ denotes the diagonal matrix whose entries are equal to the reciprocals of the diagonal entries of $\mathsfbi{A}$. 
Here, $\mathsfbi{A}$ can be either the mass or the stiffness matrix. 
Equation~\eqref{eq.Jacobi_precond} corresponds to left Jacobi preconditioning,
which improves the conditioning of~$\mathsfbi{A}$ without altering the basis functions. 
In this case, explicit normalization of the smooth center‑splines~$\widetilde{B}_l^m$ (see Sec.~\ref{Sec.orthonormal_spline}) becomes redundant.

\begin{figure}[htbp]
\centering
    \subfloat[Preconditioned mass matrix $\mathsfbi{D} \mathsfbi{M}$.]{\includegraphics[width=0.45\textwidth]{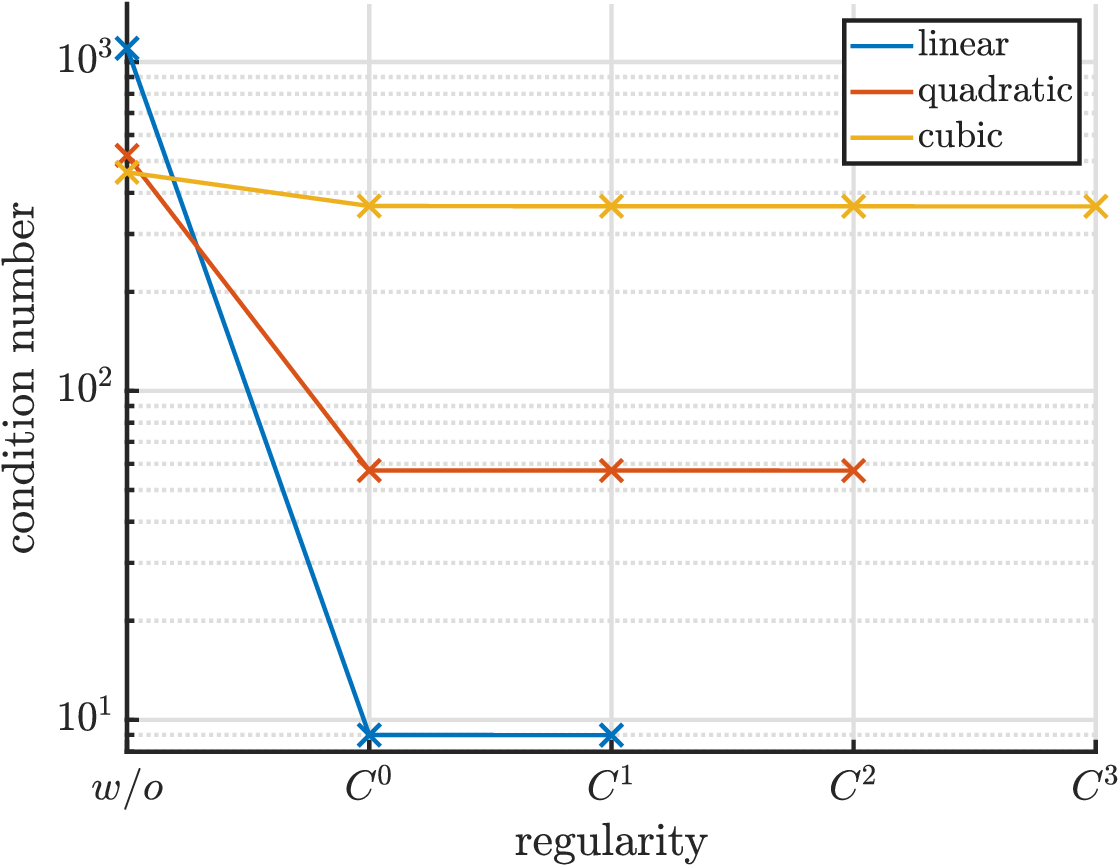}\label{fig.Mass_condition_number}}
 \hfill 	
    \subfloat[Preconditioned stiffness matrix $\mathsfbi{D} \mathsfbi{S}$.]{\includegraphics[width=0.45\textwidth]{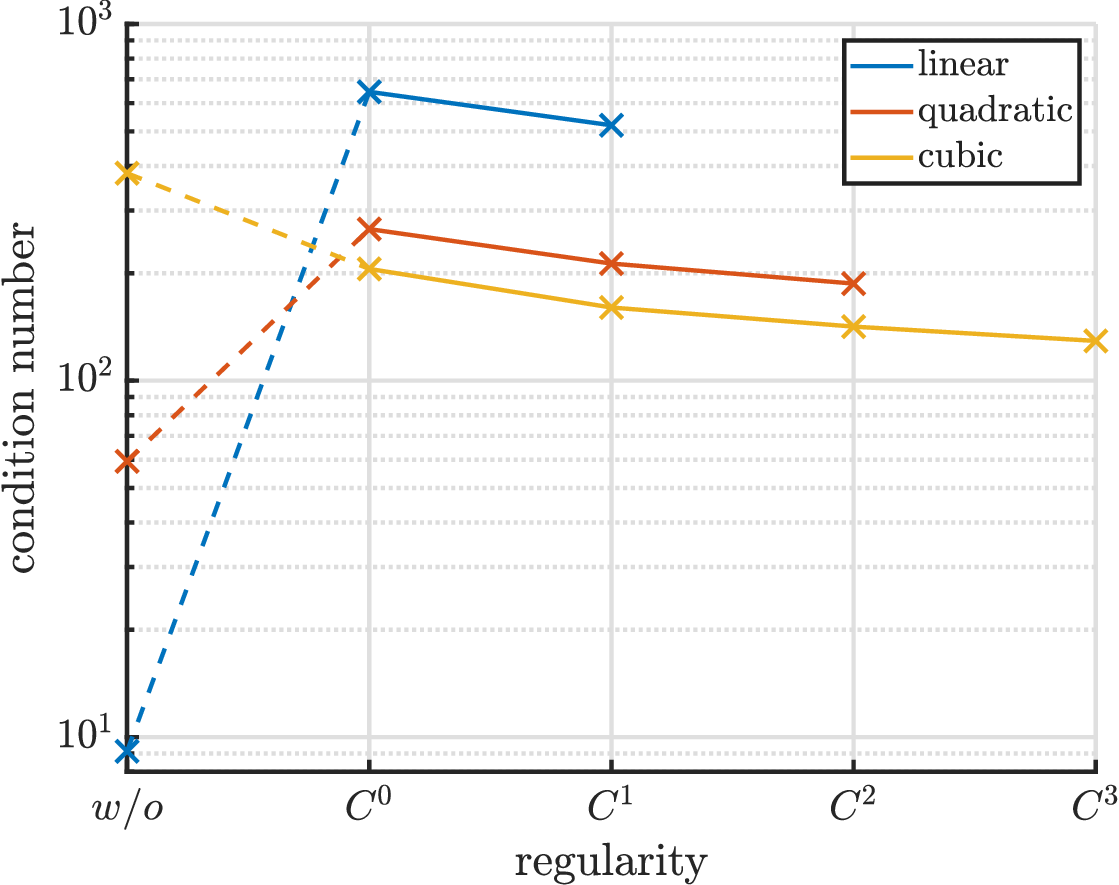}\label{fig.Poisson_condition_number}}
 \caption{Condition numbers of (a) $\mathsfbi{D} \mathsfbi{M}$ and (b) $\mathsfbi{D} \mathsfbi{S}$, 
shown on a logarithmic scale for cases without regularity and with regularity from $C^0$ to~$C^3$. 
Results are given for linear, quadratic, and cubic B‑splines. 
The radial and angular grids each contain $N_r = N_\theta = 24$ points.}
\label{fig.Condition_number}
\end{figure}

Figure~\ref{fig.Mass_condition_number} shows the condition number of the 
preconditioned mass matrix $\mathsfbi{D} \mathsfbi{M}$ for different levels of 
enforced regularity at the origin, for linear ($p{=}1$), quadratic ($p{=}2$), and 
cubic ($p{=}3$) B-splines. The horizontal axis denotes the order of regularity, ranging from without to maximum regularity $C^\infty \equiv C^p$ permitted by the spline degree~$p$, 
while the vertical axis shows the condition number on a logarithmic scale. 
The condition number initially decreases at $C^0$‑regularity and then remains constant for all higher orders. At the highest regularity level permitted by each spline degree~$p$, the condition number increases with~$p$.

Figure~\ref{fig.Poisson_condition_number} presents the condition number of the preconditioned stiffness matrix $\mathsfbi{D} \mathsfbi{S}$, 
where $a(\boldsymbol{r})=1$ and $c(\boldsymbol{r})=0$ in Eq.~\eqref{eq.qp-funcs} (Laplacian), for linear, quadratic, and cubic B‑splines, 
using the same regularity configurations as in Fig.~\ref{fig.Mass_condition_number}. 
Increasing regularity to $C^0$ produces a sharp initial jump in the condition number when normalization is applied. 
The case without regularity is notably detached from the one with $C^0$‑regularity, because the non‑regular discretization of the Laplacian contains a pole at the origin (see the discussion after Eq.~\eqref{eq.S_mat}), making it a qualitatively different problem.
The condition number decreases monotonically with increasing regularity, indicating improved matrix conditioning due to smoother basis functions. At the highest regularity level, the condition number decreases with spline degree~$p$.
Overall, these results show that, for linear, quadratic, and cubic B‑splines, 
in the case of a Jacobi preconditioner, increasing regularity at the origin typically yields smaller condition numbers.

\subsection{Eigenvalues and eigenfunctions of the Poisson equation}

On the unit disc $\Omega$, the Dirichlet eigenvalue problem for the Laplacian is given by
\begin{subequations}\label{eq.poisson_eigen}
\begin{alignat}{3}
\label{eq.poisson_interior}
    -\nabla^2 u(r,\theta) &= \lambda_\mathrm{ana} u(r,\theta)
    &&\quad \mbox{in} && \quad \Omega , \\
\label{eq.poisson_boundary}
    u(r,\theta) &= 0 
    &&\quad \mbox{on} && \quad \partial\Omega .
\end{alignat}
\end{subequations}
Its eigenfunctions are
\begin{equation} \label{eq.poisson_eigenfunctions}
    u_{m,k}(r,\theta) =
    \begin{cases}
        J_m \! \left(\alpha_{m,k} r\right)\cos(m\theta), \\
        J_m \! \left(\alpha_{m,k} r\right)\sin(m\theta),
    \end{cases}
\end{equation}
where $J_m$ denotes the Bessel function of the first kind of order $m$, 
and $\alpha_{m,k}$ is the $k$-th positive root of $J_m$. The corresponding eigenvalues are
\begin{equation}\label{eq.poisson_eigenvalues}
    \lambda_{\mathrm{ana},m,k} = \alpha_{m,k}^2 .
\end{equation}
Here, the integer $m \ge 0$ denotes the angular mode number, while $k \ge 1$ counts the radial zeros of $J_m$. 
The functions $u_{m,k}(r,\theta)$ form a complete orthogonal set on $\Omega$.

Discretizing Eq.~\eqref{eq.poisson_eigen} with a B-spline finite-element method yields the generalized eigenvalue problem
\begin{equation} \label{eq.gen_eigen}
     \mathsfbi{S} \boldsymbol{u}_{m,k} 
     = \lambda_{m,k} \mathsfbi{M} \boldsymbol{u}_{m,k},
\end{equation}
where $\mathsfbi{S}$ and $\mathsfbi{M}$ are the stiffness and mass matrices 
defined in Eqs.~\eqref{eq.S_mat} and~\eqref{eq.mass_matrix_org}, respectively.
We denote the computed eigenpairs by $\{ \lambda_{m,k}, \boldsymbol{u}_{m,k} \}$.
When higher regularity is imposed at the origin via smooth polar splines, 
Eq.~\eqref{eq.gen_eigen} is solved in the space of $C^n$-regularity. 
The stiffness and mass matrices are transformed using Eq.~\eqref{eq.stiff_regularity_trans}. 
For comparison, the corresponding eigenvectors~$\boldsymbol{u}_{\mathrm{reg}}$ are expressed in the tensor‑product B‑spline space obtained via Eq.~\eqref{eq.tildephi_phi_trans}.

We quantify the effect of increased regularity on an eigenvector by defining the relative regularity error~$\epsilon_\mathrm{reg}$ as the $L^2$-distance from the coefficient vector~$\boldsymbol{u}_{\mathrm{reg}}$ to its projection onto the subspace of highest regularity for degree~$p$:
\begin{equation}
    \epsilon_\mathrm{reg}(\boldsymbol{u}_{\mathrm{reg}}) 
    = \frac{\left \lVert \mathsfbi{\Pi}_\perp \boldsymbol{u}_{\mathrm{reg}} 
    - \boldsymbol{u}_{\mathrm{reg}} \right\rVert_{L^2}} 
    {\left \rVert \boldsymbol{u}_{\mathrm{reg}} \right\rVert_{L^2}},
\end{equation}
where $\mathsfbi{\Pi}_\perp$ denotes the $M$‑orthogonal projector onto the $C^\infty$‑regularity subspace (see Eq.~\eqref{eq.projection}).

\begin{figure}[htbp]
\centering
\includegraphics[width=0.85\linewidth]{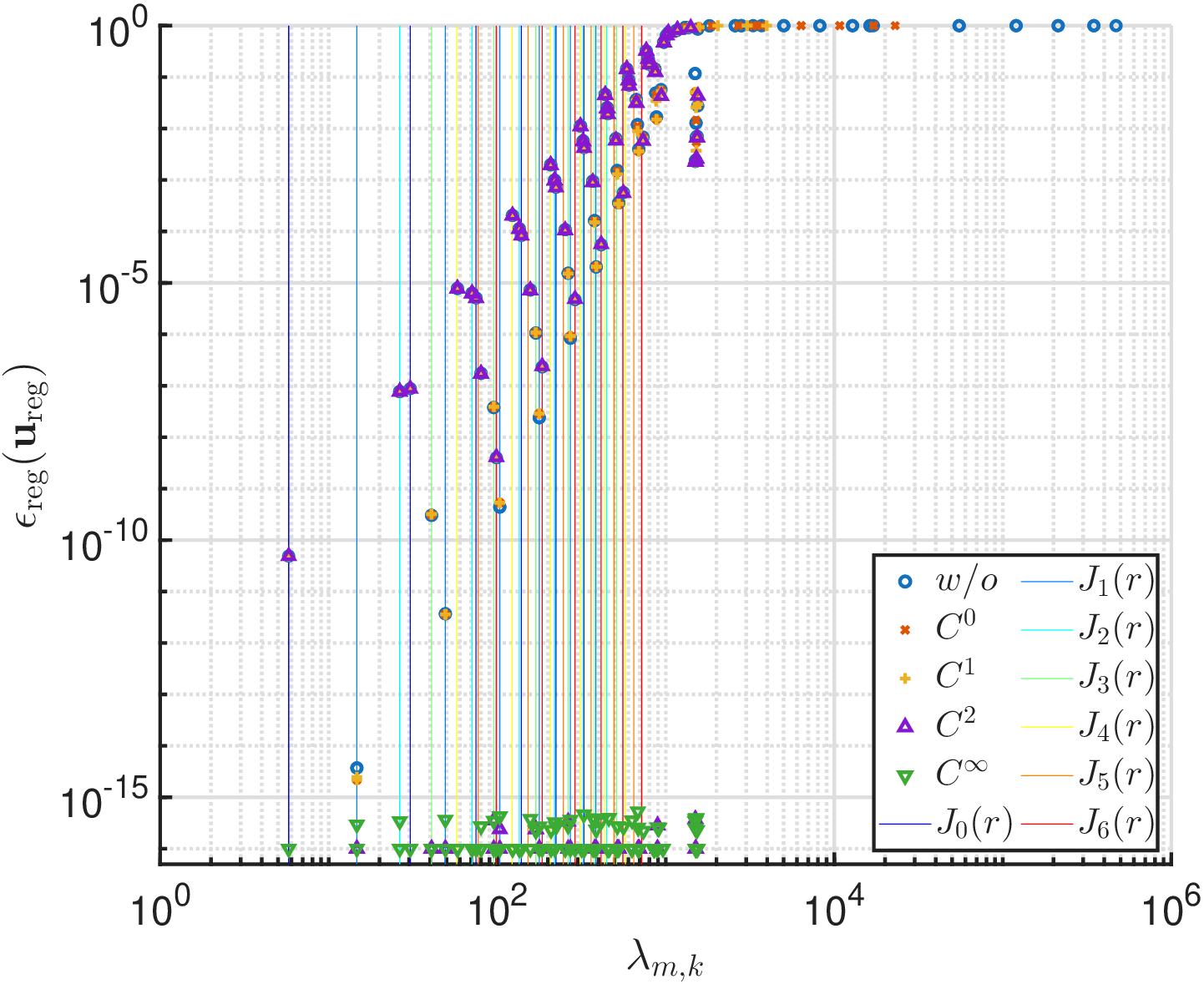}
\caption{Log--log plot of the relative regularity error 
$\epsilon_\mathrm{reg}(\boldsymbol{u}_{\mathrm{reg}})$ versus eigenvalue~$\lambda_{m,k}$ 
for several regularity levels at the origin, including the case without regularity at the origin. 
Colored vertical lines indicate the analytical eigenvalues corresponding to the first seven Bessel functions $J_m$. 
Values below $10^{-15}$ are consistent with machine precision and are clipped to a lower bound of $10^{-16}$ for plotting.}
\label{fig:eigenvalue_error}
\end{figure}

Figure~\ref{fig:eigenvalue_error} shows a log--log plot of 
$\epsilon_\mathrm{reg}(\boldsymbol{u}_{\mathrm{reg}})$ versus the corresponding eigenvalues~$\lambda_{m,k}$, 
obtained using a cubic B-spline finite-element discretization on a grid with $N_r=10$ radial and $N_\theta=12$ angular points. 
Symbols denote different orders of regularity.
Colored vertical lines indicate the analytical eigenvalues~$\lambda_{\mathrm{ana},m,k}$. 
Low-order modes, which are smooth, agree well with the analytical spectrum and exhibit very small relative regularity error.  
As $\lambda_{m,k}$ increases, the error typically grows and approaches unity unless the highest regularity ($C^\infty$) is enforced.  
For modes with small $\epsilon_\mathrm{reg}$, changing the imposed regularity affects the error but not the eigenvalue, because these modes naturally satisfy the smoothness constraints.  
In contrast, modes with $\epsilon_\mathrm{reg} \approx 1$ are spurious: they originate from insufficient continuity and are gradually removed as regularity increases, disappearing entirely from the discrete spectrum only at $C^\infty$-regularity because the approximation space no longer supports them.  
At that point, the remaining modes already possess $C^\infty$-regularity at the origin, so the $L^2$‑orthogonal projector~$\mathsfbi{\Pi}_\perp$ leaves them unchanged.  
Consequently, their relative error necessarily drops to $\epsilon_\mathrm{reg}<10^{-15}$, which is consistent with machine precision.  
Under $C^\infty$-regularity at the origin, all remaining modes satisfy $\lambda_{m,k} < 1.6 \times 10^3$. 
The removal of spurious eigenvalues -- which can reach $\approx 4\times 10^5$ without regularity -- leads to a substantial improvement in the condition number.

\begin{figure}[htbp]
\centering
\subfloat[$J_0(\alpha_{0,1}r)$, $\lambda_{0,1}=5.78$]
    {\includegraphics[width=0.45\textwidth]{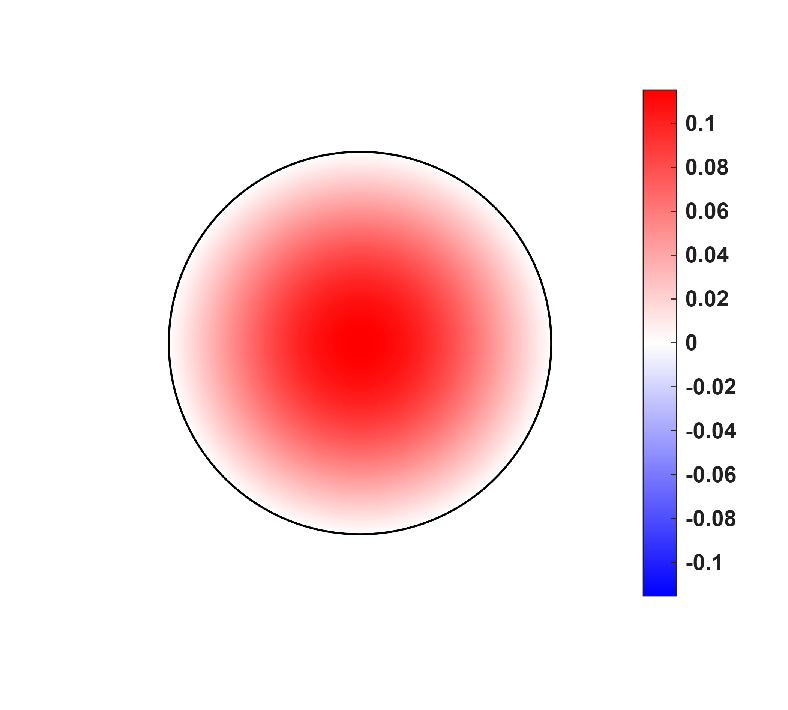}%
    \label{fig.general_eigenfunction_1}}
\hfill 	
\subfloat[$J_2(\alpha_{2,3} r) \cos\!\big( 2 (\theta - \theta_0) \big)$, $\lambda_{2,3}=135.03$]
    {\includegraphics[width=0.45\textwidth]{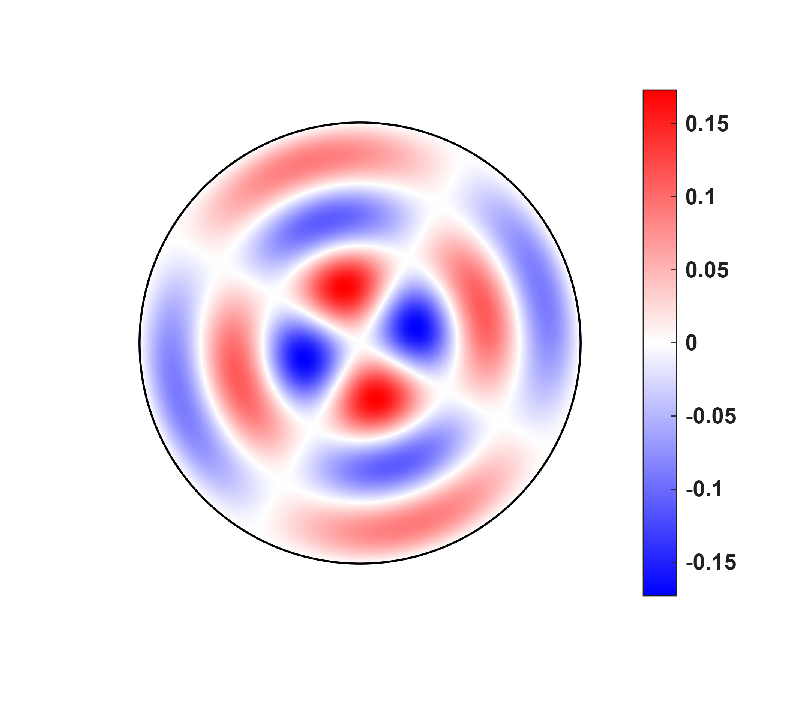}%
    \label{fig.general_eigenfunction_2}}
\hfill
\subfloat[$J_{11}(\alpha_{11,1} r)$, $\lambda_{11,1}=1498.07$]
    {\includegraphics[width=0.45\textwidth]{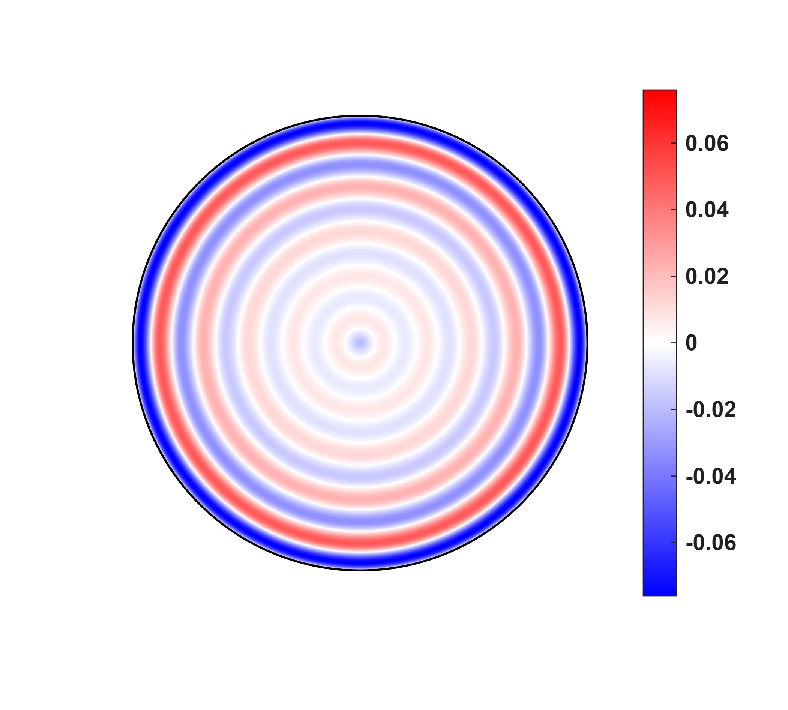}%
    \label{fig.general_eigenfunction_3}} 
\hfill	
\subfloat[Spurious mode (with $C^0$-regularity), $\lambda=10770.57$]
    {\includegraphics[width=0.45\textwidth]{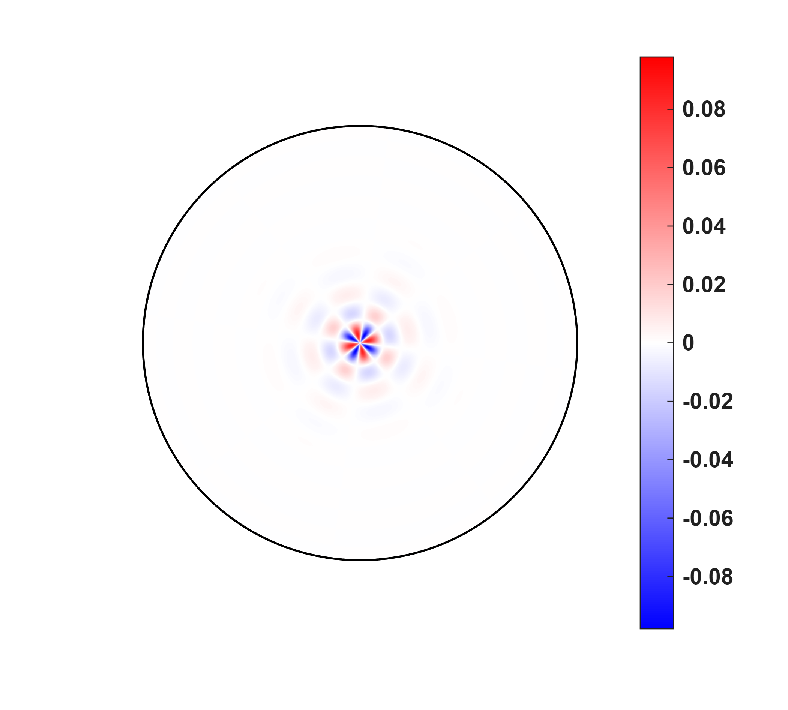}%
    \label{fig.general_eigenfunction_4}}
    \caption{Representative numerical eigenfunctions for different eigenvalues. 
The Bessel functions given in the subfigure titles are the corresponding analytical eigenfunctions and eigenvalues, listed for reference where an exact solution is available.}
\label{fig.general_eigenfunction}
\end{figure}

Figure~\ref{fig.general_eigenfunction} illustrates four representative eigenpairs of the discrete Dirichlet Laplacian.  
Figure~\ref{fig.general_eigenfunction_1} is the fundamental radially symmetric eigenfunction, in excellent agreement with the analytical form $J_0(\alpha_{0,1}r)$.  
Figure~\ref{fig.general_eigenfunction_2} corresponds to an $m=2$ mode and matches 
$J_2(\alpha_{2,3} r) \cos\!\big( 2 (\theta - \theta_0) \big)$, where $\theta_0$ is a constant.
In Figure~\ref{fig.general_eigenfunction_3}, the numerical eigenfunction resembles $J_{11}(\alpha_{11,1} r)$ but shows noticeable deviation: analytically, the radial peaks should decay in amplitude, whereas the numerical mode displays a growing envelope. 
This occurs because the mode is close to the resolution limit of the grid.  
Figure~\ref{fig.general_eigenfunction_4} presents a high-eigenvalue mode computed with $C^0$-regularity; it clearly violates the regularity constraints and is eliminated when regularity is enforced.

In summary, increasing the regularity preserves the physical part of the spectrum and its eigenfunctions, while progressively eliminating spurious components arising from insufficient regularity.  
All spurious eigenpairs vanish only when the $C^\infty$-regularity is applied.

\subsection{Convergence of smooth polar-spline basis (Bessel function test case)}

\begin{figure}[htbp]
\centering
\includegraphics[width=0.55\linewidth]{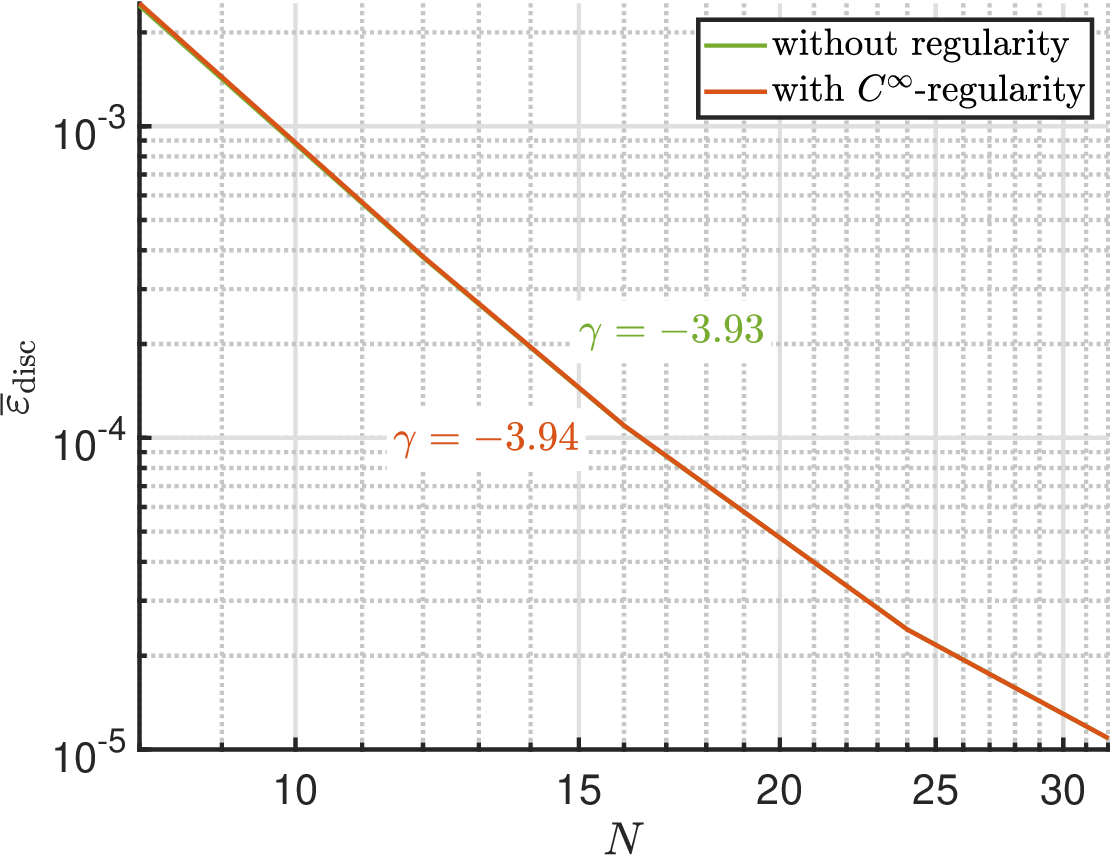}
\caption{The red/green lines represent the root mean square error~$\overline{\varepsilon}_{\mathrm{disc}}$ when approximating $f_0 = J_1(10r) \cos(\theta)$ using a cubic B-spline basis with/without $C^\infty$-regularity at the origin. The results are shown as a function of the radial and angular grids, each of which contains $N = N_r = N_\theta$ points, on a log--log scale. The parameter $\gamma$ is the slope of the log--log fit.}
\label{fig.relaerror_N}
\end{figure}

We use $\phi_h$ from Eq.~\eqref{eq.2d_bspline_phi_org}, expanded in the tensor-product B-spline basis $\boldsymbol{B}$, and $\widetilde{\phi}_h$ from Eq.~\eqref{eq.phiapprox_Cp}, expanded in the smooth polar-spline basis $\widetilde{\boldsymbol{B}}$. These representations are used to approximate the smooth target function
\begin{equation}
    f_0(r,\theta) = J_1(10r) \cos(\theta),
\end{equation}
where $J_1(r)$ is the Bessel function of the first kind, a $C^\infty$ function that cannot be exactly represented by a finite B-spline expansion. The discrete solution is achieved by solving the mass matrix equation~\eqref{eq.mass_org}. Furthermore, we define the average root mean square error (RMSE) over the unit disc as
\begin{equation}\label{eq.av_error_disc}
    \overline{\varepsilon}_{\mathrm{disc}} \coloneqq \frac{1}{\sqrt{\uppi}} \lVert \widetilde{\phi}_h(r,\theta) - f_0(r,\theta) \rVert_{L^2}
\end{equation}
which is shown in Fig.~\ref{fig.relaerror_N}. The results are obtained using a cubic B-spline basis, with and without $C^\infty$-regularity at the origin. The error is shown as a function of the radial and angular grids, each of which contains $N = N_r = N_\theta$ points, on a log--log scale. No significant difference is observed at the resolved scales considered between the results with and without $C^\infty$-regularity at the origin. The order of convergence~$\gamma$, calculated by a log--log fit, is close to four, which agrees with the expected order of convergence for a cubic B-spline basis. Therefore, adding the smoothness constraint at the origin does not change the order of convergence.

\subsection{Poisson problem with Helmholtz eigenfunction: analytic solution and numerical convergence studies}

\subsubsection{Analytic solution}

On the unit disc $\Omega$, we consider the two-dimensional Poisson problem
\begin{subequations}\label{eq.Poisson}
\begin{alignat}{3}
\label{eq.interior_Poisson}
    \nabla^2 g(r,\theta) &= f(r,\theta)
    &&\quad \mbox{in}&& \quad \Omega , \\
\label{eq.boundary_condition_Poisson}
    g(r,\theta) &= 0 &&\quad \mbox{on}&& \quad \partial\Omega.
\end{alignat}
\end{subequations}

The following functions satisfy this Poisson problem 
\begin{align}
     g(r,\theta) &= \sum_{m=0}^\infty \sum_{k=1}^\infty J_m(\alpha_{m,k} \, r) \left[ a_{m,k}\cos(m\theta) + b_{m,k}\sin(m\theta) \right] , \label{eq.y_Poisson_y}\\
     f(r,\theta) &= -\sum_{m=0}^\infty \sum_{k=1}^\infty \alpha_{m,k}^2 J_m(\alpha_{m,k} \, r) \left[ a_{m,k}\cos(m\theta) + b_{m,k}\sin(m\theta) \right], \label{eq.y_Poisson_f}
\end{align}
where $\alpha_{m,k}$ is the $k$-th positive root of the Bessel function $J_m$. The coefficients $a_{m,k}$ and $b_{m,k}$ are free parameters. 

As a simple test case, we use
\begin{align}
\label{eq.y_Poisson_y0}
    g_0(r,\theta)&= J_m(\alpha_{m,4} \, r) \cos(m\theta) , \\
\label{eq.y_Poisson_f0}
    f_0(r,\theta) &= -\alpha_{m,4}^2 g_0
\end{align}
with $m=1$.
In the following $f_0$ will be prescribed and the numerical solution will be compared to $g_0$.

\subsubsection{Grid convergence study on unit disc}
We use again $\phi_h$ from Eq.~\eqref{eq.2d_bspline_phi_org}, expanded in the tensor-product B-spline basis $\boldsymbol{B}$, and $\widetilde{\phi}_h$ from Eq.~\eqref{eq.phiapprox_Cp}, expanded in the smooth polar-spline basis $\widetilde{\boldsymbol{B}}$. These representations are used to approximate the solution~$g_0$, and the discrete solution is obtained by solving the stiffness matrix equation~\eqref{eq.stiff_org}. The average RMS error is evaluated only over the center of the unit disc, defined as
\begin{equation}\label{eq.av_error_disc_inner}
    \overline{\varepsilon}_\mathrm{center} = \sqrt{ \frac{256}{\uppi}} \lVert \widetilde{\phi}_h(r,\theta) - g_0(r,\theta) \rVert_{L^2}
\end{equation}
to quantify the difference between the numerical solutions $\phi_h$ and $\widetilde{\phi}_h$, and the analytic solution~$g_0$. The center of the unit disc, defined by $r \le 1/16$, to evaluate $\overline{\varepsilon}_\mathrm{center}$ is fixed and does not change with the grid resolution~$\Delta r$. We focus on it because the largest differences are expected to occur near the origin, where the regularity condition has the greatest influence.

\begin{figure}[htbp]
\centering
\includegraphics[width=0.55\linewidth]{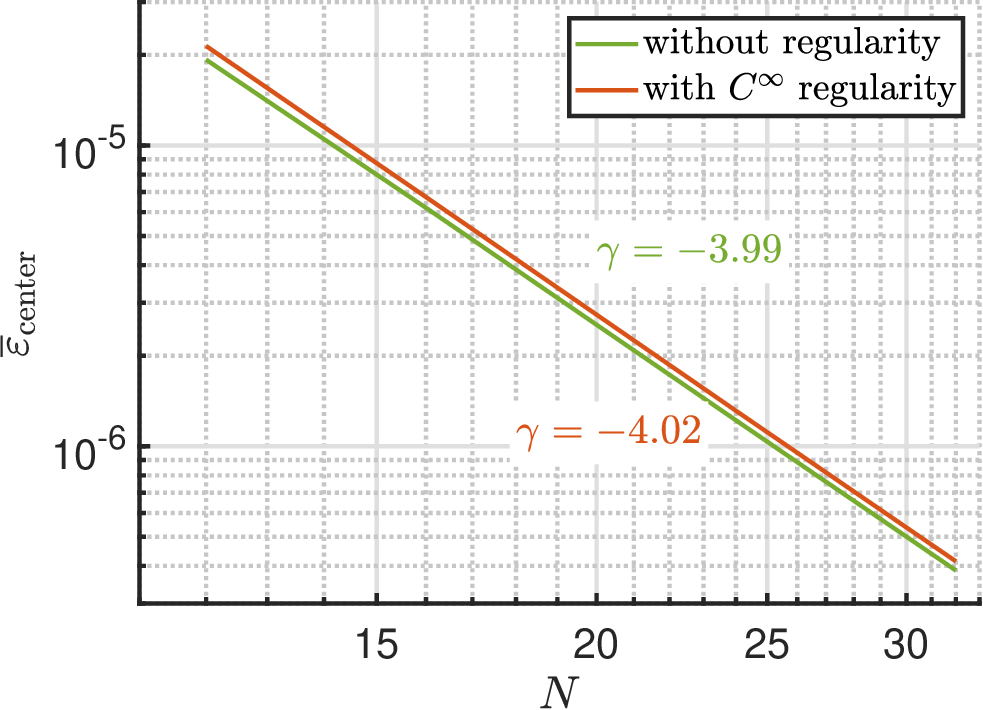}
\caption{The red/green lines represent the root mean square error~$\overline{\varepsilon}_{\mathrm{center}}$ of the solution~$g_0$~(see Eq.~\eqref{eq.y_Poisson_y0}) of the Poisson equation~\eqref{eq.Poisson} using a cubic B-spline basis with/without $C^\infty$-regularity at the origin. The error is plotted as a function of the radial and angular grids, each containing $N = N_r = N_\theta$ points, on a log--log scale. The parameter $\gamma$ is the slope of the log--log fit.}
\label{fig.Poisson_error_N}
\end{figure}

In Fig.~\ref{fig.Poisson_error_N}, the red/green lines show $\overline{\varepsilon}_{\mathrm{center}}$ of the solution~$g_0$ of the Poisson equation, calculated using a cubic B-spline basis with/without $C^\infty$-regularity at the origin. The error is plotted as a function of the radial and angular grids, each containing $N = N_r = N_\theta$ points, on a log--log scale. No significant difference is observed at the resolved scales considered between the results with and without $C^\infty$-regularity at the origin. The order of convergence $\gamma$ is close to four, which agrees with the expected order of convergence for a cubic B-spline basis. Therefore, adding the smoothness constraint at the origin does not change the order of convergence. It also appears to have little impact on the quality of the approximated solution, since the analytical solution~$g_0$ already has the correct smoothness at the origin and thus does not excite discontinuous modes.

\subsubsection{Regularity constraints on derivative continuity at the origin}

We use cubic B-splines for discretization with $N_r = N_\theta = 32$. Increasing the angular mode number~$m$ in Eqs.~\eqref{eq.y_Poisson_y0} and~\eqref{eq.y_Poisson_f0} from $m = 1$ to $m = 3$ increases the angular oscillation frequency of the solution, $\cos(m\theta)$. Whether or not $C^\infty$-regularity is enforced at the origin, the approximation error changes little, consistent with the results of the previous section. 
However, this behavior changes for $m = 4$. The angular function $\cos(4\theta)$ is excluded in the innermost region, $r \le \Delta r$, by the $C^\infty$-regularity constraint. The smooth center splines $\widetilde{B}_l^{\pm m}$, with $0 \le l \le m \le 3$, possess angular dependence only up to $\cos(3\theta)$. Consequently, modes $\cos(m\theta)$ for $m > 3$ are excluded, 
even though they could, in principle, be resolved by the angular spline resolution $N_\theta = 32$.

\begin{figure}
\centering

    \subfloat[First-order partial derivative $\partial g_0/\partial x$ of analytical solution~$g_{0,\mathrm{ana}}$.]{\includegraphics[width=0.5\textwidth]{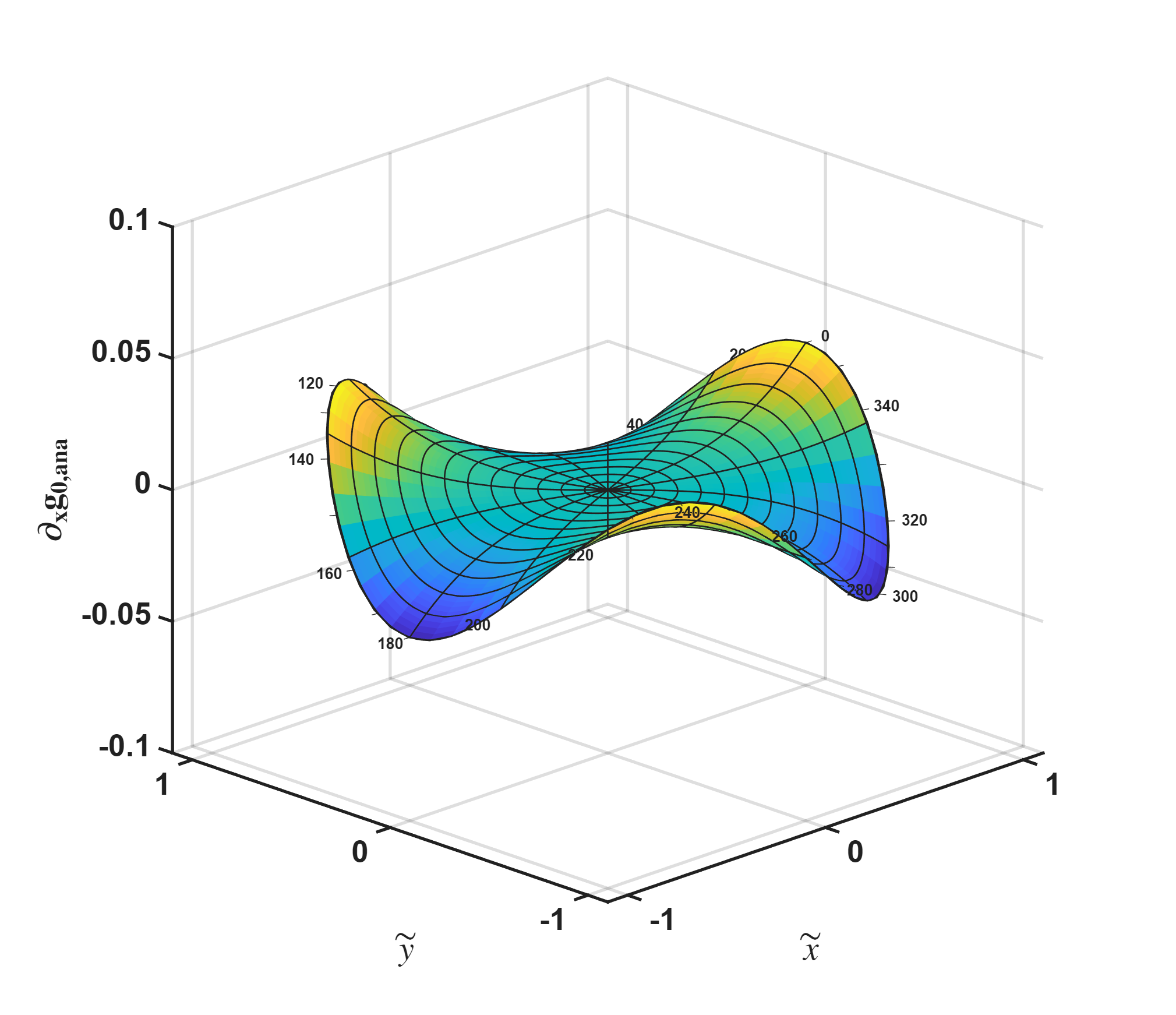}\label{fig.Poisson_Dx_ref}}
 \hfill 	
    \subfloat[$\varepsilon_{\partial g_0/\partial x}$ with $C^0$-regularity (contour plot).]{\includegraphics[width=0.5\textwidth]{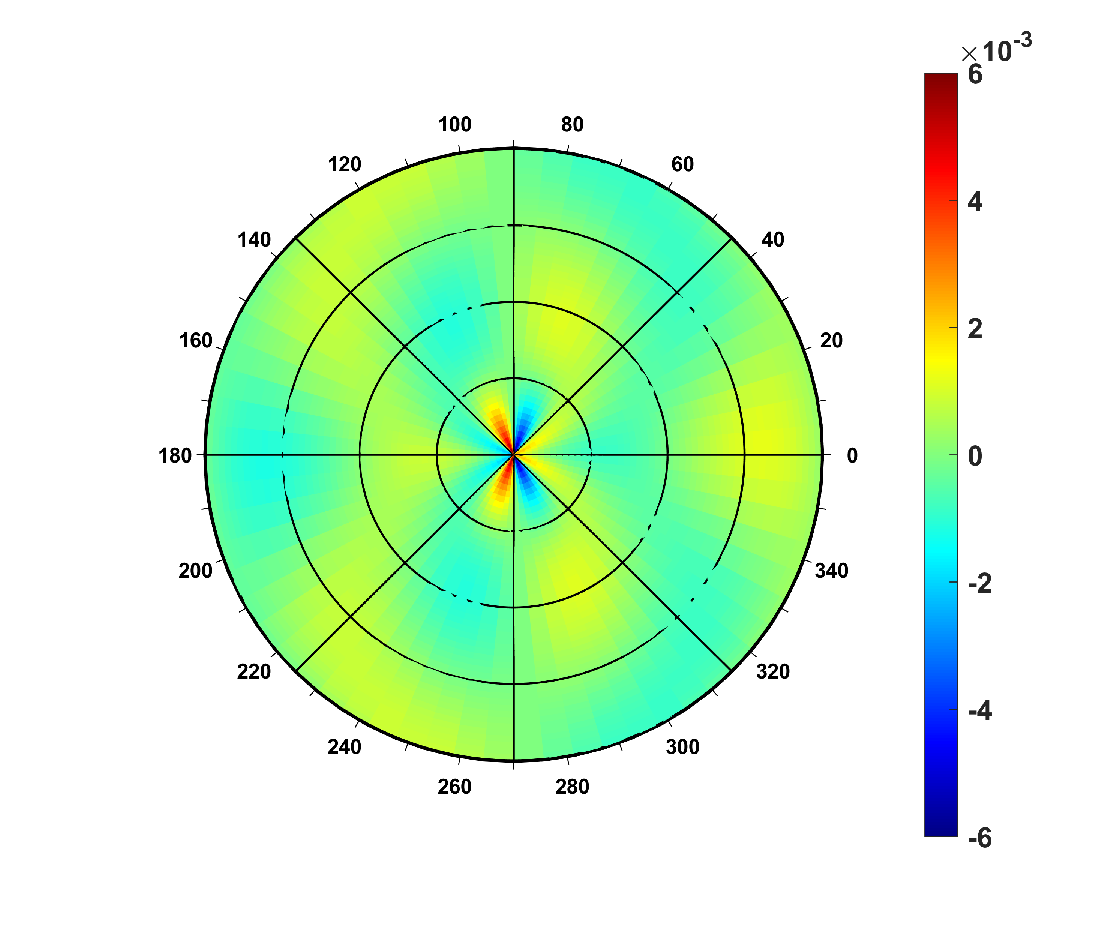}\label{fig.Poisson_Dx_C0}} 
 \hfill	
    \subfloat[$\varepsilon_{\partial g_0/\partial x}$ with $C^1$-regularity (contour plot).]{\includegraphics[width=0.5\textwidth]{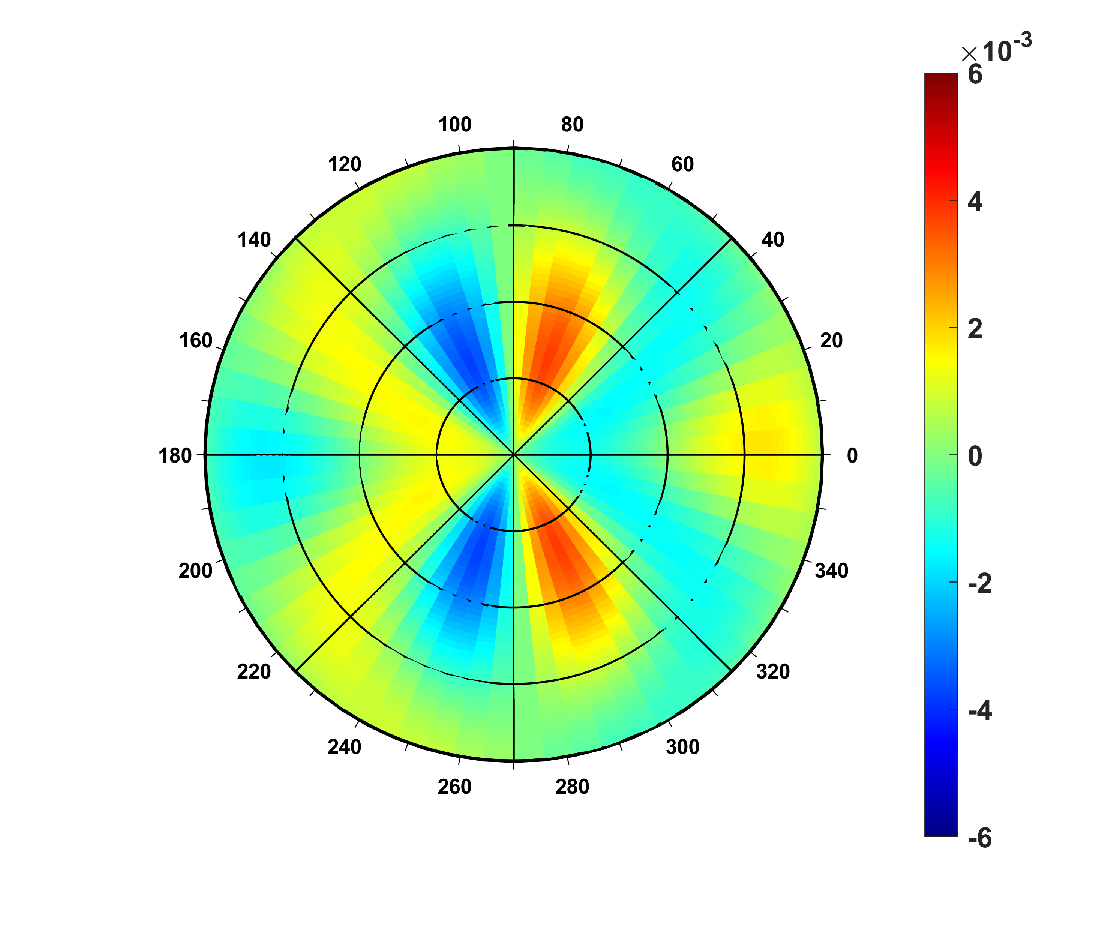}\label{fig.Poisson_Dx_C1}} 
 \hfill	
    \subfloat[First-order partial derivative $\partial g_0/\partial x$ of numerical solution $g_{0,\mathrm{num}}$ with $C^\infty$-regularity.]{\includegraphics[width=0.5\textwidth]{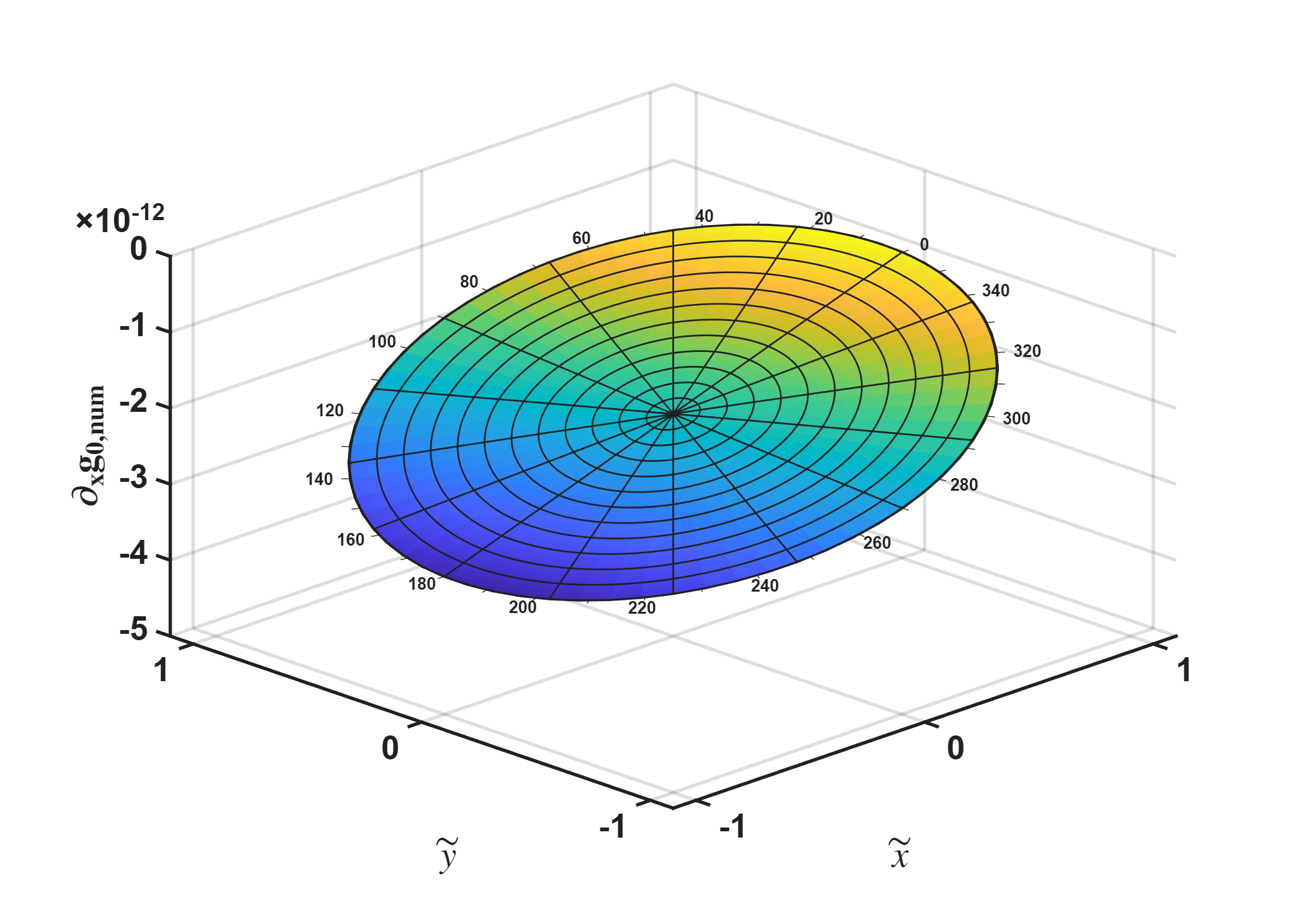}\label{fig.Poisson_Dx_C3}} 
 \caption{First-order partial derivative $\partial g_0/\partial x$ of the analytical (Eq.~\eqref{eq.y_Poisson_y0} for $m=4$) and numerical solution and its discretization error $\varepsilon_{\partial g_0/\partial x}$ for different orders of regularity within the innermost circle ($r \le \Delta r$). The radial and angular grids in the B-spline discretization each contain $N_r = N_\theta = 32$ points.}
 \label{fig.Poisson_partialx}
\end{figure}

Figure~\ref{fig.Poisson_partialx} shows the Cartesian derivative
$
\partial g_0 / \partial x 
= \cos(\theta) \partial g_0/\partial r 
- 1/r \sin(\theta) \partial g_0/\partial \theta
$
for both the analytical and numerical cubic B-spline solutions within the innermost circle.

Figure~\ref{fig.Poisson_Dx_ref} shows $\partial g_0/\partial x$ for the analytical solution.  
Figure~\ref{fig.Poisson_Dx_C0} shows the discretization error
\begin{equation} \label{eq.eps_partial}
   \varepsilon_{\partial g_0/\partial x} 
   \coloneqq \frac{\partial g_{0,\mathrm{num}}}{\partial x} 
         - \frac{\partial g_{0,\mathrm{ana}}}{\partial x}
\end{equation}
for $C^0$-regularity at the origin (contour plot), and Fig.~\ref{fig.Poisson_Dx_C1} shows the error for $C^1$-regularity.
In addition, Fig.~\ref{fig.Poisson_Dx_C3} shows $\partial g_0/\partial x$ for $C^\infty$-regularity.

For $C^0$-regularity, $\partial g_0/\partial x$ is discontinuous at the center;  
for $C^1$-regularity it is continuous but not smooth.  
Most notably, in Fig.~\ref{fig.Poisson_Dx_C3}, $\partial g_0/\partial x$ is nearly zero,  
because the $m=4$ mode of the analytic solution cannot be represented:  
it is excluded by the $C^\infty$-regularity constraint in the innermost region (see Sec.~\ref{sec:regularity}).

This confirms our expectation that the $C^\infty$-regularity at the origin filters out the high harmonics with $m>p$ in the innermost region. This is an important property, particularly in the presence of high-frequency modes induced, for example, by statistical noise in PIC codes (see next section).

\section{Impact of the high-order regularity method on particle-in-cell codes}
\label{sec:numerical_results_PIC}

In this section, we assess the impact of the high-order regularity method on PIC codes, providing a combined theoretical and numerical analysis.

\subsection{Mitigating PIC statistical noise near the origin}

Simulation markers (Monte Carlo particles) represent $\delta$-functions in phase space and therefore induce statistical noise in quantities such as the charge density. This noise is global: it affects the entire domain and spans all spatial scales. At any given radial position on the unit disc, the noise excites the entire available Fourier spectrum up to the angular grid resolution. If full $C^\infty$-regularity is not enforced at the origin, the resulting particle noise also couples into all existing non-regular modes.

We solve the two-dimensional Poisson problem (see Eq.~\eqref{eq.interior_Poisson}) with the right-hand side $f_0$~(see Eq.~\eqref{eq.y_Poisson_f0}) for $m=4$. The load vector~$\boldsymbol{f}_\mathrm{MC}$ is assembled by the PIC method for cubic B-splines. The markers are uniformly distributed, i.e.\ the sampling distribution~$g$ is constant in phase space (see Eq.~\eqref{eq.g_density}). The grid number is $N_r = N_\theta = 32$, and the average number of particles per cell is $N_\mathrm{cell} = 80$. We distinguish the cases of no smoothness constraint, and $C^0$-, up to $C^\infty$-regularity at the origin.

\begin{figure}[htbp]
\centering
\includegraphics[width=0.6\linewidth]{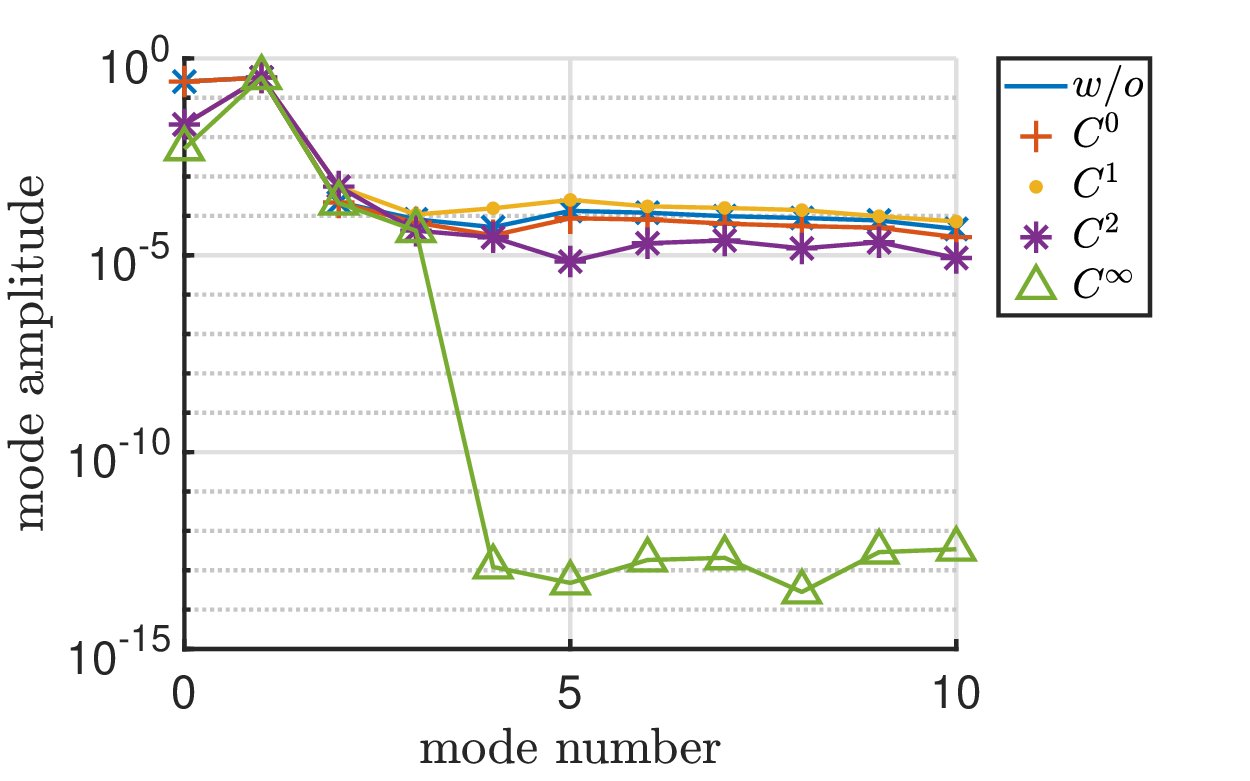}
\caption{Amplitude on a logarithmic scale of Fourier modes $m = 0$ to $10$ in the solution $\phi$ at the innermost radial grid interval, obtained by Fourier decomposition in $\theta$ at $\tilde{r}=0.5$. Results are plotted for cases without regularity (blue line) and with regularity from $C^0$ to $C^\infty$ at the origin. Particle noise dominant for $m>4$ is included in the load vector $\boldsymbol{f}_\mathrm{MC}$ during solution of the Poisson equation. The radial and angular grids in the cubic B-spline discretization each contain $N_r = N_\theta = 32$ points.}
\label{fig.Poisson_PIC_FFT}
\end{figure}

Figure~\ref{fig.Poisson_PIC_FFT} shows the amplitude of the Fourier components $0 \le m \le 10$ of the solution~$\phi$, computed using different orders of regularity within the innermost interval. The Fourier decomposition is carried out at equidistant points along the $\theta$-direction and at a fixed radial position $\tilde{r} = 0.5$ in the innermost radial interval. The smooth center-splines serve as a Fourier filter locally near the origin, allowing harmonics with $m \le p$ while suppressing those with $m > p$ within the innermost interval. This filtering effect arises only when full $C^\infty$-regularity is imposed at the origin and does not distinguish between physical signal and numerical noise.

Due to the regularity constraint, the radial dependence of each harmonic mode $m$ is proportional to $r^l$ with $l \ge |m|$ (see Eq.~\eqref{eq.crude_reg_cond}). Since for cubic B-splines~$p{=}3$, the maximum allowed radial exponent is $l_\mathrm{max} = 3$, and thus the maximum harmonic mode is $m_\mathrm{max} = 3$. Consequently, modes with $m > 3$ are expected to have significantly reduced magnitude in the innermost region. Ideally, they should vanish entirely in the continuum limit. In Figure~\ref{fig.Poisson_PIC_FFT}, we observe (green line) that modes with $m > 3$ have amplitudes approximately eight orders of magnitude smaller than those with $m \le 3$.
For $C^n$-regularities with $n < 3$, this filtering effect is absent because not all B-splines contributing to the innermost interval are subject to the regularity constraints. These B-splines retain their degrees of freedom in the $\theta$-direction.

To prevent an abrupt suppression of Fourier modes $m > p$ at the boundary between the innermost and second innermost radial intervals, the interval size $\Delta r$ should be sufficiently small. In such a case, the amplitude of the Fourier modes $m > p$, which scale radially at least as $r^m$, is already sufficiently small within the innermost interval where $r \le \Delta r$. Thus, in the limit $\Delta r \to 0$, the transition is smooth, and no numerical artifacts are expected.


\subsubsection{Theory of variance and error propagation}

In the PIC method, the particle distribution function $f$ is estimated using a Monte Carlo approach, which inherently introduces a statistical standard error~\citep{hatzky2019reduction}. Physical quantities in PIC simulations are obtained by evaluating integrals over the phase space~$\Omega_z$:
\begin{equation} \label{eq.lambda_integral}
    I(\Lambda) \coloneqq \int_{\Omega_z} \Lambda(\boldsymbol{z}) f(\boldsymbol{z}) \, \mathcal{J}_z \, \mathrm{d}^6 z.
\end{equation}
The markers have a sampling distribution function $g(\boldsymbol{z})$ that is normalized:
\begin{equation} \label{eq.g_density}
     \int_{\Omega_z} g(\boldsymbol{z}) \, \mathcal{J}_z \, \mathrm{d}^6 z = 1, \quad g(\boldsymbol{z}) > 0.
\end{equation}

We define the expectation of a random variable $X$ with respect to the probability density $g(\boldsymbol{z})$
\begin{equation}
    \mathrm{E}_g(X) \coloneqq \int_{\Omega_z} X(\boldsymbol{z}) g(\boldsymbol{z}) \, \mathcal{J}_z \, \mathrm{d}^6 z.
\end{equation}

We now define a random variable $\widetilde{X}$ so that, regardless of the choice of sampling distribution~$g$, its expected value equals the integral $I(\Lambda)$:
\begin{equation}
    I(\Lambda) = \mathrm{E}_g(\widetilde{X}) , \qquad \widetilde{X}(\boldsymbol{z}) \coloneqq \frac{\Lambda(\boldsymbol{z}) f(\boldsymbol{z})}{g(\boldsymbol{z})}.
\end{equation}
In practice, we approximate $\mathrm{E}_g(\widetilde{X})$ by an unbiased Monte Carlo estimator by sampling~$\boldsymbol{z}_p$ from~$g$ 
\begin{equation} \label{eq.MC_estimator}
\mathrm{E}_g(\widetilde{X}) \approx \frac{1}{N_p} \sum_{p=1}^{N_p} \Lambda(\boldsymbol{z}_p) w_p \pm \epsilon_E,
    \qquad w_p \coloneqq \frac{f(\boldsymbol{z}_p)}{g(\boldsymbol{z}_p)},
\end{equation}
where $N_p$ is the total number of markers.
The statistical standard error of $\mathrm{E}[\widetilde{X}]$ is given by:
\begin{equation}
    \epsilon_E \simeq \frac{\sigma_g}{\sqrt{N_p}},
\end{equation}
where the standard deviation is defined by
\begin{equation}
    \sigma_g \coloneqq \sqrt{\mathrm{Var}_g[X]}
\end{equation}
with the variance
\begin{equation}
\mathrm{Var}_g[X] \coloneqq \mathrm{E}_g(X^2) - (\mathrm{E}_g(X))^2.
\end{equation}
If we insert the specific random variable~$\widetilde{X}$ we get
\begin{equation} \label{eq.variance}
    \mathrm{Var}_g[\widetilde{X}] = \int_{\Omega_z} \Lambda^2(\boldsymbol{z}) \frac{f^2(\boldsymbol{z})}{g(\boldsymbol{z})} \mathcal{J}_z \, \mathrm{d}^6 z - I(\Lambda)^2.
\end{equation}
We use an unbiased Monte Carlo estimator to evaluate
\begin{equation}
     \mathrm{Var}_g[\widetilde{X}] \approx \frac{1}{N_p-1}\left[ \sum_{p=1}^{N_p}\big(\Lambda(\boldsymbol{z}_p) w_p\big)^2 
        - \frac{1}{N_p} \left( \sum_{p=1}^{N_p} \Lambda(\boldsymbol{z}_p) w_p \right)^2 \right] + \epsilon_{\mathrm{Var}_g}.
\end{equation}
Here $\epsilon_{\mathrm{Var}_g}$ denotes the standard error of the variance estimator itself, which scales as $\mathcal{O}(1/\sqrt{N_p})$ for independent sampling.
Unlike the expected value, the variance depends on the particle distribution~$g$.

We define the covariance by
\begin{equation} \label{eq.covariance}
    \mathrm{Cov}_g[X,Y] \coloneqq  \mathrm{E}_g(XY) - \mathrm{E}_g(X)\mathrm{E}_g(Y).
\end{equation}
Using the exact integral formulation from Eq.~\eqref{eq.lambda_integral}, the analytical covariance is
\begin{equation} \label{eq.covariance_Lambda}
\begin{aligned}
\mathrm{Cov}_g[\widetilde{X}_k,\widetilde{Y}_{k'}]
= {} & \int_{\Omega_z} \Lambda_k(\boldsymbol{z}) \Lambda_{k'}(\boldsymbol{z}) \frac{f(\boldsymbol{z})^2}{g(\boldsymbol{z})} \mathcal{J}_z \, \mathrm{d}^6 z - I(\Lambda_k) I(\Lambda_{k'}),
\end{aligned}
\end{equation}

To determine the statistical standard error of the load vector $\boldsymbol{f}$ in Eq.~\eqref{eq.stiff_org}, we introduce the covariance matrix $\Sigma_{\hat{\boldsymbol{f}}}$ of the Monte Carlo estimator (see Eq.~\eqref{eq.MC_estimator}), defined as
\begin{equation} \label{eq.MC_esti_covar}
    (\Sigma_{\hat{\boldsymbol{f}}})_{k,k'} \coloneqq \frac{1}{N_p} \mathrm{Cov}_g[\widetilde{X}_k,\widetilde{Y}_{k'}],
    \qquad
    \Lambda_k(\boldsymbol{z}) \coloneqq B_k (\boldsymbol{r}).
\end{equation}

The diagonal entries contain the square of the statistical standard error while the off-diagonal entries contain the statistical covariances between different $B_k(\boldsymbol{r})$ estimates. We compute $(\Sigma_{\hat{\boldsymbol{f}}})_{k,k'}$ as
\begin{equation}
    (\Sigma_{\hat{\boldsymbol{f}}})_{k,k'} = \frac{1}{N_p} \left(\overline{\matcomp{m}}_{k,k'} - \overline{f}_k \overline{f}_{k'} \right),
\end{equation}
where the generalized mass matrix~$\overline{\mathsfbi{M}}$ and load  vector~$\boldsymbol{f}$ are defined by
\begin{equation}
    \overline{\matcomp{m}}_{k,k'} \coloneqq \int_{\Omega_z} B_k(\boldsymbol{r}) B_{k'}(\boldsymbol{r}) \frac{f(\boldsymbol{z})^2}{g(\boldsymbol{z})} \mathcal{J}_z \, \mathrm{d}^6 z \quad \text{and} \quad
    \overline{\boldsymbol{f}}_k \coloneqq \int_{\Omega_z} B_k(\boldsymbol{r}) f(\boldsymbol{z}) \, \mathcal{J}_z \, \mathrm{d}^6 z.
\end{equation}
We explicitly distinguish $\overline{f}_k$ as the phase‑space moment (integral over $\Omega_z$), while $f_k$ in Eq.~\eqref{eq.stiff_org} is the corresponding real‑space load vector after velocity integration. For basis functions with disjoint support, e.g.\ constant B-splines, $\overline{\matcomp{m}}_{k,k'} = 0$ when $k \neq k'$, but $\overline{f}_k \overline{f}_{k'} > 0$, so the subtraction yields negative off-diagonal covariances: this results in an anti‑correlation between their estimates. In some cases, the contribution of the term~$-\overline{\boldsymbol{f}} \, \overline{\boldsymbol{f}}^\mathrm{T}$ is negligible. 

When we solve Eq.~\eqref{eq.stiff_org}, the statistical error also propagates to the solution vector $\boldsymbol{\phi} = \mathsfbi{S}^{-1} \boldsymbol{f}$.  
The covariance matrix $\Sigma_\phi$ can be evaluated by a congruence transformation:
\begin{align}\label{eq.covariance_phi}
     \Sigma_\phi &= \mathsfbi{S}^{-1} \Sigma_{\hat{\boldsymbol{f}}} \bigl(\mathsfbi{S}^{-1} \bigr)^\mathrm{T}.
\end{align}
Since the solution $\phi(\boldsymbol{r})$ is expanded in basis functions (see Eq.~\eqref{eq.2d_bspline_phi_org_single}), the random variables are the coefficients~$\phi_i$, while the basis functions~$B_k(\boldsymbol{r})$ are deterministic.
When we propagate the covariance of the coefficients $\Sigma_\phi$ to obtain the variance of the function $\phi(\boldsymbol{r})$, we use the standard linear error‑propagation formula:
\begin{equation} \label{eq.error_prop}
    \mathrm{Var}\big[\phi(\boldsymbol{r})\big] =
    \sum_{k=0}^{N-1} \sum_{k'=0}^{N-1} 
    \frac{\partial \phi(\boldsymbol{r})}{\partial \phi_k}
    (\Sigma_\phi)_{k,k'}
    \frac{\partial \phi(\boldsymbol{r})}{\partial \phi_{k'}}.
\end{equation}
Because $\phi(\boldsymbol{r})$ depends linearly on each coefficient (see Eq.~\eqref{eq.2d_bspline_phi_org_single}), we have
$\partial \phi(\boldsymbol{r}) / \partial \phi_k = B_k(\boldsymbol{r})$, where we
define the basis-function vector at $\boldsymbol r$ as
$\boldsymbol{B}(\boldsymbol r) \coloneqq \bigl(B_0(\boldsymbol r),\dots,B_{N-1}(\boldsymbol r)\bigr)^{\mathrm T}$. Inserting this into Eq.~\eqref{eq.error_prop} yields the quadratic form
\begin{equation} \label{eq.var_phi_matrix}
    \mathrm{Var}\big[\phi(\boldsymbol{r})\big]
    = \boldsymbol{B}^\mathrm{T}(\boldsymbol{r}) \, \Sigma_\phi \, \boldsymbol{B}(\boldsymbol{r}).
\end{equation}

Using equation~\eqref{eq.covariance_phi} for the smooth polar-spline basis, the covariance matrix~$\Sigma_{\widetilde\phi}$ is:
\begin{align}
     \Sigma_{\widetilde\phi} &= \mathsfbi{\widetilde S}^{-1} \Sigma_{\widetilde{\boldsymbol{f}}} \bigl(\mathsfbi{\widetilde S}^{-1} \bigr)^\mathrm{T}
   = \mathsfbi{\widetilde S}^{-1} \bigl( \mathsfbi{P} \Sigma_{\hat{\boldsymbol{f}}} \mathsfbi{P}^\mathrm{T} \bigr) \bigl(\mathsfbi{\widetilde S}^{-1}\bigr)^\mathrm{T}.
\end{align}
Finally, for the smooth polar-spline basis, the point‑wise variance of the reconstructed function $\widetilde{\phi}(\boldsymbol{r})$ is:
\begin{equation} \label{eq.var_phi_matrix_tiled}
    \mathrm{Var}\big[\widetilde{\phi}(\boldsymbol{r})\big] = \boldsymbol{\widetilde{B}}^\mathrm{T}(\boldsymbol{r}) \Sigma_{\widetilde\phi} \boldsymbol{\widetilde{B}}(\boldsymbol{r}).
\end{equation}

\subsubsection{Numerical examples for statistical error}

For simplicity, we now restrict ourselves to the special case where both the marker distribution $g$ 
and the particle distribution~$f$ are constant.
The standard deviations $\sigma_\phi$ shown in the following figures are computed from Eqs.~\eqref{eq.var_phi_matrix} and~\eqref{eq.var_phi_matrix_tiled} using the propagated covariance $\Sigma_\phi$ from Eq.~\eqref{eq.covariance_phi}, applied to the corresponding discrete system. 
In the cases shown first, $\Sigma_\phi$ is obtained from the mass-matrix solve via Eq.~\eqref{eq.covariance_phi}.

\begin{figure}[htbp]
\centering
\includegraphics[width=0.6\linewidth]{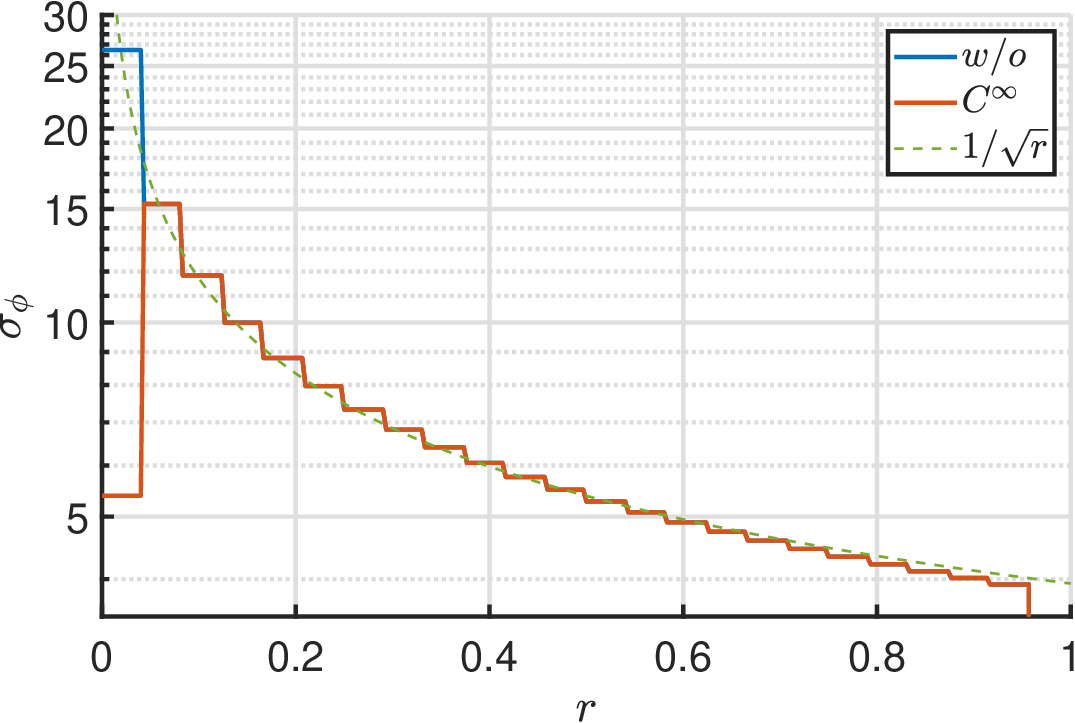}
\caption{Standard deviation $\sigma_\phi$ of the constant test function~$\phi(r)=1$, obtained from the mass matrix equation~\eqref{eq.mass_org} on the unit disc with homogeneous Dirichlet boundary conditions, plotted as a function of~$r$ along $\theta=0$. Results are plotted on a logarithmic scale for constant B-splines ($p{=}0$), both without and with $C^\infty$-regularity at the origin, on a grid with $N_r = N_\theta = 24$ points. Without regularity (blue line), $\sigma_\phi$ scales as $1/\sqrt{r}$ (dashed green line) due to the inverse dependence of the variance on the polar bin area $S(r)$. With $C^\infty$-regularity (red line), the standard deviation at the innermost B-spline drops significantly because $\widehat{B}_0^0$ spans the entire inner region.}
\label{fig:Std_phi_r_0}
\end{figure}

Figure~\ref{fig:Std_phi_r_0} shows the standard deviation~$\sigma_\phi$ for $\phi(r) = 1$ with homogeneous Dirichlet boundary conditions. The result is obtained from the mass matrix equation~\eqref{eq.mass_org} on the unit disc  using constant B‑splines, both without and with $C^0$‑regularity. The results are plotted on a logarithmic scale as a function of the radius~$r$ along $\theta=0$. The radial and angular grids each contain $N_r = N_\theta = 24$ points. 
In this case, the contribution of the term~$-\overline{\boldsymbol{f}} \, \overline{\boldsymbol{f}}^\mathrm{T}$ is negligible. 
Without regularity (blue line), $\sigma_\phi$ scales as $\sigma_\phi(r) \propto 1/\sqrt{S(r)} \propto 1/\sqrt{r}$ (dashed green line), which follows from the fact that the variance~$\mathrm{Var}_\phi$ scales inversely with the bin area $S(r)$, which in polar coordinates, for a bin of fixed $\Delta r$ and $\Delta \theta$, is given by $S(r) \approx r \, \Delta r\, \Delta\theta$. 
In the case of constant B‑splines, the standard deviation~$\sigma_\phi$ and hence the standard error increase monotonically towards the origin at $r=0$. 
With $C^0$‑regularity (red line), $\sigma_\phi$ at the innermost B‑spline drops dramatically because $\widehat{B}_0^0$ covers the whole inner region (see App.~\ref{App.C0-regularity}).

\begin{figure}[htbp]
\centering
\includegraphics[width=0.6\linewidth]{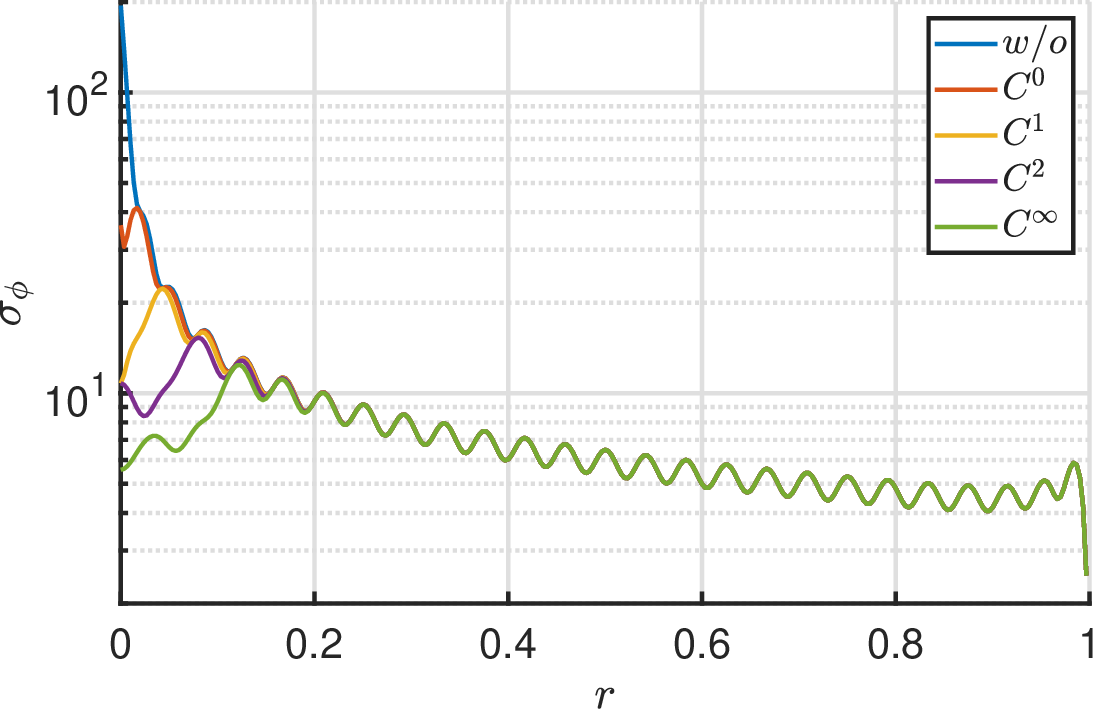}
\caption{Standard deviation $\sigma_\phi$ of the constant test function~$\phi(r) = 1$ with homogeneous Dirichlet boundary conditions to the mass matrix equation~\eqref{eq.mass_org} on the unit disc, shown as a function of $r$ along $\theta = 0$, for cubic B‑splines ($p{=}3$). 
Results are plotted on a logarithmic scale for cases without regularity (blue line) and with regularity from $C^0$ to $C^\infty$ at the origin, on a grid with $N_r = N_\theta = 24$ points. 
Without regularity, $\sigma_\phi$ scales as $1/\sqrt{r}$ across the full radial range. 
The repeated knots at $r=0$ and $r=1$ clamp the basis functions at both boundaries. At the origin, the steep $1/\sqrt{r}$ scaling combines with the maximum of the innermost basis function $B_0$ to strongly amplify $\sigma_\phi$ compared to the constant B‑spline case. 
At $r=1$, clamping modifies the outermost basis shape, but the $1/\sqrt{r}$ scaling varies little over its support, so the effect is much weaker and the drop is dominated by the Dirichlet boundary condition. Increasing regularity reduces the central peak, with $C^\infty$-regularity lowering it by a factor of~30.}
\label{fig:Std_phi_3}
\end{figure}

Figure~\ref{fig:Std_phi_3} shows the same constant–$\phi$ test case as in Fig.~\ref{fig:Std_phi_r_0}, 
but now using cubic B‑splines with regularity ranging up to~$C^\infty$. 
Without regularity, $\sigma_\phi$ follows the $1/\sqrt{r}$ scaling across the entire radial range, as expected from the bin‑area argument discussed for Fig.~\ref{fig:Std_phi_r_0}. 
Superimposed on this baseline scaling is the effect of the clamped cubic B‑spline basis at both $r=0$ and $r=1$. 
At the origin, the innermost basis function~$B_0$ attains its maximum exactly at $r=0$ due to the repeated end knots, and here the steep $1/\sqrt{r}$ scaling amplifies this maximum to produce a $\sigma_\phi$ value about seven times larger than in the constant B‑spline case. 
At the outer edge, clamping also modifies the spline shape, but the $1/\sqrt{r}$ scaling varies only weakly over the support of the outermost spline, so its contribution is small and the observed drop is dominated by the Dirichlet boundary condition. 
Oscillations in the curves occur because the local standard deviation peaks whenever two or more B‑splines have maximal overlap, leading to additive variances. 
Introducing $C^0$‑regularity significantly reduces the central peak, and increasing the regularity up to~$C^\infty$ continues this trend. 
With $C^\infty$‑regularity at the origin, $\sigma_\phi$ is reduced by a factor of about~30 compared to the case without regularity, and the first four innermost radial splines are likewise affected, as the smooth center‑splines up to $\widehat{B}_3^{\pm 3}$ possess far fewer degrees of freedom than their tensor‑product B‑spline counterparts.

\begin{figure}[htbp]
\centering
\includegraphics[width=0.6\linewidth]{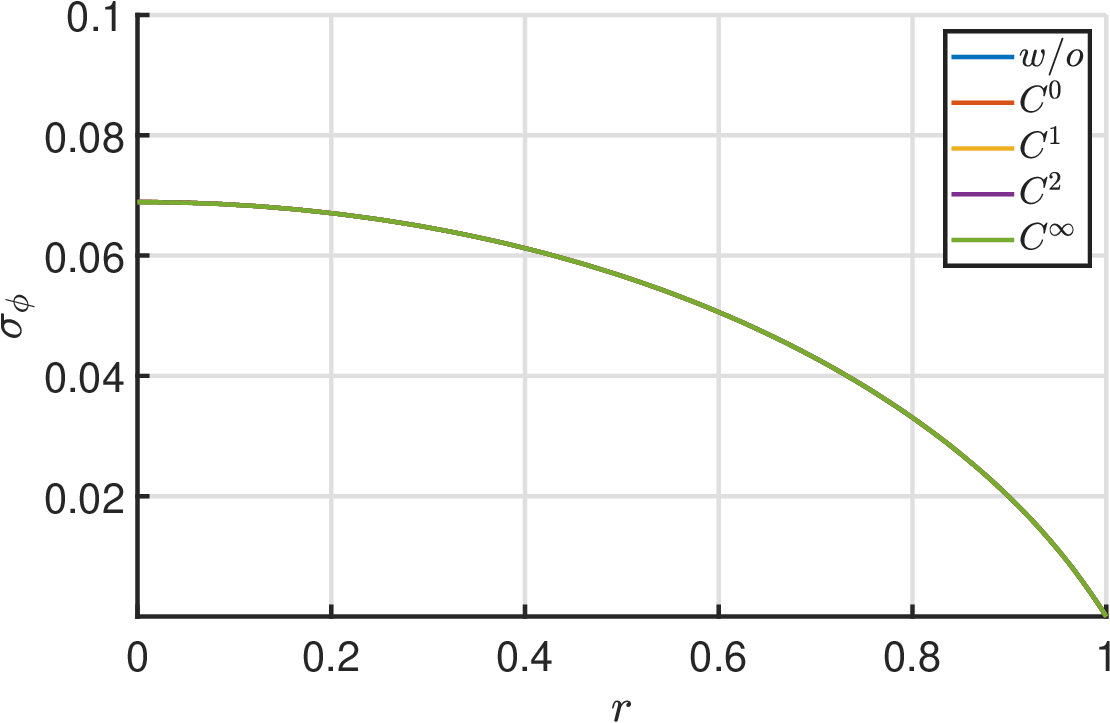}
\caption{Standard deviation $\sigma_\phi$ of the analytic solution $\phi(r) = (1-r^2)/4$ to the Poisson equation with homogeneous Dirichlet boundary conditions on the unit disc, discretized using cubic B‑splines without and with various regularity degrees ($C^0$–$C^\infty$) at the origin. The $\sigma_\phi$ profile is dominated by the analytic parabola shape and is nearly identical for all regularity cases.}
\label{fig:Std_phi_poisson}
\end{figure}

Figure~\ref{fig:Std_phi_poisson} shows the standard deviation~$\sigma_\phi$ of the analytic solution~$\phi(r) = (1-r^2)/4$ with homogeneous Dirichlet boundary conditions to the stiffness matrix equation~\eqref{eq.stiff_org} on the unit disc for cubic B‑splines without and with regularity ranging from $C^0$ up to $C^\infty$. 
Here, the standard deviation is likewise obtained from Eqs.~\eqref{eq.var_phi_matrix} and~\eqref{eq.var_phi_matrix_tiled}. 
In this case, $\Sigma_\phi$ comes from the stiffness‑matrix solve via Eq.~\eqref{eq.covariance_phi}. 
The results are plotted as a function of the radius~$r$ along $\theta=0$. 
The radial and angular grids each contain $N_r = N_\theta = 24$ points. 
The~$\sigma_\phi$ profile has its maximum at the center~$r=0$ and decreases smoothly to zero at the boundary~$r=1$, closely following the analytic solution shape $(1-r^2)/4$.
In this constant‑source Poisson case, the stiffness‑matrix inversion of a constant load produces a smooth profile whose shape is dominated by the analytic solution, so regularity of the basis at the origin has no visible effect on the standard deviation. All regularity cases ($C^0$--$C^\infty$) yield overlapping curves, and no oscillations are visible because the PDE solve smooths out local variations from overlaps of the basis elements.

\subsection{Charge conservation}

\subsubsection{Tensor-product B-spline basis}
\label{Sec.charge_cons_Bspline}

Charge conservation is one of the most important conditions in particle-in-cell (PIC) simulations, and for simplicity, we assume that all particles have the same charge. The charge density is given by
\begin{equation}\label{eq.rho_exact}
    \rho(\boldsymbol{r}) = q \sum_{p=1}^{N_p} \delta(\boldsymbol{r} - \boldsymbol{r}_p),
\end{equation}
where $\boldsymbol{r}_p$ is the position of marker $p$, and $N_p$ is the total number of markers. For the tensor-product B-spline basis $\boldsymbol{B}$ the approximated charge distribution function $\rho_h$ is
\begin{equation}\label{eq.rho_approx}
    \rho_h(\boldsymbol{r}) = \sum_{k'=0}^{N-1} \rho_{k'} B_{k'}(\boldsymbol{r}),
\end{equation}
where $k'$ is a single index.

The coefficients $\rho_{k'}$ are obtained via Galerkin projection, i.e.\ we require that the projections of $\rho_h$ and $\rho$ onto each basis function $B_k$ are equal:
\begin{equation}\label{eq.rho_projected}
    \int_\Omega \rho_h(\boldsymbol{r}) B_k(\boldsymbol{r}) \, \mathrm{d} \boldsymbol{r} = \int_\Omega \rho(\boldsymbol{r}) B_k(\boldsymbol{r}) \, \mathrm{d} \boldsymbol{r},
    \qquad (k = 0, \ldots, N-1).
\end{equation}
Summing Eq.~\eqref{eq.rho_projected} over all values of $k$, and exchanging the summation and integration, we derive
\begin{equation}\label{eq.charge_cons_aux1}
     \int_\Omega \rho_h(\boldsymbol{r}) \left( \sum_{k=0}^{N-1} B_k(\boldsymbol{r}) \right) \mathrm{d} \boldsymbol{r} = \int_\Omega \rho(\boldsymbol{r}) \left( \sum_{k=0}^{N-1} B_k(\boldsymbol{r}) \right) \mathrm{d} \boldsymbol{r}.
\end{equation}
Using the partition of unity, Eq.~\eqref{eq.unity_tensor}, we can further simplify,
\begin{equation}\label{eq.charge_conservation}
    \int_\Omega \rho_h(\boldsymbol{r}) \, \mathrm{d} \boldsymbol{r} = \int_\Omega \rho(\boldsymbol{r}) \, \mathrm{d} \boldsymbol{r}.
\end{equation}
This is the proof of charge conservation, showing that the total charge of the particles is conserved after discretization by the tensor-product B-spline basis, since the constant function (the harmonic $m=0$ mode) is exactly representable due to the partition-of-unity property.

\subsubsection{Smooth polar-spline basis}
\label{Sec.charge_cons_regularity_spline}
Unfortunately, for the smooth polar-spline basis $\widehat{\boldsymbol{B}}$, the partition of unity is not satisfied. Nevertheless, we can generalize Eq.~\eqref{eq.charge_cons_aux1} by summing Eq.~\eqref{eq.rho_projected} for some values of $k$ after multiplying it by weighting factors $c_k$:
\begin{equation}\label{eq.charge_cons_aux2}
     \int_\Omega \widehat{\rho}_h(\boldsymbol{r}) \left( \sum_k c_k \widehat{B}_k(\boldsymbol{r}) \right) \mathrm{d} \boldsymbol{r} = \int_\Omega \rho(\boldsymbol{r}) \left( \sum_k c_k \widehat{B}_k(\boldsymbol{r}) \right) \mathrm{d} \boldsymbol{r}.
\end{equation}
As we can see, it is a sufficient condition for charge conservation whenever the basis is able to express the constant function even with only a subset of the basis functions; that is, there exists a subset $\mathcal{S}$ and weights $c_k$ such that
\begin{equation}
    \sum_{k\in\mathcal{S}} c_k \widehat{B}_k(\boldsymbol{r}) = 1, \qquad \forall \boldsymbol{r} \in \Omega
\end{equation}

For the smooth polar-spline basis, we use the fact that it is possible to constitute unity by a linear combination of the constant smooth center-spline basis function~$\widehat{B}_{0,C^n}^0$ together with the original basis functions~$B_k(r,\theta)$, by choosing $c_k = 1$ for all remaining original basis functions:
\begin{equation}
   c_0 \widehat{B}_{0,C^n}^0 + \sum_{i=n+1}^{N_r-1} \sum_{j=0}^{N_\theta-1} B_{r,i}(r)B_{\theta,j}(\theta) = \sum_{i=0}^{N_r-1} \sum_{j=0}^{N_\theta-1} B_{r,i}(r)B_{\theta,j}(\theta) = 1,
\end{equation}
where we have used 
\begin{equation}
    \widehat{B}_{0,C^n}^0 \coloneqq \frac{1}{c_0} \sum_{i=0}^{n} \sum_{j=0}^{N_\theta-1} B_{r,i}(r)B_{\theta,j}(\theta),
\end{equation}
and the partition of unity, Eq.~\eqref{eq.unity_tensor}, of the original B-spline basis. Thus, we have proven that the smooth polar-spline basis conserves charge, provided that the constant mode is part of the combination. Note that the factor~$1/{c_0}$ stems from the normalization of the basis function~$\widehat{B}_{0,C^n}^0$.

\subsection{Application of high-order regularity in the gyrokinetic PIC code \textsc{EUTERPE}}

\textsc{EUTERPE}~\citep{kleiber2024euterpe} is a global (full-volume) gyrokinetic PIC code for plasma turbulence simulations in stellarator geometry. 
It solves the multi-species electromagnetic gyrokinetic equations in the parallel magnetic potential~$\delta A_{\shortparallel}$ approximation, where $\delta A_{\shortparallel}$ is the component of the perturbed vector potential along the magnetic field.
The code employs the pullback mixed-variables method~\citep{Kleiber2016} to allow for larger time-step sizes and to avoid the well-known ``cancellation problem''~\citep{hatzky2019reduction}. 
For noise reduction, the $\delta f$ method is adopted to avoid resolving the background distribution with markers. 
As boundary conditions, it applies the so-called ``unicity'' inner boundary condition, ensuring that the potentials are continuous and single-valued at the magnetic axis; this corresponds to a $C^0$-regularity condition. 
At the outer boundary, \textsc{EUTERPE} applies a homogeneous Dirichlet condition.

In \textsc{EUTERPE}, the flux-surface label
\begin{equation}
s \coloneqq \sqrt{\frac{\psi_\mathrm{T}}{\psi_\mathrm{T}(a)}}
\end{equation}
is defined via the normalized toroidal flux~$\psi_\mathrm{T}$, with $a$ denoting the minor radius at the plasma edge. 
For a smooth, regular MHD equilibrium (no magnetic islands or singularities near the axis), the toroidal flux satisfies $\psi_\mathrm{T} \propto r^{2}$ to leading order, implying $s \approx r$ as $r \to 0$.
In the special case of a perfect cylinder, one has $s = r$ exactly. 
Near the magnetic axis, the poloidal angle $\vartheta$ in flux coordinates approaches the geometric polar angle~$\theta$ due to the locally cylindrical metric. 
Therefore, the local coordinate system $(s,\vartheta)$ is asymptotically equivalent to polar coordinates $(r,\theta)$, 
and the regularity condition~\eqref{eq.crude_reg_cond} derived for the unit disc applies directly in the neighborhood of the magnetic axis.

We chose the \textsc{EUTERPE} code for the following reasons: 
First, it employs tensor-product B-spline basis functions in magnetic flux-surface coordinates, which exhibit discontinuities near the magnetic axis -- a key issue addressed by our method. 
Second, being a PIC code, the statistical noise inherent in PIC simulations can produce high numerical noise levels at the origin, which our method can effectively mitigate.

\subsubsection{Simulation of toroidal Alfv\'en eigenmode}

The fast-particle-driven Toroidal Alfv\'en Eigenmode (TAE) instability arises when a TAE,
usually damped, is destabilized by a fast-ion species. 
We simulate this process using the well-known ITPA TAE tokamak benchmark parameters~\citep{Koenies_2018}. 
The main parameters are: the major radius $R_0 = 10~\mathrm{m}$, the minor radius $a = 1~\mathrm{m}$ of a circular tokamak, and its toroidal magnetic field on the magnetic axis $B_\mathrm{T} = B_\ast = 3~\mathrm{T}$. 
The bulk ions are hydrogen with a uniform density $n_0 = 2\times 10^{19}~\mathrm{m}^{-3}$ and a constant pressure of $6408\ \mathrm{Pa}$. 
The safety factor profile is
$
q(r) = 1.71 + 0.16(r/a)^2,
$
with $q(0.5a) = 1.75$ at mid-radius. 
The fast ions are modeled as deuterons with a Maxwellian velocity distribution and a uniform temperature $T_\mathrm{f} = 400~\mathrm{keV}$. 
The density profile of the energetic particles is
\begin{equation}
    n(s) = n_0 c_3 \exp{\bs{-\frac{c_2}{c_1} \tanh{\bk{\frac{s - c_0}{c_2}}}}},
\end{equation}
where $n_0 = 1.4431\times 10^{17}\ \mathrm{m}^{-3}$ is the energetic particle density at $s = 0$, and the coefficients are $c_0 = 0.49123$, $c_1 = 0.298228$, $c_2 = 0.198739$, and $c_3 = 0.521298$. 
This density profile has a large gradient at the mid-radius where $q = 1.75$, which drives the TAE instability.

In this case, we simulate only the linear TAE for the $n = 6$ mode. 
Quadratic B-splines are used with grid numbers $N_s = 100$, $N_\vartheta = 64$, and $N_\varphi = 16$. 
The number of markers is $N_p = 2.4\times 10^6$ for both bulk ions and fast ions, and $4.8\times 10^6$ for electrons.
A $C^0$-regularity condition and a homogeneous Dirichlet boundary condition are applied at the axis and outer boundary, respectively.

\begin{figure}[htbp]
\centering
    \subfloat[with $C^0$-regularity at the magnetic axis]{%
        \includegraphics[width=0.48\textwidth]{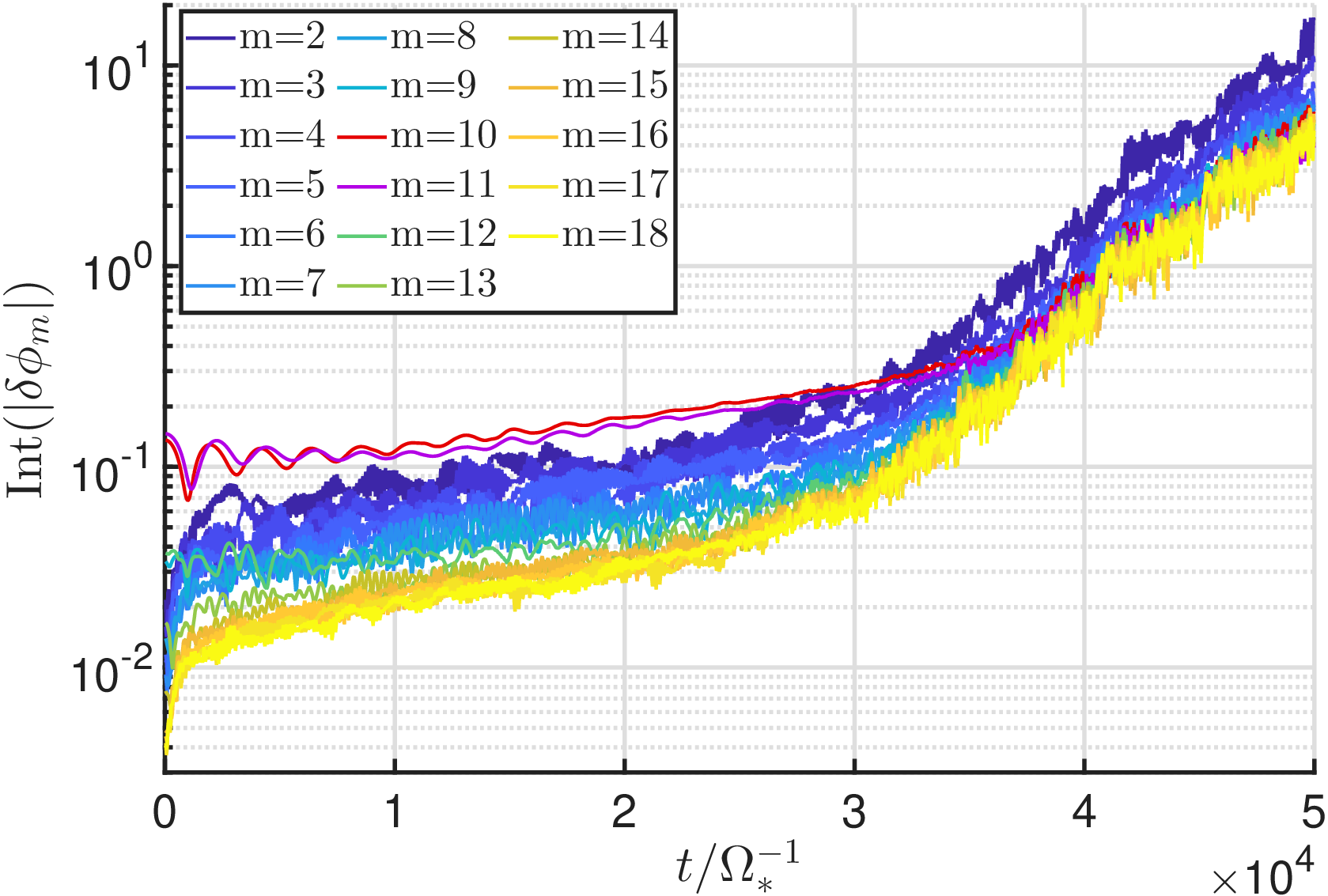}%
    \label{fig.ITPA_modes_evolution_a}}%
    \hfill
    \subfloat[with $C^\infty$-regularity at the magnetic axis]{%
        \includegraphics[width=0.48\textwidth]{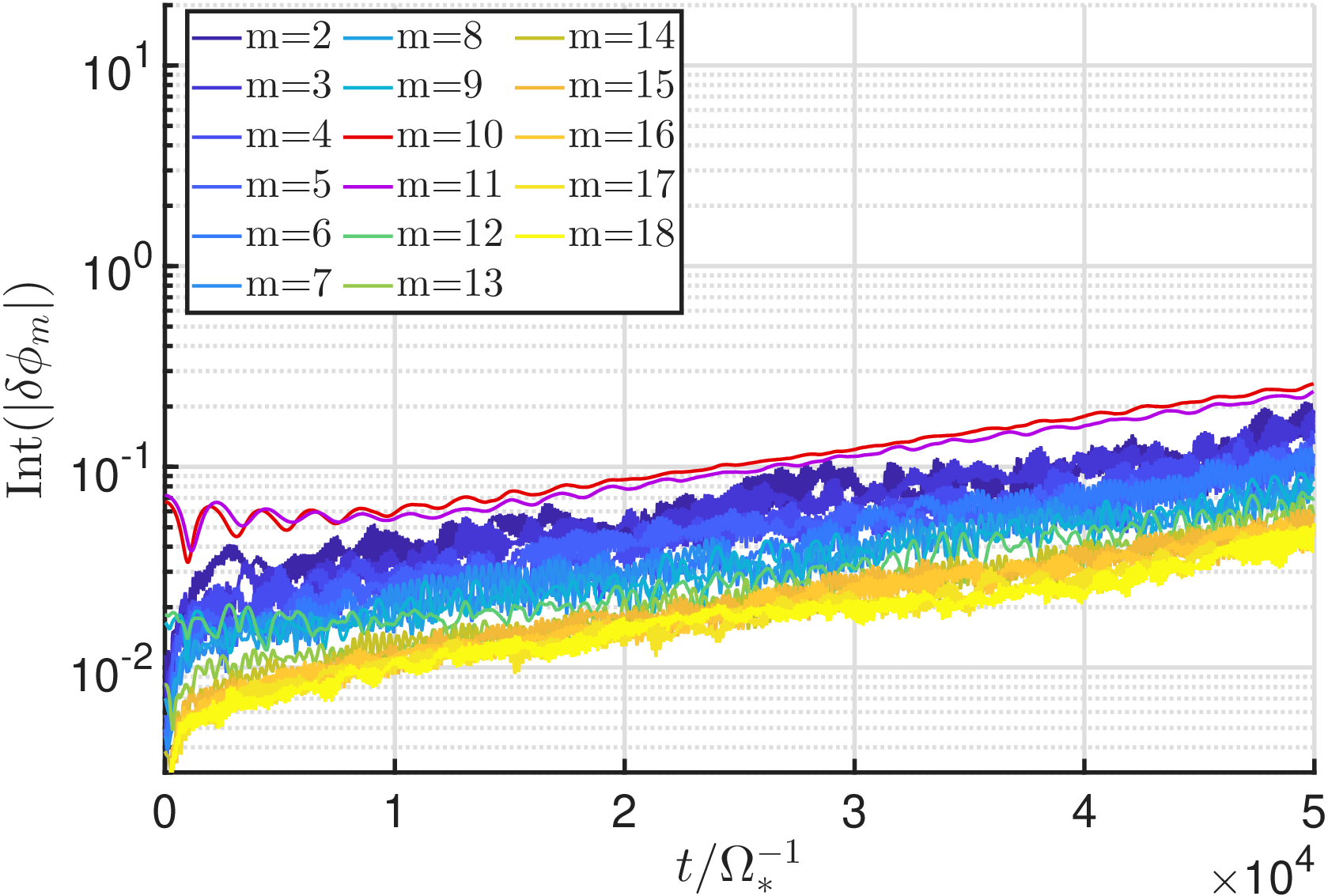}%
    \label{fig.ITPA_modes_evolution_b}}%
\caption{PIC simulation of a fast-particle-driven toroidal Alfv\'en eigenmode (TAE) instability using quadratic B-splines. The perturbed electrostatic potential, $|\delta \phi_m|$, integrated over the plasma volume and plotted on a logarithmic scale, is shown as a function of time normalized to the ion cyclotron frequency~$\Omega_\ast$. Results are shown for poloidal mode numbers in the range $m = 2$ to $18$.}
\label{fig.ITPA_modes_evolution}
\end{figure}

Unlike the \textsc{EUTERPE} results in~\citep{Koenies_2018}, which retain only the poloidal mode numbers $m \in [9,\dots,12]$ for benchmark purposes, here we include a broader range of Fourier harmonics $m \in [2,\dots,18]$ to highlight the numerical instability that develops at the magnetic axis. 
Figure~\ref{fig.ITPA_modes_evolution} shows the time evolution of the absolute value of the Fourier mode amplitude~$|\delta \phi_m|$, integrated over the plasma volume, as a function of the normalized time $t/\Omega_\ast^{-1}$, where $\Omega_\ast^{-1} \coloneqq m_\mathrm{p} / (e B_\ast)$ is the inverse ion cyclotron frequency. Figure~\ref{fig.ITPA_modes_evolution_a} is computed imposing $C^0$-regularity at the axis, while Fig.~\ref{fig.ITPA_modes_evolution_b} enforces $C^\infty$-regularity. 
Initially, the dominant modes are $m = 10$ and $m = 11$, which -- after some oscillations -- settle into steady exponential growth. 
In the case with $C^0$-regularity, from $t = 3\times 10^4\ \Omega_\ast^{-1}$ onward, the lower-$m$ modes begin to dominate, and their growth rate increases significantly. 
This behavior is attributed to numerical instability.

\begin{figure}[htbp]
\centering
    \subfloat[$|\delta \phi_m|$ as a function of~$s$]{%
        \includegraphics[width=0.5\textwidth]{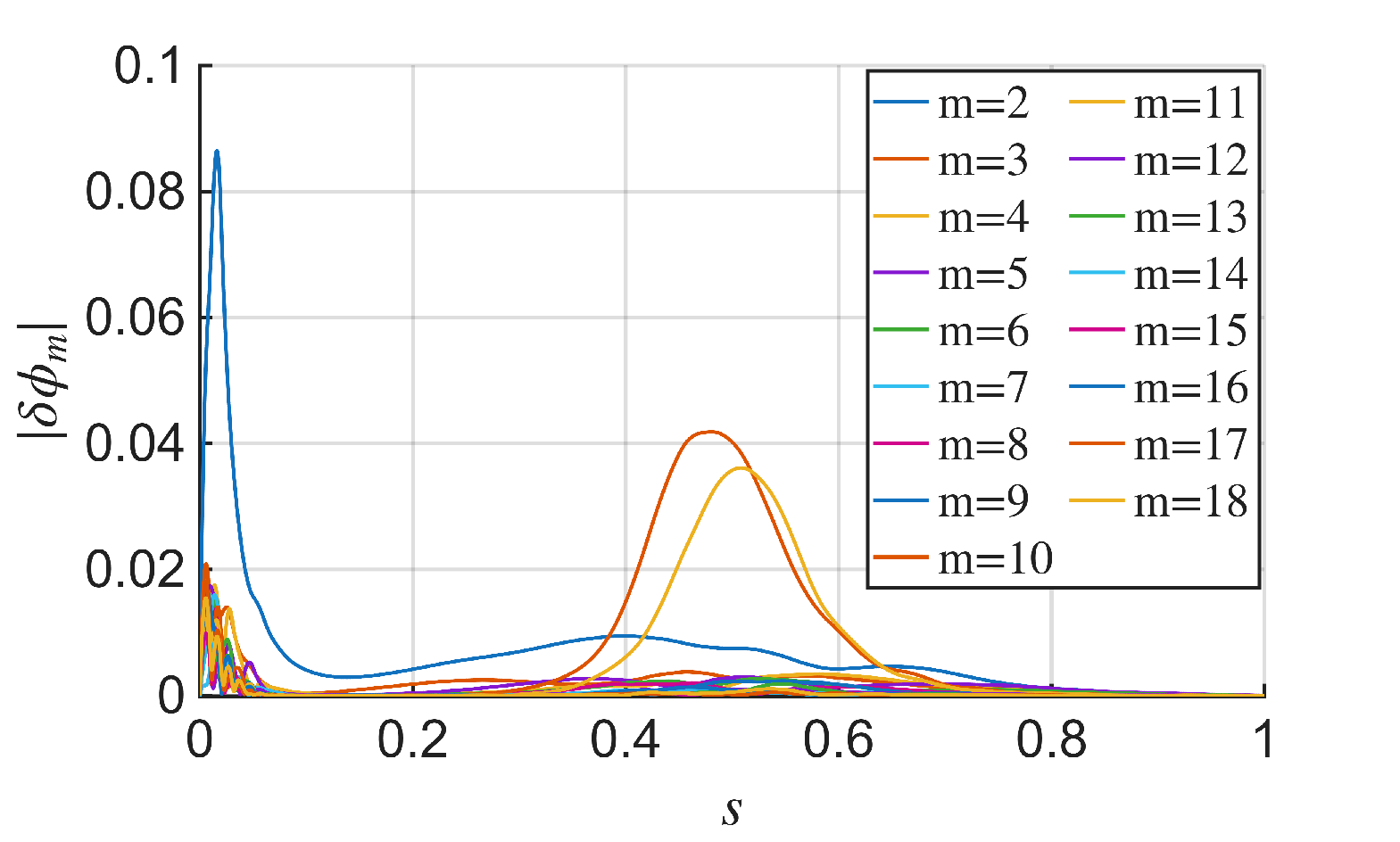}%
        \label{fig.ITPA_modes_after}} 
    \hfill
    \subfloat[Enlarged view of panel~(a)]{%
        \includegraphics[width=0.5\textwidth]{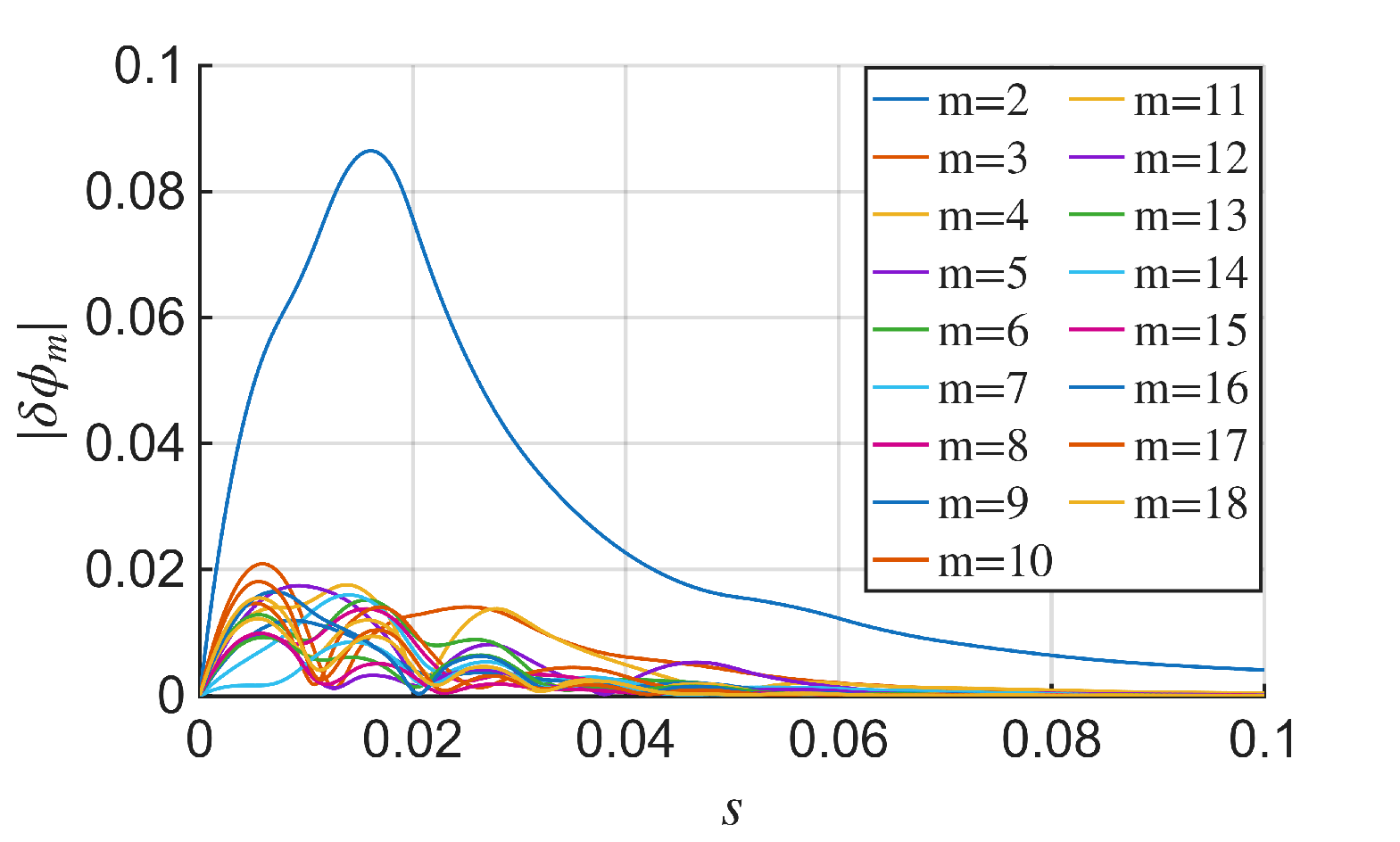}%
        \label{fig.ITPA_modes_after_enlarge}}
\caption{Radial mode structure of the perturbed electrostatic potential $|\delta \phi_m|$ as a function of the flux-surface label~$s$ at $t = 2.8\times10^4\ \Omega_\ast^{-1}$ normalized to the ion cyclotron frequency~$\Omega_\ast$, for poloidal modes $m = 2$ to $18$, with a $C^0$-regularity condition at the axis.}
\label{fig.ITPA_modes_wo}
\end{figure}

Figure~\ref{fig.ITPA_modes_after} shows the radial mode structure of the perturbed electric potential $|\delta \phi|$ as a function of the flux-surface label~$s$ at $t = 2.8\times 10^4\ \Omega_\ast^{-1}$.  
The physical mode is localized near flux-surface label~$s=0.5$ and is characterized by the poloidal harmonics $m = 10$ and $m = 11$.  
This agrees with the results reported in~\citep{Koenies_2018}.  
In our simulation, where a broader spectrum of poloidal harmonics is included, a numerical instability appears close to the axis.  
As the simulation progresses, this instability grows rapidly, exceeding the physical mode amplitude.  
Figure~\ref{fig.ITPA_modes_after_enlarge} provides an enlarged view of the numerical artifact near the axis for $s \in [0,0.1]$.  
The behavior of the mode structure as $r \rightarrow 0$ is inconsistent with the regularity condition in Eq.~\eqref{eq.crude_reg_cond}; for example, the $m=2$ mode should decrease as $r^2$, confirming that a $C^0$-regularity condition is not sufficient at the axis.

\begin{figure}[htbp]
\centering
    \subfloat[$|\delta \phi_m|$ as a function of~$s$ at $t = 2.8\times 10^4\ \Omega_\ast^{-1}$]{%
        \includegraphics[width=0.5\textwidth]{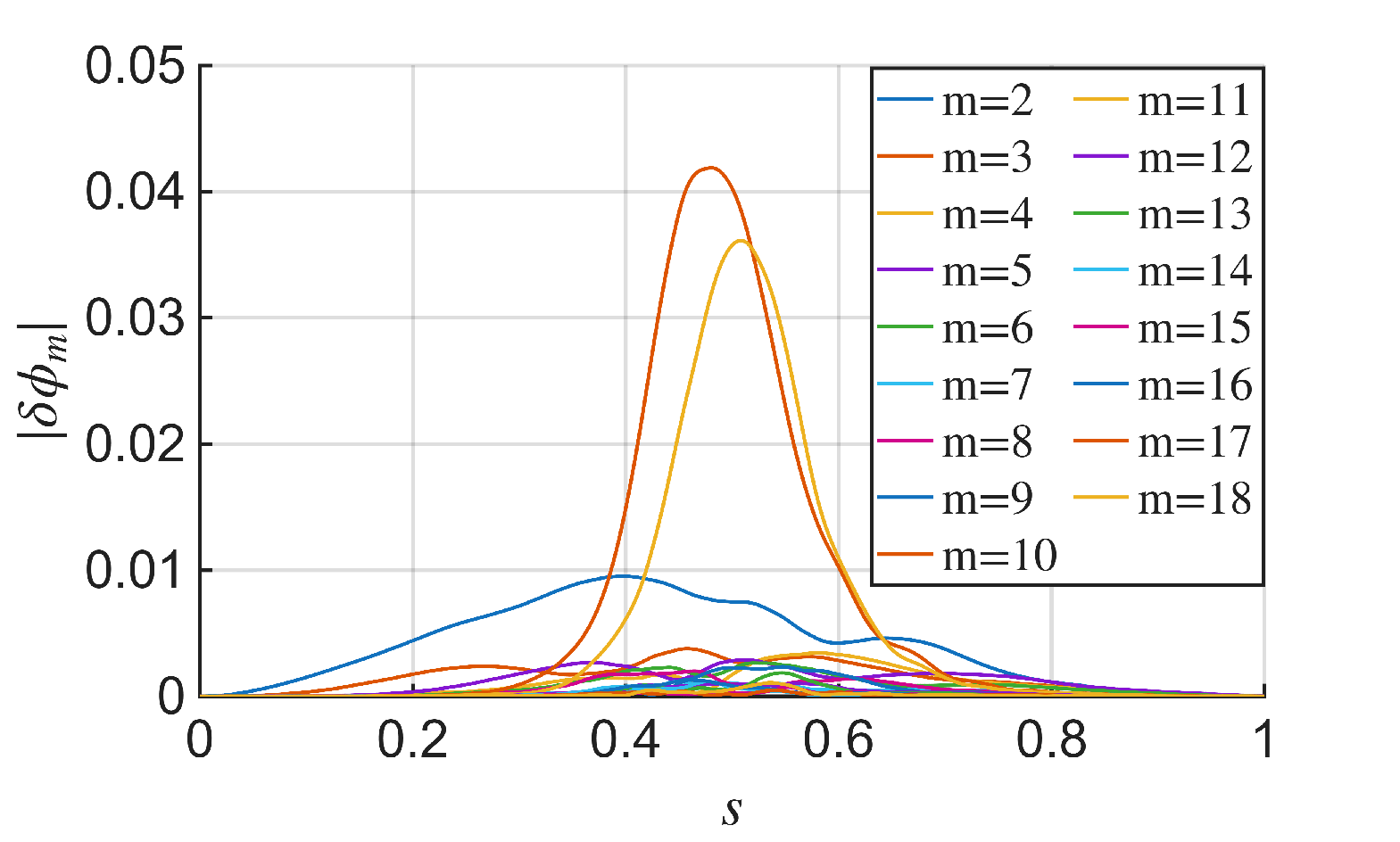}%
        \label{fig.ITPA_modes_C2_after}} 
    \hfill
    \subfloat[$|\delta \phi_m|$ as a function of~$s$ at $t = 7\times 10^4\ \Omega_\ast^{-1}$]{%
        \includegraphics[width=0.5\textwidth]{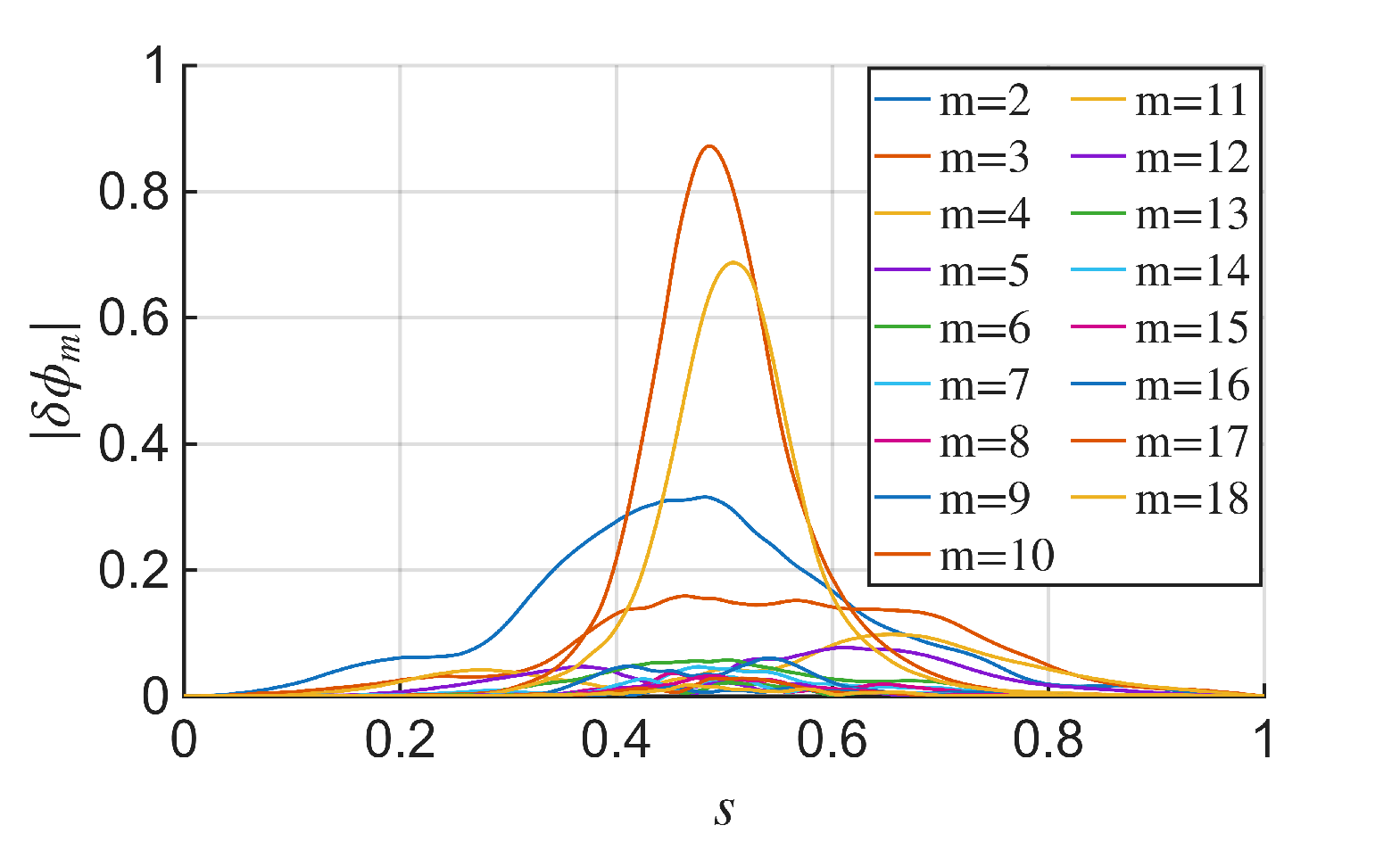}%
        \label{fig.ITPA_modes_C2_after_long}}
\caption{Radial mode structure of the perturbed electrostatic potential $|\delta \phi_m|$ as a function of the flux-surface label~$s$ at different times normalized to the ion cyclotron frequency~$\Omega_\ast$, for poloidal modes $m = 2$ to $18$, with a $C^\infty$-regularity condition at the axis.}
\label{fig.ITPA_modes_C2}
\end{figure}

To eliminate the near-axis numerical instability, we now impose $C^\infty$-regularity at the axis. 
Figure~\ref{fig.ITPA_modes_evolution}(b) shows that now the $m = 10$ and $m = 11$ modes consistently remain dominant, i.e.\ no numerical instability arises. 
Figure~\ref{fig.ITPA_modes_C2} shows the radial mode structure $|\delta \phi_m|$ as a function of the flux-surface label~$s$ for different simulation times. 
At $t = 2.8\times 10^4\ \Omega_\ast^{-1}$, shown in Fig.~\ref{fig.ITPA_modes_C2_after}, there is no instability close to the axis.
Figure~\ref{fig.ITPA_modes_C2_after_long} demonstrates that the mode structure remains smooth even after a long simulation time ($t = 7\times 10^4\ \Omega_\ast^{-1}$), confirming the effectiveness of the $C^\infty$-regularity method.

\section{Conclusions}
\label{sec:conclusion}

We have introduced smooth polar splines of polynomial degree~$p$ to address regularity constraints in discretizations of domains naturally formulated in polar coordinates, such as the unit disc. In particular, for a given poloidal Fourier mode~$m$, regularity at the origin requires $\Phi_m(r) = \mathcal{O}(r^{|m|})$ as $r \to 0$, where $|m|$ is the minimal radial power needed to satisfy the regularity condition. By enforcing these polar regularity conditions, the origin is treated as an interior point of the domain, and the resulting solution space consists of functions that are compatible with Cartesian‑sense $C^\infty$-regularity at $r=0$. Our procedure projects the harmonic polar functions~$S_l^m(r,\theta) = r^l h_{\theta,m}(\theta)$ onto the tensor-product B-spline space on the innermost region, thereby recovering the maximal $C^\infty$‑compatibility in the algebraic sense achievable with degree‑$p$ B‑splines. At the discrete level, Cartesian‑sense $C^\infty$‑regularity at the origin is recovered asymptotically as $\Delta\theta \to 0$, while exact $C^0$-regularity at the polar origin is independent of $\Delta \theta$. The approach extends naturally to three-dimensional geometries, including toroidal topologies relevant to plasma physics in tokamaks and stellarators, and it retains the optimal $(p{+}1)$‑order accuracy characteristic of standard degree‑$p$ B‑splines globally, without degradation near the origin.

The structure-preserving property of the proposed construction relies on two
essential ingredients. First, for $0 \le l \le p$, the radial B-spline basis
reproduces the monomials $r^l$ exactly on the innermost radial interval,
ensuring that the radial part of the admissible harmonic polar functions
$S_l^m(r,\theta)$ is represented exactly. Second, owing to the circulant
structure of the angular mass matrix, the discrete $L^2$-projections of
distinct harmonics~$h_m$ remain mutually $L^2$-orthogonal at finite
angular resolution.
As a consequence, for each admissible radial index $l$, the discrete formulation
decomposes algebraically into invariant subspaces indexed by the angular
harmonic wavenumber $|m|$, with the axisymmetric mode $m=0$ corresponding to a
one-dimensional subspace and $|m|>0$ to two-dimensional subspaces. The
corresponding angular components associated with different $|m|$ are mutually
$L^2$-orthogonal, and no algebraic coupling occurs between different angular
wavenumbers, nor between different radial indices~$l$ within the
finite-dimensional approximation space. This preservation of this algebraic modal structure provides a robust and practically advantageous mechanism for enforcing polar regularity and preventing the excitation of non‑physical modes.

The center-splines are linear combinations of the original basis functions and provide the correct regularity properties at the axis. Smooth polar splines are
constructed by replacing the innermost tensor-product basis functions with
orthonormal center-splines of identical radial support, covering the first
$p{+}1$ intervals. They span a regularized subspace embedded in the full tensor-product space, with an exact prolongation operator and a corresponding restriction operator for mapping load vectors between the regularized subspace and the original discretization space. In contrast to conventional tensor-product B-splines, the proposed basis does not satisfy partition of unity; nevertheless, total charge is exactly preserved in PIC simulations because the constant mode is representable within the regularized polar-spline space.

For Galerkin discretizations, matrix equations are solved in the regularized subspace. The right-hand side, i.e.\ load vector, is assembled in the original tensor-product space and then transformed to the regularized space. The solution is subsequently transformed back to the tensor-product space by the exact prolongation operator. This has the advantage that, in existing implementations, only the matrix solving step needs to be modified, without the need to alter the underlying tensor-product space.

An additional perpendicular $L^2$-projector can act directly in the tensor-product B-spline space as a regularity filter, removing irregular modes that would otherwise produce nonphysical discontinuities in a function or its derivatives at the polar origin.

Imposing high-order regularity effectively reduces the number of degrees of freedom by eliminating, e.g.\ basis functions whose radial part does not satisfy the regularity condition. This, in turn, leads to several benefits: improved conditioning of system matrices, removal of spurious eigenfunctions, and a substantial reduction of statistical noise in PIC simulations near the origin.

In PIC simulations, particle noise excites all available degrees of freedom in the solution, including non-regular modes; removing these significantly improves numerical robustness. As a practical example, in the \textsc{EUTERPE} code, employing smooth polar splines with quadratic B-splines to guarantee $C^\infty$-regularity eliminated a numerical instability at the magnetic axis in simulations of Toroidal Alfv\'en Eigenmodes (ITPA--TAE benchmark) that persisted despite $C^0$-regularity.

In summary, smooth polar splines provide a high-order, efficient, and minimally intrusive modification of tensor-product B-splines for use in polar coordinates. They allow the enforcement of compatibility with the desired $C^\infty$-regularity at the origin, improve the conditioning of system matrices, suppress nonphysical modes, and enhance noise resilience in PIC simulations, making them highly suitable for demanding computational physics applications.

\section*{CRediT authorship contribution statement}

{\bf Peiyou Jiang}: Writing -- original draft, Writing -- Review \& Editing, Validation, Software, Conceptualization; 
{\bf Roman Hatzky}: Writing -- original draft, Writing -- Review \& Editing, Validation, Conceptualization; 
{\bf Zhixin Lu}: Writing -- original draft, Writing -- Review \& Editing, Validation, Conceptualization;
{\bf Eric Sonnendr\"ucker}: Writing -- Review \& Editing, Validation, Resources, Funding acquisition, Conceptualization;
{\bf Matthias Borchardt}: Writing -- Review \& Editing, Validation, Software, Conceptualization;
{\bf Ralf Kleiber}: Writing -- Review \& Editing, Validation, Software, Resources, Funding acquisition, Conceptualization;
{\bf Martin Campos Pinto}: Writing -- Review \& Editing, Validation, Conceptualization;
{\bf Ronald Remmerswaal}: Writing -- Review \& Editing, Validation, Conceptualization.

\section*{Data availability}
No data was used for the research described in the article.

\section*{Declaration of competing interest}

The authors declare that they have no known competing financial interests or personal relationships that could have appeared to influence the work reported in this paper.

\section*{Declaration of generative AI and AI-assisted technologies in the manuscript preparation process}

During the preparation of this work the author(s) used the ChatGPT language model (OpenAI) in order to assist in improving the clarity and readability of portions of the manuscript. After using this tool/service, the author(s) reviewed and edited the content as needed and take(s) full responsibility for the content of the published article.

\section*{Acknowledgments}
We thank Alberto Bottino, Florian Hindenlang, Axel K\"onies, Omar Maj, Stefan Possanner, and Christoph Slaby for helpful discussions.

This work has been carried out within the framework of the EUROfusion Consortium, funded by the European Union via the Euratom Research and Training Programme (Grant Agreement No~101052200--EUROfusion). Views and opinions expressed are however those of the author(s) only and do not necessarily reflect those of the European Union or the European Commission. Neither the European Union nor the European Commission can be held responsible for them.

\begin{appendix}

\section{Cox--de~Boor recursion for a uniform open knot vector}
\label{App.Bspline_rec}

For uniformly spaced interior knots in an open uniform knot vector with spacing $\Delta r$ in the interior,  
the B-splines are defined recursively via the Cox--de~Boor formula~\cite[p.~90]{deBoor2001splines}.

The index $i$ denotes the index of the basis function relative to the left boundary knot $t_0 = 0$.  
Thus, $B_{0,p}$ is the first nonzero basis on $[0,(p{+}1)\Delta r)$, $B_{1,p}$ the second, and so on.

Let $k \in \mathbb{N}_0$ denote the recursion level, which also equals the current polynomial degree.  
We start from $k = 0$ for degree-0 (piecewise constant) splines; each recursion step increases $k$ by one.  
A B‑spline of degree $p$ is obtained at recursion level $k=p$, corresponding to $p{+}1$ knot intervals in its support:

\begin{subequations}
\begin{align}
B_{i,0}(r) &=
\begin{cases}
1, & t_i \le r < t_{i+1},\\
0, & \text{otherwise},
\end{cases} \\[0.5em]
B_{i,k}(r) &=
\omega_{i,k}(r)\, B_{i,k-1}(r) +
\big( 1 - \omega_{i+1,k}(r) \big)\, B_{i+1,k-1}(r),
\label{eq.CoxBoor}
\end{align}
\end{subequations}
where $k \ge 1$ is the current recursion level.

The Cox--de~Boor weights are defined as
\begin{equation}
\omega_{i,k}(r) =
\begin{cases}
\frac{r - i\Delta r}{k\Delta r}, & \text{if } t_{i+k} \ne t_i,\\
0, & \text{otherwise},
\end{cases}
\label{eq.omega_def}
\end{equation}
where $t_i = i\Delta r$ for interior knot positions (i.e.\ excluding the repeated knots at the ends in an open uniform knot vector). Introducing the normalized coordinate $\tilde{r} \coloneqq r / \Delta r$ gives
\begin{equation}
\omega_{i,k}(\tilde{r}) = \frac{\tilde{r} - i}{k}.
\label{eq.omega_norm}
\end{equation}
These forms apply for interior basis functions;  
at the ends, repeated knots are handled via the zero-case rule in Eq.~\eqref{eq.omega_def}.

\section{Generalizing the radial part $R_l$ of the basis functions~$S_l^{\pm m}$}
\label{App.smooth_constraint}

Building on the harmonic polar basis functions~$S_l^{\pm m}$ defined in
Eq.~\eqref{eq.smooth_basis}, we now generalize the construction to allow a more general form for the radial parts while preserving the smoothness constraints in polar coordinates. This generalization prepares the basis for the normalization and orthogonalization procedure described in Sec.~\ref{Sec.orthonormal}.


We define a generalized radial part $\overline{R}_l^m(r)$ as a polynomial of total radial degree~$l$:
\begin{equation}\label{eq.bar_Rlm_def}
    \overline{R}_l^m(r) \coloneqq r^m \left[ \big(\overline{\boldsymbol{c}}_l^m\big)_m + \big(\overline{\boldsymbol{c}}_l^m\big)_{m+2} \, r^2 + \ldots + \big(\overline{\boldsymbol{c}}_l^m\big)_l \, r^{l-m} \right],
\end{equation}
where $0 \le m \le l \le n, \, m \equiv l \ (\mathrm{mod} \ 2)$. 
Here $\overline{\boldsymbol{c}}_l^m$ denotes the vector of coefficients for the polynomial describing~$\overline{R}_l^m$. The prefactor $r^m$ enforces the correct near‑origin behavior for Fourier mode $m$. The subscript of each coefficient denotes the total power of $r$ after including the $r^m$ prefactor, increasing in steps of two due to the $r^2$ dependence of the bracketed part. The last term $r^{l-m}$ ensures that the radial degree matches~$l$. The generalized radial part $\overline{R}_l^m$ includes $R_l$ as a special case.


Compared to~$R_l$, the generalized radial part~$\overline{R}_l^m$ has additional degrees of freedom, which are encoded in the entries of the coefficient vector~$\overline{\boldsymbol{c}}_l^m$. This vector can be uniquely determined by normalizing and orthogonalizing the radial parts~$\overline{R}_l^m$.

We define the generalized basis functions~$\overline{S}_l^{\pm m}$ for $0 \le m \le l$ and $m \equiv l \ (\mathrm{mod} \ 2)$ by the ansatz
\begin{equation}
 \label{eq.gen_harmon_ansatz}
  \overline{S}_l^m(r,\theta) \coloneqq \overline{R}_l^{|m|}(r) h_m( \theta).
 \end{equation}
For $m=0$, the sine term vanishes, i.e.\ $\overline{S}_l^{-0}=0$, and only the cosine term is present.
A real bivariate monomial of total degree~$n$ can be expressed as
\begin{equation}\label{eq.general_ansatz}
    \widetilde{\mathcal{T}}_n(r,\theta) =
    \sum_{l=0}^{n} 
    \sum_{\substack{m = -l \\ m \equiv l \ (\mathrm{mod} \ 2)}}^{l}
    \overline{s}_l^m \, \overline{S}_l^m(r,\theta).
\end{equation}

\section{Exact representation of monomials by radial B-splines}
\label{App.exact_poly_r}

In the following, we show explicitly that
$\widetilde{R}_l(r) = r^l$ on $[0,\Delta r]$ for $0 \le l \le p$,
because $r^l \in V_{r,h}([0,\Delta r])$ (due to the clamped knot vector at $r=0$, see Sec.~\ref{Sec.radial_boundary}), and therefore
Eq.~\eqref{eq.FEM_r} is satisfied identically. 

The radial B-splines $\boldsymbol{B}_{r,\mathrm{c}}(r)$ can be expressed as linear combinations of the corresponding monomials within the innermost radial interval $r \in [0,\Delta r]$.
\begin{equation}
     \boldsymbol{B}_{r,\mathrm{c}} = \mathsfbi{D} \boldsymbol{m},
\end{equation}
where $\boldsymbol{m}$ is the $(p{+}1)$-component monomial vector 
$(1,r,r^2, \ldots, r^p)^\mathrm{T}$, and $\mathsfbi{D}$ is the $(p{+}1) \times (p{+}1)$ upper triangular coefficient matrix.
Explicitly, each row of $\mathsfbi{D}$ gives the representation of a B-spline in terms of monomials on the radial interval $[0,\Delta r]$.
The matrix $\mathsfbi{D}$ is invertible because its diagonal elements are nonzero. 

As an example, for cubic ($p{=}3$) B-splines restricted to $\tilde{r} \in [0,1]$ (where $\tilde{r} \coloneqq r / \Delta r$ is the normalized coordinate), the local polynomial forms are (see Eq.~\eqref{eq.Express_cubic}):
\begin{equation}\label{eq.cubic_C3_radial_matrix}
    \left[
    \begin{array}{c}
       B_{r,0}  \\[6pt] B_{r,1} \\[6pt] B_{r,2} \\[6pt] B_{r,3} \\
    \end{array}
    \right]
    =
    \left[
     \begin{array}{c c c c}
        1  & -3 & 3 & -1 \\[6pt]
        0  &  3 & -\frac{9}{2} & \frac{7}{4} \\[6pt]
        0  & 0 & \frac{3}{2} & -\frac{11}{12} \\[6pt]
        0  & 0 & 0 & \frac{1}{6} \\
     \end{array}
    \right]
    \left[
    \begin{array}{c}
       1  \\[6pt] \tilde{r} \\[6pt] \tilde{r}^2 \\[6pt] \tilde{r}^3 \\
    \end{array}
    \right] .
\end{equation}
This results in:
\begin{equation}\label{eq.cubic_C3_m_matrix}
    \left[
    \begin{array}{c}
     1 \\[6pt] \tilde{r} \\[6pt] \tilde{r}^2 \\[6pt] \tilde{r}^3 \\
    \end{array}
    \right]
    =
    \left[
    \begin{array}{c c c c}
       1  & 1 & 1 & 1 \\[6pt]
       0  &  \frac{1}{3} & 1 & 2 \\[6pt]
       0  & 0 & \frac{2}{3} & \frac{11}{3} \\[6pt]
       0  & 0 & 0 & 6 \\
    \end{array}
    \right]
    \left[
    \begin{array}{c}
     B_{r,0} \\[6pt] B_{r,1} \\[6pt] B_{r,2} \\[6pt] B_{r,3} \\
    \end{array}
    \right] .
\end{equation}

The $l$-th row of $\mathsfbi{D}^{-1}$, denoted $\boldsymbol{c}_{r,l}$, gives the coefficients for expressing the monomial~$\tilde{r}^l$ 
in terms of the radial center B-spline basis functions $\boldsymbol{B}_{r,\mathrm{c}}$:
\begin{equation} \label{m_eq}
    \boldsymbol{m} = [\boldsymbol{c}_{r,0},\ldots, \boldsymbol{c}_{r,p}]^{\mathrm T} \boldsymbol{B}_{r,\mathrm{c}}.
\end{equation}
Note that the elements of $\boldsymbol{c}_{r,0}$ must be equal to one on $[0,\Delta r]$, as this follows from the partition of unity property.

\section{B-spline approximation of the harmonic functions}
\label{App.trigono_approx}

The harmonic polar basis functions~$S_l^{\pm m}$ are expressed as a product of functions in the radial and angular coordinates. Therefore, the one-dimensional approximation of the angular part by B-splines is decoupled from the radial part. For the approximation of the harmonics~$h_m(\theta)$ (see Eq.~\eqref{eq.h_basis}) we use
\begin{equation}
    f(\theta) = \sum_{j=0}^{N_\theta-1} (\boldsymbol{c}_{\theta,m})_j B_{\theta,j}(\theta).
\end{equation}
Owing to the periodic B‑spline discretization and the resulting circulant structure, the angular grid can be interpreted in terms of discrete Fourier modes (see App.~\ref{App.M_ortho_angular_coeff}). Consequently, resolving two harmonics $\sin(m\theta)$ and $\cos(m\theta)$ of index~$m$ on a discrete angular grid requires a sufficient number of degrees of freedom in~$\theta$. 
By analogy with the Nyquist sampling criterion, the number of angular grid points~$N_\theta$ -- and thus the number of B-splines in~$\theta$ -- must satisfy
\begin{equation} \label{eq.theta_condition}
    N_\theta \ge 2|m| + 1
\end{equation}
to represent the mode without aliasing.
In the special case $m=0$, the angular dependence is constant, and a single periodic B‑spline degree of freedom is sufficient.

\subsection{Sampling method}
The simplest choice is to sample the function $h_m(\theta)$ at the center positions of each B-spline basis function and set the corresponding spline coefficient to this sampled value (sampling method). Namely,
\begin{equation}
    f(\theta) \approx \sum_{j=0}^{N_\theta-1} h_m(\theta_j) B_{\theta,j}(\theta) ,
\end{equation}
where $\theta_j \coloneqq j \Delta\theta$ is the sampling coordinate associated with the angular basis function $B_{\theta,j}$,  and $\Delta \theta$ is the grid resolution. This method is accurate only for constant and linear B-splines. For higher-order B-splines, the overlap between different B-spline basis functions is ignored, which leads to poor convergence (e.g.\ for cubic B-splines of $\mathcal{O}((\Delta\theta)^2)$).

\subsection{Collocation method}
The collocation method uses the sampling points $h_m(\theta_i)$ as the right-hand side of a coupled system of equations
\begin{equation}
    \sum_{j=0}^{N_\theta-1} (\boldsymbol{c}_{\theta,m})_j B_{\theta,j}(\theta_i) = h_m(\theta_i) ,
 \qquad (i = 0, \ldots, N_\theta-1).
\end{equation}
This can be written in matrix form
\begin{equation}
    \mathsfbi{K}\boldsymbol{c}_{\theta,m} = \boldsymbol{f}_{\theta,m},
\end{equation}
where $\mathsfbi{K}$ is a circulant matrix with a bandwidth of $p$ -- the degree of the B-splines. For cubic B-splines, it results in the circulant matrix
\newcommand{\Rm}{\frac{2}{3}}
\newcommand{\Rl}{\frac{1}{6}}
\begin{equation}
\mathsfbi{K}=
\begin{bmatrix}
\Rm&\Rl& &&\Rl\\[6pt]
\Rl&\Rm&\Rl&& \\[6pt]
&\ddots&\ddots&\ddots&\\[6pt]
&&\Rl&\Rm&\Rl\\[6pt]
\Rl&&&\Rl&\Rm
\end{bmatrix}  ,
\end{equation}
and the column vector~$\boldsymbol{f}_{\theta,m}$ with the entries $h_m(\theta_i)$. The elements of the matrix and the column vector can be evaluated without performing any integration. The coefficients are calculated as follows:
\begin{equation}
    \boldsymbol{c}_{\theta,m}=\mathsfbi{K}^{-1}\boldsymbol{f}_{\theta,m}.
\end{equation}

\subsection{Finite element method using the Galerkin orthogonality}
\label{App.trigono_approx_FEM}

The finite element method (see Eq.~\eqref{eq.FEM_theta}) leads to a system of coupled equations
\begin{equation}
    \sum_{j=0}^{N_\theta-1} (\boldsymbol{c}_{\theta,m})_j \int_0^{2 \uppi} B_{\theta,i}(\theta)B_{\theta,j}(\theta) \, \mathrm{d} \theta  = \int_0^{2 \uppi} h_m(\theta) B_{\theta,i}(\theta) \, \mathrm{d} \theta , \qquad
    (i = 0, \ldots, N_\theta-1) ,
\end{equation}
which can be written as a matrix equation
\begin{equation} \label{eq.theta_fitting}
    \mathsfbi{M}_\theta \boldsymbol{c}_{\theta,m}=\boldsymbol{b}_{\theta,m}.
\end{equation}
The circulant mass matrix $\mathsfbi{M}_\theta$ and the load vector $\boldsymbol{b}_{\theta,m}$ are defined as
\begin{align}
\label{eq.M_theta}
    (\mathsfbi{M}_{\theta})_{i,j} &\coloneqq \int_0^{2 \uppi} B_{\theta,i}(\theta)B_{\theta,j}(\theta) \, \mathrm{d} \theta , \\
\label{eq.rhs_h_m}
    (\boldsymbol{b}_{\theta,m})_i &\coloneqq \int_0^{2 \uppi} h_m(\theta) B_{\theta,i}(\theta) \, \mathrm{d} \theta .
\end{align}
Note that the mass matrix $\mathsfbi{M}_\theta$ has a bandwidth of $2p{+}1$, which is larger than the bandwidth of $\mathsfbi{K}$.
The mass matrix $\mathsfbi{M}_\theta$ and load vector are typically calculated using a numerical integration method such as, for example, Gauss--Legendre. The coefficients are calculated as follows:
\begin{equation}\label{eq.c_coeff_trigono}
   \boldsymbol{c}_{\theta,m} =\mathsfbi{M}_\theta^{-1}\boldsymbol{b}_{\theta,m}.
\end{equation}

\subsection{Numerical results}
\begin{figure}[htbp]
\centering
\includegraphics[width=0.55\linewidth]{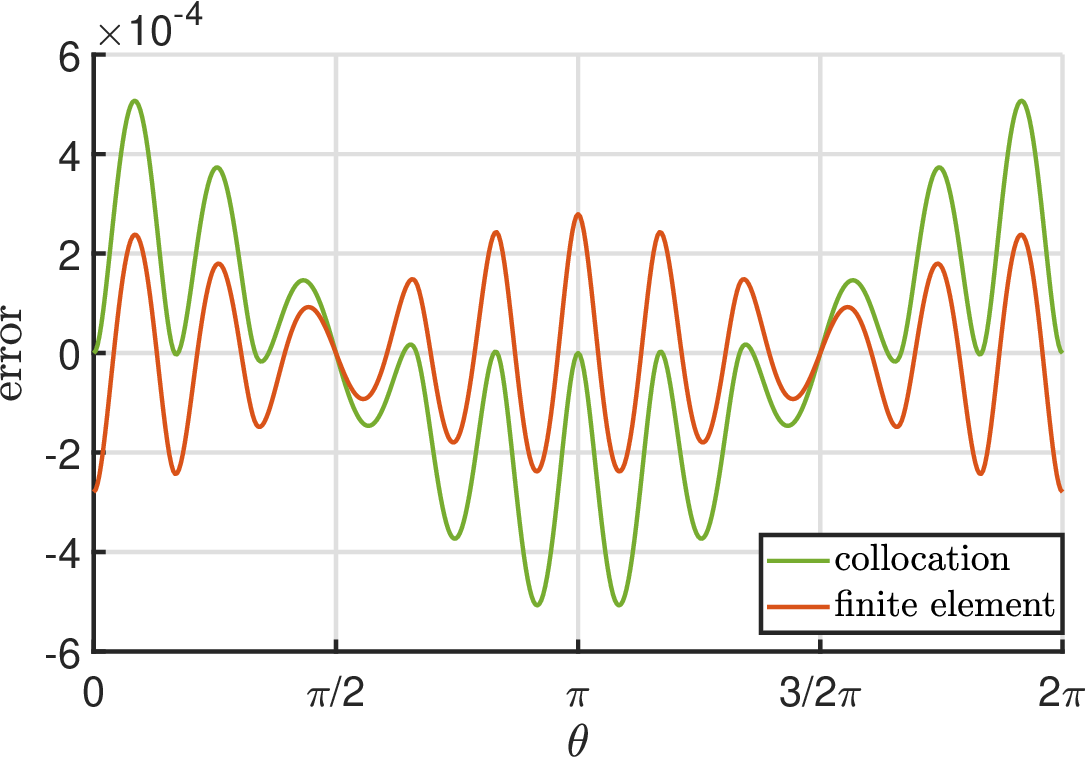}
\caption{Error of the collocation and finite element methods in approximating the function $h_1=\cos(\theta)$ by a cubic B-spline discretization with $N_\theta = 12$.}
\label{fig.collocation_Vs_finiteelement}
\end{figure}

Figure~\ref{fig.collocation_Vs_finiteelement} shows the error $f(\theta)-h_1(\theta)$ between both the collocation and finite element methods and the analytic function $h_1=\cos(\theta)$ as a function of $\theta$ for a cubic B-spline discretization with $N_\theta = 12$. The absolute errors are less than $5 \times 10^{-4}$. The finite element method has a smaller error compared to the collocation method.

To quantify the difference between the analytic function $h_m$ and its approximation~$f$, we define the average error over $\theta$ as
\begin{equation}\label{eq.av_error_theta}
    \overline{\varepsilon}_{\theta} \coloneqq \frac{1}{\sqrt{2\uppi}} \lVert f(\theta) - h_m(\theta) \rVert_{L_2}.
\end{equation}

\begin{figure}[htbp]
\centering
\includegraphics[width=0.55\linewidth]{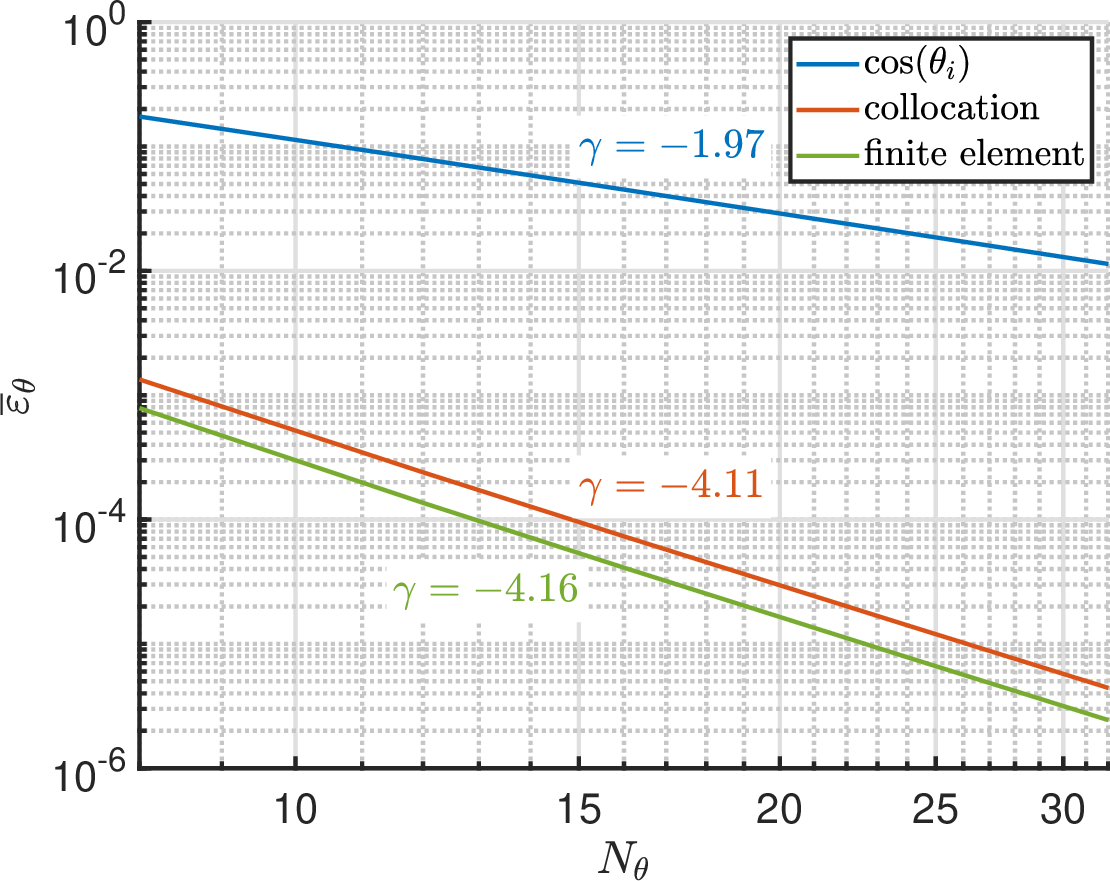}
\caption{Average error~$\overline{\varepsilon}_{\theta}$ of the approximation of $h_1=\cos(\theta)$ in a log--log plot using the sampling, collocation, and finite element method as a function of the number of B-splines $N_\theta$ in the $\theta$-direction. All three methods use a cubic B-spline discretization. The variable $\gamma$ is the slope of the log--log fit.}
\label{fig.aver_error_Ntheta}
\end{figure}

Figure~\ref{fig.aver_error_Ntheta} shows the average error, $\overline{\varepsilon}_{\theta}$, of approximating $h_1 = \cos(\theta)$ in a log--log plot, obtained using the sampling, collocation, and finite element methods, as a function of the number of B-splines $N_\theta$ used in the $\theta$-direction. All three methods employ a cubic B-spline discretization. 

As shown in Fig.~\ref{fig.aver_error_Ntheta}, the sampling method has a convergence rate of $\mathcal{O}((\Delta\theta)^2)$. Both the collocation and finite element methods achieve a much higher convergence rate of $\mathcal{O}((\Delta\theta)^4)$. The average error of the finite element method is slightly smaller than that of the collocation method. However, it results in a matrix with larger bandwidth, and the entries of the matrix and load vector typically need to be evaluated via numerical integration, which increases the computational cost.

\begin{figure}[htbp]
\centering
\includegraphics[width=0.55\linewidth]{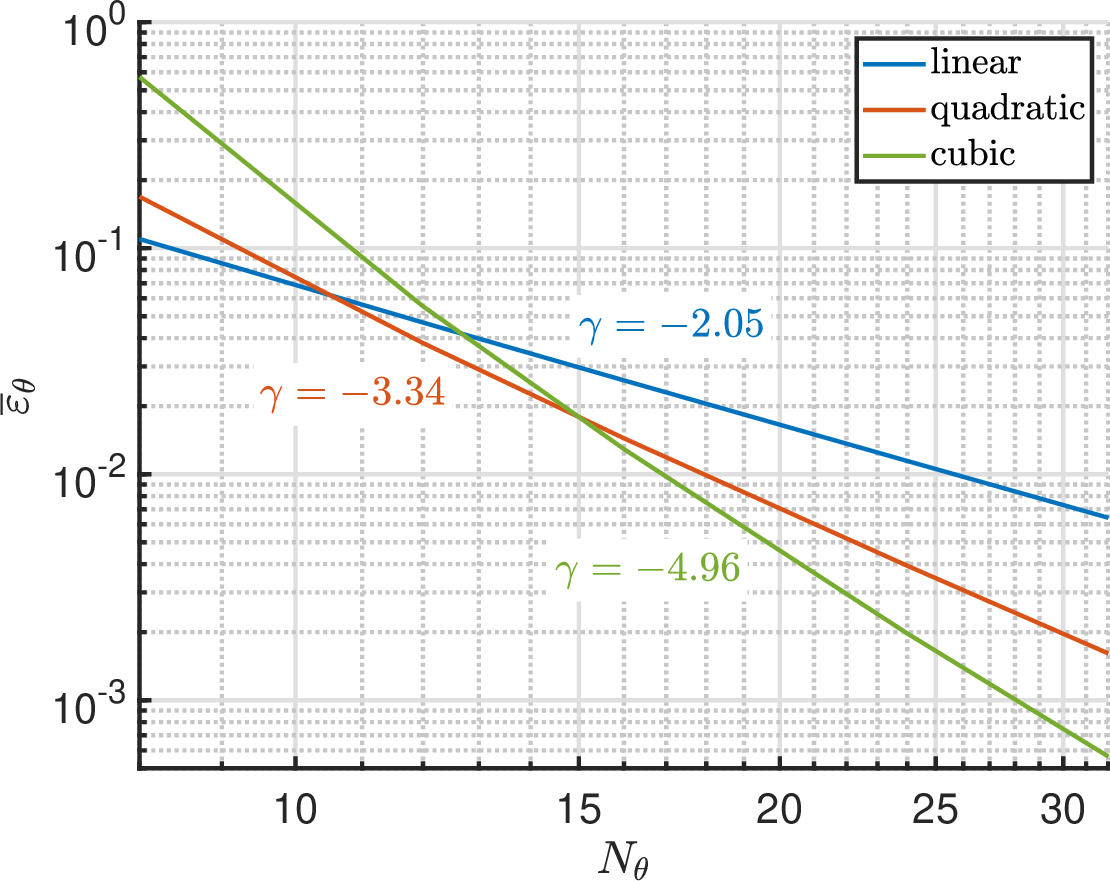}
\caption{Average error~$\overline{\varepsilon}_{\theta}$ of the B-spline approximation of $h_p=\cos(p\theta)$ in a log--log plot as a function of the number of B-splines $N_\theta$, where $p$ is the degree of the B-spline. The finite element method (see App.~\ref{App.trigono_approx_FEM}) is applied for linear (p{=}1), quadratic (p{=}2), and cubic (p{=}3) B-splines. The variable $\gamma$ is the slope of the log--log fit.}
\label{fig.aver_error_Ntheta_maxm}
\end{figure}

Figure~\ref{fig.aver_error_Ntheta_maxm} shows the average error, $\varepsilon_{\theta}$, of the B-spline approximation of $h_p = \cos(p\theta)$ in a log--log plot as a function of the number of B-splines $N_\theta$. Here, the harmonic index~$m$ of $\cos(m\theta)$ is set equal to the spline degree~$p$ due to the regularity constraint of the harmonic basis: for B-splines of degree~$p$, the highest regular harmonic mode is $|m|=p$. It has the largest approximation error compared to other harmonics~$h_m$ for $|m|<p$ when approximated by the angular B-spline basis functions.   As expected, the higher the degree of the B-splines, the faster the average error decreases. Surprisingly, for cubic B-splines, the convergence rate of the average error is $\mathcal{O}((\Delta\theta)^5)$ rather than the expected $\mathcal{O}((\Delta\theta)^4)$. This higher observed rate is likely due to the symmetry and smoothness of the cosine test function.

\subsection{Analytic mode-wise expression for the angular coefficient vectors}
\label{App.angularmodecoeff}

Since the angular discretization is uniform and periodic, all discrete
quantities depend on $m$ only modulo $N_\theta$, and the discretization admits
exactly $N_\theta$ independent Fourier modes.

Starting from Eq.~\eqref{eq.rhs_h_m}, we define for a Fourier mode~$m$ the modified
load vector
\begin{equation}
(\overline{\boldsymbol{b}}_{\theta,m})_j
\coloneqq
\int_0^{2\uppi} B_{\theta,j}(\theta)\,
\exp(\mathrm{i} m\theta)\,\mathrm{d}\theta
=
\exp(\mathrm{i} \Delta\theta \, jm)
\int_0^{2\uppi} B_0(\theta')\,
\exp(\mathrm{i} m \theta') \,\mathrm{d} \theta'
=
\omega^{jm} I_{0,m}.
\end{equation}
Here we used the translation invariance of the angular basis functions (see
Eq.~\eqref{eq.Btheta_translate}). The load vector thus admits the compact
representation
$
\overline{\boldsymbol{b}}_{\theta,m} = I_{0,m} \boldsymbol{\omega}_m ,
$
where $\omega \coloneqq \exp(\mathrm{i}\Delta\theta)$ and
\begin{equation} \label{eq.DF_vector}
\boldsymbol{\omega}_m
\coloneqq
\bigl[1,\omega^m,\omega^{2m},\ldots,\omega^{(N_\theta-1)m}\bigr]^{\mathrm{T}}
\end{equation}
is the (unnormalized) discrete Fourier basis vector corresponding to mode~$m$.
The integral~$I_{0,m}$ is given by
\begin{equation} \label{eq.B-spline_Fourier}
I_{0,m}
=
\int_0^{2\uppi} B_0(\theta)\,\exp(\mathrm{i}m\theta)\,\mathrm{d}\theta
=
\Delta\theta\,
\operatorname{sinc}^{\,p+1}\!\left(\frac{m\Delta\theta}{2}\right)
\exp\!\left(-\frac{\mathrm{i} m\Delta\theta p}{2}\right),
\qquad m\in\mathbb{Z},
\end{equation}
where $\operatorname{sinc}(x)\coloneqq \sin(x)/x$.  This expression follows from the Fourier transform of uniform cardinal B‑splines of degree~$p$. With the centering convention of the angular B-splines from Sec.~\ref{Sec.1d_basis_angular}, the phase factor in Eq.~\eqref{eq.B-spline_Fourier} is uniquely fixed.

The angular mass matrix $\mathsfbi{M}_\theta$ is an
$N_\theta\times N_\theta$ real circulant matrix, since the angular grid is
uniform and the basis functions are translationally invariant. Any circulant
matrix is diagonalizable by the discrete Fourier transform (DFT),
\begin{equation}
\mathsfbi{M}_\theta
=
\mathsfbi{F}^\mathrm{H}
\mathsfbi{\Lambda}
\mathsfbi{F},
\end{equation}
where $\mathsfbi{\Lambda}=\mathrm{diag}(\lambda_0,\ldots,\lambda_{N_\theta-1})$
and the normalized unitary DFT matrix is defined by
\begin{equation} \label{eq.F_kl}
(\mathsfbi{F})_{kl}
\coloneqq
\frac{1}{\sqrt{N_\theta}}
\exp\!\left(-\mathrm{i}\Delta\theta\,kl\right),
\qquad k,l=0,\ldots,N_\theta-1 .
\end{equation}
Since $\mathsfbi{M}_\theta$ is real symmetric positive definite,
$\mathsfbi{\Lambda}$ is a real diagonal matrix with positive entries, and the
eigenvalues satisfy the symmetry $\lambda_{(N_\theta-k)\bmod N_\theta}=\lambda_k$.

The coefficient vectors are obtained from
$\boldsymbol{c}_{\theta,m}
= \mathsfbi{M}_\theta^{-1}\boldsymbol{b}_{\theta,m}$. For the constant mode $m=0$, corresponding to $h_0=1$, the
partition‑of‑unity property of the angular B‑spline basis implies that the
$L^2(0,2\uppi)$‑projection is exact, yielding
\begin{equation} \label{eq.vecofones}
\boldsymbol{c}_{\theta,0} = \boldsymbol{1}.
\end{equation}
For $m\in\mathbb{Z}$ with
$0<|m|\le\lfloor N_\theta/2\rfloor$, the coefficient vectors admit the explicit
expressions
\begin{equation} \label{eq.a_mp_b_mp}
\boldsymbol{c}_{\theta,m}
=
\left\{
\begin{aligned}
\displaystyle
\frac{1}{2}\!\left(
\frac{I_{0,m}}{\lambda_m}\boldsymbol{\omega}_m
+
\frac{I_{0,m}^*}{\lambda_m}\boldsymbol{\omega}_m^*
\right)
&=
a_{m,p}
\begin{bmatrix}
\cos(m\theta_0)\\
\cos(m\theta_1)\\
\vdots\\
\cos(m\theta_{N_\theta-1})
\end{bmatrix},
\qquad & m>0,
\\[2ex]
\displaystyle
\frac{1}{2\mathrm{i}}\!\left(
\frac{I_{0,|m|}}{\lambda_{|m|}}\boldsymbol{\omega}_{|m|}
-
\frac{I_{0,|m|}^*}{\lambda_{|m|}}\boldsymbol{\omega}_{|m|}^*
\right)
&=
b_{|m|,p}
\begin{bmatrix}
\sin(|m|\theta_0)\\
\sin(|m|\theta_1)\\
\vdots\\
\sin(|m|\theta_{N_\theta-1})
\end{bmatrix},
\qquad & m<0 ,
\end{aligned}
\right.
\end{equation}
where $\theta_j \coloneqq j\,\Delta\theta$, $j=0,\ldots,N_\theta-1$.
Moreover, $a_{m,p}$ and $b_{m,p}$ are mode- and degree-dependent scaling factors arising from the complex ratio $I_{0,m}/\lambda_m$:
\begin{equation} \label{eq.ampl_factors}
\begin{aligned}
a_{m,p}
&\coloneqq
\frac{1}{\Delta\theta}\,
\operatorname{sinc}^{-(p+1)}\!\left(\frac{m\Delta\theta}{2}\right)
\cos\!\left(\frac{p\,m\Delta\theta}{2}\right),
\\
b_{m,p}
&\coloneqq
\frac{1}{\Delta\theta}\,
\operatorname{sinc}^{-(p+1)}\!\left(\frac{m\Delta\theta}{2}\right)
\sin\!\left(\frac{p\,m\Delta\theta}{2}\right).
\end{aligned}
\end{equation}
The restriction
$0<|m|\le\lfloor N_\theta/2\rfloor$
excludes the zero mode and, for even $N_\theta$, the Nyquist mode
$m=\pm N_\theta/2$, which require separate treatment.

\subsection{Analytic expression for the eigenvalues $\lambda_m$}
\label{App.lambda}

As $\mathsfbi{M}_\theta$ is circulant, its eigenvalues are given by the discrete
Fourier transform of its first row,
\begin{equation}
\lambda_m
= \sum_{j=0}^{N_\theta-1} (\mathsfbi{M}_\theta)_{0j} \, \omega^{-mj},
\qquad m=0,\ldots,N_\theta-1 .
\end{equation}
Using translational invariance of the angular B‑splines,
$
(\mathsfbi{M}_\theta)_{0j}
= \int_0^{2\uppi} B_0(\theta)\,B_0(\theta-j\Delta\theta)\,\mathrm{d}\theta,
$
and applying the discrete Fourier transform, one obtains
\begin{equation} \label{eq.eigenvalues}
\lambda_m
= \bigl| I_{0,m} \bigr|^2
= \Delta\theta^2\,
\operatorname{sinc}^{\,2(p+1)}\!\left(\frac{m\Delta\theta}{2}\right),
\qquad m = 0,\ldots, N_\theta-1 ,
\end{equation}
where we have used Eq.~\eqref{eq.B-spline_Fourier}. Thus the eigenvalues depend
only on the discrete Fourier mode~$m$, as required by periodicity.

In agreement with the properties of $\mathsfbi{M}_\theta$, these eigenvalues are
real, positive, and symmetric in the sense that
$\lambda_{(-m)\bmod N_\theta}=\lambda_m$.

\section{Proof of the $M_\theta$-orthogonality of the angular coefficient vectors}
\label{App.M_ortho_angular_coeff}

We prove that the projections of the Fourier modes
$\varphi_m(\theta)=\exp(\mathrm{i}m\theta)$, $m\in\mathbb{Z}$,
onto the angular B-splines $B_{\theta,j}$ produce coefficient vectors
$\boldsymbol{c}_{\theta,m}$ that are mutually orthogonal under the
$M_\theta$-inner product. The indices $m,m'$ are taken modulo $N_\theta$,
corresponding to the $N_\theta$ independent discrete Fourier modes. We use several properties shown in App.~\ref{App.angularmodecoeff}.

To test $M_\theta$-orthogonality, we consider:
\begin{subequations}
\begin{equation}
\overline{\boldsymbol{c}}_{\theta,m}^\mathrm{H} \mathsfbi{M}_\theta \, \overline{\boldsymbol{c}}_{\theta,m'} 
= \left( \mathsfbi{M}_\theta^{-1} \overline{\boldsymbol{b}}_{\theta,m} \right)^\mathrm{H}  \mathsfbi{M}_\theta \, \mathsfbi{M}_\theta^{-1} \overline{\boldsymbol{b}}_{\theta,m'} = \overline{\boldsymbol{b}}_{\theta,m}^\mathrm{H} \mathsfbi{M}_\theta^{-1} \overline{\boldsymbol{b}}_{\theta,m'},
\end{equation}
where we have used $(\mathsfbi{M}_\theta^{-1})^\mathrm{H} = \mathsfbi{M}_\theta^{-1}$.  
Inserting $\mathsfbi{M}_\theta^{-1} = \mathsfbi{F}^\mathrm{H} \mathsfbi{\Lambda}^{-1} \mathsfbi{F}$ yields
\begin{equation}
\overline{\boldsymbol{c}}_{\theta,m}^\mathrm{H} \,\mathsfbi{M}_\theta \,\overline{\boldsymbol{c}}_{\theta,m'} 
= I_{0,m}^* I_{0,m'} \,\boldsymbol{\omega}_m^\mathrm{H} \,\mathsfbi{F}^\mathrm{H} \mathsfbi{\Lambda}^{-1}  \mathsfbi{F} \,\boldsymbol{\omega}_{m'} 
= I_{0,m}^* I_{0,m'} \,\boldsymbol{w}_m^\mathrm{H} \,\mathsfbi{\Lambda}^{-1} \,\boldsymbol{w}_{m'},
\end{equation}
where $\boldsymbol{w}_m \coloneqq \mathsfbi{F} \boldsymbol{\omega}_m$.
\end{subequations}

Evaluating $\boldsymbol{w}_m$ componentwise:
\begin{equation}
(\boldsymbol{w}_m)_k = \sum_{l=0}^{N_\theta-1} \frac{1}{\sqrt{N_\theta}} \,\omega^{-kl} \,\omega^{ml} 
= \frac{1}{\sqrt{N_\theta}} \sum_{l=0}^{N_\theta-1} \omega^{(m - k)l} = \sqrt{N_\theta}\,\,\delta_{(m-k)\bmod N_\theta,\,0},
\end{equation}
where we have used the standard discrete orthogonality relation of Fourier modes on a periodic uniform grid. Note that the above orthogonality result relies crucially on the discrete orthogonality of Fourier modes on a uniform grid. For a non-equidistant angular discretization, the discrete Fourier vectors $\boldsymbol{\omega}_m$ are no longer orthogonal.

Substituting back, we obtain the $M_\theta$-orthogonality
\begin{equation} \label{eq:Mtheta_ortho_general}
\begin{aligned}
\overline{\boldsymbol{c}}_{\theta,m}^\mathrm{H} \, \mathsfbi{M}_\theta \, \overline{\boldsymbol{c}}_{\theta,m'}
= I_{0,m}^* I_{0,m'} \, \lambda_m^{-1} \, N_\theta \, \delta_{(m-m')\bmod N_\theta,\,0} =
N_\theta \, \delta_{(m-m')\bmod N_\theta,\,0},
\end{aligned}
\end{equation}
where we have used Eqs.~\eqref{eq.B-spline_Fourier} and \eqref{eq.eigenvalues}. The scalar factor $N_\theta$ originates from the use of unnormalized discrete Fourier
modes. Consequently, Eq.~\eqref{eq:Mtheta_ortho_general} shows that the Galerkin projections of distinct angular Fourier modes produce coefficient vectors that are exactly orthogonal with respect to the
$M_\theta$‑inner product.

\section{Orthonormalization of the radial part of the smooth center-splines}
\label{App.M_ortho_radial_coeff}

As described in Sec.~\ref{Sec.orthonormal}, the orthogonalization of the radial part
of the smooth center-spline basis functions~$\widetilde{B}_{r,l}$ is carried out
with respect to the $M_r$-inner product applied to their coefficient vectors
$\boldsymbol{c}_{r,l}$. As a result, we obtain orthogonal functions
$\widetilde{B}_{r,l}^m$ that now also depend on the $m$-index:
\begin{equation}
    \widetilde{B}_{r,l}^m(r)
    = \left(\boldsymbol{c}_{r,l}^m\right)^\mathrm{T}
      \boldsymbol{B}_{r,\mathrm{c}}(r).
\end{equation}
Here $\boldsymbol{B}_{r,\mathrm{c}}(r)$ denotes the original radial B-spline basis
at the center.

The $M_r$-inner product is defined as
\begin{equation}
(\boldsymbol{c}_{r,l}^m)^\mathrm{T} \mathsfbi{M}_r \boldsymbol{c}_{r,l'}^m
    = \sum_{i,j=0}^p (\boldsymbol{c}_{r,l}^m)_i
      (\mathsfbi{M}_{r})_{i,j}
      (\boldsymbol{c}_{r,l'}^m)_j,
\end{equation}
where $\mathsfbi{M}_r$ is the radial mass matrix, computed from the $L^2$-overlaps
of the first $p{+}1$ radial B-spline basis functions and including the Jacobian
factor $r$ from polar coordinates:
\begin{equation}
(\mathsfbi{M}_{r})_{i,j}
\coloneqq \int_0^{(p+1)\Delta r}
B_{r,i}(r) B_{r,j}(r)\, r\, \mathrm{d} r,
\qquad (i,j=0,\ldots,p).
\end{equation}
The radial mass matrix $\mathsfbi{M}_r$ is symmetric and positive definite.

We assemble the matrix
\begin{equation}
\mathsfbi{E}
\coloneqq
[\boldsymbol{c}_{r,m}, \boldsymbol{c}_{r,m+2}, \ldots,
 \boldsymbol{c}_{r,\widetilde{l}}],
\end{equation}
from the subset of the original (non‑orthonormal) monomial coefficient vectors $\boldsymbol{c}_{r,l}$
satisfying $l \equiv m \ (\mathrm{mod}\ 2)$, where $\widetilde{l}$ is the largest
index such that $\widetilde{l} \equiv m \ (\mathrm{mod}\ 2)$ and
$\widetilde{l} \le p$.

For orthonormalization we use a ``metric-aware'' QR decomposition. First, the mass
matrix is factored as
\begin{equation}
\mathsfbi{M}_r = \mathsfbi{W}^\mathrm{T}\mathsfbi{W},
\end{equation}
for example via a Cholesky decomposition. We then apply a standard QR decomposition
to the weighted matrix
\begin{equation}
\mathsfbi{W}\mathsfbi{E} = \mathsfbi{Q}\mathsfbi{R},
\end{equation}
where the columns of $\mathsfbi{Q}$ are orthonormal with respect to the Euclidean
inner product. The $M_r$-orthonormal coefficient matrix is obtained by solving the triangular system
\begin{equation}
\mathsfbi{W}\,\widehat{\mathsfbi{C}} = \mathsfbi{Q}.
\end{equation}
We write the resulting coefficient matrix columnwise as
\begin{equation}
\widehat{\mathsfbi{C}}
\coloneqq
[\widehat{\boldsymbol{c}}_{r,m}^m,
 \widehat{\boldsymbol{c}}_{r,m+2}^m,
 \ldots,
 \widehat{\boldsymbol{c}}_{r,\widetilde{l}}^m].
\end{equation}
By construction, $\widehat{\mathsfbi{C}}$ satisfies
\begin{equation}
\widehat{\mathsfbi{C}}^\mathrm{T}\mathsfbi{M}_r\widehat{\mathsfbi{C}}=\mathsfbi{I}.
\end{equation}
This procedure is equivalent to a Gram--Schmidt orthonormalization in the weighted
discrete $L^2$-space defined by $\mathsfbi{M}_r$.

\section{Quadratic smooth center-spline basis functions}
\label{App.quad_splines}

This appendix presents the quadratic smooth center-spline basis functions. They are constructed using the same $M_r$-orthogonalization and normalization procedure 
as described in Sec.~\ref{Sec.orthonormal_spline}. These basis functions are used when $C^2$-regularity at the origin is required.

The radial parts of the orthonormal smooth center-splines~$\widehat{B}_{r,l}^m$ are given by
\begin{equation}\label{eq.quadratic_C2_radial_projection}
    \begin{bmatrix}
     \widehat{B}_{r,0}^0 \\[6pt] 
     \widehat{B}_{r,2}^0 \\[6pt] 
     \widehat{B}_{r,1}^1 \\[6pt] 
     \widehat{B}_{r,2}^2
    \end{bmatrix}
    =
    \mathrm{diag}\!\left(
    \begin{bmatrix}
    2\sqrt{\frac{15}{97}} \\[4pt] 
    2\sqrt{\frac{15}{1340831}} \\[4pt]
    \sqrt{\frac{15}{134}} \\[4pt] 
    2\sqrt{\frac{10}{33}} 
    \end{bmatrix}
    \right)
    \begin{bmatrix}
       1  &  1  &  1 \\[6pt]
      -251  & -251  &  137 \\[6pt]
       0    &   1   &  3 \\[6pt]
       0    &   0   &  1
    \end{bmatrix}
    \begin{bmatrix}
     B_{r,0} \\[6pt] B_{r,1} \\[6pt] B_{r,2}
    \end{bmatrix}.
\end{equation}

\begin{figure}[htbp]
\centering
\includegraphics[width=0.55\linewidth]{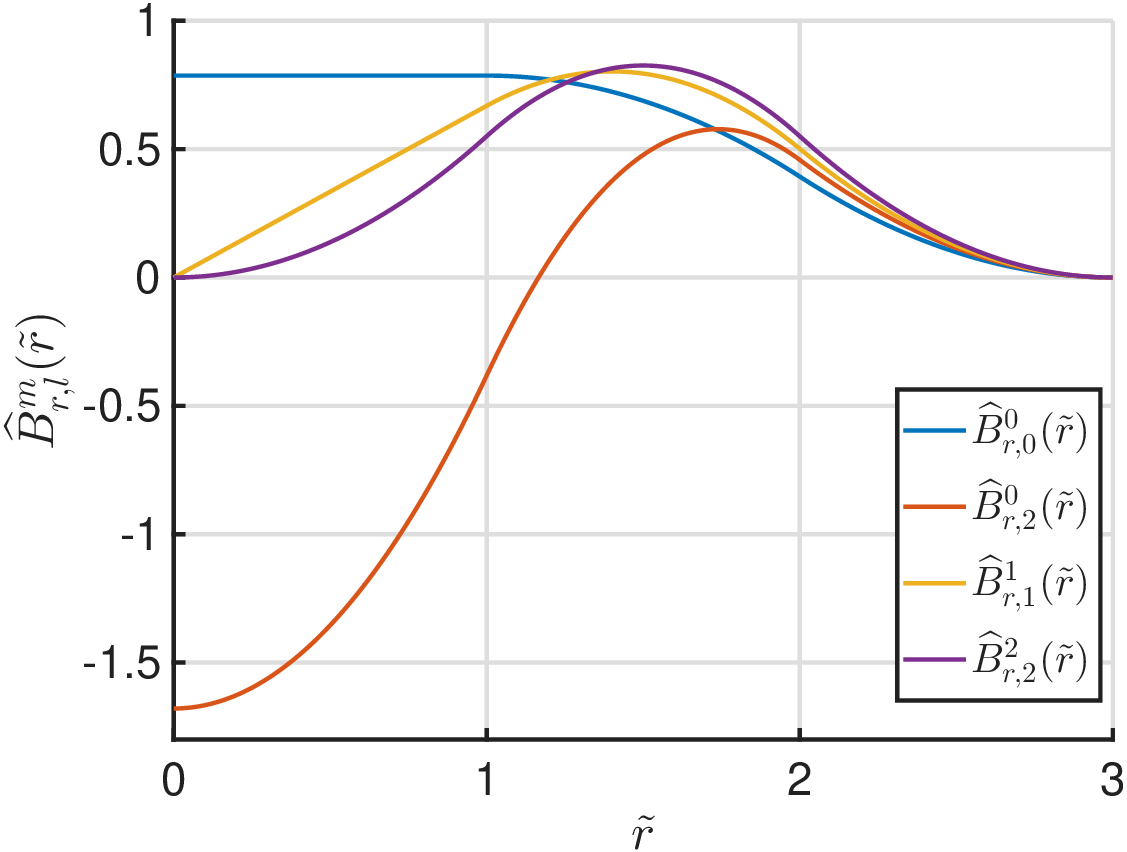}
\caption{Normalized radial parts $\widehat{B}_{r,l}^m(r)$ of the smooth center-splines for quadratic B-splines. The functions $\widehat{B}_{r,0}^0$ and $\widehat{B}_{r,2}^0$ are orthogonal with respect to the radial $L^2$-inner product (see Eq.~\eqref{eq.innerprod_r}).}
\label{fig.radial_orthogonal_quadratic}
\end{figure}

Figure~\ref{fig.radial_orthogonal_quadratic} shows these normalized radial parts. Note that $\widehat{B}_{r,0}^0$ and $\widehat{B}_{r,2}^0$ are orthogonal with respect to the radial $L^2$-inner product (see Eq.~\eqref{eq.L2_scalar_prod}).

\begin{figure*}[ht]
\centering
    \subfloat[\rbasis{0}{0}]{\includegraphics[width = 0.24\linewidth]{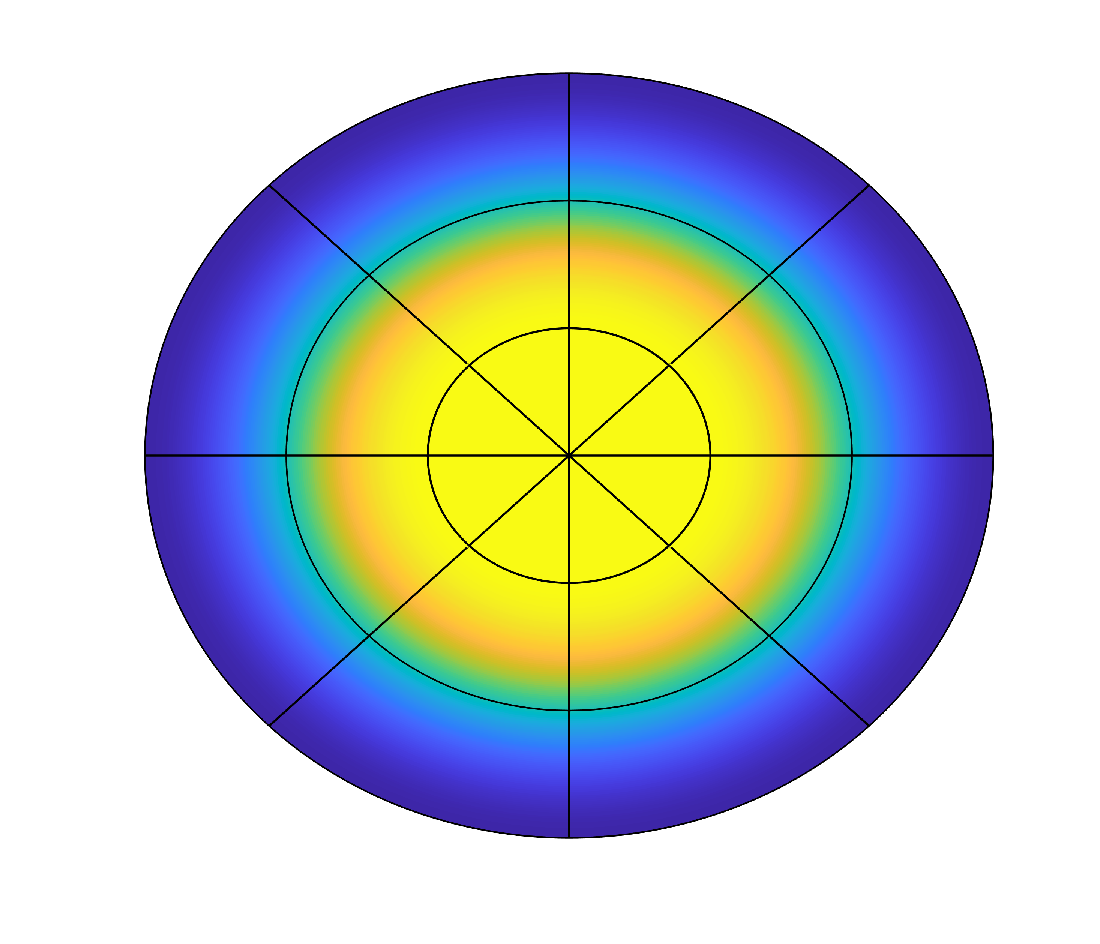}} \\
	\subfloat[\rbasis{1}{-1}]{\includegraphics[width = 0.24\linewidth]{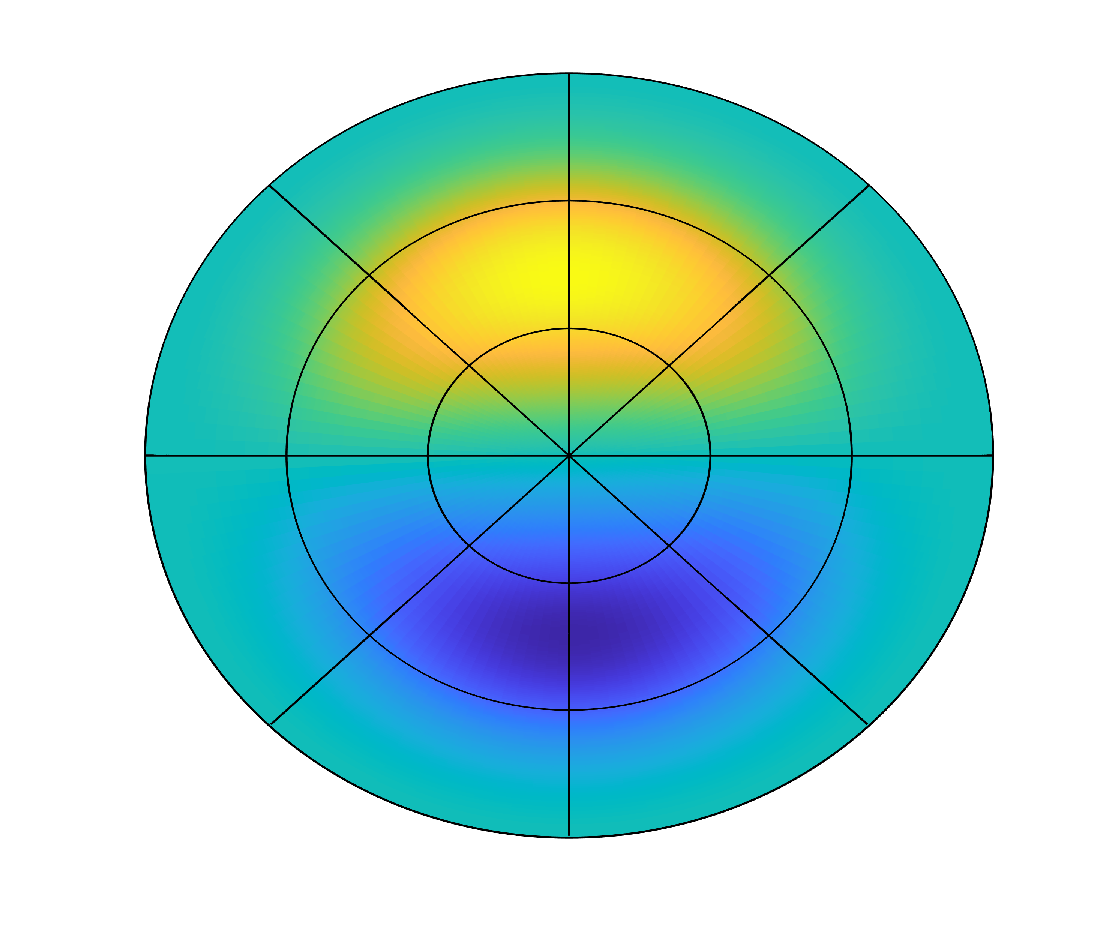}}
	\subfloat[\rbasis{1}{1}]{\includegraphics[width = 0.24\linewidth]{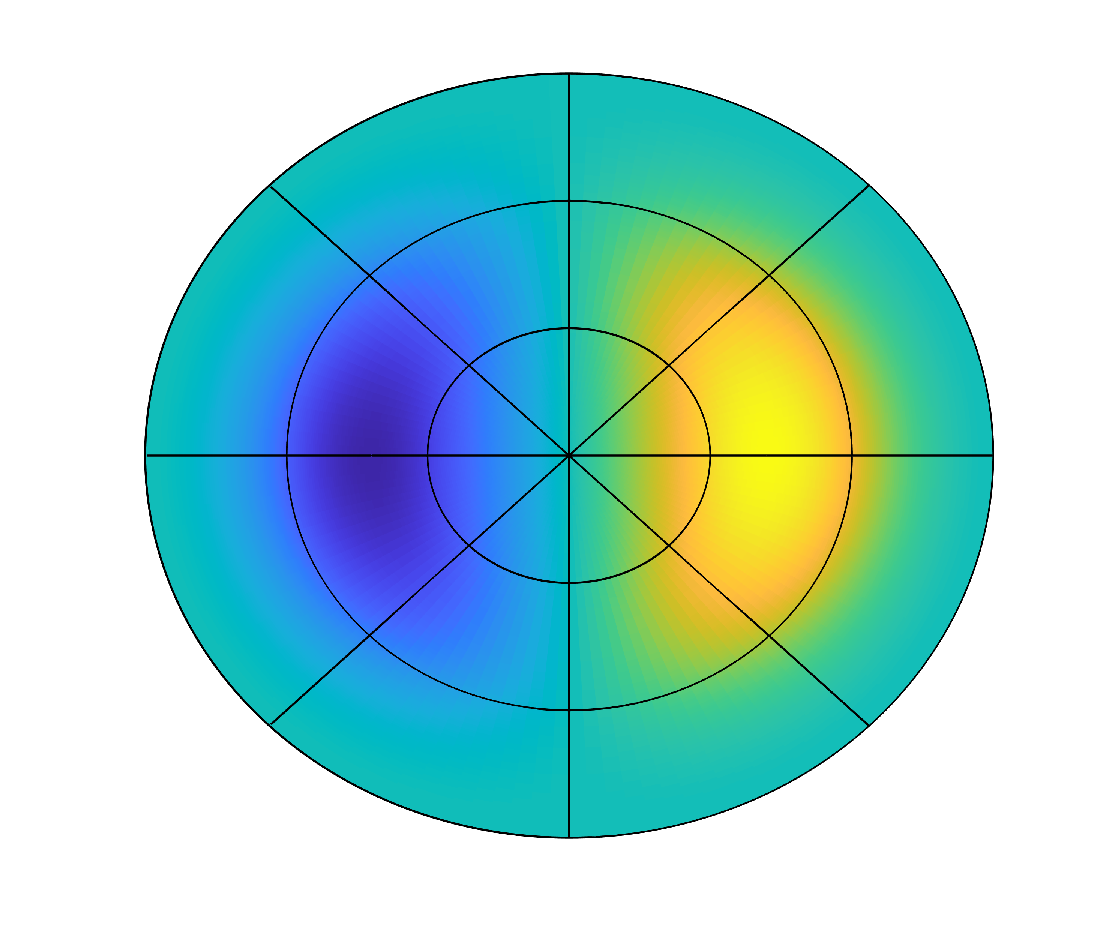}} \\
	\subfloat[\rbasis{2}{-2}]{\includegraphics[width = 0.24\linewidth]{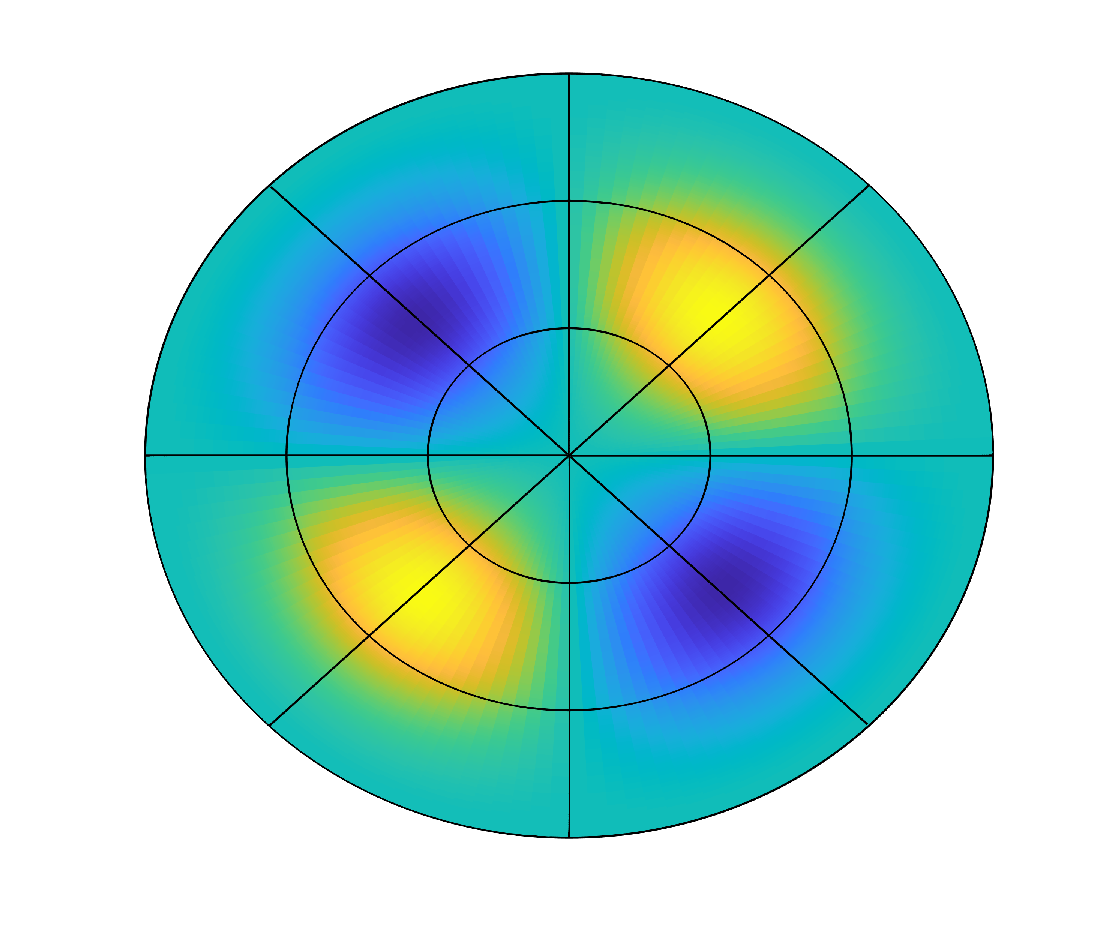}}
	\subfloat[\rbasis{2}{0}]{\includegraphics[width = 0.24\linewidth]{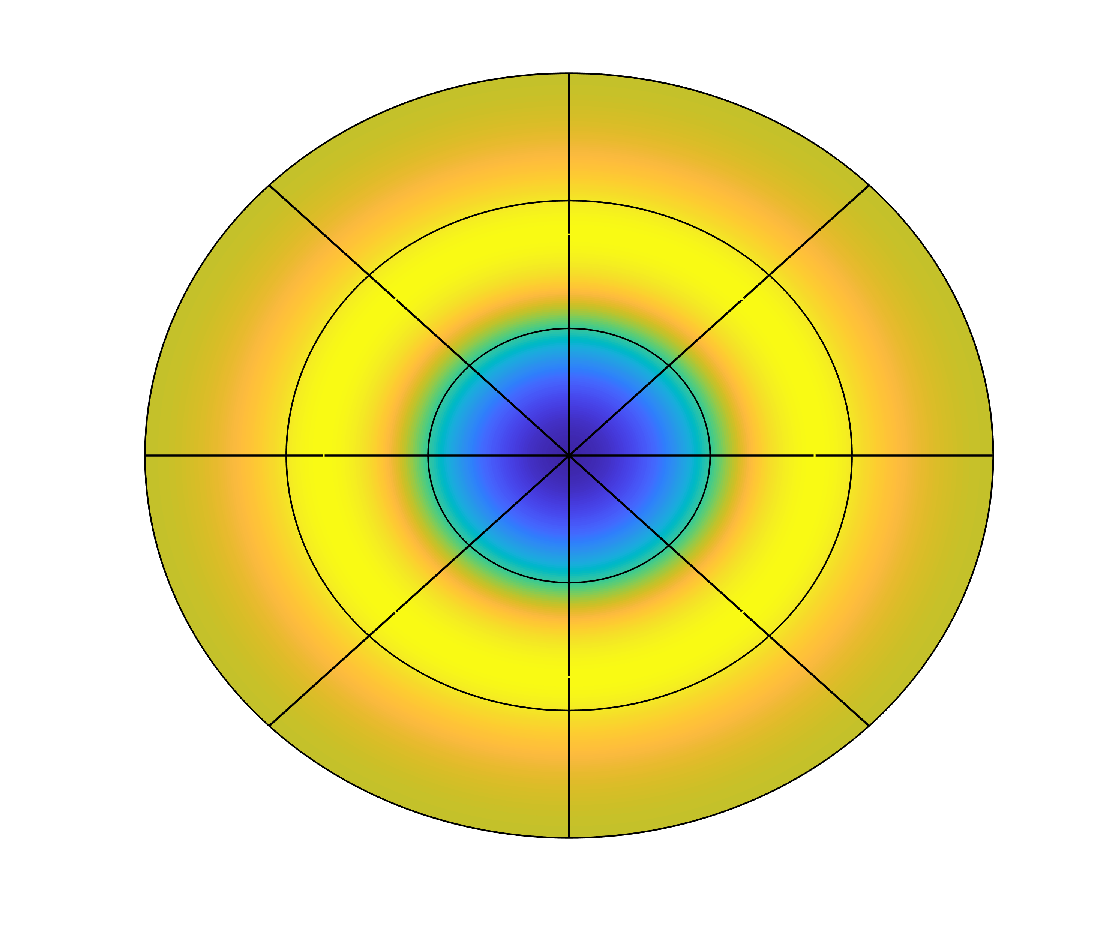}}
    \subfloat[\rbasis{2}{2}]{\includegraphics[width = 0.24\linewidth]{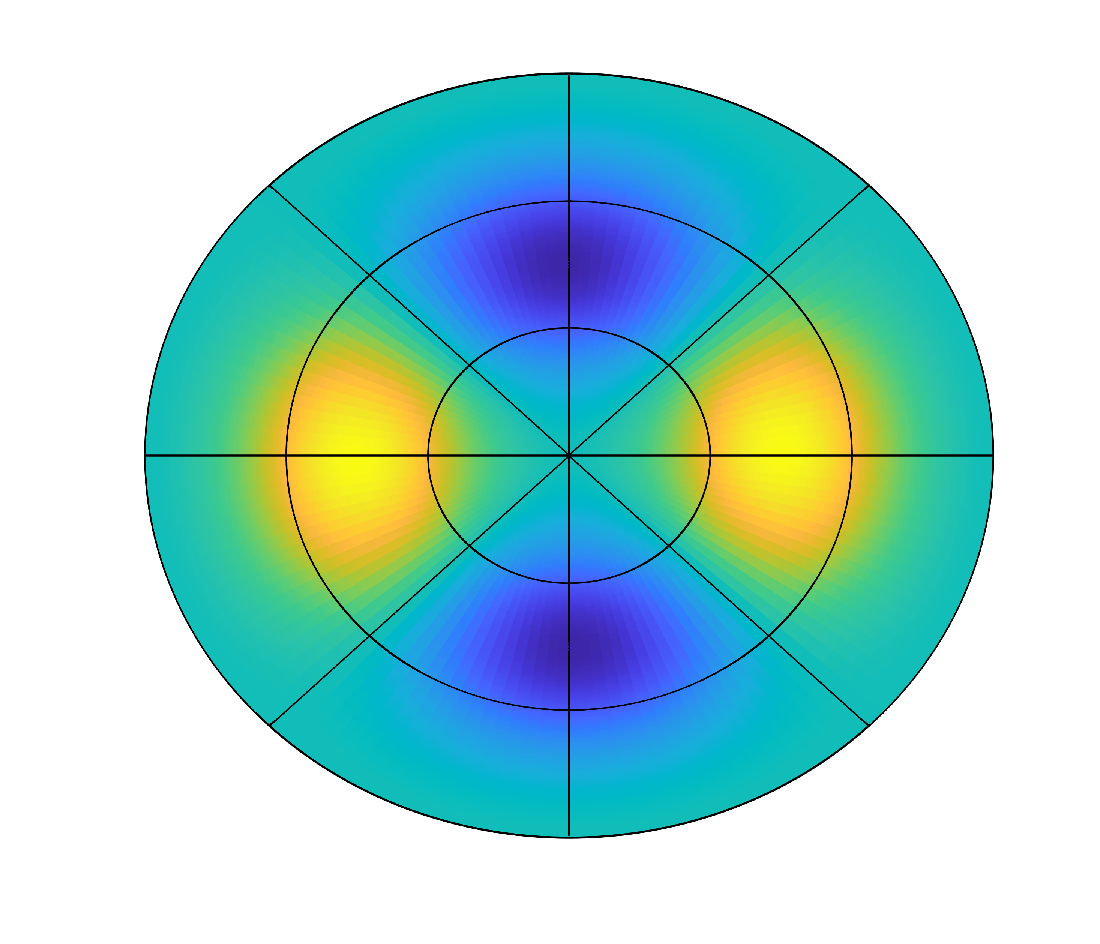}}
\caption{Contour plots of the smooth center-splines $\widehat{B}_l^m(\tilde{r},\theta)$,
for $0 \leq l \leq 2$, $-l \leq m \leq l$, and $m \equiv l \ (\mathrm{mod}\ 2)$.  
Radial part: quadratic B-splines; angular part: normalized harmonic functions.  
Circles at fixed $\tilde{r}$ mark boundaries of the three innermost radial intervals.}
\label{fig.quadratic_basis_c2_all}
\end{figure*}

Figure~\ref{fig.quadratic_basis_c2_all} presents the corresponding quadratic smooth center-spline basis functions $\widehat{B}_l^m(\tilde{r},\theta)$ as contour plots.  
These are obtained by combining the above orthonormalized radial functions with the normalized harmonics~$\widehat{h}_m$ from Sec.~\ref{Sec.orthonormal_spline}.  

\section{Examples of smooth polar-spline bases of $C^0$- and $C^1$-regularity}
\label{App.C0andC1examples}

\subsection{$C^0$-regularity}
\label{App.C0-regularity}

\begin{figure}[htbp]
\centering
\includegraphics[width=0.5\linewidth]{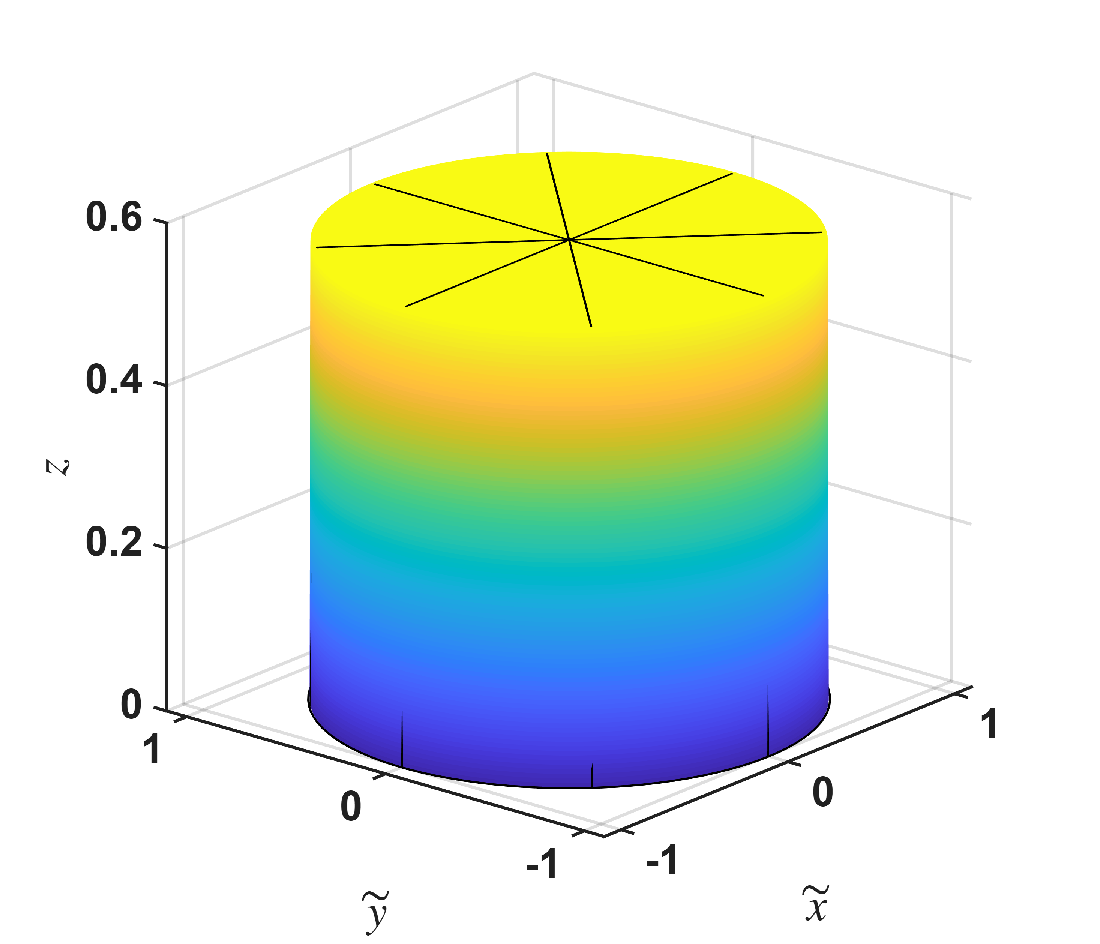}
\caption{The basis function $\widehat{B}_0^0$ with the radial part represented by constant B-splines and the angular part given by normalized harmonic function~$\widehat{h}_0$.}
\label{fig.gamma_c0}
\end{figure}

The $C^0$-regularity condition, based on the smooth center-spline basis functions~$\widehat{\boldsymbol{B}}_\mathrm{c}$, is equivalent to the ``unity'' (i.e.\ uniqueness) boundary condition~\citep{Fivaz1998gygles,lanti2020orb5,kleiber2024euterpe}. Since only the basis functions~$B_{r,0}(r)B_{\theta,j}(\theta)$ have a non-zero value at the origin, we can derive the expression
\begin{equation}
\begin{aligned}
    \widehat{B}_0^0(r,\theta) &= \sum_{j=0}^{N_\theta-1} \widehat{q}_{r,0}^0 B_{r,0}(r) (\widehat{\boldsymbol{c}}_{\theta,0})_j B_{\theta,j}(\theta) = \widehat{q}_{r,0}^0(\widehat{\boldsymbol{c}}_{\theta,0})_0  B_{r,0}(r) \sum_{j=0}^{N_\theta-1} B_{\theta,j}(\theta) \\
    &= \widehat{q}_{r,0}^0 (\widehat{\boldsymbol{c}}_{\theta,0})_0 B_{r,0}(r),
\end{aligned}
\end{equation}
where the scalar $\widehat{q}_{r,0}^0$ is the normalization factor of $B_{r,0}(r)$ and $\widehat{\boldsymbol{c}}_{\theta,0}$ is the normalized coefficient vector of length~$N_\theta$ (see Eq.~\eqref{eq.c_theta_m_norm}) representing the harmonic~$h_0$ in the angular B-spline basis. Note that the normalization factor of $\widehat{B}_0^0$ is $\widehat{q}_{r,0}^0 (\widehat{\boldsymbol{c}}_{\theta,0})_0$ as every component of $\widehat{\boldsymbol{c}}_{\theta,0}$ is the same (see Eq.~\eqref{eq.vecofones}). In addition, we use the partition of unity (see Eq.~\eqref{eq.unity_r}). Hence, only a single $\theta$-independent degree of freedom is retained at the polar origin, corresponding to $C^0$-regularity. Figure~\ref{fig.gamma_c0} shows, for constant B-splines, the shape of the basis function $\widetilde{B}_0^0(r)$, which is independent of the $\theta$-direction.
 
The matrix $\widehat{\mathsfbi{P}}_0$ (see Eq.~\eqref{eq.restriction_op}) is a $(N_r N_\theta) \times [N_r (N_\theta{-}1){+}1]$ matrix
\begin{equation}\label{eq.restriction_op_C0}
    \widehat{\mathsfbi{P}}_0 \coloneqq \left[
    \begin{array}{c|c}
        \widehat{\boldsymbol{L}}_0 & \boldsymbol{0} \\
        \hline
        \boldsymbol{0} & \mathsfbi{I}
    \end{array}
    \right],
\end{equation}
where the vector $\widehat{\boldsymbol{L}}_0 \coloneqq \widehat{q}_{r,0}^0(\widehat{\boldsymbol{c}}_{\theta,0})_0$ is of length $N_\theta$, and the matrix $\mathsfbi{I}$ is the $[N_r (N_\theta{-}1)] \times [N_r (N_\theta{-}1)]$ unit matrix.

\subsection{$C^1$-regularity}
\label{App.C1-regularity}
For $C^1$-regularity at the origin, there are three degrees of freedom: one for the constant function, and two for the monomials $x$ and $y$ (see Eq.~\eqref{eq.deg_freedom}). Thus, the three smooth center-spline basis functions~$\widehat{B}_0^0$, $\widehat{B}_1^{-1}$ and $\widehat{B}_{1}^1$ are contributing in the innermost radial interval.
\begin{figure*}[ht]
\centering
    \subfloat[$\widehat{B}_0^0(\tilde{r},\theta)$]{\includegraphics[width = 0.4\linewidth]{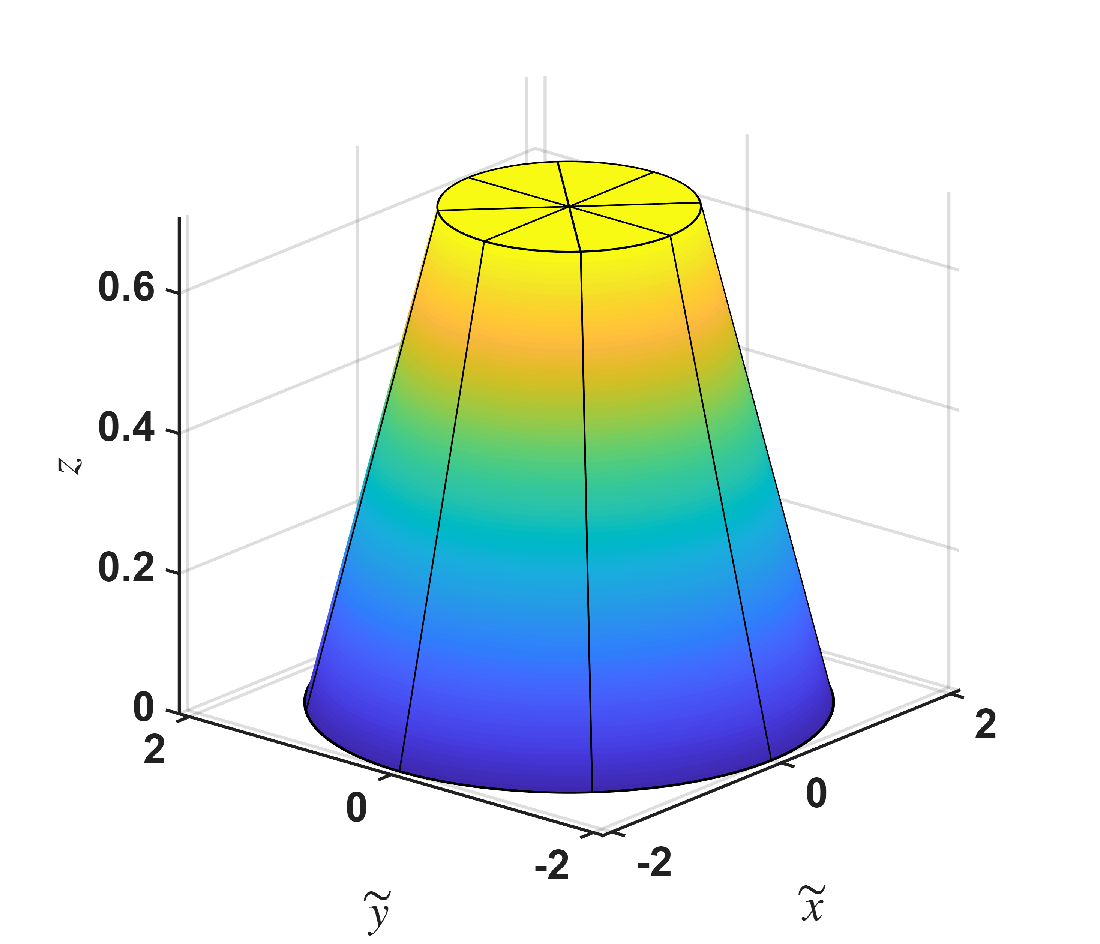}} \\
	\subfloat[$\widehat{B}_{1}^{-1}(\tilde{r},\theta)$]{\includegraphics[width = 0.4\linewidth]{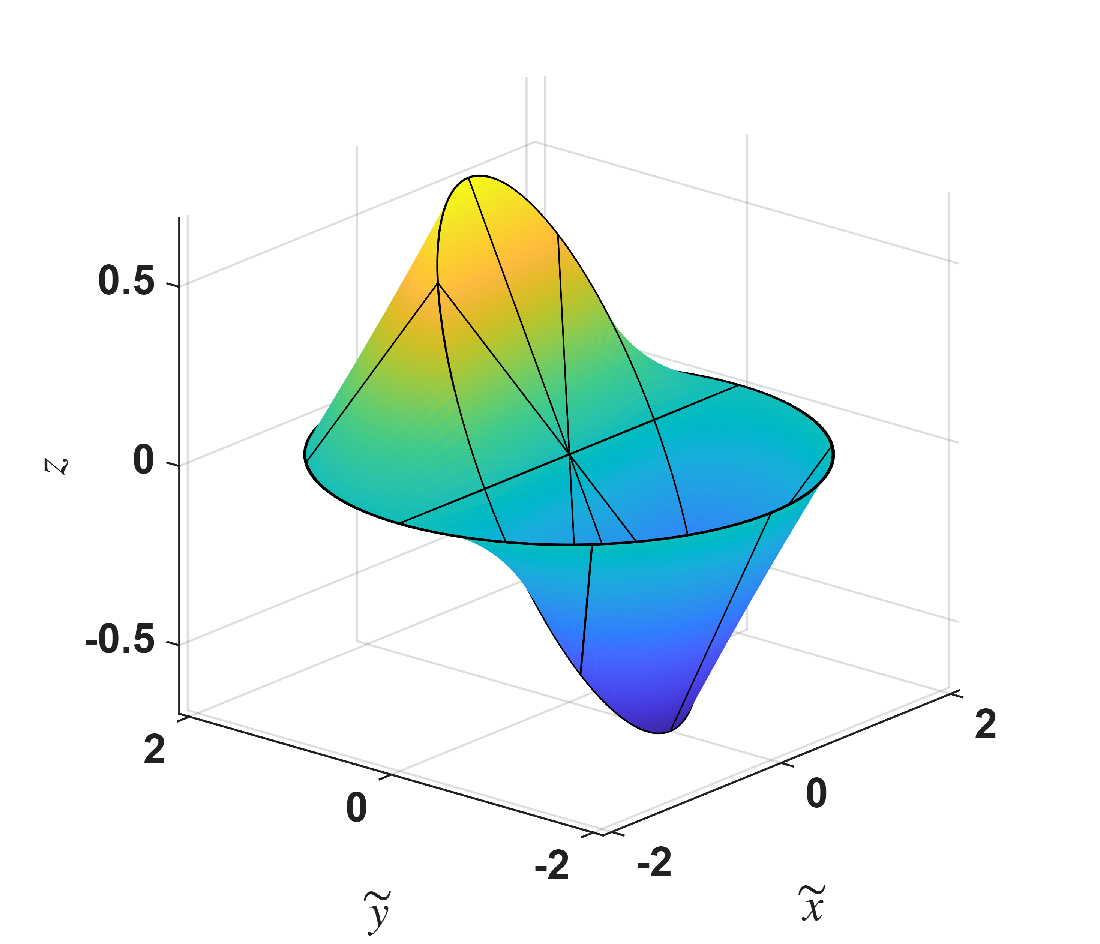}}
	\subfloat[$\widehat{B}_1^1(\tilde{r},\theta)$]{\includegraphics[width = 0.4\linewidth]{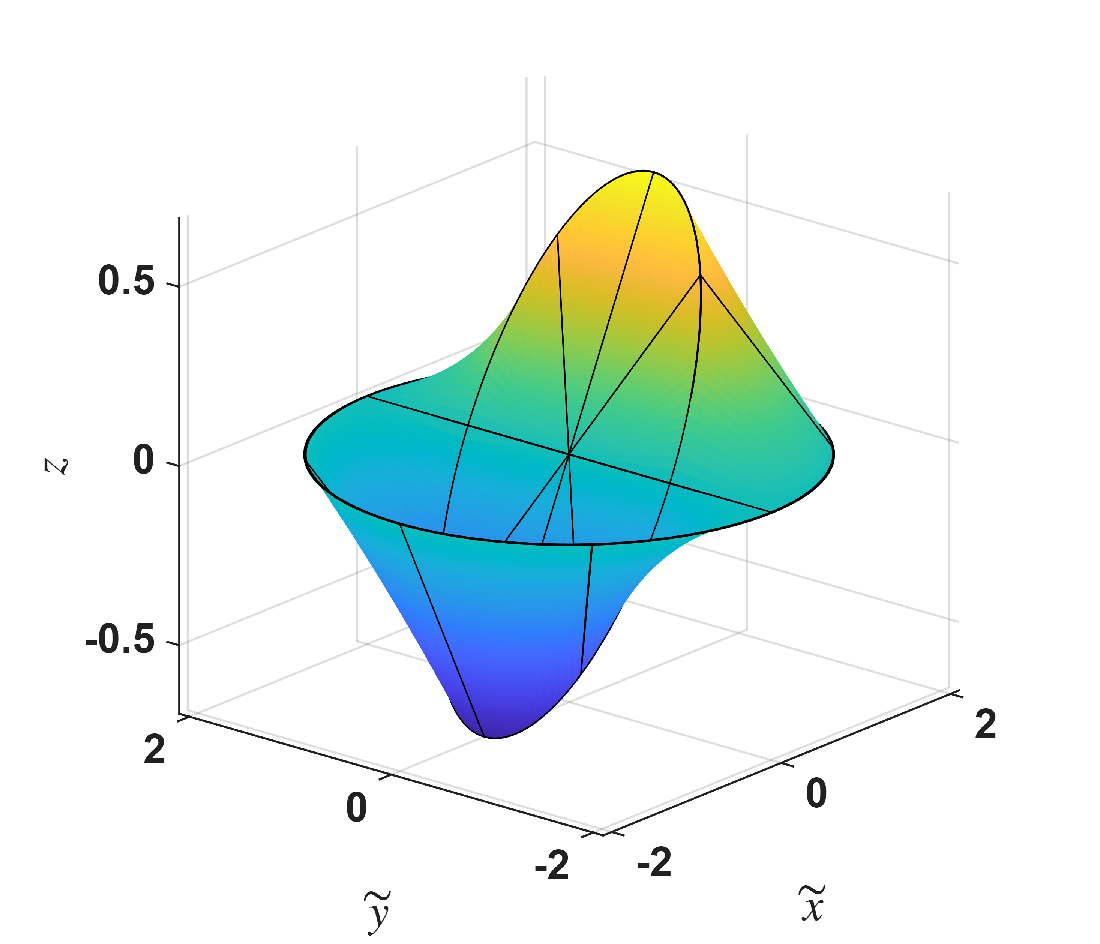}}
\caption{The basis functions $\widehat{B}_0^0(\tilde{r},\theta)$, $\widehat{B}_{1}^{-1}(\tilde{r},\theta)$ and $\widehat{B}_1^1(\tilde{r},\theta)$ with the radial part represented by linear B-splines and the angular part given by normalized harmonic functions~$\widehat{h}_m$.}
\label{fig.gamma1}
\end{figure*}
Figure~\ref{fig.gamma1} shows the three basis functions expressed by linear B-splines in the limit $N_\theta \rightarrow \infty$. In the innermost radial interval, the shape matches that of the corresponding bivariate monomials, while in the second radial interval it results from the declining profile of the linear B-spline basis function~$\widehat{B}_{1,j}$.

The corresponding matrix~$\widehat{\mathsfbi{P}}_1$ with the size $(N_r N_\theta) \times [N_r(N_\theta{-}2){+}3]$ is
\begin{equation}\label{eq.projection_C1}
    \widehat{\mathsfbi{P}}_1 \coloneqq \left[
    \begin{array}{c|c}
       \widehat{\mathsfbi{L}}_1 & \boldsymbol{0} \\
        \hline
       \widehat{\mathsfbi{Z}}_1 & \boldsymbol{0} \\
        \hline
        \boldsymbol{0} & \mathsfbi{I}
    \end{array}
    \right].
\end{equation}
The matrix $\widehat{\mathsfbi{L}}_1 \coloneqq  \big[ \big(\widehat{\boldsymbol{c}}_{r,0}^0\big)_0 \widehat{\boldsymbol{c}}_{\theta,0}, \boldsymbol{0}, \boldsymbol{0} \big]$ is a $N_\theta \times 3$ matrix, where $\widehat{\boldsymbol{c}}_{r,0}^0$ is the normalized coefficient vector (see Eq.~\eqref{eq.hat_tilde_B_r_l_m}) representing $\widehat{B}_{r,0}^0$ in the radial B-spline basis~$\boldsymbol{B}_{r,\mathrm{c}}$. In addition, $\widehat{\boldsymbol{c}}_{\theta,0}$ is the normalized coefficient vector (see Eq.~\eqref{eq.c_theta_m_norm}) representing the harmonic~$h_0$ in the angular B-spline basis. The matrix $\widehat{\mathsfbi{Z}}_1 \coloneqq \big[ \big(\widehat{\boldsymbol{c}}_{r,0}^0\big)_1 \widehat{\boldsymbol{c}}_{\theta,0}, \big(\widehat{\boldsymbol{c}}_{r,1}^1\big)_1 \widehat{\boldsymbol{c}}_{\theta,-1}, \big(\widehat{\boldsymbol{c}}_{r,1}^1\big)_1 \widehat{\boldsymbol{c}}_{\theta,1} \big]$ is a $N_\theta \times 3$ matrix, where $\widehat{\boldsymbol{c}}_{\theta,0},\widehat{\boldsymbol{c}}_{\theta,-1}$ and $\widehat{\boldsymbol{c}}_{\theta,1}$ are the coefficient vectors representing the harmonics for $m=\pm 1$, $\sin(\theta)$ and $\cos(\theta)$ in the angular B-spline basis (see Sec.~\ref{Sec.angular_basis}). The matrix $\mathsfbi{I}$ is a $[N_r (N_\theta{-}2)] \times [N_r (N_\theta{-}2)]$ unit matrix.

A function $\phi_h$ can be expressed in the basis $\widehat{\boldsymbol{B}}_1$ by
\begin{equation}\label{eq.phiapprox_C1}
    \phi_h(r,\theta) = \widehat{b}_0^0  \widehat{B}_0^0(r,\theta) + \widehat{b}_1^{-1}  \widehat{B}_{1}^{-1}(r,\theta) + \widehat{b}_1^1  \widehat{B}_1^1(r,\theta) + \sum_{i=2}^{N_r-1} \sum_{j=0}^{N_\theta-1} \widehat{\phi}_{i,j}\widehat{B}_{i,j}(r,\theta),
\end{equation}
where the coefficient vector is $\widehat{\boldsymbol{\phi}} = [\widehat{b}_0^0, \widehat{b}_1^{-1}, \widehat{b}_1^1, \widehat{\phi}_{2,0}, \ldots,\widehat{\phi}_{i,j}, \ldots]^{\mathrm T}$. We use Eq.~\eqref{eq.tildephi_phi_trans} to write the coefficients in the original basis
\begin{subequations}
\begin{align}
     \phi_{0,j} &= \widehat{b}_0^0 \big(\widehat{\boldsymbol{c}}_{r,0}^0\big)_0 \big(\widehat{\boldsymbol{c}}_{\theta,0}\big)_j \,, \\
     \phi_{1,j} &=
\widehat{b}_0^0 (\widehat{\boldsymbol{c}}_{r,0}^0)_1 (\widehat{\boldsymbol{c}}_{\theta,0})_j
+ \widehat{b}_1^{-1} (\widehat{\boldsymbol{c}}_{r,1}^1)_1 (\widehat{\boldsymbol{c}}_{\theta,-1})_j
+ \widehat{b}_1^1 (\widehat{\boldsymbol{c}}_{r,1}^1)_1 (\widehat{\boldsymbol{c}}_{\theta,1})_j \,, \\
     \phi_{i,j} &= \widehat{\phi}_{i,j} \,, \qquad (i \geq 2) .
\end{align}   
\end{subequations}
The coefficient $\big(\widehat{\boldsymbol{c}}_{r,1}^1\big)_0 = 0$ because $B_0(r)$ contains a constant component (see Eq.~\eqref{eq.boundary_poly_order}), and therefore it can not contribute to $r$.

\end{appendix}

\newpage

\bibliographystyle{elsarticle-num-names} 
\bibliography{references}

@book{deBoor2001splines,
  author       = {de Boor, C.},
  title        = {A Practical Guide to Splines},
  series       = {Applied Mathematical Sciences},
  volume       = {27},
  year         = {2001},
  publisher    = {Springer},
  address      = {New York, NY}
}

@article{Mishchenko2008AlfvenicPIC,
  author  = {Mishchenko, A. and Hatzky, R. and K{\"o}nies, A.},
  title   = {Global particle-in-cell simulations of {Alfv\'enic} modes},
  journal = {Physics of Plasmas},
  volume  = {15},
  number  = {11},
  pages   = {112106},
  year    = {2008}
}

@article{Nuehrenberg_JPP_2025,
  title   = {Gyrokinetic simulations of magnetohydrodynamic modes in stellarator plasmas},
  author  = {N{\"u}hrenberg, C. and Kleiber, R. and Mishchenko, A. and K{\"o}nies, A. and Borchardt, M. and Hatzky, R.},
  journal = {Journal of Plasma Physics},
  volume  = {91},
  number  = {4},
  pages   = {E93},
  year    = {2025}
}

@article{Toshniwal2017isogeometric,
  author    = {Toshniwal, D. and Speleers, H. and Hiemstra, R.R. and Hughes, T. J.},
  title     = {Multi-degree smooth polar splines: A framework for geometric modeling and isogeometric analysis},
  journal   = {Computer Methods in Applied Mechanics and Engineering},
  volume    = {316},
  pages     = {1005--1061},
  year      = {2017},
}

@article{Guclu2025BrokenFEEC,
  title   = {A broken-{FEEC} framework for structure-preserving discretizations of polar domains with tensor-product splines},
  author  = {G{\"u}{\c{c}}l{\"u}, Y. and Patrizi, F. and Campos Pinto, M.},
  year    = {2025},
  month   = {May},
  journal = {arXiv preprint arXiv:2505.15996},
  archivePrefix = {arXiv}
}

@book{Boyd2000,
  author    = {Boyd, J.P.},
  title     = {Chebyshev and Fourier Spectral Methods},
  edition   = {2},
  publisher = {Dover Publications},
  address   = {New York},
  year      = {2000},
}

@article{Lewis1990constraints,
   author = {Lewis, H.R. and Bellan, P.M.},
   title = {Physical constraints on the coefficients of {F}ourier expansions in cylindrical coordinates},
   journal = {Journal of Mathematical Physics},
   volume = {31},
   number = {11},
   pages = {2592--2596},
   year = {1990},
   issn = {0022-2488}
}

@article{hatzky2002energy,
  title={Energy conservation in a nonlinear gyrokinetic particle-in-cell code for ion-temperature-gradient-driven modes in $\theta$-pinch geometry},
  author={Hatzky, R. and Tran, T.M. and K{\"o}nies, A. and Kleiber, R. and Allfrey, S.J.},
  journal={Physics of Plasmas},
  volume={9},
  number={3},
  pages={898--912},
  year={2002},
  publisher={American Institute of Physics}
}

@article{wan2012global,
  title={Global gyrokinetic simulation of tokamak edge pedestal instabilities},
  author={Wan, W. and Parker, S.E. and Chen, Y. and Yan, Z. and Groebner, R.J. and Snyder, P.B.},
  journal={Physical Review Letters},
  volume={109},
  number={18},
  pages={185004},
  year={2012},
  publisher={APS}
}

@article{Niu2022,
  title={Zernike polynomials and their applications},
  author={Niu, K. and Tian, C.},
  journal={Journal of Optics},
  volume={24},
  number={},
  pages={123001},
  year={2022},
  publisher={IOP Publishing}
}

@article{Zernike1934,
  title={Beugungstheorie des {S}chneidenverfahrens und seiner verbesserten {F}orm, der {P}hasenkontrastmethode},
  author={Zernike, F.},
  journal={Physica},
  volume={1},
  number={7--12},
  pages={689--704},
  year={1934},
  publisher={Elsevier}
}

@article{Lin98zonalflow,
  title={Turbulent Transport Reduction by Zonal Flows: Massively Parallel Simulations},
  author={Lin, Z. and Hahm, T.S. and Lee, W.W. and Tang, W.M. and White, R.B.},
  journal={Science},
  volume={281},
  pages={1835--1837},
  year={1998}
}

@article{Fivaz1998gygles,
  title={Finite element approach to global gyrokinetic Particle-In-Cell simulations using magnetic coordinates},
  author={Fivaz, M. and Brunner, S. and de Ridder, G. and Sauter, O. and Tran, T.M. and Vaclavik, J. and Villard, L. and Appert, K.},
  journal={Computer Physics Communications},
  volume={111},
  number={1--3},
  pages={27--47},
  year={1998},
  publisher={Elsevier}
}

@article{lanti2020orb5,
  title={{ORB5}: A global electromagnetic gyrokinetic code using the {PIC} approach in toroidal geometry},
  author={Lanti, E. and Ohana, N. and Tronko, N. and Hayward-Schneider, T. and Bottino, A. and McMillan, B.F. and Mishchenko, A. and Scheinberg, A. and Biancalani, A. and Angelino, P. and Brunner, S. and Dominski, J. and Donnel, P. and Gheller, C. and Hatzky, R. and Jocksch, A. and Jolliet, S. and Lu, Z.X. and Martin Collar, J.P. and Novikau, I. and Villard, L.},
  journal={Computer Physics Communications},
  volume={251},
  pages={107072},
  year={2020},
  publisher={Elsevier BV}
}

@article{hatzky2019reduction,
  title={Reduction of the statistical error in electromagnetic gyrokinetic particle-in-cell simulations},
  author={Hatzky, R. and Kleiber, R. and K{\"o}nies, A. and Mishchenko, A. and Borchardt, M. and Bottino, A. and Sonnendr{\"u}cker, E.},
  journal={Journal of Plasma Physics},
  volume={85},
  number={1},
  pages={905850112},
  year={2019},
  publisher={Cambridge University Press}
}

@article{cole2014fluid,
  title={Fluid electron, gyrokinetic ion simulations of linear internal kink and energetic particle modes},
  author={Cole, M. and Mishchenko, A. and K{\"o}nies, A. and Kleiber, R. and Borchardt, M.},
  journal={Physics of Plasmas},
  volume={21},
  number={7},
  year={2014},
  pages={072123},
  publisher={AIP Publishing}
}

@article{eisen1991Spectra,
	author = {Eisen, H. and Heinrichs, W. and Witsch, K.},
	journal = {Journal of Computational Physics},
	number = {2},
	pages = {241-257},
	title = {Spectral collocation methods and polar coordinate singularities},
	volume = {96},
	year = {1991}}

@article{lu2021development,
  title={The development of an implicit full f method for electromagnetic particle simulations of {A}lfv{\'e}n waves and energetic particle physics},
  author={Lu, Z.X. and Meng, G. and Hoelzl, M. and Lauber, P.},
  journal={Journal of Computational Physics},
  volume={440},
  pages={110384},
  year={2021},
  publisher={Elsevier}
}

@article{holderied2022magneto,
  title={Magneto-hydrodynamic eigenvalue solver for axisymmetric equilibria based on smooth polar splines},
  author={Holderied, F. and Possanner, S.},
  journal={Journal of Computational Physics},
  volume={464},
  pages={111329},
  year={2022},
  publisher={Elsevier}
}

@article{lee1983gyrokinetic,
	title={Gyrokinetic approach in particle simulation},
	author={Lee, W.W.},
	journal={Physics of Fluids},
	volume={26},
	number={2},
	pages={556--562},
	year={1983},
	publisher={American Institute of Physics}
}

@article{chen2007electromagnetic,
  title={Electromagnetic gyrokinetic $\delta$f particle-in-cell turbulence simulation with realistic equilibrium profiles and geometry},
  author={Chen, Y. and Parker, S.E.},
  journal={Journal of Computational Physics},
  volume={220},
  number={2},
  pages={839--855},
  year={2007},
  publisher={Elsevier}
}

@article{mishchenko2014pullback,
  title={Pullback transformation in gyrokinetic electromagnetic simulations},
  author={Mishchenko, A. and K{\"o}nies, A. and Kleiber, R. and Cole, M.},
  journal={Physics of Plasmas},
  volume={21},
  number={9},
  pages={092110},
  year={2014},
  publisher={AIP Publishing}
}

@article{kraus2017gempic,
  title={{GEMPIC}: geometric electromagnetic particle-in-cell methods},
  author={Kraus, M. and Kormann, K. and Morrison, P.J. and Sonnendr{\"u}cker, E.},
  journal={Journal of Plasma Physics},
  volume={83},
  number={4},
  pages={905830401},
  year={2017},
  publisher={Cambridge University Press}
}

@article{kleiber2024euterpe,
  title={{EUTERPE}: A global gyrokinetic code for stellarator geometry},
  author={Kleiber, R. and Borchardt, M. and Hatzky, R. and K{\"o}nies, A. and Leyh, H. and Mishchenko, A. and Riemann, J. and Slaby, C. and Garc\'ia-Rega\~na, J.M. and S\'anchez, E. and Cole, M.},
  journal={Computer Physics Communications},
  volume={295},
  pages={109013},
  year={2024},
  publisher={Elsevier}
}

@article{Kleiber2016,
    title   = {An explicit large time step particle-in-cell scheme for nonlinear gyrokinetic simulations in the electromagnetic regime},
    author  = {Kleiber, R. and Hatzky, R. and K{\"o}nies, A. and Mishchenko, A. and Sonnendr{\"u}cker, E.},
    journal = {Physics of Plasmas},
    volume  = {23},
    number  = {3},
    pages   = {032501},
    year    = {2016},
    issn    = {1070-664X}
}

@article{Koenies_2018,
author = {K{\"o}nies, A. and Briguglio, S. and Gorelenkov, N. and Feh{\'e}r, T. and Isaev, M. and Lauber, P. and Mishchenko, A. and Spong, D.A. and Todo, Y. and Cooper, W.A. and Hatzky, R. and Kleiber, R. and Borchardt, M. and Vlad, G. and Biancalani, A. and Bottino, A. and {ITPA EP TG}},
title = {Benchmark of gyrokinetic, kinetic {MHD} and gyrofluid codes for the linear calculation of fast particle driven {TAE} dynamics},
year = {2018},
publisher = {IOP Publishing},
journal = {Nuclear Fusion},
volume = {58},
number = {12},
pages = {126027}
}

@article{zoni2019solving,
  title={Solving hyperbolic-elliptic problems on singular mapped disk-like domains with the method of characteristics and spline finite elements},
  author={Zoni, Edoardo and G{\"u}{\c{c}}l{\"u}, Yaman},
  journal={Journal of Computational Physics},
  volume={398},
  pages={108889},
  year={2019},
  publisher={Elsevier}
}

\end{document}